\documentclass[longauth]{aa}  

\usepackage{graphicx}
\usepackage{txfonts}

\usepackage{hyperref}

\newcommand{\bdmuESPRESSOeighteen}{\ensuremath{-20490.87^{+0.66}_{-0.71}}}
\newcommand{\bdsigmawESPRESSOeighteen}{\ensuremath{2.10^{+0.86}_{-0.64}}}
\newcommand{\bdmuESPRESSOnineteen}{\ensuremath{-20494.89^{+0.40}_{-0.46}}}
\newcommand{\bdsigmawESPRESSOnineteen}{\ensuremath{0.023^{+0.223}_{-0.020}}}
\newcommand{\bdmuHARPS}{\ensuremath{-20337.12^{+0.59}_{-0.65}}}
\newcommand{\bdsigmawHARPS}{\ensuremath{0.31^{+1.51}_{-0.30}}}
\newcommand{\bdPpone}{\ensuremath{41.68547^{+0.00097}_{-0.00094}}}

\newcommand{\bdtzeropone}{\ensuremath{2457151.9014^{+0.0034}_{-0.0035}}}
\newcommand{\bdKpone}{\ensuremath{1.08^{+0.46}_{-0.44}}}
\newcommand{\bdrho}{\ensuremath{1758.5^{+905.8}_{-1201.6}}}
\newcommand{\bdppone}{\ensuremath{0.0188^{+0.0017}_{-0.0009}}}
\newcommand{\bdbpone}{\ensuremath{0.50^{+0.31}_{-0.38}}}
\newcommand{\bdqoneKtwofour}{\ensuremath{0.34^{+0.28}_{-0.22}}}
\newcommand{\bdqtwoKtwofour}{\ensuremath{0.37^{+0.35}_{-0.25}}}
\newcommand{\bdmfluxKtwofour}{\ensuremath{(-0.86^{+0.18}_{-0.20})\times10^{-5}}}
\newcommand{\bdsigmawKtwofour}{\ensuremath{0.97^{+4.28}_{-0.75}}}
\newcommand{\bdGPsigmaKtwofour}{\ensuremath{(7.341^{+0.090}_{-0.00000.089})\times10^{-5}}}
\newcommand{\bdGPrhoKtwofour}{\ensuremath{0.01575^{+0.00044}_{-0.00046}}}
\newcommand{\bdeccpone}{\ensuremath{0}}
\newcommand{\bdomegapone}{\ensuremath{90}}
\newcommand{\bdmdilutionKtwofour}{\ensuremath{1}}
\newcommand{\ktwosixtytwomuESPRESSOeighteen}{\ensuremath{4363.7^{+2.3}_{-2.4}}}
\newcommand{\ktwosixtytwosigmawESPRESSOeighteen}{\ensuremath{1.8\pm1.0}}
\newcommand{\ktwosixtytwomuESPRESSOnineteen}{\ensuremath{4359.9^{+1.8}_{-2.5}}}
\newcommand{\ktwosixtytwosigmawESPRESSOnineteen}{\ensuremath{1.81^{+0.65}_{-0.50}}}
\newcommand{\ktwosixtytwoPpone}{\ensuremath{6.67188^{+0.00024}_{-0.00026}}}

\newcommand{\ktwosixtytwotzeropone}{\ensuremath{2456982.6856115987^{+0.0014}_{-0.0013}}}
\newcommand{\ktwosixtytwoKpone}{\ensuremath{1.77^{+0.69}_{-0.61}}}
\newcommand{\ktwosixtytworho}{\ensuremath{7824.4^{+1458.1}_{-2736.5}}}
\newcommand{\ktwosixtytwoppone}{\ensuremath{0.0243^{+0.0027}_{-0.0016}}}
\newcommand{\ktwosixtytwobpone}{\ensuremath{0.45^{+0.20}_{-0.16}}}
\newcommand{\ktwosixtytwoPptwo}{\ensuremath{16.19752\pm0.00081}}

\newcommand{\ktwosixtytwotzeroptwo}{\ensuremath{2456991.5437\pm0.0015}}
\newcommand{\ktwosixtytwoKptwo}{\ensuremath{2.19^{+0.89}_{-0.92}}}
\newcommand{\ktwosixtytwopptwo}{\ensuremath{0.0299^{+0.0022}_{-0.0018}}}
\newcommand{\ktwosixtytwobptwo}{\ensuremath{0.795^{+0.063}_{-0.042}}}
\newcommand{\ktwosixtytwoqoneKtwothree}{\ensuremath{0.87^{+0.09}_{-0.13}}}
\newcommand{\ktwosixtytwoqtwoKtwothree}{\ensuremath{0.78^{+0.14}_{-0.21}}}
\newcommand{\ktwosixtytworvintercept}{\ensuremath{10.1^{+2.5}_{-2.1}}}
\newcommand{\ktwosixtytworvslope}{\ensuremath{-0.0627\pm0.0023}}
\newcommand{\ktwosixtytwomfluxKtwothree}{\ensuremath{(-2.8^{+4.2}_{-0.000004.0})\times 10^{-6}}}
\newcommand{\ktwosixtytwosigmawKtwothree}{\ensuremath{65.8^{+2.0}_{-1.9}}}
\newcommand{\ktwosixtytwoGPsigmaKtwothree}{\ensuremath{(5.55^{+0.20}_{-0.21})\times 10^{-5}}}
\newcommand{\ktwosixtytwoGPrhoKtwothree}{\ensuremath{0.16^{+0.016}_{-0.015}}}
\newcommand{\ktwosixtytwoeccpone}{\ensuremath{0}}
\newcommand{\ktwosixtytwoomegapone}{\ensuremath{90}}
\newcommand{\ktwosixtytwoeccptwo}{\ensuremath{0}}
\newcommand{\ktwosixtytwoomegaptwo}{\ensuremath{90}}
\newcommand{\ktwosixtytwomdilutionKtwothree}{\ensuremath{1}}
\newcommand{\ktwooneofivemuESPRESSOeighteen}{\ensuremath{-32534.7\pm1.2}}
\newcommand{\ktwooneofivesigmawESPRESSOeighteen}{\ensuremath{5.42^{+1.01}_{-0.79}}}
\newcommand{\ktwooneofivemuESPRESSOnineteen}{\ensuremath{-32532.63\pm0.83}}
\newcommand{\ktwooneofivesigmawESPRESSOnineteen}{\ensuremath{3.77^{+0.79}_{-0.61}}}
\newcommand{\ktwooneofivemuSubaru}{\ensuremath{0.5^{+2.5}_{-2.7}}}
\newcommand{\ktwooneofivesigmawSubaru}{\ensuremath{0.24^{+3.76}_{-0.23}}}
\newcommand{\ktwooneofivePpone}{\ensuremath{5.02173^{+0.00042}_{-0.00065}}}

\newcommand{\ktwooneofivetzeropone}{\ensuremath{2458437.97438^{+0.0096}_{-0.0102}}}
\newcommand{\ktwooneofivePptwo}{\ensuremath{8.2669721^{+0.0000082}_{-0.0000077}}}

\newcommand{\ktwooneofivetzeroptwo}{\ensuremath{2457147.99182^{+0.00091}_{-0.00093}}}
\newcommand{\ktwooneofiveKpone}{\ensuremath{4.36^{+0.99}_{-1.04}}}
\newcommand{\ktwooneofiveKptwo}{\ensuremath{1.7^{+1.2}_{-1.0}}}
\newcommand{\ktwooneofiveppone}{\ensuremath{0.0057^{+0.0032}_{-0.0034}}}
\newcommand{\ktwooneofivebpone}{\ensuremath{0.66^{+0.24}_{-0.39}}}
\newcommand{\ktwooneofiverho}{\ensuremath{928.9^{+50.7}_{-85.6}}}
\newcommand{\ktwooneofivepptwo}{\ensuremath{0.03252^{+0.00069}_{-0.00059}}}
\newcommand{\ktwooneofivebptwo}{\ensuremath{0.591^{+0.036}_{-0.023}}}
\newcommand{\ktwooneofiveqoneKtwofiveKtwoeighteen}{\ensuremath{0.84^{+0.12}_{-0.16}}}
\newcommand{\ktwooneofiveqtwoKtwofiveKtwoeighteen}{\ensuremath{0.26^{+0.14}_{-0.13}}}
\newcommand{\ktwooneofivemfluxKtwofive}{\ensuremath{(0.58^{+0.79}_{-0.73})\times10^{-5}}}
\newcommand{\ktwooneofivesigmawKtwofive}{\ensuremath{120.7^{+4.0}_{-4.1}}}
\newcommand{\ktwooneofiveGPsigmaKtwofive}{\ensuremath{(21.00^{+0.47}_{-0.45})\times10^{-5}}}
\newcommand{\ktwooneofiveGPrhoKtwofive}{\ensuremath{0.0454^{+0.0026}_{-0.0027}}}
\newcommand{\ktwooneofivemfluxKtwoeighteen}{\ensuremath{(-0.3^{+1.4}_{-1.5})\times10^{-5}}}
\newcommand{\ktwooneofivesigmawKtwoeighteen}{\ensuremath{86.5^{+2.4}_{-2.5}}}
\newcommand{\ktwooneofiveGPsigmaKtwoeighteen}{\ensuremath{(23.68^{+0.60}_{-0.62})\times10^{-5}}}
\newcommand{\ktwooneofiveGPrhoKtwoeighteen}{\ensuremath{0.0910^{+0.0037}_{-0.0036}}}
\newcommand{\ktwooneofiveeccpone}{\ensuremath{0.0}}
\newcommand{\ktwooneofiveomegapone}{\ensuremath{90.0}}
\newcommand{\ktwooneofiveeccptwo}{\ensuremath{0.0}}
\newcommand{\ktwooneofiveomegaptwo}{\ensuremath{90.0}}
\newcommand{\ktwooneofivemdilutionKtwofive}{\ensuremath{1}}
\newcommand{\ktwooneofivemdilutionKtwoeighteen}{\ensuremath{1}}
\newcommand{\ktwooneeightfourmuESPRESSOeighteen}{\ensuremath{-33344.0\pm1.6}}
\newcommand{\ktwooneeightfoursigmawESPRESSOeighteen}{\ensuremath{2.47^{+0.45}_{-0.38}}}
\newcommand{\ktwooneeightfourmuESPRESSOnineteen}{\ensuremath{-33342.5^{+1.6}_{-1.9}}}
\newcommand{\ktwooneeightfoursigmawESPRESSOnineteen}{\ensuremath{0.11^{+1.23}_{-0.10}}}
\newcommand{\ktwooneeightfourPpone}{\ensuremath{16.977959^{+0.000089}_{-0.000090}}}

\newcommand{\ktwooneeightfourtzeropone}{\ensuremath{2457148.5219^{+0.0023}_{-0.0020}}}
\newcommand{\ktwooneeightfoursesinomegapone}{\ensuremath{0.02^{+0.34}_{-0.30}}}
\newcommand{\ktwooneeightfoursecosomegapone}{\ensuremath{-0.71^{+0.14}_{-0.05}}}
\newcommand{\ktwooneeightfoureccpone}{\ensuremath{0.574^{+0.083}_{-0.090}}}
\newcommand{\ktwooneeightfouromegapone}{\ensuremath{93.6^{+75.8}_{-263.4}}}
\newcommand{\ktwooneeightfourKpone}{\ensuremath{4.4^{+1.5}_{-1.6}}}
\newcommand{\ktwooneeightfourrho}{\ensuremath{1748.3^{+1811.5}_{-967.5}}}
\newcommand{\ktwooneeightfourppone}{\ensuremath{0.01467^{+0.00074}_{-0.00068}}}
\newcommand{\ktwooneeightfourbpone}{\ensuremath{0.407^{+0.093}_{-0.088}}}
\newcommand{\ktwooneeightfourqoneKtwofiveKtwoeighteen}{\ensuremath{0.896^{+0.071}_{-0.106}}}
\newcommand{\ktwooneeightfourqtwoKtwofiveKtwoeighteen}{\ensuremath{0.930^{+0.050}_{-0.078}}}
\newcommand{\ktwooneeightfourGPsigmarv}{\ensuremath{4.1^{+1.0}_{-0.8}}}
\newcommand{\ktwooneeightfourGPalpharv}{\ensuremath{0.026^{+0.023}_{-0.017}}}
\newcommand{\ktwooneeightfourGPGammarv}{\ensuremath{5.2^{+3.1}_{-2.4}}}
\newcommand{\ktwooneeightfourGPProtrv}{\ensuremath{47.9^{+19.9}_{-17.3}}}
\newcommand{\ktwooneeightfourmfluxKtwofive}{\ensuremath{(2.23^{+0.22}_{-0.20})\times10^{-5}}}
\newcommand{\ktwooneeightfoursigmawKtwofive}{\ensuremath{49.88^{+0.78}_{-0.73}}}
\newcommand{\ktwooneeightfourGPsigmaKtwofive}{\ensuremath{(3.55\pm0.11)\times10^{-5}}}
\newcommand{\ktwooneeightfourGPrhoKtwofive}{\ensuremath{0.137\pm0.012}}
\newcommand{\ktwooneeightfourmfluxKtwoeighteen}{\ensuremath{(-0.85^{+0.30}_{-0.3})\times10^{-5}}}
\newcommand{\ktwooneeightfoursigmawKtwoeighteen}{\ensuremath{60.8\pm1.1}}
\newcommand{\ktwooneeightfourGPsigmaKtwoeighteen}{\ensuremath{(4.63^{+0.17}_{-0.16})\times10^{-5}}}
\newcommand{\ktwooneeightfourGPrhoKtwoeighteen}{\ensuremath{0.0944\pm0.0085}}
\newcommand{\ktwooneeightfourmdilutionKtwofive}{\ensuremath{1}}
\newcommand{\ktwooneeightfourmdilutionKtwoeighteen}{\ensuremath{1}}
\newcommand{\bdab}{\ensuremath{0.2226 \pm 0.0002}}
\newcommand{\bdincb}{\ensuremath{89.35 \pm 0.49}}
\newcommand{\bdTdurb}{\ensuremath{6.42 \pm 1.55}}
\newcommand{\bdmassb}{\ensuremath{<12.0}}

\newcommand{\bdradb}{\ensuremath{2.23 \pm 0.20}}

\newcommand{\bdrhoplb}{\ensuremath{5.91 \pm 1.95}}
\newcommand{\bdteqb}{\ensuremath{613.0 \pm 4.0}}
\newcommand{\bdSb}{\ensuremath{23.9 \pm 1.0}}
\newcommand{\ktwosixtytwoab}{\ensuremath{0.0606 \pm 0.0002}}
\newcommand{\ktwosixtytwoincb}{\ensuremath{88.68 \pm 0.59}}
\newcommand{\ktwosixtytwoTdurb}{\ensuremath{2.4 \pm 0.26}}
\newcommand{\ktwosixtytwomassb}{\ensuremath{<8.6}}

\newcommand{\ktwosixtytworadb}{\ensuremath{1.77 \pm 0.20}}

\newcommand{\ktwosixtytworhoplb}{\ensuremath{8.63 \pm 3.27}}
\newcommand{\ktwosixtytwoteqb}{\ensuremath{745.0 \pm 3.0}}
\newcommand{\ktwosixtytwoSb}{\ensuremath{51.4 \pm 1.0}}
\newcommand{\ktwosixtytwoac}{\ensuremath{0.1095 \pm 0.0003}}
\newcommand{\ktwosixtytwoincc}{\ensuremath{88.72 \pm 0.1}}
\newcommand{\ktwosixtytwoTdurc}{\ensuremath{2.31 \pm 0.25}}
\newcommand{\ktwosixtytwomassc}{\ensuremath{<14.7}}

\newcommand{\ktwosixtytworadc}{\ensuremath{2.17 \pm 0.16}}

\newcommand{\ktwosixtytworhoplc}{\ensuremath{7.88 \pm 2.29}}
\newcommand{\ktwosixtytwoteqc}{\ensuremath{554.0 \pm 2.0}}
\newcommand{\ktwosixtytwoSc}{\ensuremath{15.8 \pm 0.0}}
\newcommand{\ktwooneofiveab}{\ensuremath{0.0559 \pm 0.0002}}

\newcommand{\ktwooneofivemassb}{\ensuremath{11.1 \pm 2.7}}

\newcommand{\ktwooneofiveteqb}{\ensuremath{1037.0 \pm 5.0}}
\newcommand{\ktwooneofiveSb}{\ensuremath{191.6 \pm 4.0}}
\newcommand{\ktwooneofiveac}{\ensuremath{0.078 \pm 0.0002}}
\newcommand{\ktwooneofiveincc}{\ensuremath{88.21 \pm 0.11}}
\newcommand{\ktwooneofiveTdurc}{\ensuremath{2.83 \pm 0.09}}
\newcommand{\ktwooneofivemassc}{\ensuremath{<15.5}}

\newcommand{\ktwooneofiveradc}{\ensuremath{3.145 \pm 0.073}}

\newcommand{\ktwooneofiverhoplc}{\ensuremath{2.73 \pm 0.64}}
\newcommand{\ktwooneofiveteqc}{\ensuremath{878.0 \pm 5.0}}
\newcommand{\ktwooneofiveSc}{\ensuremath{98.5 \pm 2.0}}
\newcommand{\ktwooneeightyfourab}{\ensuremath{0.1179 \pm 0.0004}}
\newcommand{\ktwooneeightyfourincb}{\ensuremath{88.36 \pm 0.54}}
\newcommand{\ktwooneeightyfourTdurb}{\ensuremath{3.61 \pm 0.16}}
\newcommand{\ktwooneeightyfourmassb}{\ensuremath{<29.8}}

\newcommand{\ktwooneeightyfourradb}{\ensuremath{1.216 \pm 0.062}}

\newcommand{\ktwooneeightyfourrhoplb}{\ensuremath{90.99 \pm 20.68}}
\newcommand{\ktwooneeightyfourteqb}{\ensuremath{671.0 \pm 13.0}}
\newcommand{\ktwooneeightyfourSb}{\ensuremath{33.9 \pm 3.0}}
\newcommand{\toitwosixtysixab}{\ensuremath{0.0899 \pm 0.0001}}
\newcommand{\toitwosixtysixincb}{\ensuremath{88.35 \pm 0.2}}
\newcommand{\toitwosixtysixTdurb}{\ensuremath{3.44 \pm 0.2}}
\newcommand{\toitwosixtysixmassb}{\ensuremath{7.39 \pm 0.92}}

\newcommand{\toitwosixtysixradb}{\ensuremath{2.57 \pm 0.12}}

\newcommand{\toitwosixtysixrhoplb}{\ensuremath{2.39 \pm 0.44}}
\newcommand{\toitwosixtysixteqb}{\ensuremath{884.0 \pm 3.0}}
\newcommand{\toitwosixtysixSb}{\ensuremath{101.6 \pm 1.0}}
\newcommand{\toitwosixtysixac}{\ensuremath{0.1341 \pm 0.0002}}
\newcommand{\toitwosixtysixincc}{\ensuremath{89.33 \pm 0.17}}
\newcommand{\toitwosixtysixTdurc}{\ensuremath{4.79 \pm 0.17}}
\newcommand{\toitwosixtysixmassc}{\ensuremath{9.0 \pm 1.2}}

\newcommand{\toitwosixtysixradc}{\ensuremath{2.48 \pm 0.14}}

\newcommand{\toitwosixtysixrhoplc}{\ensuremath{3.26 \pm 0.7}}
\newcommand{\toitwosixtysixteqc}{\ensuremath{723.0 \pm 3.0}}
\newcommand{\toitwosixtysixSc}{\ensuremath{45.6 \pm 1.0}}
\newcommand{\toioneeightysixab}{\ensuremath{0.0691 \pm 0.0002}}
\newcommand{\toioneeightysixincb}{\ensuremath{89.05 \pm 0.29}}
\newcommand{\toioneeightysixTdurb}{\ensuremath{2.64 \pm 0.12}}
\newcommand{\toioneeightysixmassb}{\ensuremath{<3.83}}

\newcommand{\toioneeightysixradb}{\ensuremath{0.984 \pm 0.038}}

\newcommand{\toioneeightysixrhoplb}{\ensuremath{22.08 \pm 5.29}}
\newcommand{\toioneeightysixteqb}{\ensuremath{728.0 \pm 10.0}}
\newcommand{\toioneeightysixSb}{\ensuremath{46.5 \pm 3.0}}
\newcommand{\toioneeightysixac}{\ensuremath{0.1904 \pm 0.0005}}
\newcommand{\toioneeightysixincc}{\ensuremath{89.38 \pm 0.33}}
\newcommand{\toioneeightysixTdurc}{\ensuremath{4.05 \pm 0.26}}
\newcommand{\toioneeightysixmassc}{\ensuremath{25.9 \pm 1.3}}

\newcommand{\toioneeightysixradc}{\ensuremath{2.809 \pm 0.062}}

\newcommand{\toioneeightysixrhoplc}{\ensuremath{6.43 \pm 0.53}}
\newcommand{\toioneeightysixteqc}{\ensuremath{436.0 \pm 2.0}}
\newcommand{\toioneeightysixSc}{\ensuremath{6.0 \pm 0.0}}
\newcommand{\TeffTOIonetwozerothree}{\ensuremath{5617\pm113}}
\newcommand{\loggTOIonetwozerothree}{\ensuremath{4.1\pm0.02}}
\newcommand{\FeHTOIonetwozerothree}{\ensuremath{-0.53\pm0.02}}
\newcommand{\VbroadTOIonetwozerothree}{\ensuremath{1.55\pm0.03}}
\newcommand{\TeffTOIonetwothreethree}{\ensuremath{5707\pm115}}
\newcommand{\loggTOIonetwothreethree}{\ensuremath{4.49\pm0.02}}
\newcommand{\FeHTOIonetwothreethree}{\ensuremath{-0.33\pm0.01}}
\newcommand{\VbroadTOIonetwothreethree}{\ensuremath{1.58\pm0.01}}
\newcommand{\TeffTOIsixninesix}{\ensuremath{3536\pm73}}
\newcommand{\loggTOIsixninesix}{\ensuremath{4.81\pm0.08}}
\newcommand{\FeHTOIsixninesix}{\ensuremath{-0.5\pm0.04}}
\newcommand{\VbroadTOIsixninesix}{\ensuremath{\leq1.0}}
\newcommand{\TeffKtwoonefivefive}{\ensuremath{4173\pm89}}
\newcommand{\loggKtwoonefivefive}{\ensuremath{4.13\pm0.12}}
\newcommand{\FeHKtwoonefivefive}{\ensuremath{-0.73\pm0.04}}
\newcommand{\VbroadKtwoonefivefive}{\ensuremath{1.4\pm0.05}}
\newcommand{\TeffTOIfourthreeeight}{\ensuremath{5186\pm105}}
\newcommand{\loggTOIfourthreeeight}{\ensuremath{4.47\pm0.03}}
\newcommand{\FeHTOIfourthreeeight}{\ensuremath{0.04\pm0.01}}
\newcommand{\VbroadTOIfourthreeeight}{\ensuremath{1.42\pm0.02}}
\newcommand{\TeffTOIoneonethreezero}{\ensuremath{4436\pm89}}
\newcommand{\loggTOIoneonethreezero}{\ensuremath{4.72\pm0.02}}
\newcommand{\FeHTOIoneonethreezero}{\ensuremath{-0.04\pm0.01}}
\newcommand{\VbroadTOIoneonethreezero}{\ensuremath{1.45\pm0.02}}
\newcommand{\TeffTOIeightsevenone}{\ensuremath{4975\pm101}}
\newcommand{\loggTOIeightsevenone}{\ensuremath{4.49\pm0.04}}
\newcommand{\FeHTOIeightsevenone}{\ensuremath{-0.15\pm0.01}}
\newcommand{\VbroadTOIeightsevenone}{\ensuremath{1.29\pm0.03}}
\newcommand{\TeffTOIonezerosixthree}{\ensuremath{5295\pm107}}
\newcommand{\loggTOIonezerosixthree}{\ensuremath{4.74\pm0.02}}
\newcommand{\FeHTOIonezerosixthree}{\ensuremath{-0.04\pm0.01}}
\newcommand{\VbroadTOIonezerosixthree}{\ensuremath{2.73\pm0.04}}
\newcommand{\TeffTOIonezeroseveneight}{\ensuremath{3474\pm73}}
\newcommand{\loggTOIonezeroseveneight}{\ensuremath{4.87\pm0.05}}
\newcommand{\FeHTOIonezeroseveneight}{\ensuremath{-0.17\pm0.06}}
\newcommand{\VbroadTOIonezeroseveneight}{\ensuremath{2.0\pm0.04}}
\newcommand{\TeffTOIsevenzerofour}{\ensuremath{3599\pm74}}
\newcommand{\loggTOIsevenzerofour}{\ensuremath{4.54\pm0.07}}
\newcommand{\FeHTOIsevenzerofour}{\ensuremath{-0.28\pm0.05}}
\newcommand{\VbroadTOIsevenzerofour}{\ensuremath{2.21\pm0.04}}
\newcommand{\TeffTOIonezeronineseven}{\ensuremath{5941\pm120}}
\newcommand{\loggTOIonezeronineseven}{\ensuremath{4.62\pm0.03}}
\newcommand{\FeHTOIonezeronineseven}{\ensuremath{0.0\pm0.04}}
\newcommand{\VbroadTOIonezeronineseven}{\ensuremath{5.13\pm0.03}}
\newcommand{\TeffTOIonetwothreezero}{\ensuremath{5910\pm120}}
\newcommand{\loggTOIonetwothreezero}{\ensuremath{4.68\pm0.04}}
\newcommand{\FeHTOIonetwothreezero}{\ensuremath{-0.43\pm0.03}}
\newcommand{\VbroadTOIonetwothreezero}{\ensuremath{1.78\pm0.02}}
\newcommand{\TeffTOIonefoursixeight}{\ensuremath{3438\pm69}}
\newcommand{\loggTOIonefoursixeight}{\ensuremath{4.69\pm0.05}}
\newcommand{\FeHTOIonefoursixeight}{\ensuremath{-0.19\pm0.05}}
\newcommand{\VbroadTOIonefoursixeight}{\ensuremath{2.05\pm0.08}}
\newcommand{\TeffTOIonezerofiveeight}{\ensuremath{5834\pm122}}
\newcommand{\loggTOIonezerofiveeight}{\ensuremath{3.71\pm0.06}}
\newcommand{\FeHTOIonezerofiveeight}{\ensuremath{0.0\pm0.02}}
\newcommand{\VbroadTOIonezerofiveeight}{\ensuremath{4.74\pm0.05}}
\newcommand{\TeffTOInineonezero}{\ensuremath{3401\pm72}}
\newcommand{\loggTOInineonezero}{\ensuremath{4.85\pm0.08}}
\newcommand{\FeHTOInineonezero}{\ensuremath{-0.12\pm0.07}}
\newcommand{\VbroadTOInineonezero}{\ensuremath{2.11\pm0.05}}
\newcommand{\TeffBDtwozerofiveninefour}{\ensuremath{5748\pm116}}
\newcommand{\loggBDtwozerofiveninefour}{\ensuremath{4.32\pm0.03}}
\newcommand{\FeHBDtwozerofiveninefour}{\ensuremath{-0.24\pm0.02}}
\newcommand{\VbroadBDtwozerofiveninefour}{\ensuremath{1.45\pm0.01}}
\newcommand{\logRhkBDtwozerofiveninefour}{\ensuremath{-5.682\pm0.18}}
\newcommand{\ProtFullMedianBDtwozerofiveninefour}{\ensuremath{114.0\pm42.0}}

\newcommand{\TeffKtwosixtwo}{\ensuremath{4662\pm94}}
\newcommand{\loggKtwosixtwo}{\ensuremath{4.69\pm0.02}}
\newcommand{\FeHKtwosixtwo}{\ensuremath{-0.21\pm0.01}}
\newcommand{\VbroadKtwosixtwo}{\ensuremath{1.13\pm0.05}}
\newcommand{\logRhkKtwosixtwo}{\ensuremath{-4.983\pm0.046}}
\newcommand{\ProtFullMedianKtwosixtwo}{\ensuremath{31.0\pm5.0}}

\newcommand{\TeffTOIoneeightsix}{\ensuremath{4675\pm95}}
\newcommand{\loggTOIoneeightsix}{\ensuremath{4.5\pm0.04}}
\newcommand{\FeHTOIoneeightsix}{\ensuremath{0.04\pm0.02}}
\newcommand{\VbroadTOIoneeightsix}{\ensuremath{1.23\pm0.04}}
\newcommand{\logRhkTOIoneeightsix}{\ensuremath{-4.921\pm0.028}}
\newcommand{\ProtFullMedianTOIoneeightsix}{\ensuremath{28.0\pm4.0}}

\newcommand{\TeffTOItwosixsix}{\ensuremath{5613\pm113}}
\newcommand{\loggTOItwosixsix}{\ensuremath{4.38\pm0.02}}
\newcommand{\FeHTOItwosixsix}{\ensuremath{-0.15\pm0.01}}
\newcommand{\VbroadTOItwosixsix}{\ensuremath{1.39\pm0.01}}
\newcommand{\logRhkTOItwosixsix}{\ensuremath{-5.024\pm0.122}}
\newcommand{\ProtFullMedianTOItwosixsix}{\ensuremath{33.0\pm9.0}}

\newcommand{\TeffKtwooneeightfour}{\ensuremath{5214\pm105}}
\newcommand{\loggKtwooneeightfour}{\ensuremath{4.6\pm0.02}}
\newcommand{\FeHKtwooneeightfour}{\ensuremath{-0.2\pm0.02}}
\newcommand{\VbroadKtwooneeightfour}{\ensuremath{1.17\pm0.03}}
\newcommand{\logRhkKtwooneeightfour}{\ensuremath{-5.117\pm0.339}}
\newcommand{\ProtFullMedianKtwooneeightfour}{\ensuremath{40.0\pm26.0}}
\newcommand{\TeffKtwoonezerofive}{\ensuremath{5400\pm109}}
\newcommand{\loggKtwoonezerofive}{\ensuremath{4.49\pm0.03}}
\newcommand{\FeHKtwoonezerofive}{\ensuremath{0.16\pm0.02}}
\newcommand{\VbroadKtwoonezerofive}{\ensuremath{1.5\pm0.03}}
\newcommand{\logRhkKtwoonezerofive}{\ensuremath{-5.038\pm0.035}}
\newcommand{\ProtFullMedianKtwoonezerofive}{\ensuremath{34.0\pm5.0}}
\newcommand{\toioneeightysixmuESPRESSO}{\ensuremath{59604.41^{+0.76}_{-0.73}}}
\newcommand{\toioneeightysixsigmawESPRESSO}{\ensuremath{2.43^{+0.74}_{-0.50}}}
\newcommand{\toioneeightysixmuHARPS}{\ensuremath{-0.12^{+0.57}_{-0.56}}}
\newcommand{\toioneeightysixsigmawHARPS}{\ensuremath{4.44^{+0.46}_{-0.41}}}
\newcommand{\toioneeightysixmuPFSF}{\ensuremath{0.10\pm0.68}}
\newcommand{\toioneeightysixsigmawPFSF}{\ensuremath{4.24^{+0.53}_{-0.42}}}
\newcommand{\toioneeightysixmuPFSS}{\ensuremath{0.51^{+0.25}_{-0.24}}}
\newcommand{\toioneeightysixsigmawPFSS}{\ensuremath{3.01^{+0.18}_{-0.17}}}
\newcommand{\toioneeightysixPpone}{\ensuremath{7.789787^{+0.000018}_{-0.000013}}}

\newcommand{\toioneeightysixtzeropone}{\ensuremath{2458371.2286^{+0.0016}_{-0.0013}}}
\newcommand{\toioneeightysixsesinomegapone}{\ensuremath{-0.37^{+0.17}_{-0.12}}}
\newcommand{\toioneeightysixsecosomegapone}{\ensuremath{0.29^{+0.27}_{-0.34}}}
\newcommand{\toioneeightysixeccpone}{\ensuremath{0.25^{+0.20}_{-0.13}}}
\newcommand{\toioneeightysixomegapone}{\ensuremath{-49.0^{+20.3}_{-47.3}}}
\newcommand{\toioneeightysixKpone}{\ensuremath{0.57^{+0.32}_{-0.29}}}
\newcommand{\toioneeightysixrho}{\ensuremath{4922.8^{+1324.6}_{-1072.6}}}
\newcommand{\toioneeightysixppone}{\ensuremath{0.01272^{+0.00049}_{-0.00047}}}
\newcommand{\toioneeightysixbpone}{\ensuremath{0.401^{+0.087}_{-0.096}}}
\newcommand{\toioneeightysixPptwo}{\ensuremath{35.613442\pm0.000016}}

\newcommand{\toioneeightysixtzeroptwo}{\ensuremath{2458385.92562^{+0.00054}_{-0.00052}}}
\newcommand{\toioneeightysixsesinomegaptwo}{\ensuremath{0.08\pm 0.14}}
\newcommand{\toioneeightysixsecosomegaptwo}{\ensuremath{-0.319^{+0.050}_{-0.034}}}
\newcommand{\toioneeightysixeccptwo}{\ensuremath{0.122\pm0.022}}
\newcommand{\toioneeightysixomegaptwo}{\ensuremath{146.0^{+23.4}_{-313.7}}}
\newcommand{\toioneeightysixKptwo}{\ensuremath{6.24^{+0.30}_{-0.31}}}
\newcommand{\toioneeightysixpptwo}{\ensuremath{0.03630^{+0.00078}_{-0.00072}}}
\newcommand{\toioneeightysixbptwo}{\ensuremath{0.584^{+0.066}_{-0.077}}}
\newcommand{\toioneeightysixqoneTESS}{\ensuremath{0.29^{+0.11}_{-0.07}}}
\newcommand{\toioneeightysixqtwoTESS}{\ensuremath{0.62^{+0.23}_{-0.27}}}

\newcommand{\toitwosixsixmuESPRESSO}{\ensuremath{41406.16 \pm 0.31}}
\newcommand{\toitwosixsixsigmawESPRESSO}{\ensuremath{1.42^{+0.26}_{-0.20}}}
\newcommand{\toitwosixsixmuHIRES}{\ensuremath{-1.42^{+0.36}_{-0.35}}}
\newcommand{\toitwosixsixsigmawHIRES}{\ensuremath{3.17^{+0.28}_{-0.25}}}
\newcommand{\toitwosixsixPpone}{\ensuremath{10.751014^{+0.000053}_{-0.000058}}}

\newcommand{\toitwosixsixtzeropone}{\ensuremath{2458393.0855^{+0.0031}_{-0.0026}}}
\newcommand{\toitwosixsixKpone}{\ensuremath{2.42\pm0.30}}
\newcommand{\toitwosixsixrho}{\ensuremath{1901.5^{+357.3}_{-309.7}}}
\newcommand{\toitwosixsixppone}{\ensuremath{0.0246\pm0.0011}}
\newcommand{\toitwosixsixbpone}{\ensuremath{0.583^{+0.065}_{-0.069}}}
\newcommand{\toitwosixsixPptwo}{\ensuremath{19.60550\pm0.00018}}

\newcommand{\toitwosixsixtzeroptwo}{\ensuremath{2458398.2917^{+0.0057}_{-0.0048}}}
\newcommand{\toitwosixsixKptwo}{\ensuremath{2.41^{+0.33}_{-0.32}}}
\newcommand{\toitwosixsixpptwo}{\ensuremath{0.0237459057^{+0.0012}_{-0.0013}}}
\newcommand{\toitwosixsixbptwo}{\ensuremath{0.354^{+0.081}_{-0.090}}}
\newcommand{\toitwosixsixqoneTESS}{\ensuremath{0.34^{+0.37}_{-0.22}}}
\newcommand{\toitwosixsixqtwoTESS}{\ensuremath{0.39^{+0.35}_{-0.27}}}
\newcommand{\toitwosixsixmfluxTESSthree}{\ensuremath{(-1.6^{+3.4}_{-3.1})\times10^{-5}}}
\newcommand{\toitwosixsixsigmawTESSthree}{\ensuremath{1.6^{+8.7}_{-1.3}}}
\newcommand{\toitwosixsixGPsigmaTESSthree}{\ensuremath{(3.4^{+2.8}_{-2.2})\times10^{-5}}}
\newcommand{\toitwosixsixGPrhoTESSthree}{\ensuremath{222.9^{+199.4}_{-134.8}}}
\newcommand{\toitwosixsixmfluxTESSthirty}{\ensuremath{(-1.0\pm-1.3)\times10^{-5}}}
\newcommand{\toitwosixsixsigmawTESSthirty}{\ensuremath{144.9^{+22.0}_{-29.3}}}
\newcommand{\toitwosixsixGPsigmaTESSthirty}{\ensuremath{(10.25^{+0.76}_{-0.74})\times10^{-5}}}
\newcommand{\toitwosixsixGPrhoTESSthirty}{\ensuremath{0.114^{+0.022}_{-0.019}}}
\newcommand{\toitwosixsixeccpone}{\ensuremath{0}}
\newcommand{\toitwosixsixomegapone}{\ensuremath{90}}
\newcommand{\toitwosixsixeccptwo}{\ensuremath{0}}
\newcommand{\toitwosixsixomegaptwo}{\ensuremath{90}}

\usepackage[dvipsnames]{xcolor}

\begin{document}

   \title{The small transiting planet population revealed by ESPRESSO with extreme precision radial velocities}

   \author{M. J. Hobson
          \inst{\ref{unige}}
          \and
           B. Lavie
          \inst{\ref{unige}}
          \and
           F. Bouchy
          \inst{\ref{unige}}
          \and
          C. Lovis
          \inst{\ref{unige}}
          \and 
          F. Pepe
          \inst{\ref{unige}}
          \and         
          S. G. Sousa
          \inst{\ref{IAporto}, \ref{uniporto}}
          \and  
          H. M. Tabernero
          \inst{\ref{IEEC}, \ref{ICE}}
          \and
          S. E. van Terwisga
          \inst{\ref{IWF}}
          \and
          S. Oe
          \inst{\ref{unige2}, \ref{leuven}}
          \and  
          V. Adibekyan
          \inst{\ref{IAporto}, \ref{uniporto}}
          \and
          C. Allende Prieto
          \inst{\ref{iac}, \ref{laguna}}
          \and 
          Y. Alibert
          \inst{\ref{unibe1}, \ref{unibe2}}
          \and
          S. C. C. Barros
          \inst{\ref{IAporto}, \ref{uniporto}}
          \and
          A. Castro-González
          \inst{\ref{unige}}
          \and
          S. Cristiani
          \inst{\ref{inaf-trieste}, \ref{IFPU}}
          \and
          V. D'Odorico
          \inst{\ref{inaf-trieste}}
          \and 
          O. D. S. Demangeon
          \inst{\ref{IAporto}}
          \and
          X. Dumusque
          \inst{\ref{unige}}
          \and
          D. Ehrenreich
          \inst{\ref{unige}}
          \and
          P. Figueira
          \inst{\ref{CSIC}}
          \and
          R. Génova Santos
          \inst{\ref{iac}}
          \and
          J. I. Gonz\'alez Hern\'andez
          \inst{\ref{iac}, \ref{laguna}}
          \and
          J. Lillo-Box
          \inst{\ref{cab}}
          \and
          G. Lo Curto
          \inst{\ref{ESO}}
          \and
          C. J. A. P. Martins
          \inst{\ref{astro-porto}, \ref{IAporto}}
          \and
          P. Di Marcantonio
          \inst{\ref{inaf-trieste}}
          \and
          A. Mehner
          \inst{\ref{ESO}}
          \and
          G. Micela
          \inst{\ref{inaf-palermo}}
          \and
          P. Molaro
          \inst{\ref{inaf-trieste}}
          \and
          N. J. Nunes
          \inst{\ref{lisboa}}
          \and
          E. Palle
          \inst{\ref{iac}, \ref{laguna}}
          \and
          R. Rebolo
          \inst{\ref{iac}\ref{laguna}\ref{consejo-sup-madrid}}
          \and
          J. Rodrigues
          \inst{\ref{IAporto}, \ref{uniporto}, \ref{unige}}
          \and
          N. Santos
          \inst{\ref{IAporto}, \ref{uniporto}}
          \and
          A. Sozzetti
          \inst{\ref{inaf-torino}}
          \and
          S. Udry
          \inst{\ref{unige}}
          \and
          A. Suárez Mascareño
          \inst{\ref{iac}, \ref{laguna}}
          \and
          M.-R. Zapatero Osorio
          \inst{\ref{cab}}
          }

   \institute{Observatoire de Genève, Département d'Astronomie, Université de Genève, Chemin Pegasi 51b, 1290 Versoix, Switzerland \label{unige}\\
              \email{melissa.hobson@unige.ch}
    \and
    Instituto de Astrofísica e Ciências do Espaço, CAUP, Universidade do Porto, Rua das Estrelas, 4150-762, Porto, Portugal\label{IAporto}
    \and
    Departamento de Física e Astronomia, Faculdade de Ciências, Universidade do Porto, Rua do Campo Alegre, 4169-007, Porto, Portugal\label{uniporto}
    \and
    Institut d'Estudis Espacials de Catalunya (IEEC), Edifici RDIT, Campus UPC, 08860 Castelldefels (Barcelona), Spain\label{IEEC}
    \and
    Institut de Ciències de l’Espai (ICE, CSIC), Campus UAB, c/ de Can Magrans s/n, 08193 Cerdanyola del Vallès, Barcelona, Spain\label{ICE}
    \and
    Space Research Institute, Austrian Academy of Sciences, Schmiedlstrasse 6, A-8042 Graz, Austria\label{IWF}
    \and
    Institute of Astronomy, KU Leuven, Celestĳnlaan 200D, 3001 Leuven, Belgium\label{leuven}
    \and
    University of Geneva, Faculty of Science, Geneva, Switzerland\label{unige2}
    \and
    Instituto de Astrof\'{\i}sica de Canarias, c/ V\'ia L\'actea s/n, 38205 La Laguna, Tenerife, Spain\label{iac}
    \and
    Departamento de Astrof\'{\i}sica, Universidad de La Laguna, 38206 La Laguna, Tenerife, Spain\label{laguna}
    \and
    Physics Institute, University of Bern, Gesellsschaftstrasse 6, CH-3012 Bern, Switzerland\label{unibe1}
    \and
    Center for Space and Habitability, University of Bern, Gesellsschaftstrasse 6, CH-3012 Bern, Switzerland\label{unibe2}
    \and
    INAF- Osservatorio Astronomico di Trieste, via G. B. Tiepolo 11, I-34143, Trieste, Italy\label{inaf-trieste}
    \and
    IFPU–Institute for Fundamental Physics of the Universe, via Beirut 2, I-34151 Trieste, Italy\label{IFPU}
    \and
    Instituto de Astrof\'{i}sica de Andaluc\'{i}a-CSIC, Glorieta de la Astronom\'{i}a s/n, E-18008 Granada, Spain\label{CSIC}
    \and
    Centro de Astrobiolog\'{i}a, CSIC-INTA, Camino Bajo del Castillo s/n, 28692 Villanueva de la Ca\~{n}ada, Madrid, Spain\label{cab}    
    \and
    European Southern Observatory, Av. Alonso de Cordova, 3107, Vitacura, Santiago de Chile, Chile\label{ESO}
    \and
    Centro de Astrof\'{\i}sica da Universidade do Porto, Rua das Estrelas, 4150-762 Porto, Portugal\label{astro-porto}
    \and
    INAF - Osservatorio Astronomico di Palermo, Piazza del Parlamento 1, 90134, Palermo, Italy\label{inaf-palermo}
    \and
    Instituto de Astrof\'isica e Ci\^encias do Espa\c{c}o, Faculdade de Ci\^encias da Universidade de Lisboa, 1749-016 Lisboa, Portugal\label{lisboa}
    \and
    Consejo Superior de Investigaciones Científicas,  28006 Madrid, Spain\label{consejo-sup-madrid}
    \and
    INAF - Osservatorio Astrofisico di Torino, Via Osservatorio 20, 10025 Pino Torinese, Italy\label{inaf-torino}
             }

   \date{Submitted 17 March 2026; accepted 22 May 2026}

  \abstract
   {Small planets are extremely common in the Galaxy, including planets with masses and radii between those of Earth and Neptune. Characterizing these planets' masses requires ultra-precise radial velocities. The ESPRESSO spectrograph was designed and built for this purpose.}
   {We present an overview of the ESPRESSO Guaranteed Time Observations transit follow-up sub-program, aimed at confirming and characterizing small transiting planet candidates from the K2 and TESS space missions.}
   {We analyse the global stellar and planetary properties of the sample of 65 planets in 30 systems characterized by this sub-program. This includes six systems presented in this paper, for which we either obtain only upper mass limits, or provide updates to previously published parameters. We also place this sample in the context of the overall population of precisely characterized small planets.}
   {Separating the population into insolation regimes, we find a tentative mass threshold at $\mathrm{\simeq6\,M_\oplus}$ for the rocky to volatile-rich composition transition in the medium-insolation regime, and a population of likely stripped massive rocky planets in the high-insolation regime. We likewise find a correlation between planet mass and stellar metallicity, with more massive planets being hosted by more metal-rich stars. We also explore the radius valley, finding that planets below the gap have a tighter mass distribution. We compare planet masses with typical protoplanetary disk masses and draw tentative conclusions about likely formation conditions. Finally, we discuss the impact of our observing strategy on our results.}
   {The ESPRESSO transit follow-up sub-program has been highly productive, characterizing a diverse population of small planets that allows us to identify population-level features. Likewise, the lessons learned from this sub-program will be valuable for PLATO follow-up planning.}


   \maketitle

\section{Introduction}

The most common planets in the Galaxy are super-Earths and sub-Neptunes \citep{Mayor2011, Fressin2013}. These planets, with masses and radii between those of Earth and Neptune, are intriguing as we do not find such planets in the Solar System, raising the question of what their likely compositions are; scaled-up Earths, scaled-down Neptunes, gas dwarfs, water worlds, have all been proposed as options \citep[e.g.][and references therein]{Valencia2007, Fortney2007, Rogers2015, Zeng2019, Bean2021, Luo2024, Parc2024}. Likewise, they are among the most promising candidates for habitability \citep{Heller2014}. However, due to their small size and low mass, they are also some of the hardest planets to detect and characterize, requiring highly precise instrumentation.

The shallow transits from small planets are difficult to catch with ground-based (and thus atmosphere-limited) facilities, and the technique is thus relatively inefficient for large blind surveys from the ground. Instead, they are primarily identified by space-based missions such as \textit{Kepler}, K2, and TESS. \textit{Kepler} \citep{Borucki2010} was active from April 2009 to May 2013, observing the same sector of the northern sky for four years. While it identified thousands of transiting planet candidates, most orbit faint stars, making radial velocity (RV) follow-up difficult or impossible. Likewise, the Kepler field is inaccessible to southern RV facilities. \textit{Kepler}'s successor mission, K2 \citep{Howell2014}, was active from June 2014 to October 2018, and observed 19 campaigns of around three months each along the ecliptic, covering a field of 115 square degrees in each campaign. Meanwhile, TESS \citep{Ricker2015}, which began operations in 2018 and is still observing today, observes 3200 square degree sectors for 27 days each, covering most of the sky in sequence. Both K2 and TESS observed brighter stars than the \textit{Kepler} mission, and observed regions of the sky accessible to southern velocimetric facilities.

The premier southern extreme precision radial velocity (EPRV) facility is the Echelle SPectrograph for Rocky Exoplanets and Stable Spectroscopic Observations \citep[ESPRESSO,][]{Pepe2013, Pepe2021} at ESO’s Very Large Telescope (VLT), Paranal, Chile. ESPRESSO is a fibre-fed echelle spectrograph covering the $\mathrm{378.2 - 788.7\,nm}$ wavelength range, with a simultaneous reference fibre that can observe either the sky or a calibration lamp. It can observe with any of the four Unit Telescopes (UTs) of the VLT, reaching a resolving power of $140\,000$ (high resolution mode) or $190\,000$ (ultra-high resolution mode). It can also observe with all four UTs together, with a resolving power of $70\,000$. Being designed for ultra-high RV precision, ESPRESSO is thus ideally suited to precisely determining the masses of low-mass planets. It has been shown to achieve an RV precision better than 25 $\mathrm{cm\, s^{-1}}$ during a single night \citep{Pepe2021} and of 40 $\mathrm{cm\, s^{-1}}$ over 3.5 years \citep{Figueira2025}. 

Since October 2018, the ESPRESSO consortium has been carrying out an ambitious program of low-mass planet characterization, through its Guaranteed Time Observations \citep[GTO,][]{Pepe2021}. One of the major sub-programs, known as Working Group 3 (WG3), is dedicated to the follow-up of candidate transiting small planets from TESS and K2. In this paper, we present an overview of the entire WG3 sub-program. Section \ref{s:wg3-presentation} describes the WG3 goals and observing strategy, and summarises the resulting discoveries. In Section \ref{s:unpublished}, we analyse previously unpublished ESPRESSO datasets for six systems for which we either obtain only upper mass limits on the transiting planets, or provide parameter updates and refinements to previously published planets using other facilities. We also provide stellar parameters for 15 stars for which less than 10 measurements were obtained with ESPRESSO, and whose RVs we provide but do not analyse further. Section \ref{s:overview} explores the population of planets from the WG3 sub-program, including all published and in prep planet detections; we explore population-level features such as the radius valley, the mass-metallicity correlation, and the comparison of planet masses to typical disk masses, and discuss the impact of the observing strategy and stellar activity. Finally, we conclude in Section \ref{s:conclusions}.

\section{Overview of the WG3 sub-program}\label{s:wg3-presentation}

\subsection{Goals and sample definition}

The main goals of the ESPRESSO GTO WG3 sub-program are: 
\begin{enumerate}
    \item The confirmation and detailed characterization of small rocky transiting planet candidates;
    \item The probing of the rocky-to-gaseous transition.
\end{enumerate}

For the first goal, the sample was selected from K2 and TESS validated or vetted candidates with $\mathrm{R_p \lesssim 2\, R_\oplus}$, orbiting bright stars ($\mathrm{V \lesssim 14.5}$), and with no mass measurement available. For the second, \cite{Fulton2017} suggested that the insolation range of $\mathrm{50\,S_\oplus\lesssim S_p \lesssim200\,S_\oplus}$ was particularly interesting, as sub-Neptunes and rocky planets coexist there. The sample for this goal was thus constructed to include all the more massive companions to the rocky planet sample, plus other sub-Neptune and Neptune candidates with $\mathrm{R_p \lesssim 4\, R_\oplus}$ within this insolation range. 

\begin{figure}[hbt!]
    \centering
    \includegraphics[width=\linewidth]{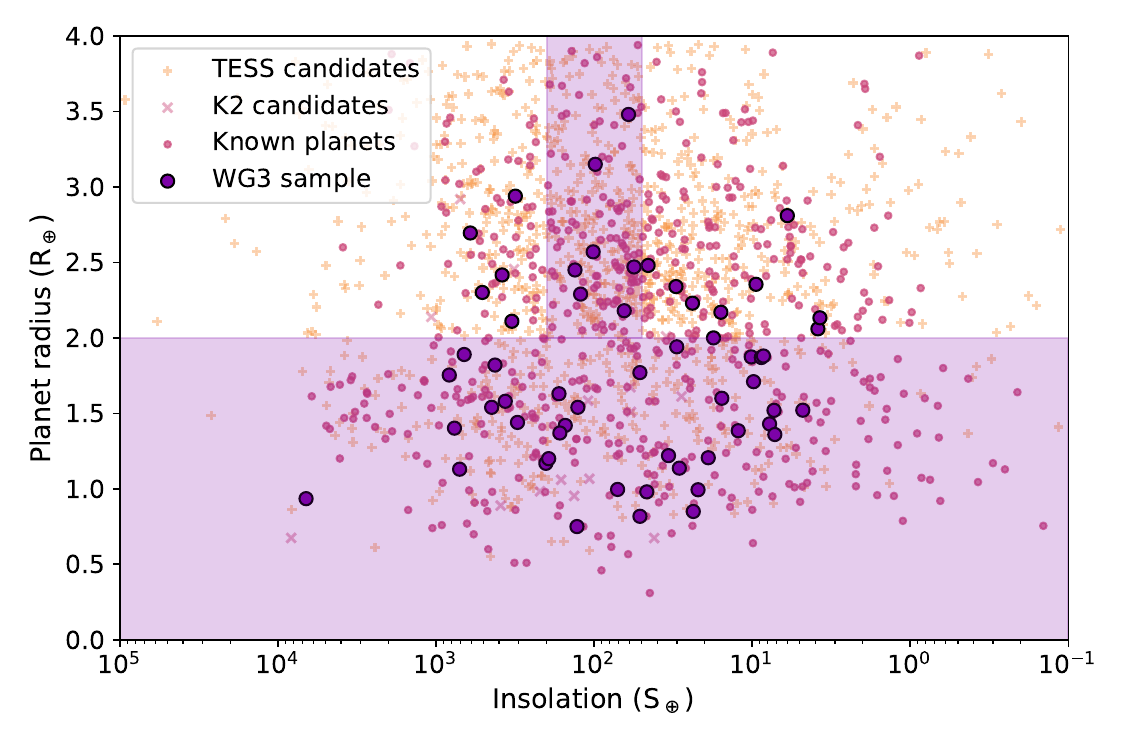}
    \caption{WG3 sample region (purple shaded areas) and individual target planets (purple circles), together with the population of confirmed and validated small planets (pink circles) and TESS and K2 candidates (orange plusses and pink crosses respectively). Target planets outside the sample regions are either part of multiplanetary systems with planets within the sample regions, or were initially thought to fall within them.}
    \label{fig:WG3-sample}
\end{figure}

Figure \ref{fig:WG3-sample} shows the sample regions and the planets characterized by the WG3, together with the overall population of small planets, and the TESS and K2 candidates from which our targets were drawn. Most of the WG3 planets outside the target regions are in multiplanetary systems with companions within them. The exceptions are $\pi$ Men c \citep{Damasso2020}, TOI-283 c \citep{Murgas2025}, and BD+20954 b (this work), all of which were initially announced as having radii of $\mathrm{\sim2\,R_\oplus}$ and thus considered as falling within the sample region, but whose final radii are slightly above $\mathrm{2\,R_\oplus}$.

\subsection{Observations and results}

The observations for the WG3 were carried out between September 2018 and March 2023, spanning four and a half years. There was, however, a nine-month interruption in 2020, when the Paranal observatory was closed in response to the COVID-19 pandemic. The targets to be observed were drawn from candidates in our sample region, with additional considerations such as observability from Paranal, stellar brightness, or stellar activity level. Bright targets were particularly prioritized, as they require less observing time for equal RV precision and thus enable more systems to be followed. We did not observe the same targets over the entire duration, but dynamically updated the target list as new candidates were released by TESS. A total of 50 stars were observed, though 15 of these were dropped after a handful of observations for varied reasons, such as being already intensively followed by other teams, or the RV photon-noise uncertainties and/or level of stellar activity being too large for precise mass characterization. 

Individual WG3 targets were observed over time spans of 2 to 36 months, with the average baseline being of 16 months. We used the 1-UT, high resolution mode for all observations. The binning and readout mode, exposure time, and use of the calibration fibre were selected according to the target magnitude, as summarized in Table \ref{tab:obs_strategy}. Regarding observing strategy, we generally requested measurements every $2-4$ nights depending on the orbital period of the planet candidate(s). 

\begin{table}[htb]
\begin{center} 
    \caption{Observing strategy summary for the WG3 targets by target magnitude.}
    \label{tab:obs_strategy}
\resizebox{\columnwidth}{!}{  
    \centering
    \begin{tabular}{llll}
    \hline \hline
    V magnitude & binning and readout & exposure time & calibration fibre \\ 
    \hline
    $\Pi$ Men C (V=5.7) &  1x1, FAST &  6x120 s  & FP  \\
    V < 11  & 1x1, FAST &   900 s  &    FP \\        
    11 < V < 12 & 2x1, SLOW & 900 s  & Sky \\
    12 < V < 13 & 2x1, SLOW & 1200 s  &   Sky\\
    V > 13 & 2x1, SLOW & 1800 s &  Sky  \\
    \hline
    \end{tabular}
    }
\end{center}
\end{table}

The WG3 sub-program has been highly productive. From these observations, 54 planets around 23 stars have been, or will shortly be, published by the ESPRESSO GTO consortium. In Table \ref{tab:WG3-planets} we list the planets, their main characteristics, and the reference WG3 publication. A further 10 planets around six stars are presented in this paper. For six of them we can provide only upper mass limits based on the existing data. The remaining four were previously published by other groups; we use the ESPRESSO RVs, together with the published data, to provide parameter updates. Combining the two groups, this comprises a final total of 65 planets in 30 systems. We note that some of these planets are non-transiting companions to transiting planets, that were revealed by the RV data. For these planets, we report the minimum mass. 

\begin{table*}[]
\caption{ESPRESSO WG3 planets.}
\label{tab:WG3-planets}
\resizebox{\textwidth}{!}{  
\begin{tabular}{llllllllll}
\hline \hline
Name       & K [m s$^{-1}$]     & Mass [M$_\oplus$]                               & Radius [R$_\oplus$] & density [g cm$^{-3}$]                              & Period [d]                & eccentricity   & S [S$_\oplus$]    & RV jitter [m s$^{-1}$]  & Reference                       \\
\hline
GJ 9827 b   & $3.53  \pm 0.22 $     & $4.28^{+0.35}_{-0.33}$          & $1.44  ^{+0.09  }_{-0.07 }$       &  $7.9\pm 1.6$ & 1.208974         & $0$     & $306 \pm 14   $  & 0.42     & \cite{Passegger2024}            \\
GJ 9827 c   & $1.06  \pm 0.21 $     & $1.86^{+ 0.37}_{- 0.39}$       & $1.13  ^{+0.07  }_{-0.05 }$       & $7.1\pm 2.0$ &3.648103         & $0$     & $711.0  \pm 35.4   $  & 0.42     & \cite{Passegger2024}            \\
GJ 9827 d   & $1.44  \pm 0.27 $     & $3.02 ^{+0.58 }_{-0.57 }$       & $1.89  ^{+0.16  }_{-0.14 }$       & $2.46 \pm0.78$ & 6.201812         & $0$     & $665.5 \pm 31.3   $  & 0.42     & \cite{Passegger2024}            \\
K2-38 b     & $2.95  ^{+0.44  }_{-0.39 }$     & $7.3  ^{+1.1  }_{-1    }$       & $1.54  \pm 0.14 $  & $11.0 \pm 3.4$     & 4.01593          & $0.197  ^{+0.067   }_{-0.060    }$     & $445.8 \pm 13.9   $  & 0.96     & \cite{Toledo2020}              \\
K2-38 c     & $2.41  ^{+0.39  }_{-0.37 }$     & $8.3  \pm 1.3 $       & $2.29  \pm 0.26 $   & $3.8 \pm1.4$    & 10.56103         & $0.161  ^{+0.096   }_{-0.078   }$     & $122.0 \pm 3.9  $  & 0.96     & \cite{Toledo2020}              \\
K2-111 b    & $2.21  \pm 0.32 $     & $5.29 ^{+0.76 }_{-0.77 }$       & $1.82  ^{+0.11  }_{-0.09 }$ & $4.8\pm1.1$      & 5.3518           & $0.13   ^{+0.13    }_{-0.09    }$     & $424.7 \pm 21.1   $  & 1.57     & \cite{Mortier2020}             \\
K2-111 c    & $3.27  ^{+0.31  }_{-0.32 }$     & $^{\dagger}11.3 \pm 1.1  $       & $-$       & $-$ & 15.6785          & $0.07   ^{+0.07    }_{-0.05    }$     & $100.9 \pm 4.8    $  & 1.57     & \cite{Mortier2020}             \\
K2-157 b    & $1.1   ^{+0.39  }_{-0.41 }$     & $1.14 ^{+0.41 }_{-0.42 }$       & $0.935 \pm 0.09 $    & $7.7\pm3.6$   & 0.3652575        & $0$     & $6660             \pm 200                $  & 1.43     & \cite{Castro2025}              \\
K2-157 c    & $7.17  \pm 0.42$     & $^{\dagger}30.8 \pm 1.9  $       & $-$   & $-$    & 25.942           & $<0.2$     & $22.65            \pm 0.68               $  & 1.43     & \cite{Castro2025}              \\
K2-157 d    & $3.96  ^{+0.42  }_{-0.41 }$     & $^{\dagger}23.3 \pm 2.5  $       & $-$   & $-$     & 66.5             & $<0.5$     & $6.46             \pm 0.21               $  & 1.43     & \cite{Castro2025}              \\
TOI-130 b   & $1.93  ^{+0.334 }_{-0.35 }$     & $7.8  ^{+1.5  }_{-1.4  }$       & $2.45  \pm 0.05 $   & $2.91\pm 0.59$   & 14.339156        & $0.08   ^{+0.09    }_{-0.05    }$     & $132.4\pm 4.0   $  & 1.17     & \cite{Sozzetti2021}            \\
TOI-130 c   & $3.17  ^{+0.34  }_{-0.32 }$     & $^{\dagger}18.4 ^{+1.8  }_{-1.9  }$       & $-$  & $-$    & 40.87            & $0.07   ^{+0.08    }_{-0.04    }$     & $32.8 \pm 1.0  $  & 1.17     & \cite{Sozzetti2021}            \\
TOI-134 b   & $3.2   ^{+0.29  }_{-0.3  }$     & $4.07 \pm 0.45 $       & $1.63  \pm 0.14 $   & $5.2\pm 1.5$    & 1.40152604       & $0$     & $167.1 \pm 10.2   $  & 2.145    & \cite{Hobson2024}              \\
TOI-144 b   & $196.1 \pm 0.7  $     & $^{\dagger}9.89 \pm 0.25$       & $-$   & $-$   & 2088.8           & $0$     & $0.1427\pm 0.0045$  & 1.2      & \cite{Damasso2020}             \\
TOI-144 c   & $1.5   \pm 0.2 $     & $4.3  \pm 0.7  $       & $2.11 \pm 0.05 $ & $2.52\pm0.45$     & 6.267852         & $0$     & $332.0 \pm 10.3   $  & 1.2      & \cite{Damasso2020}             \\
TOI-174 b   & $2.68  ^{+0.24  }_{-0.25 }$     & $8.32 ^{+0.78 }_{-0.79 }$       & $2     ^{+0.11  }_{-0.10  }$ & $5.7 \pm1.1$      & 17.667087        & $0.072  ^{+0.039   }_{-0.040    }$     & $17.60 \pm 0.65  $  & 0.617    & \cite{Barros2022}              \\
TOI-174 c   & $0.92 \pm 0.23 $     & $3.41 ^{+0.88 }_{-0.81 }$       & $1.87  ^{+0.12  }_{-0.11 }$  & $2.87\pm 0.92$    & 29.79749         & $0.063  ^{+0.054   }_{-0.043   }$     & $8.77 \pm 0.31  $  & 0.617    & \cite{Barros2022}              \\
TOI-174 d   & $0.29 \pm 0.11 $     & $0.55 ^{+0.21 }_{-0.20  }$       & $0.75  ^{+0.067 }_{-0.057}$  & $7.2\pm3.3$     & 3.97664          & $0.07   ^{+0.050    }_{-0.047   }$     & $128.6 \pm 4.7   $  & 0.617    & \cite{Barros2022}              \\
TOI-174 e   & $0.3   ^{+0.12  }_{-0.11 }$     & $0.72 ^{+0.28 }_{-0.27 }$       & $0.818 ^{+0.080  }_{-0.065}$ & $7.2\pm3.5$     & 7.90754          & $0.07   ^{+0.052   }_{-0.047   }$     & $51.4 \pm 1.8   $  & 0.617    & \cite{Barros2022}              \\
TOI-174 f   & $0.29  \pm 0.11 $     & $0.77 ^{+0.44 }_{-0.40  }$       & $1.137 ^{+0.084 }_{-0.077}$   & $2.9\pm1.8$    & 12.1621839       & $0.07   ^{+0.048   }_{-0.051   }$     & $28.9 \pm 1.1   $  & 0.617    & \cite{Barros2022}              \\
TOI-175 b   & $0.46  ^{+0.20   }_{-0.17 }$     & $0.4  ^{+0.16 }_{-0.15 }$       & $0.85  ^{+0.061 }_{-0.047}$  & $3.6\pm1.6$     & 2.2531136        & $0.103  ^{+0.117   }_{-0.045   }$     & $23.6  \pm 1.9   $  & 0.895    & \cite{Demangeon2021}           \\
TOI-175 c   & $2.19  ^{+0.17  }_{-0.20  }$     & $2.22 ^{+0.26 }_{-0.25 }$       & $1.385 ^{+0.095 }_{-0.075}$ & $4.6\pm1.1$      & 3.6906777        & $0.103  ^{+0.045   }_{-0.058   }$     & $12.3 \pm 1.0   $  & 0.895    & \cite{Demangeon2021}           \\
TOI-175 d   & $1.5   ^{+0.22  }_{-0.19 }$     & $1.94 \pm 0.28 $       & $1.521 ^{+0.119 }_{-0.098}$   & $3.03\pm 0.83$    & 7.4507245        & $0.074  ^{+0.057   }_{-0.046   }$     & $4.79 \pm 0.39  $  & 0.895    & \cite{Demangeon2021}           \\
TOI-175 e   & $2.01  ^{+0.16  }_{-0.20  }$     & $^{\dagger}3.06 ^{+0.33 }_{-0.37 }$       & $-$   & $-$    & 12.796           & $0.128  ^{+0.108   }_{-0.076   }$     & $2.21 \pm 0.37  $  & 0.895    & \cite{Demangeon2021}           \\
TOI-178 b   & $1.05  ^{+0.25  }_{-0.30  }$     & $0.96 ^{+0.70  }_{-0.65 }$       & $1.2   ^{+0.041 }_{-0.037}$  & $3.1\pm2.3$     & 1.9145601        & $0$     & $194.2 \pm 12.3   $  & 0.51     & \cite{Leleu2021}               \\
TOI-178 c   & $2.77  ^{+0.22  }_{-0.33 }$     & $4.64 ^{+0.52 }_{-0.53 }$       & $1.754 ^{+0.032 }_{-0.040 }$  & $4.72 \pm0.63$     & 3.238486         & $0.00032^{+0.00065 }_{-0.00020  }$     & $826.9  \pm 51.9   $  & 0.51     & \cite{Leleu2021}               \\
TOI-178 d   & $1.34  ^{+0.31  }_{-0.39 }$     & $5.2  ^{+0.39 }_{-0.43 }$       & $2.695 ^{+0.041 }_{-0.046}$   & $1.46\pm 0.14$    & 6.557569         & $0.0068 \pm 0.0016  $     & $608.4 \pm 38.9   $  & 0.51     & \cite{Leleu2021}               \\
TOI-178 e   & $1.62  ^{+0.41  }_{-0.34 }$     & $3.48 \pm 0.29 $       & $2.301 ^{+0.038 }_{-0.039}$  & $1.57\pm 0.15$     & 9.96318          & $0.00038^{+0.00067 }_{-0.00025 }$     & $510.9 \pm 32.9   $  & 0.51     & \cite{Leleu2021}               \\
TOI-178 f   & $2.76  ^{+0.46  }_{-0.42 }$     & $5.63 ^{+0.45 }_{-0.41 }$       & $2.417 ^{+0.041 }_{-0.048}$  & $2.19\pm0.22$     & 15.23335         & $0.00045^{+0.00070  }_{-0.00026 }$     & $382.8 \pm 24.1   $  & 0.51     & \cite{Leleu2021}               \\
TOI-178 g   & $1.3   ^{+0.38  }_{-0.59 }$     & $4.4  ^{+0.39 }_{-0.37 }$       & $2.939 ^{+0.057 }_{-0.055}$   & $0.95\pm0.10$    & 20.71663         & $0.00043^{+0.00065 }_{-0.00028 }$     & $315.7 \pm 20.3   $  & 0.51     & \cite{Leleu2021}               \\
TOI-198 b   & $1.673 ^{+0.207 }_{-0.240 }$     & $3.17 ^{+0.64 }_{-0.65 }$       & $1.36  \pm 0.13$ & $6.9\pm 2.4$      & 10.215213        & $0$     & $7.2              \pm 2.4                $  & 0.005    & \cite{Zapatero2026} \\
TOI-214 b   & $0.49  ^{+0.17}_{-0.16}$     & $1.35_{-0.46}^{+0.46};<2.73$                         & $0.996 \pm 0.050$  & $7.5\pm 2.8; <15.2$     & 9.695931          & $0.035^{+0.027}_{0.018}$     & $71.2 \pm 2.4    $  & 1.53      & Benatti et al. in prep          \\
TOI-214 c   & $2.3   \pm 0.17 $     & $7.88^{+0.61}_{-0.66}$       & $1.941  \pm 0.074$   & $5.92 \pm0.84$   & 18.5537114  & $0.095^{+0.035}_{-0.023}$     & $30.1 \pm 1.0   $  & 1.53      & Benatti et al. in prep          \\
TOI-238 b   & $2.36  \pm 0.32 $     & $3.4  \pm 0.46 $       & $1.402 \pm 0.086$  & $6.8\pm 1.6$    & 1.2730991        & $<0.18$     & $766.4 \pm 40.6   $  & 1.99     & \cite{Suarez2024}              \\
TOI-238 c   & $2.46  \pm 0.41 $     & $6.7  \pm 1.1  $       & $2.18 \pm 0.18 $   & $3.6 \pm1.1$   & 8.465651         & $<0.36$     & $64.6  \pm 10.0   $  & 1.99     & \cite{Suarez2024}              \\
TOI-244 b   & $1.55 \pm 0.16 $     & $2.68 \pm 0.3  $       & $1.52  \pm 0.12 $   & $4.2\pm1.1$    & 7.397225         & $0$     & $7.29 \pm 0.37  $  & 0.19     & \cite{Castro2023}              \\
TOI-260 b   & $1.57  ^{+0.59  }_{-0.57 }$     & $4.23 \pm 1.6  $       & $1.71  \pm 0.08 $   & $4.7\pm1.9$    & 13.475853        & $0$     & $9.82 \pm 0.39  $  & 1.51     & \cite{Hobson2024}              \\
TOI-262 b & $2.01^{+0.20}_{-0.21}$ & $6.59\pm0.81$ & $2.47\pm0.13$ & $2.40\pm 0.48$ & 11.145334 & 0 & $56^{+8}_{-7}$ & 0.008 & Zapatero Osorio et al. in prep \\
TOI-270 b   & $1.27  \pm 0.21 $     & $1.58 \pm 0.26$       & $1.206 \pm 0.039$  & $4.95\pm 0.95$    & 3.3601538        & $0.034  \pm 0.025   $     & $18.99 \pm 0.46 $  & 0.68     & \cite{VanEylen2021}            \\
TOI-270 c   & $4.16  \pm 0.24 $     & $6.15 \pm 0.37 $       & $2.355 \pm 0.064$ & $2.59 \pm0.26$      & 5.6605731        & $0.027  \pm 0.021   $     & $9.47 \pm 0.23  $  & 0.68     & \cite{VanEylen2021}            \\
TOI-270 d   & $2.56  \pm 0.23 $     & $4.78 \pm 0.43 $       & $2.133 \pm 0.058$  & $2.71 \pm 0.33$     & 11.379573        & $0.032  \pm 0.023   $     & $3.734 \pm 0.091 $  & 0.68     & \cite{VanEylen2021}            \\
TOI-283 b   & $1.86 ^{+0.58 }_{-0.56}$        & $6.54 \pm 2.04$          & $2.34  \pm 0.09 $   & $2.80\pm0.93$    & 17.61745         & $0$     & $30.4     \pm 2.5  $  & 5.11     & \cite{Murgas2025} \\ 
TOI-286 b   & $1.98 \pm 0.33 $     & $4.53 \pm 0.78 $       & $1.42  \pm 0.1  $    & $8.7\pm 2.4$   & 4.5117244        & $0$     & $152.5 \pm 6.8    $  & 1.56     & \cite{Hobson2024}              \\
TOI-286 c   & $0.79  ^{+0.64  }_{-0.47 }$     & $3.72 \pm 2.22$       & $1.88  \pm 0.12 $  & $3.1\pm 1.9$     & 39.361826        & $0$     & $8.50 \pm 0.38  $  & 1.56     & \cite{Hobson2024}              \\
TOI-455 b   & $2.15  \pm 0.19 $     & $2.32 \pm 0.25 $       & $1.43  \pm 0.09 $   & $4.36 \pm0.95$    & 5.358764         & $0$     & $7.8       \pm 2.9   $  & 0.88     & \cite{Lavie2023}               \\
TOI-455 c   & $1.11  \pm 0.2  $     & $1   \pm 0.19 $       & $1.6   ^{+0.67  }_{-0.34 }$   & $1.3\pm 1.7$    & 3.123898         & $0$     & $15.6 \pm 5.5   $  & 0.88     & \cite{Lavie2023}               \\
TOI-455 d   & $1.53  ^{+0.40   }_{-0.41 }$     & $^{\dagger}2.72 ^{+0.75 }_{-0.75 }$       & $-$   & $-$    & 24.3             & $0$     & $0.98 \pm 0.44   $  & 0.88     & \cite{Lavie2023}               \\
TOI-469 b   & $2.8   \pm 0.2  $     & $9.6  \pm 0.8  $       & $3.48  ^{+0.07  }_{-0.08 }$ & $1.25\pm0.14$     & 13.63083         & $0$     & $60.7 \pm 2.2   $  & 0.7      & \cite{Damasso2023}             \\
TOI-469 c   & $2.1   \pm 0.1  $     & $4.5  \pm 0.3  $       & $1.58  ^{+0.10   }_{-0.11 }$    & $6.3 \pm 1.4$   & 3.53796          & $0$     & $365.6 \pm 13.8  $  & 0.7      & \cite{Damasso2023}             \\
TOI-469 d   & $1.9   \pm 0.1  $     & $5.1  \pm 0.4  $       & $1.37  \pm 0.11 $   & $10.9\pm 2.8$   & 6.42975          & $0$     & $165.0\pm 6.1   $  & 0.7      & \cite{Damasso2023}             \\
TOI-512 b   & $1.54  \pm 0.21 $     & $3.57 ^{+0.53 }_{-0.55 }$       & $1.54  \pm 0.1  $  & $5.4 \pm 1.3$    & 7.18886          & $0.02   ^{+0.06    }_{-0.03    }$     & $127              \pm 9                  $  & 0.79     & \cite{Rodrigues2025}           \\
TOI-713 b   & $1.062 ^{+0.186 }_{-0.179}$     & $1.754^{+0.321}_{-0.326}$       & $1.169 ^{+0.097 }_{-0.099}$  & $6.0 \pm 1.9$   & 1.87152392922476 & $0$     & $202              \pm 33                 $  & 0.47     & Barros et al. in prep           \\
TOI-713 c   & $1.548 ^{+0.513 }_{-0.548}$     & $6.521^{+2.191}_{-2.522}$       & $2.06  ^{+0.170  }_{-0.176}$  &  $4.1\pm 1.9$   & 35.9994728661654 & $0.26   ^{+0.16    }_{-0.17    }$     & $3.84     \pm 0.84  $  & 0.47     & Barros et al. in prep \\    
TOI-2322 b  & $0.84  ^{+0.97  }_{-0.56 }$     & $2.31 ^{+2.67 }_{-1.54 }$       & $0.994 ^{+0.057 }_{-0.059}$   & $12.9 \pm 8.9$    & 11.30717         & $0$     & $22.04            \pm 5.34               $  & 4.07     & \cite{Hobson2025}              \\
TOI-2322 c  & $5.37  ^{+1.29  }_{-1.6  }$     & $18.1 ^{+4.34 }_{-5.36 }$       & $1.874 ^{+0.066 }_{-0.057}$   & $15.1\pm 4.8$    & 20.225528        & $0$     & $10.15            \pm 2.5                $  & 4.07     & \cite{Hobson2025}              \\
\hline
BD+20594 b  & $1.08  ^{+0.46  }_{-0.44 }$     & $5.2 \pm 2.2;<12.0$                        & $2.23  \pm 0.20  $   & $2.6\pm1.3; <5.9$    & 41.68547         & $0$     & $23.9             \pm 0.5                $  & 2.1      & this work                       \\
K2-62 b     & $1.77  ^{+0.69  }_{-0.61 }$     &  $4.0 \pm 1.6;<8.6   $                       & $1.77  \pm 0.20  $     & $4.0 \pm 2.1; <8.7$ & 6.67188          & $0$     & $51.4             \pm 0.9                $  & 1.805    & this work                       \\
K2-62 c     & $2.19  ^{+0.89  }_{-0.92 }$     &  $6.6 \pm 2.8;<14.7   $                      & $2.17 \pm 0.16 $   & $3.6\pm 1.7; <8.1$    & 16.19752         & $0$     & $15.8             \pm 0.3                $  & 1.805    & this work                       \\
K2-105 b    & $4.36  ^{+0.99  }_{-1.04 }$     & $^{\dagger}11.1\pm 2.7 $       &  $- $   & $-$    & 5.0217295799     & $0$     & $191.6            \pm 3.7                $  & 4.595    & this work                       \\
K2-105 c    & $1.7   ^{+1.2   }_{-1.0    }$     & $5.0 \pm 3.5;<15.5$                        & $3.145  \pm 0.073 $    & $0.88 \pm 0.62; <2.74$   & 8.2669720877     & $0$     & $98.5             \pm 1.9                $  & 4.595    & this work                       \\
K2-184 b    & $4.4   ^{+1.5   }_{-1.6  }$     & $10.9 \pm 4.7;<29.8$                        & $1.216  \pm 0.062 $ & $33\pm 15;< 76$     & 16.977959        & $\ktwooneeightfoureccpone$     & $33.9             \pm 2.6                $  & 2.47     & this work                       \\
TOI-186 b   & $6.24  ^{+0.30   }_{-0.31 }$     & $25.94 \pm 1.3  $       & $2.809  \pm 0.062 $       & $6.42\pm 0.52$ & 35.613442        & $0.122   \pm 0.022    $     & $6.0                \pm 0.1                $  & 2.43     & this work                       \\
TOI-186 c   & $0.57  ^{+0.32  }_{-0.39 }$     & $1.43 \pm 0.80;<3.83$                         & $0.98 \pm 0.04 $   & $8.4\pm4.8; 22<.4$  & 7.789787         & $0.25 _{-0.13}^{+0.20}$     & $46.5             \pm 2.6                $  & 2.43     & this work                       \\
TOI-266 b   & $2.42  \pm 0.30$     & $7.39 \pm 0.92 $       & $2.57  \pm 0.12 $    & $2.39 \pm 0.45$ & 10.751014        & $0$     & $101.6            \pm 1.2                $  & 1.42     & this work                       \\
TOI-266 c   & $2.41  ^{+0.33  }_{-0.32 }$     & $9.0 \pm 1.2 $       & $2.48 \pm 0.14 $ &  $3.25\pm  0.71$   & 19.60550         & $0$     & $45.6             \pm 0.5                $  & 1.42     & this work                       \\
\hline
\end{tabular}
}
\tablefoot{Top: planets confirmed and characterized by the ESPRESSO WG3 sub-program, listed alphabetically. For non-transiting planets, we provide the minimum mass $\mathrm{m \sin i}$, indicated by a prepended $\dagger$. Bottom: planets for which we present either upper mass limits or parameter updates in this work. For those where we do not reach a $3\sigma$ measurement of the semi-amplitude, we give both the best-fit mass and density and the $3\sigma$ upper mass and density limits.}
\end{table*}

\subsection{Some remarkable individual systems}

There are many noteworthy individual planets and planetary systems among the WG3 harvest. Here, we highlight a small selection. Considering the sample as a whole, the lowest-mass planet detected is TOI-175 b \citep[][also known as L 98-59 b]{Demangeon2021}, with less than half the mass of the Earth. This is not, however, the smallest RV signal detected, as the RV semi-amplitude is a combination of the stellar mass, planetary mass, and orbital period; that honour goes to TOI-174 d, e, and f \citep{Barros2022}, all at $\mathrm{29-30\,cm\,s^{-1}}$. The densest planet of the sample is TOI-2322 c \citep{Hobson2025}, a remarkably massive super-Earth, while the least dense planet is TOI-178 g \citep{Leleu2021}, a mini-Neptune with a large gas fraction. Regarding density, we also highlight TOI-175 d \citep{Demangeon2021} and TOI-244 b \citep{Castro2023}, two of the first low-density super-Earths to be discovered; and K2-38 b \citep{Toledo2020}, with an extraordinarily dense, Mercury-like composition. 
While around half the planets have periods shorter than 10 days, there are a handful of transiting planets with periods longer than 35 days: BD+20594 b (this work), TOI-286 c \citep{Hobson2024}, TOI-713 c (Barros et al. in prep), and TOI-186 b \citep[][this work]{Dragomir2019}. Insolation-wise, the planets range from K2-157 b \citep{Castro2025} at over 6000 times the Earth's insolation, to TOI-270 d \citep{VanEylen2021}, the lowest insolation transiting planet in the sample at less than $\mathrm{4\,S_\oplus}$; there are also non-transiting outer companions to transiting planets at insolations down to a tenth of the Earth's \citep[TOI-1444 b,][]{Damasso2020}. The TOI-178 system \citep{Leleu2021, Leleu2024}, comprised of six small planets in which all but the innermost are in a chain of Laplace resonances, has the largest number of planets out of the systems in the sample, closely followed by the TOI-174 system \citep{Barros2022} with five planets.

\section{Analysis of six unpublished WG3 systems}\label{s:unpublished}

In this section, we analyse six previously unpublished ESPRESSO datasets. These stars were observed in the context of the WG3, but either the data do not suffice for a robust mass measurement, or the planets were previously confirmed by teams using other facilities. Sub-section \ref{s:obs} describes the observations, sub-section \ref{s:stellar-params} the stellar parameter computation, and sub-section \ref{s:analysis} the planet modelling.

\subsection{Observations}\label{s:obs}

\subsubsection{ESPRESSO radial velocities}\label{s:ESPRESSO-RVs}

All our targets were observed with ESPRESSO in the context of the GTO WG3 transiting candidate follow-up. Table \ref{tab:espresso-summary} lists the number of observations, time span, exposure time, and CCF mask used for the reduction for each of the targets presented in this paper. All the spectra were processed with the ESPRESSO data reduction software (DRS, \citealt{Pepe2021}) v.3.0.0, which computes RVs through the cross-correlation function (CCF) method \citep{Baranne1996}. The DRS also computes activity indicators: the Mount-Wilson S-index ($\mathrm{S_{MW}}$, \citealt{Vaughan1978}) and $\log R'_{\rm hk}$ \citep{Noyes1984}, which measure chromospheric emission in the cores of the Ca II H and K lines; the $\mathrm{H_\alpha}$ index \citep{Cincunegui2007,Bonfils2007}, which measures it in the $\mathrm{H_\alpha}$ line; the Na index \citep{Diaz2007}, which measures it in the Na I D1 and D2 lines; and the full width at half maximum (FWHM), bisector inverse slope \citep[BIS,][]{Queloz2001}, and contrast of the CCF. The RVs and activity indicators were obtained from the Data \& Analysis Center for Exoplanets (DACE) platform\footnote{Available online at \url{https://dace.unige.ch}}.

For some targets, observations were performed both before and after ESPRESSO underwent a major intervention in June-July 2019, when the fibre link was replaced \citep{Pepe2021}. When this is the case, we split the ESPRESSO data into two sets pre- and post-upgrade, labelled ESPRESSO18 and ESPRESSO19 respectively. We treat these as independent data sets in the fitting.

\begin{table*}[htb]
\caption{ESPRESSO observations summary for the systems presented in this work.}
\label{tab:espresso-summary}
\begin{tabular}{llllll}
\hline \hline
Target   & Observations & Time span & Exp. time [s] & CCF mask & median RV error [m/s] \\
\hline
K2-184   & 46 & 6 Feb. 2019 - 27 Dec. 2020 & $600-900$ & K2 & 0.85 \\
K2-105   & 40 & 14 Nov. 2018 - 27. Dec. 2020 & $900-1500$ & G9 & 1.11 \\  
TOI-266  & 26 & 13 Sep. 2021 - 19 Dec. 2022 & $900-1200$ & G8 & 0.51 \\
K2-62    & 24 & 3 May 2019 - 14 Aug. 2021 & $1200-1800$ & K2 & 1.13 \\
BD+20594 & 15 & 30 Oct. 2018 - 15 Sep. 2019 & 900 & G9 & 1.09 \\
TOI-186  & 10 & 30 Nov. 2019 -  9 Aug. 2021 & 900 & K2 & 0.14 \\
\hline
\end{tabular}
\end{table*}

\subsubsection{TESS photometry}\label{s:TESS-data}

We use TESS Presearch Data Conditioning Simple Aperture Photometry (PDC-SAP) \citep{Stumpe2012,Stumpe2014,Smith2012} photometry, which we obtained from the Mikulski Archive for Space Telescopes (MAST) archive, for TOI-266 and TOI-186. This photometry was processed by the \textit{TESS} Science Processing Operations Center \citep[SPOC,][]{Jenkins2016} at NASA Ames Research Center. The resulting light curves were also searched for transit signals by SPOC \citep{Jenkins2002, Jenkins2010, Jenkins2020}.

TOI-266 was observed in sector 3 (24 September - 14 October 2018) during the TESS prime mission, and in sector 30 (23 September - 20 October 2020) during the first extended mission. In both sectors it was observed in camera 1, CCD 2, in 2-minute cadence. Two potential planet candidates at $\mathrm{\sim 11\,d}$ and $\mathrm{\sim 20\,d}$ were identified by SPOC and designated as TESS Objects of Interest TOI-266.01 and TOI-266.02 by the TESS Science Office \citep{Guerrero2021} on 30 November 2018. 

TOI-186 was observed in sectors 1-4 (25 July - 14 November 2018) during the TESS prime mission; in sectors 28-30 and 34 (31 July - 20 October 2020 and 14 January - 8 February 2021) during the first extended mission; and sectors 61, 64, 68, and 69 (18 January - 12 February 2023, 6 April - 3 May 2023, and 29 July - 20 September 2023) during the second extended mission. All sectors were observed in 2-minute cadence. Sectors 1, 2, 28, 29, 69, and 69 were observed with camera 3, CCD 4; sectors 3 and 30 with camera 3, CCD 3; sector 4 with camera 4, CCD 2; sectors 34 and 61 with camera 4, CCD 3; sector 64 with camera 4, CCD 4. Two potential planet candidates at $\mathrm{\sim 8\,d}$ and $\mathrm{\sim 36\,d}$ were identified by SPOC and designated by the TESS Science Office as TOI-186.01 on 4 December 2018 and TOI-186.02 on 9 July 2020, respectively.

\subsubsection{K2 photometry}\label{s:K2-data}

We use K2 photometry for K2-105, K2-184, K2-62, and BD+20594. While all these targets were also observed with TESS, the photometric precision is insufficient to clearly detect the transits, so we do not include the TESS photometry in the analysis. Both K2-105 and K2-184 were observed in campaign 5, from 27 April to 10 July 2015, and in campaign 18, from 13 May to 2 July 2018. K2-62 was observed in campaign 3, from 15 November 2014 to 23 January 2015. BD+20594 was observed in campaign 5, from 8 February to 20 April 2015. For K2-62 and BD+20594, only 30-minute cadence data is available. For K2-105 and K2-184, 1-minute cadence data is available only for campaign 18; for consistency, we use the 30-minute cadence photometry for both campaigns.

\subsection{Stellar parameters}\label{s:stellar-params}

The stellar parameters for all our targets are listed in Table \ref{tab:starparams}. We also list in Table \ref{tab:starparams_fewmeas} atmospheric stellar parameters for WG3 targets which were observed less than ten times, and which we do not analyse further due to the small number of RVs. 

We obtained the stellar coordinates, proper motions, and parallaxes from the \textit{Gaia} Data Release 3 \citep{GAIA2016, GaiaDR3}. We used the co-added ESPRESSO spectra to compute the atmospheric parameters via spectral synthesis using the {\sc SteParSyn} code\footnote{https://github.com/hmtabernero/SteParSyn/} \citep{Tabernero2022}. This code computes the effective temperature $\mathrm{T_{eff}}$, metallicity [Fe/H], surface gravity $\log{g}$, and a broadening parameter $\mathrm{v_{broad}}$ which includes both the macroturbulence $\zeta$ and the projected rotational velocity $\mathrm{v \sin i}$.  We use the $\mathrm{T_{eff}}$ to estimate spectral types from the tables of \cite{Pecaut2013}. Since the error bars reported by {\sc SteParSyn} represent only the internal errors, for the $\mathrm{T_{eff}}$ we add in quadrature the 2\% error floor found by \cite{Tayar2022}. 

As an independent comparison and validation, we also obtained the atmospheric parameters via the measurement of the equivalent width of specific spectral lines, employing either (for FGK stars) the combined ARES+MOOG approach of \cite{Sousa2014}, or (for M dwarfs) the \texttt{ODUSSEAS} tool \citep{Antoniadis2020}. These provide $\mathrm{T_{eff}}$, [Fe/H], and $\log{g}$. We obtain generally compatible parameters with those computed using {\sc SteParSyn}. Regarding $\mathrm{T_{eff}}$, all values are compatible at the $1\sigma$ level save for K2-155, which lies at the edge of the range of validity for \texttt{ODUSSEAS}. Regarding [Fe/H], all values are compatible at $2\sigma$ save, again, K2-155, and TOI-1203, which is the most metal-poor FGK star in the sample. Regarding $\log{g}$, half the stars are compatible at the $2\sigma$ level. We also note that while the systematic errors for [Fe/H] and $\log{g}$ are less well studied, a conservative estimate is at the 0.1 dex level \citep{Tabernero2022}. If we add this in quadrature to the {\sc SteParSyn} errors, most [Fe/H] and $\log{g}$ values become compatible at the $1\sigma$ level. We adopt the {\sc SteParSyn} values for all our targets in order to have a uniform methodology across all spectral types. 

To compute the masses, radii, bolometric luminosities, and stellar ages, we used the approach described in \cite{Brahm2019PARSEC}. Briefly, the broadband photometric measurements from \textit{Gaia} \citep{GAIA2016} and 2MASS \citep{2MASS} are converted via the \textit{Gaia} DR3 \citep{GaiaDR3} parallaxes to absolute magnitudes and subsequently compared to the \texttt{PARSEC} stellar evolutionary models \citep{Bressan2012}. We add the error floors of \cite{Tayar2022} - 5\% in mass, 4\% in radius, and 20\% in age - in quadrature to the internal model errors. 

Finally, we report the median and standard deviation of the $\log R'_{\rm hk}$ activity indicator computed from the ESPRESSO spectra. We use this to estimate the stellar rotation period $\mathrm{P_{rot}}$, following the $\log R'_{\rm hk} - \mathrm{P_{rot}}$ relation of \cite{Suarez2015}.

\begin{table*}[htb]
\caption{Stellar parameters of the targets under analysis.}
\label{tab:starparams}
\resizebox{\textwidth}{!}{  
\begin{tabular}{llllllll}
\hline \hline
Parameter     & K2-184 & K2-105 & TOI-266 & K2-62 & BD+20594 & TOI-186 & Reference \\
\hline
Names         & BD+14 1931 & K2-105 & BD-19 302 & K2-62 & BD+20 594 & HD 21749 & Simbad \\
              & TIC 21244210 & TIC 6892385 & TIC 164767175 & TIC 434094657 & TIC 26123781 & TIC 279741379 & TESS \\
              & $-$ & TOI-5164 & TOI-266 & $-$ & $-$ & TOI-186 & TESS \\
              & K2-184 & K2-105 & $-$ & K2-62 &  K2-56 & $-$ & K2 \\
              & 651522834880015872 & 651907079835937280 & 5140454049422547200 & 2612830229101544704 & 58200934326315136 & 4673947174316727040 & \textit{Gaia} DR3 \\
RA \dotfill (J2000) & $08^h36^m33^s.6296967664$ & $08^h21^m40^s.8642457671$ & $01^h44^m50^s.3157306405$ & $22^h17^m27^s.4023572863$ & $03^h34^m36^s.2293244264$ & $03^h26^m59^s.2226450279$ & \textit{Gaia} DR3 \\
DEC \dotfill (J2000) & $+14{\degr}27{\arcmin}42{\arcsec}.967623427$ & $+13{\degr}29{\arcmin}51{\arcsec}.119454268$ & $-18{\degr}24{\arcmin}03{\arcsec}.266681201$ & $-12{\degr}11{\arcmin}15{\arcsec}.010119450$ & $+20{\degr}35{\arcmin}57{\arcsec}.269730747$ & $-63{\degr}29{\arcmin}56{\arcsec}.762921534$ & \textit{Gaia} DR3 \\
pm$^{\rm RA}$ \hfill [mas yr$^{-1}$] & 89.722 & 6.963 & 140.976 & 75.462 & 36.440 & 355.195 & \textit{Gaia} DR3 \\
pm$^{\rm DEC}$ \hfill [mas yr$^{-1}$] & -12.271 & -4.878 & -87.185 & 5.831 & -51.358 & -247.388 & \textit{Gaia} DR3 \\
$\pi$ \dotfill [mas] & 13.2175 & 5.0222 & 9.8339 & 8.8865 & 5.6384 & 61.2271 & \textit{Gaia} DR3 \\
\hline
T \dotfill [mag] & 9.6071 & 11.2089 & 9.45661 & 11.3633 & 10.4021 & 6.9934 & TESS \\
B \dotfill [mag] & 11.23$^\dagger$ & 12.68$^\dagger$ & 10.84$^\dagger$ & 13.51$^\star$ & 11.77$^\dagger$ & 9.302$^\ddagger$ & Various$^a$ \\
V \dotfill [mag] & 10.34$^\dagger$ & 11.75$^\dagger$ & 10.03$^\dagger$ & 12.40$^\star$ & 10.85$^\dagger$ & 8.143$^\ddagger$ & Various \\
J \dotfill [mag] & 8.920 & 10.541 & 8.847 & 10.374 & 9.770 & 6.081 & 2-MASS \\
H \dotfill [mag] & 8.513 & 10.173 & 8.542 & 9.811 & 9.432 & 5.52 & 2-MASS \\
K \dotfill [mag] & 8.380 & 10.091 & 8.449 & 9.688 & 9.368 & 5.375 & 2-MASS \\
\hline
$\mathrm{T_{eff}}$ \dotfill [K] & \TeffKtwooneeightfour & \TeffKtwoonezerofive & \TeffTOItwosixsix & \TeffKtwosixtwo & \TeffBDtwozerofiveninefour & \TeffTOIoneeightsix & this work \\
Fe/H \dotfill [dex] &  \FeHKtwooneeightfour & \FeHKtwoonezerofive & \FeHTOItwosixsix & \FeHKtwosixtwo & \FeHBDtwozerofiveninefour & \FeHTOIoneeightsix & this work \\
$\log{g}$ \dotfill [dex] & \loggKtwooneeightfour & \loggKtwoonezerofive & \loggTOItwosixsix & \loggKtwosixtwo & \loggBDtwozerofiveninefour & \loggTOIoneeightsix & this work \\
V$_{\rm broad}$$^b$ \dotfill [km s$^{-1}$] & \VbroadKtwooneeightfour & \VbroadKtwoonezerofive & \VbroadTOItwosixsix & \VbroadKtwosixtwo & \VbroadBDtwozerofiveninefour & \VbroadTOIoneeightsix & this work \\
R$_\star$ \dotfill [R$_\odot$] & $0.760\pm0.031$ & $0.887\pm0.036$ & $0.958\pm0.039$ & $0.666\pm0.027$ & $1.088\pm0.045$ & $0.709\pm0.029$ & this work \\
M$_\star$ \dotfill [M$_\odot$] & $0.759\pm0.039$ & $0.925\pm0.047$ & $0.838\pm0.042$ & $0.667\pm0.034$ & $0.847\pm0.042$ & $0.726\pm0.037$ & this work \\
L$_{bol,\star}$ \dotfill [L$_\odot$] & $0.3857_{-0.0031}^{+0.0038}$ & $0.5991_{-0.0088}^{+0.0109}$ & $0.8203_{-0.0073}^{+0.0090}$ & $0.1889_{-0.0032}^{+0.0033}$ & $1.183\pm0.027$ & $0.2151_{-0.0032}^{+0.0033}$ & this work \\
Age \dotfill [Gyr] & $10.7_{-2.6}^{+2.6}$ & $5.1\pm1.4$ & $12.6\pm2.6$ & $12.1_{-2.9}^{+2.6}$ & $13.2\pm2.6$ & $10.1_{-2.3}^{+2.6}$ & this work \\
$\log R'_{\rm hk}$ \dotfill & \logRhkKtwooneeightfour & \logRhkKtwoonezerofive & \logRhkTOItwosixsix & \logRhkKtwosixtwo & \logRhkBDtwozerofiveninefour & \logRhkTOIoneeightsix & this work \\
$\mathrm{P_{rot}}$ \dotfill [d] & \ProtFullMedianKtwooneeightfour & \ProtFullMedianKtwoonezerofive & \ProtFullMedianTOItwosixsix & \ProtFullMedianKtwosixtwo & \ProtFullMedianBDtwozerofiveninefour & \ProtFullMedianTOIoneeightsix & this work \\
Spectral type \dotfill & K1V & G9V & G6V & K4V & G2V & K4V & PM13 \\
\hline
\end{tabular}
}
\tablefoot{$^a$: B and V magnitude sources: $\dagger$: the Tycho-2 Catalogue \citep{Tycho-2}, $\star$: The Fourth US Naval Observatory CCD Astrograph Catalog \citep[UCAC4,][]{Zacharias2013}, $\ddagger$: \cite{Koen2010}. \\
$^b$: Broadening parameter including both the macroturbulence $\zeta$ and the projected rotational velocity $\mathrm{v \sin i}$, as described in \cite{Tabernero2022}.\\
Other references: Simbad: Simbad astronomical database \citep{Wenger2000}; \textit{TESS}: \textit{TESS} Input Catalog \citep{Stassun2019}; 2MASS: Two-micron All Sky Survey \citep{2MASS}; \textit{Gaia} DR3: \textit{\textit{Gaia}} Data Release 3 \citep{GAIA2016, GaiaDR3}; PM13: using the tables of \cite{Pecaut2013}. \\
The mass, radius, and $\mathrm{T_{eff}}$ have systematic errors added in quadrature; the error bars on [Fe/H] and $\log{g}$ represent only the internal errors, as the systematic errors are less well studied (see section \ref{s:stellar-params}).}
\end{table*}

\subsection{Data analysis}\label{s:analysis}

For each of our targets, we fit the RV and photometric data simultaneously with \texttt{juliet}\footnote{Available at \url{https://github.com/nespinoza/juliet}} \citep{Espinoza2019juliet}. This software enables the joint fitting of RVs through the \texttt{radvel} package \citep{Fulton2018} and photometry through the \texttt{batman} package \citep{Kreidberg2015}, incorporating Gaussian processes (GPs) via the \texttt{celerite} \citep{Foreman-Mackey2017} and \texttt{george} \citep{Ambikasaran2015} packages. The parameter space is explored via importance nested sampling, using the \texttt{dynesty} package \citep{Speagle2020}. 

For one-planet systems, we test circular models (henceforward $1c$) and models with free eccentricity (henceforward $1e$). For two-planet systems, we test circular models for the inner and outer planet individually (henceforward $in$ and $out$ respectively), and circular and free-eccentricity models for both planets together (henceforward $2c$ and $2e$ respectively). We also test the same models with the addition of GP detrending on the RVs (henceforward $model\_gp$). To select the number of signals to fit, we generally consider only the transiting planet candidates, as there are no clear signals in the RV periodograms suggesting the presence of further planets. The exception is K2-105, with a strong signal in the RV periodogram at $\mathrm{\sim5\,d}$ that does not correspond to the transiting candidate at $\mathrm{\sim8\,d}$.

The following input parameters are required in \texttt{juliet} for joint RV and transit fits, for each planet $\mathrm{i}$: period $\mathrm{P_{i}}$, time of transit $\mathrm{t_{0,i}}$, planet-to-star radius ratio $\mathrm{p_{i}}$, impact parameter $\mathrm{b_{i}}$, RV semi-amplitude $\mathrm{K_{i}}$, and either eccentricity $\mathrm{e_{i}}$ and angle of periastron $\mathrm{\omega_{i}}$ or a derived parametrisation such as $\mathrm{\sqrt{e_{i}} \sin \omega_{i}, \sqrt{e_{i}} \cos \omega_{i}}$. We adopt the latter parametrization, as recommended by \cite{Eastman2013}. We take priors for $\mathrm{P_{i}}$, $\mathrm{t_{0,i}}$, $\mathrm{p_{i}}$, and $\mathrm{b_{i}}$ from, in order, previous confirmation papers, validation papers, or Exofop. For $\mathrm{K_{i}}$ we take a broad uniform prior of $\mathrm{\mathcal{U}(0,10)\, ms^{-1}}$, and for $\mathrm{\sqrt{e_{i}} \sin \omega_{i}, \sqrt{e_{i}} \cos \omega_{i}}$ broad uniform priors of $\mathrm{\mathcal{U}(-1,1)}$.

The following instrumental parameters are also required: for each RV instrument, the systemic radial velocity $\mathrm{\mu_{instrument}}$ and the jitter $\mathrm{\sigma_{w,instrument}}$; for each photometric instrument, the dilution factor $\mathrm{m_{dilution,instrument}}$, the flux offset $\mathrm{m_{flux,instrument}}$, the jitter $\mathrm{\sigma_{w,instrument}}$, and the limb-darkening parameters $\mathrm{q_{1,instrument}}$ and $\mathrm{q_{2,instrument}}$ for the quadratic law parametrisation of \cite{Kipping2013}.  
For $\mathrm{\mu_{instrument}}$ we take a uniform prior of $\mathrm{\mathcal{U}(RV_{min}, RV_{max})}$. We fix $\mathrm{m_{dilution,instrument}}$ to 1 as the TESS and K2 light curves are dilution corrected.
For the remaining instrumental parameters we take broad priors.

In all fits, we apply GP detrending to the photometry, for which we employ the (approximate) Matern 3/2 kernel \citep{Foreman-Mackey2017}, as implemented in \texttt{celerite}. We use a two-step process on the photometry itself, where we first mask the transits and fit GPs to the out-of-transit data only on a sector-by-sector basis, with broad log-uniform priors of $\mathcal{J}(1\times10^{-6}, 1\times10^6)$ for $\mathrm{\sigma_{GP,sector}}$ and $\mathcal{J}(0.001, 1\times10^{3})$ for $\mathrm{\rho_{GP,sector}}$. We then use the resulting parameters as normal priors for the photometry GPs in the joint full fits. This two-step process improves the exploration of the parameter space and allows for faster convergence, as described in \cite{Patel2022}.

In the models including GP detrending for the RVs, we use the quasi-periodic kernel (called exp-sine-squared kernel in \texttt{juliet}) of \cite{Haywood2014}. To constrain the priors, we first fit a GP to the FWHM values, with broad log-uniform or Jeffreys priors of $\mathcal{J}(0.001,1000)$ for $\mathrm{\sigma_{GP,FWHM}}$, $\mathcal{J}(1\times10^{-6}, 0.1)$ for $\mathrm{\alpha_{GP,FWHM}}$, and $\mathcal{J}(1\times10^{-6}, 10)$ for $\mathrm{\Gamma_{GP,FWHM}}$; and a normal prior of $\mathcal{N}(P_{rot}, \sigma_{Prot})$ for the rotation period GP hyperparameter $\mathrm{P_{rot,GP, FWHM}}$, taking the value and error of the stellar rotation period derived in Sect \ref{s:stellar-params} as the mean and standard deviation respectively. We then used the resulting parameters as normal (or, in cases where a pure normal led to sampling values below zero and thus caused the fit to fail, truncated positive normal) priors for the RV GP in the joint full fits. We also applied a scaling factor to the $\mathrm{\sigma_{GP,FWHM}}$ posterior, computed as the ratio of the standard deviation of the two time series, $\mathrm{{\sigma_{RV}}/{\sigma_{FWHM}}}$.

The ESPRESSO RVs were obtained from DACE, as described in section \ref{s:ESPRESSO-RVs}. Other RVs for previously confirmed planets were taken from the respective confirmation papers. The TESS PDCSAP light curves were downloaded using the built-in \texttt{juliet} TESS MAST archive query function. The K2 light curves were downloaded using the \texttt{Lightkurve} package\footnote{\url{https://lightkurve.github.io/lightkurve/index.html}} \citep{Lightkurve, astropy, astroquery}. We also used the built-in \texttt{Lightkurve} utilities to remove long-term trends and correct for the spacecraft motion in the K2 light curves. 

\subsubsection{K2-184}

This system has a single transiting planet, which was validated in \cite{Mayo2018} and \cite{Livingston2018}. The K2 transits show significant TTVs, as reported by \cite{Castro2022}. However, we see no significant signals in the RVs that could help constrain the body causing the TTVs, so we do not analyse them further. We note that \cite{Castro2022} were unable to derive a TTV period due to an insufficient coverage, with only linear trends being found in the two K2 campaigns, which were observed three years apart. The RV data, meanwhile, span a little under two years, though the majority of it (38 of the 46 data points) was obtained in the first four months.

We tested the $1c$, $1e$, $1c\_gp$, and $1e\_gp$ models on the ESPRESSO RVs and K2 photometry. Both with and without GP detrending, the eccentric model is significantly favoured over the circular model, with $\mathrm{\Delta \log Z_{1e-1c} \approx 2-3}$. Likewise, the GP detrending is significantly favoured, with $\mathrm{\Delta \log Z_{1c\_gp-1c} \approx 8}$, $\mathrm{\Delta \log Z_{1e\_gp-1e} \approx 7}$. We thus adopt the $1e\_gp$ model as our final model, and report the priors and posteriors in Table \ref{tab:K2-184}. The RVs and phase-folded light curves are shown in Figs. \ref{fig:K2-184_RV} and \ref{fig:K2-184_K2} respectively. K2-184 b has a $\mathrm{\ktwooneeightfourPpone \,d}$ period, $\mathrm{\ktwooneeightyfourradb \, R_\oplus}$ radius, and $\mathrm{\ktwooneeightfoureccpone}$ eccentricity. We caution that this high-eccentricity fit may be driven by a residual RV signal from the body causing the TTVs, although no significant signals emerge in the residual RV periodogram. Regarding the planet's mass, we find a best-fit value of $\mathrm{10.9 \pm 4.7\,M_\oplus}$. However, we only reach a $\sim 2.7\sigma$ precision on the semi-amplitude, so we also provide an upper mass limit of $\mathrm{M_b\ktwooneeightyfourmassb \, M_\oplus}$.

\subsubsection{K2-105}

This system has a single transiting planet, validated in \cite{Narita2017}. There is no signal in the RVs at the $\sim 8\,d$ period of this planet, but there is a strong signal at $\sim 5\,d$. We therefore considered this as a potential two-planet system, and tested the $in$, $out$, $2c$, and $2e$ models on the ESPRESSO RVs and K2 photometry, and with the addition of the Subaru HDS RVs from \cite{Narita2017}. To constrain the priors of the potential inner planet, we ran a circular fit with \texttt{juliet} on the RVs alone, and allowed its impact parameter to go above 1 in the full fits to reflect the fact it does not appear to transit, with no signal at $\sim 5\,d$ in the K2 photometry BLS. Given the already complex nature of the system, we decided not to include GP detrending. 

The $2c$ model is favoured in both cases, with $\mathrm{\Delta \log Z_{2c-in} \approx 372}$, $\mathrm{\Delta \log Z_{2c-out} \approx 6}$, and $\mathrm{\Delta \log Z_{2c-2e} \approx 26}$ for the ESPRESSO+K2 dataset, and $\mathrm{\Delta \log Z_{2c-in} \approx 370}$, $\mathrm{\Delta \log Z_{2c-out} \approx 3}$ , and $\mathrm{\Delta \log Z_{2c-2e} \approx 20}$ for the full dataset. We therefore adopt the $2c$ model from the joint dataset as our final model, and report the priors and posteriors in Table \ref{tab:K2-105}. The RVs and phase-folded photometry are shown in Figs. \ref{fig:K2-105_RVs} and \ref{fig:K2-105_TESS_stacked} respectively. The inner planet K2-105 b does not transit, with the best-fit transit model being compatible with a flat line. It has a $\mathrm{\ktwooneofivePpone\,d}$ period, and a $\mathrm{\ktwooneofivemassb \, M_\oplus}$ minimum mass. The outer planet K2-105 c has a $\mathrm{\ktwooneofivePptwo\,d}$ period, a $\mathrm{\ktwooneofiveradc\,R_\oplus}$ radius, and a $\mathrm{5.0 \pm 3.5\,M_\oplus}$. As we only reach a $\sim 1.7\sigma$ precision on the semi-amplitude of this planet, we also report an upper mass limit of $\mathrm{M_b\ktwooneofivemassc \, M_\oplus}$.

\subsubsection{TOI-266}

This system has two transiting planets that were previously confirmed in \cite{Akana2023}, using 95 RVs from the Keck Observatory High Resolution Echelle Spectrometer (HIRES, \citealt{Vogt1994}). We first tested the $in$, $out$, $2c$, $2e$, and corresponding $\_gp$ models on the ESPRESSO RVs and TESS photometry. Both with and without detrending, the best model is that with two circular planets, and the addition of the GP is not favoured, with $\mathrm{\Delta \log Z_{2c\_gp-2c} \approx 0.8}$. For the models without detrending, we have $\mathrm{\Delta \log Z_{2c-in} \approx 51}$, $\mathrm{\Delta \log Z_{2c-out} \approx 70}$ , and $\mathrm{\Delta \log Z_{2c-2e} \approx 4}$. We then tested the $in$, $out$, $2c$, and $2e$ models on the joint ESPRESSO and HIRES RV dataset together with the test photometry. Once again, the best model is that with two circular planets, with $\mathrm{\Delta \log Z_{2c-in} \approx 63}$, $\mathrm{\Delta \log Z_{2c-out} \approx 87}$ , and $\mathrm{\Delta \log Z_{2c-2e} \approx 3}$. We adopt the $2c$ model for the joint dataset as our final model, and report the priors and posteriors in Table \ref{tab:TOI-266}. The RVs and phase-folded light curves are shown in Figs. \ref{fig:TOI-266_RVs} and \ref{fig:TOI-266_TESS_stacked} respectively. TOI-266 b has a $\mathrm{\toitwosixsixPpone\,d}$ period, $\mathrm{\toitwosixtysixradb\,R_\oplus}$ radius, and $\mathrm{\toitwosixtysixmassb\,M_\oplus}$ mass, while TOI-266 c has a $\mathrm{\toitwosixsixPptwo\,d}$ period, $\mathrm{\toitwosixtysixradc\,R_\oplus}$ radius, and $\mathrm{\toitwosixtysixmassc\,M_\oplus}$ mass. Our results are comparable to those of \cite{Akana2023}, with compatible RV semi-amplitude values with improved precision, and thus improved precision on the planetary masses, as shown in Table \ref{tab:TOI-266-comp}. Our masses are systematically smaller as a result of a lower stellar mass of $\mathrm{0.838\pm0.042\,M_\odot}$, compared to $\mathrm{0.91\pm0.06\,M_\odot}$ from \cite{Akana2023}.

\begin{table}[htb]
\begin{center} 
\caption{TOI-266 semi-amplitudes and masses, from \cite{Akana2023} and this work.} 
\label{tab:TOI-266-comp} 
\centering 
\begin{tabular}{lll}
\hline \hline
Parameter & \cite{Akana2023} & this work \\ 
\hline
$\mathrm{K_b}$ \dotfill [$\mathrm{m \, s^{-1}}$] & $2.42\pm0.53$ & \toitwosixsixKpone \\
$\mathrm{M_b}$ \dotfill [$\mathrm{M_\oplus}$] & $7.8\pm1.8$ & \toitwosixtysixmassb \\
$\mathrm{K_c}$  \dotfill [$\mathrm{m \, s^{-1}}$] & $2.40\pm0.55$ & \toitwosixsixKptwo \\
$\mathrm{M_c}$ \dotfill [$\mathrm{M_\oplus}$] & $9.40\pm2.20$ & \toitwosixtysixmassc \\
\hline
\end{tabular}
\end{center}
\end{table}

\subsubsection{K2-62}

This system has two validated transiting planets, published in \cite{Mayo2018}. We tested the $in$, $out$, $2c$, $2e$, and corresponding $\_gp$ models on the ESPRESSO RVs and K2 photometry. The RVs show a strong long-term trend, which we include in all models as an additional linear term given by the $\mathrm{rv_{intercept}}$ and $\mathrm{rv_{slope}}$ parameters. The two-planet models are significantly favoured over both one-planet models, with $\mathrm{\Delta \log Z_{2c-in} \approx 167}$ and $\mathrm{\Delta \log Z_{2c-out} \approx 404}$. Without detrending, the $2e$ model is marginally but not significantly favoured ($\mathrm{\Delta \log Z_{2e-2c} \approx 0.4}$). With GP detrending, the best model is $2c\_gp$, but it is not favoured over the no-detrending model, with $\mathrm{\Delta \log Z_{2c-2c\_gp} \approx 6}$. We therefore adopt the $2c$ model as our final model. The priors and posteriors are given in Table \ref{tab:K2-62}, and the RVs and phase-folded photometry shown in Figs. \ref{fig:K2-62_RVs} and \ref{fig:K2-62_K2_stacked} respectively. K2-62 b has a $\mathrm{\ktwosixtytwoPpone\,d}$ period, a $\mathrm{\ktwosixtytworadb\,R_\oplus}$ radius, and a  $\mathrm{3.98 \pm 1.55\,M_\oplus}$ best-fit mass, while K2-62 c has a $\mathrm{\ktwosixtytwoPptwo\,d}$ period, a $\mathrm{\ktwosixtytworadc\,R_\oplus}$ radius, and a $\mathrm{6.6 \pm 2.8\,M_\oplus}$ best-fit mass. Since we do not reach a $3\sigma$ measurement of the semi-amplitude for either planets, we also report upper limits for the planetary masses of $\mathrm{\ktwosixtytwomassb \, M_\oplus}$ and $\mathrm{\ktwosixtytwomassc \, M_\oplus}$ respectively.

\subsubsection{BD+20954}

This system has one confirmed transiting planet \citep[][as K2-56 b]{Espinoza2016}, with a mass measurement using 23 HARPS \citep{Mayor2003} RVs. We tested $1c$ and $1e$ models for both the ESPRESSO RVs and K2 photometry alone, and including the HARPS RVs. We removed three ESPRESSO points with large error bars of $>2\,ms^{-1}$. In both cases, the eccentric models are marginally but not significantly favoured, with $\mathrm{\Delta \log Z_{1e-1c} \approx 0.8}$ for the fit with only ESPRESSO RVs, and $\mathrm{\Delta \log Z_{1e-1c} \approx 0.6}$ when the HARPS RVs are included. We also tested the $1c\_gp$ and $1e\_gp$ models on the ESPRESSO RVs and K2 photometry, but the inclusion of the GP detrending was not favoured, with $\mathrm{\Delta \log Z_{1c-1c\_gp} \approx 0.4}$, $\mathrm{\Delta \log Z_{1e-1e\_gp} \approx 1}$. 

We adopted as our final model the $1c$ model from the fit to the K2 photometry and joint ESPRESSO and HARPS RVs. The priors and posteriors are reported in Table \ref{tab:BD+20954}, and the RVs and phase-folded photometry shown in Figs. \ref{fig:BD+20594_RV} and \ref{fig:BD+20594_K2_stacked} respectively. BD+20954 b has a $\mathrm{\bdPpone\,d}$ period and a $\mathrm{\bdradb}$ radius. We find a notably smaller semi-amplitude than was previously reported, with $\mathrm{K = \bdKpone \,ms^{-1}}$, compared to the $\mathrm{3.1\pm 1.1\,ms^{-1}}$ found by \cite{Espinoza2016}, leading to a $\mathrm{5.2 \pm 2.2\,M_\oplus}$ mass, much lower than the previous $\mathrm{16.3^{+6.0}_{-6.1}\,M_\oplus}$ measurement. As we do not reach a $3\sigma$ significance on the semi-amplitude, we also give an upper mass limit of $\mathrm{M_b\bdmassb \, M_\oplus}$.

\subsubsection{TOI-186}
This system has two transiting planets, for which \cite{Dragomir2019} obtained a mass measurement for the outer planet (we stress that this is planet b, having been previously validated in \citealt{Trifonov2019}), and an upper mass limit for the inner (planet c, being detected later) with 55 HARPS and 225 PFS RVs \citep{Crane2010}. We first tested the $in$, $out$, $2c$, and $2e$ on the ESPRESSO RVs and TESS photometry. Given the quantity of TESS sectors observed for this target, fitting their instrumental parameters and GPs together with the planet(s) becomes computationally unfeasible. Therefore, we adopted a variant of our two-step process, in which the values obtained for the fit to the out-of-transit data are fixed in the full fit, which is performed on the in-transit data only. The two-planet models are significantly favoured over both one-planet models, with $\mathrm{\Delta \log Z_{2c-in} \approx 1011}$ and $\mathrm{\Delta \log Z_{2c-out} \approx 71}$. The $2c$ model is favoured over the $2e$ model, with , with $\mathrm{\Delta \log Z_{2c-2e} \approx 1}$. We do not apply GP detrending on the RVs in this case, as the ESPRESSO RVs are few.

We then tested the same four models on the TESS photometry and joint RV dataset including the new ESPRESSO RVs and the published HARPS and PFS RVs. We likewise do not apply GP detrending, as the joint dataset has only one common activity indicator, $\mathrm{H_\alpha}$, which shows systematic offsets and scaling differences between the three instruments. In this case, the $2e$ model was favoured, with $\mathrm{\Delta \log Z_{2e-2c} \approx 11}$. We adopt this model as our final model; the priors and posteriors are given in Table \ref{tab:TOI-186}, and the RVs and phase-folded photometry shown in Figs. \ref{fig:TOI-186_RVs} and \ref{fig:TOI-186_TESS_stacked} respectively. TOI-186 b has a $\mathrm{\toioneeightysixPptwo\,d}$ period, $\mathrm{\toioneeightysixradc\,R_\oplus}$ radius, and $\mathrm{\toioneeightysixeccptwo}$ eccentricity, while TOI-186 c has a $\mathrm{\toioneeightysixPpone\,d}$ period, $\mathrm{\toioneeightysixradb\,R_\oplus}$ radius, and $\mathrm{\toioneeightysixeccpone}$ eccentricity. Our mass values are in agreement with the previous results; we improve the mass error bars on the outer planet TOI-186 b slightly, going from a literature value of $\mathrm{M_b = 22.7^{+2.2}_{-1.9}\, M_\oplus}$ to $\mathrm{M_b = \toioneeightysixmassc\, M_\oplus}$, but similarly to \cite{Dragomir2019} do not reach a $3\sigma$ measurement of the mass of TOI-186 c (best-fit value: $\mathrm{1.43 \pm 0.80\,M_\oplus}$), and so give an upper mass limit of $\mathrm{M_c\toioneeightysixmassb\, M_\oplus}$.

\section{Population analysis}\label{s:overview}

In this section, we analyse the entire population of planets discovered by the WG3 sub-program as presented in Section \ref{s:wg3-presentation}, including both published systems, and systems with dedicated publications in preparation. We also include the six systems presented in Section \ref{s:analysis}. We highlight that of the 65 planets listed in Table \ref{tab:WG3-planets}, seven have only upper mass limits, and are therefore excluded from any analysis involving the planetary masses. Likewise, eight are non-transiting companions to transiting planets, and are thus excluded from any analysis involving the planetary radii. This leaves a final total of 51 transiting planets with well-determined masses and radii. 

In order to place our systems in a global population context among other well-characterized small planets, we use the PlanetS catalogue \citep{Otegi2020, Parc2024} to select all published planets with precisely measured masses and radii (mass measured to better than 25\%, radius to better than 8\%, as per the catalogue criteria), and set upper limits of $\mathrm{M_p \leq 40 M_\oplus}$ and $\mathrm{R_p \leq 4 R_\oplus}$, recovering a total of 182 planets. We then remove 23 WG3-detected planets that fall within the PlanetS criteria from the catalogue sample to avoid duplication. These planets correspond to $\mathrm{\approx13\%}$ of the PlanetS small planet sample. We also note that 75\% of the WG3 transiting planets fulfil the PlanetS catalogue criteria on mass and radius precision. 

\subsection{Planetary composition and insolation}\label{s:composition}

Thanks to the precisely determined masses and radii derived from the ESPRESSO RVs, complementary RVs, and transit photometry, we can analyse the composition of the WG3 population. These planets span a large range of compositions, from Mercury-like through rocky to likely possessed of extended atmospheres. One of the goals of the WG3 was to examine specifically the $\mathrm{50-200\,S_\oplus}$ insolation range, in order to study the transition from rocky super-Earths to volatile-rich sub-Neptunes and search for mass thresholds at which significant evaporation of the planetary envelopes occurs. This insolation regime corresponds to that at which the radius valley was initially detected \citep{Fulton2017}; we explore the radius valley further in Section \ref{s:rad-valley}.

\begin{figure}[htb!]
    \centering
    \includegraphics[width=\linewidth]{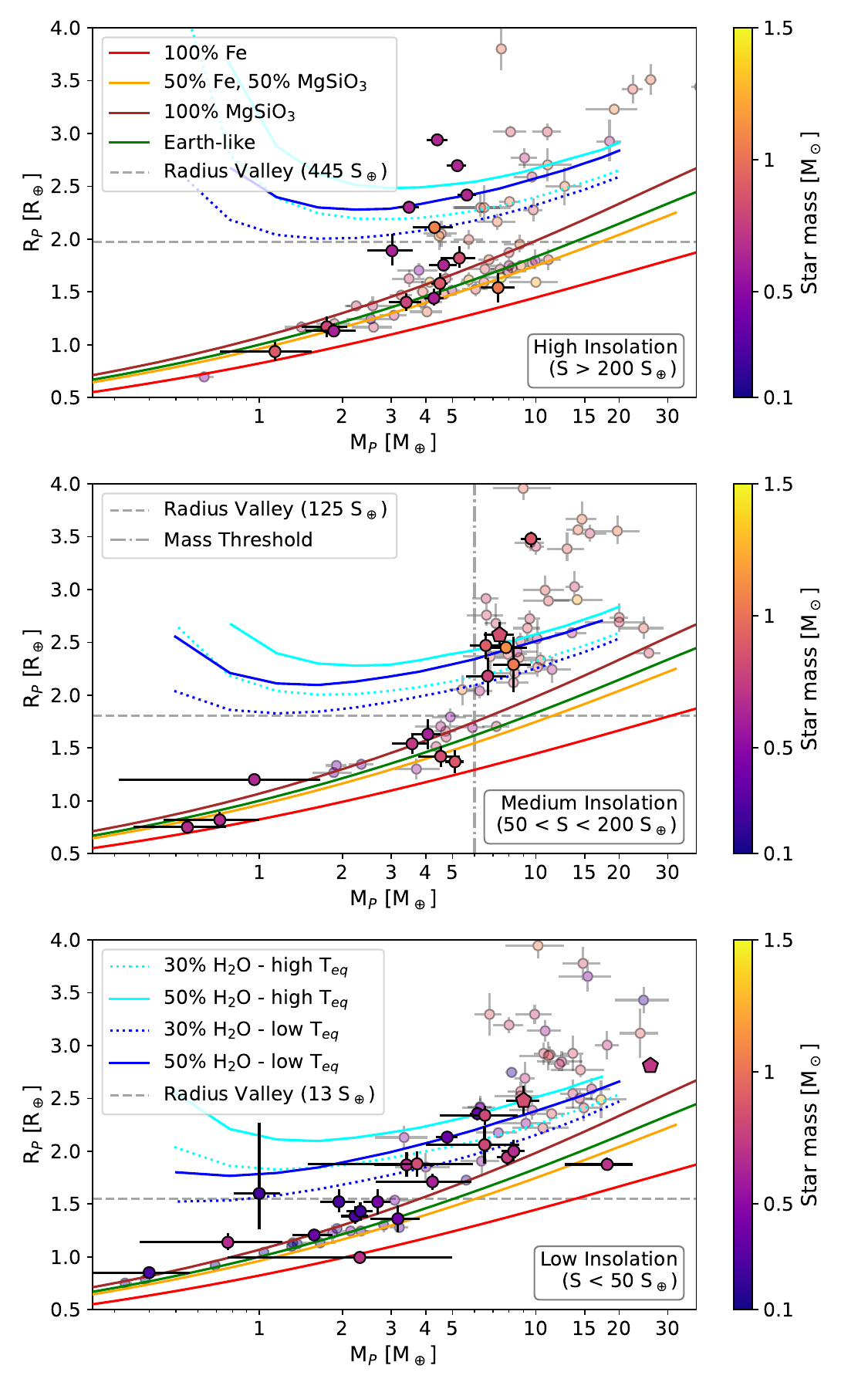}
    \caption{Mass-radius diagrams across three insolation regimes for the WG3 planets (solid points) and small planets from the PlanetS catalogue (semi-transparent points). Planets presented in this paper are shown as pentagons. All the planets are coloured by host star mass. The coloured curves show the composition models of \cite{Zeng2019} (rocky planets) and \cite{Luo2024} (volatile-rich planets). For the latter, we show the models for the temperatures above and below the mean $\mathrm{T_{eq}}$ for each panel, labelled as high and low $\mathrm{T_{eq}}$; these correspond to $\mathrm{1400\,K}$ and $\mathrm{1000\,K}$ for the top panel, $\mathrm{1000\,K}$ and $\mathrm{700\,K}$ for the middle panel, and $\mathrm{700\,K}$ and $\mathrm{400\,K}$ for the bottom panel. Dashed horizontal grey lines indicate the location of the radius valley at the median insolation for each population, as defined by \cite{Ho2023}. The dash-dotted vertical grey line in the central plot shows a tentative threshold mass between rocky and volatile-rich planets in the medium insolation regime.}
    \label{fig:mass-radius}
\end{figure}

We plot the planets on mass-radius diagrams, together with the composition curves of \cite{Zeng2019} (for rocky planets) and \cite{Luo2024} (for volatile-rich planets), in Fig. \ref{fig:mass-radius} for three insolation regimes: $\mathrm{S_p < 50\,S_\oplus}$, $\mathrm{S_p = 50-200\,S_\oplus}$, and $\mathrm{S_p> 200\,S_\oplus}$. We also show the planets from the PlanetS catalogue. As can be seen from the subplots in Fig. \ref{fig:mass-radius}, while in the high-insolation and low-insolation regimes there is no clear mass threshold for the rocky to volatile-rich transition, there appears to be one at $\mathrm{M_p\approx6\,M_\oplus}$ in the medium-insolation regime, with planets below this mass generally falling on rocky composition curves, and planets above this mass being volatile-rich. We caution that this observation is based on a small number of planets; a larger sample would be needed for a more robust detection. While all our sub-samples are inherently biased, as they consist of transiting planets, the upper mass threshold for rocky planets in the medium insolation regime should be robust, as for the same radius and period (and thus same transit probability, and same insolation for the same host star) a more massive planet is easier to detect than a low-mass one.

The lack of a clear threshold mass in the low-insolation regime, where planets of both composition types can be found essentially across the entire planetary mass range, is likely due to more of these planets being hosted by low-mass stars: 51\% are hosted by stars with $\mathrm{M_\star\leq 0.6\,M_\odot}$, compared to 6\% in the high-insolation regime and 9\% in the mid-insolation regime. It has been shown \citep{Venturini2024, Parc2024} that the radius valley (generally considered a gap between rocky super-Earths and volatile-rich sub-Neptunes) fades for M dwarfs, so this population is difficult to compare directly to that hosted by FGK stars. Alternatively, it could be related to the insolation itself; there is an emerging population of volatile-rich low-density super-Earths, that are more frequently detected in low-irradiation environments \citep{Castro2023}. We also note that many of the WG3 planets in this regime have comparatively large uncertainties on their masses, due to their longer orbital periods and later-type (and thus fainter) host stars. 

For the high-insolation regime, there is likely a combination of host-star and irradiation effects at play. On the one hand, most of the low-mass volatile-rich planets in this regime are hosted by low-mass stars - e.g., the four WG3 planets at or above the 50\% $\mathrm{H_2O}$ line; these are TOI-178 d through g, orbiting a late K dwarf \citep{Leleu2021, Leleu2024}. On the other, in this high-insolation regime we can expect both stripping of primordial atmospheres from more massive cores than in the medium-insolation case, leading to more massive planets with rocky compositions, and puffing up of volatile atmospheres, leading to larger radii for similar masses. A particularly notable feature in this regime is the massive rocky planet population, reaching to $\mathrm{\approx10\,M_\oplus}$. We do not find a similar population in the lower-insolation regimes, suggesting that rather than having a direct formation pathway, these planets are stripped cores: they accrete a significant envelope during formation, which is retained in the lower-insolation environments, but can be lost due to photo-evaporation in high-irradiation environments. 

For many of the planets presented in this paper specifically, we do not achieve $3\sigma$ measurements of the semi-amplitudes and thus reported in Section \ref{s:analysis} the standard $3\sigma$ upper mass limits. Once we take into account the radii measured from the photometry, however, we find that in the case of K2-184 b and TOI-186 c, these mass upper limits lead to unphysically dense planets with densities higher than pure iron. To obtain more physically motivated upper mass limits for these planets, we calculate the mass value of the \cite{Zeng2019} pure-iron model for the $3\sigma$ upper limit on their radii. Table \ref{tab:upper-mass-limits} lists these upper mass limits, together with the initial upper mass limits derived from the \texttt{juliet} fits for comparison.

\begin{table}[htb]
\begin{center} 
\caption{Physically motivated upper mass limits.} 
\label{tab:upper-mass-limits} 
\centering 
\begin{tabular}{llll}
\hline \hline
Planet & best-fit mass & $3\sigma$ limit & pure iron limit \\ 
\hline
K2-184 b & $\mathrm{10.87 \pm 4.68\,M_\oplus}$ & $\mathrm{\ktwooneeightyfourmassb\,M_\oplus}$ & $\mathrm{<8.59\,M_\oplus}$ \\
TOI-186 c & $\mathrm{1.43 \pm 0.80\,M_\oplus}$ & $\mathrm{\toioneeightysixmassb\,M_\oplus}$ & $\mathrm{<3.04\,M_\oplus}$ \\
\hline
\end{tabular}
\end{center}
\end{table}

Another way to study the planetary compositions is through interior structure modelling. We use the ExoMDN code \citep{Baumeister2023}\footnote{Available at \url{https://github.com/philippbaumeister/ExoMDN}} to carry out rapid uniform interior structure modelling for the full WG3 sample. ExoMDN employs mixture density networks to perform interior inference modelling, fitting a four-layer model consisting of an iron core, a silicate mantle, a water layer, and a H/He atmosphere from the planet's mass, radius, and equilibrium temperature. The outputs are mass fractions $w_{layer}$, radius fractions $d_{layer}$, and posterior distributions of these values, for each of the four layers. Figures \ref{fig:ExoMDN_mass} and \ref{fig:ExoMDN_radius} show the output mass and radius fraction posterior distribution histograms for the WG3 sample, ordered by core mass fraction and core radius fraction respectively. We also show the results for the Earth and Neptune as comparison points. Our outputs range from dense, core-rich planets, through Earth-like compositions, intermediate cases with large water fractions, to Neptune-like gas-rich planets.

To compare the mass and radius layer fractions to the results of the same model for Earth-like and Neptune-like planets, as presented in \cite{Baumeister2023}\footnote{Earth: $w_{core}\approx0.74$, $w_{mantle}\approx0.19$, $w_{water}\approx0.06$, $w_{gas}\approx0$; $d_{core}\approx0.69$, $d_{mantle}\approx0.14$, $d_{water}\approx0.09$, $d_{gas}\approx0.05$. Neptune: $w_{core}\approx0.18$, $w_{mantle}\approx0.21$, $w_{water}\approx0.26$, $w_{gas}\approx0.27$; $d_{core}\approx0.22$, $d_{mantle}\approx0.14$, $d_{water}\approx0.16$, $d_{gas}\approx0.46$}, we compute the four-dimensional Euclidean distances between each planet's mass (radius) layer fractions and the mass (radius) layer fractions of Earth and Neptune respectively.
Around 56\% of the sample (28 planets) are closer to Earth-like than Neptune-like values in both mass and radius, while 30\% (15 planets) are closer to Neptune-like values. The remaining 14\% (7 planets) are an intermediate case, closer to the Neptune model in mass fractions but to the Earth model in radius fractions. All these planets have relatively low core fractions ($w_{core}\approx0.21-0.40$, $d_{core}\approx0.33-0.42$), high mantle ($w_{mantle}\approx0.31-0.37$, $d_{mantle}\approx0.21-0.29$) and water ($w_{water}\approx0.21-0.40$, $d_{water}\approx0.21-0.33$) fractions, and low gas fractions ($w_{gas}\approx0$, $d_{core}\approx0.03-0.14$). The high water fractions suggest these planets may be water worlds. Inspecting the position of the planets in each of the three groups compared to composition tracks on the mass-radius diagram, we find that those closer to Earth-like values correspond well to rocky models, while the intermediate and Neptune-like planets fall on volatile-rich tracks. 

\subsection{The radius valley}\label{s:rad-valley}

\begin{figure}
    \centering
    \includegraphics[width=\linewidth]{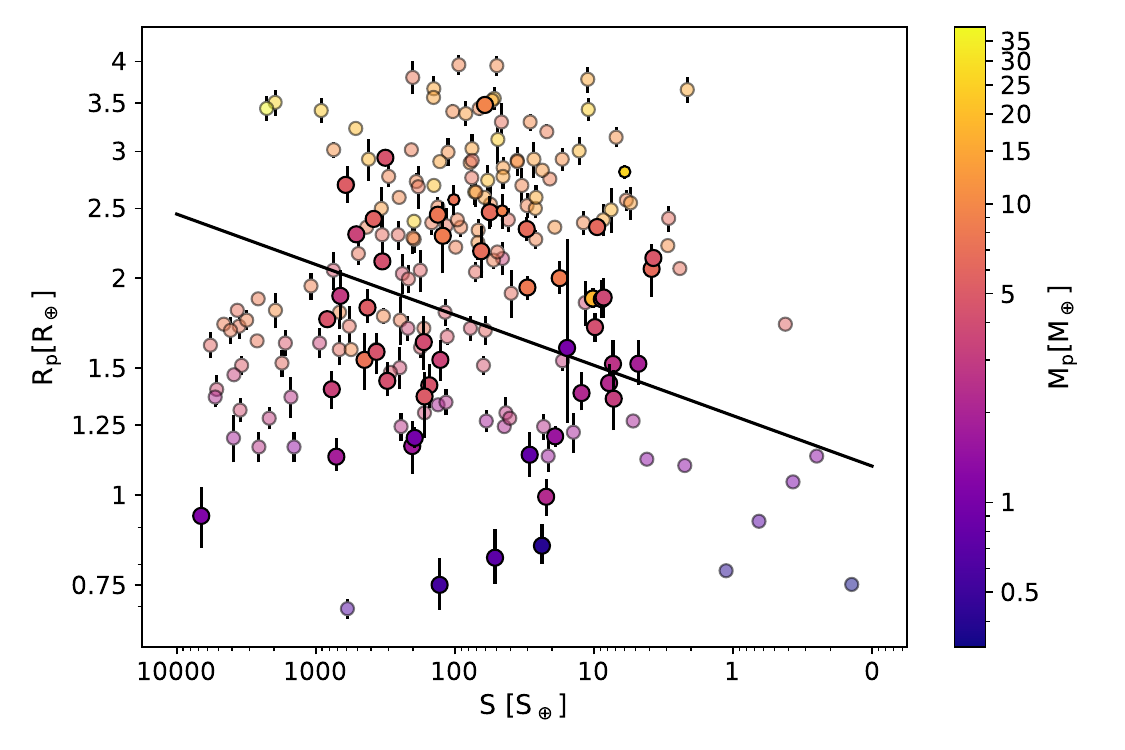}
    \caption{Insolation-radius diagram for the WG3 planets (solid circles) and PlanetS catalogue small planets (semi-transparent circles). The planets are coloured by planet mass. The black line shows the insolation-dependent radius valley from \cite{Ho2023}.}
    \label{fig:isolation-radius}
\end{figure}

One of the most remarkable features of the small planet population is the radius valley or radius gap, an underdensity of planets in the $\mathrm{1.5 - 2\, R_\oplus}$ region, first identified by \cite{Fulton2017}. Many mechanisms have been invoked to explain the valley, from core-powered or XUV-driven mass loss \citep{Petigura2022, Ho2023}, to in-situ rocky planet versus ex-situ water-rich planet formation \citep{Venturini2020, Burn2024}. The location and depth of the valley also depends on parameters such as stellar host mass \citep{Petigura2022, Luque2022, Venturini2024, Parc2024} and planet insolation \citep{Ho2023}. We show the planetary radius distribution as a function of insolation for the WG3 and PlanetS planets in Fig. \ref{fig:isolation-radius}, overplotting the insolation-dependent gap derived by \cite{Ho2023}. We do not attempt to re-estimate the location of the valley because our sample has non-uniformly-derived radii and insolation, and it has been shown that uniformly deriving the radii increases the depth of the valley and improves its localization \citep{VanEylen2018}. The separation between the two populations is clearest at mid-to-high insolations, while at low insolations several planets overlap the gap. This is also consistent with the mass-radius diagrams for different insolation regimes shown in Fig. \ref{fig:mass-radius}, where for high and medium insolations the radius valley generally separates the rocky and volatile-rich populations, while for low insolation the separation is less clear.

The planets below the radius gap appear to be less massive and span a narrower mass range than those above the gap, a feature of this population previously noted in \cite{Armstrong2025}. To better examine this mass bimodality, we compute the kernel density estimations (KDE) for the masses of the planets below and above the gap. Since the planet masses have associated uncertainties, in order to take these into account we adopt a similar methodology to \cite{Parc2024}. We create 1000 samples where we take a random value from a Gaussian with parameters $\mu = M_p,\, \sigma = \sigma_{M_p}$ for each planet, and recompute the KDE for each sample. We then take the median and standard deviation of these KDEs as our final distributions and their associated errors. 

\begin{figure}
    \centering
    \includegraphics[width=\linewidth]{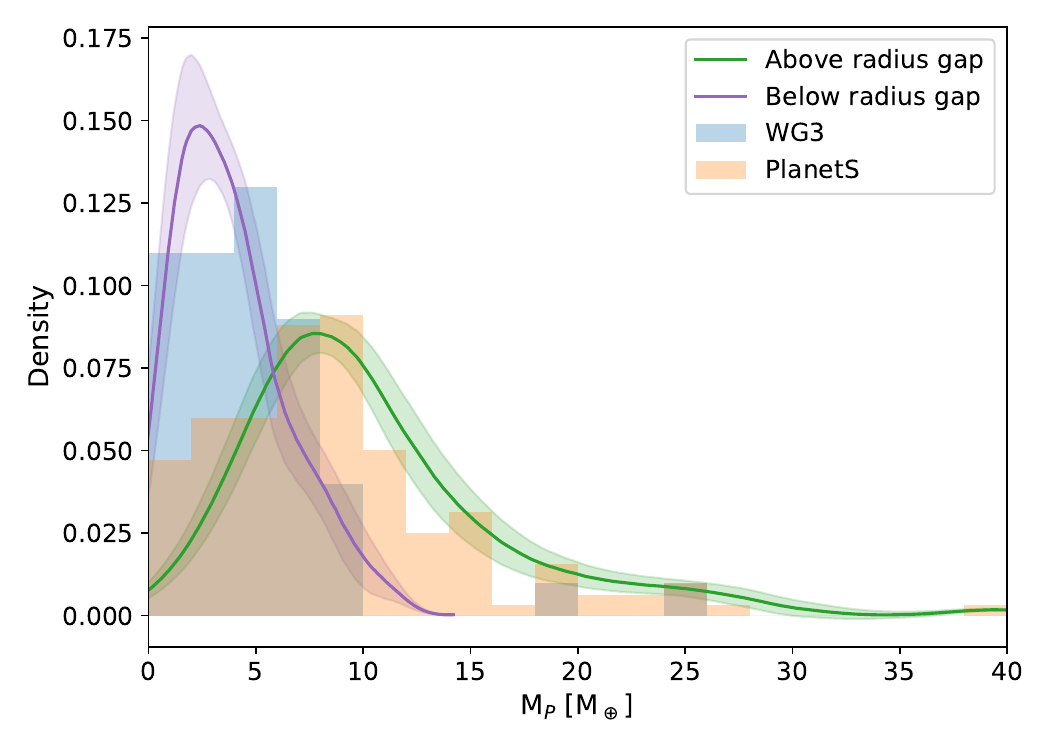}
    \caption{Kernel density estimations of the mass distributions for the planets above (green) and below (purple) the radius gap. The shaded purple and green regions show the $1\sigma$ region. The blue and orange histograms show the mass distributions for the WG3 and PlanetS samples respectively.}
    \label{fig:kde-mass}
\end{figure}

Figure \ref{fig:kde-mass} shows these final distributions for the masses of the planets below and above the gap, together with the mass histograms for the WG3 and PlanetS samples. It can be seen that the planets below the gap span a narrow $\mathrm{\approx0.3-10\, M_\oplus}$ mass range. In agreement with \cite{Armstrong2025}, we also find a sharp mass cut-off at $\mathrm{\approx10\,M_\oplus}$ for the planets below the gap; our overall distribution for these planets, however, appears more sharply peaked, and peaks at a lower value of $\mathrm{2.5\,M_\oplus}$ compared to their broad peak at $\mathrm{4\,M_\oplus}$. For the planets above the gap, meanwhile, we find a similar broad distribution to \cite{Armstrong2025}, with a $\mathrm{\approx1-40\, M_\oplus}$ mass range, and peaking at $\mathrm{7\,M_\oplus}$. We note that the PlanetS sample has more restrictive criteria on the errors in mass and radius than that applied by \cite{Armstrong2025}, who took all planets in the Exoplanet Archive with mass and radius determinations better than $3\sigma$; conversely, our sample extends out to planets with masses up to $\mathrm{40\,M_\oplus}$, while they cut off at $\mathrm{20\,M_\oplus}$. The two KDEs intersect at $\mathrm{\approx6\,M_\oplus}$, in line with our tentative rocky-volatile transition threshold presented in Section \ref{s:composition}. It is also clear from Fig. \ref{fig:kde-mass} that ESPRESSO is a major contributor to the precise mass measurements of the below-the-gap planets, which is only to be expected as their small masses require ESPRESSO's exquisite RV precision. 

\subsection{Exploring the mass-metallicity correlation}

The correlation between stellar metallicity and planet occurrence has been greatly explored in the literature. It is generally accepted that there is a strong giant planet occurrence - metallicity correlation \citep[][and references therein]{Gonzalez1997, Santos2001, Udry2007, Sousa2021}, and a weaker superEarth occurrence - metallicity correlation \citep[][and references therein]{Zhu2021, Sousa2021}. This correlation is due to the stellar metallicity acting as a proxy for the amount of solids in the protoplanetary disk; to form large planets, more solids - and thus a higher stellar metallicity - are required \citep[e.g.][]{Ida2004, Chen2025}. The question then arises of whether planetary mass also correlates with metallicity, as more massive planets would require more solids available for their formation. 

\begin{figure}[htb!]
    \centering
    \includegraphics[width=\linewidth]{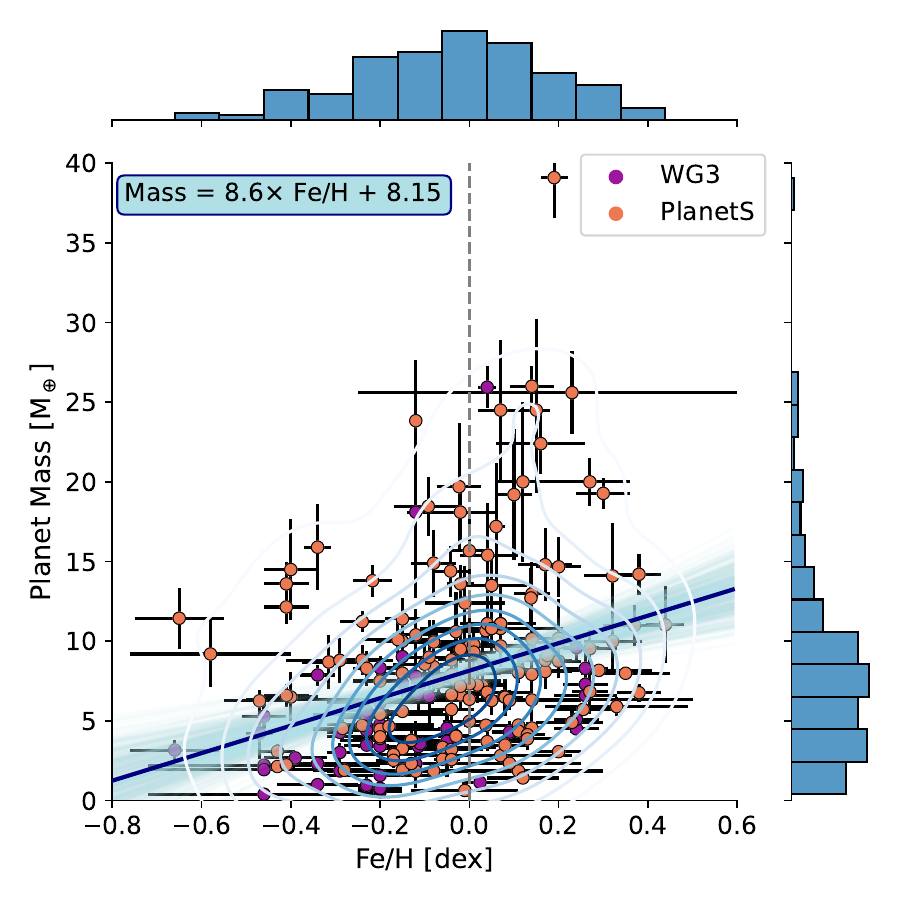}
    \caption{Stellar metallicity versus planet mass for the WG3 planets (purple points) and small planets from the PlanetS catalogue (orange points), with contour lines in blue. Marginalized histograms of the metallicity and mass are shown on the top and right respectively. The translucent light blue lines show random draws from the \texttt{linmmix} MCMC sampling, and the solid navy line shows the median model, whose values are annotated in the blue box. The vertical dashed grey line shows the solar metallicity.}
    \label{fig:mass-metallicity}
\end{figure}

In Fig. \ref{fig:mass-metallicity} we show the planet mass - stellar metallicity distribution for the combined WG3 and PlanetS sample. The overall metallicity distribution is centred on solar values, with a mean of $\mathrm{\mu_{Fe/H}} = -0.05$ and a standard deviation of $\mathrm{\sigma_{Fe/H}} = 0.22$. The shape of this distribution is similar to that predicted by core-accretion models for small planets \citep{Chen2025}, providing empirical support for a core-accretion formation. We note a scarcity of planets at very low metallicities ($\mathrm{Fe/H\lesssim-0.5}$), consistent with the observational metallicity cut-off seen both for giant planets \citep{Mortier2012}, and more recently in the TESS small planet sample \citep{Boley2024}. The maximum mass likewise increases with metallicity, an effect first noted by \cite{Courcol2016}; ten years later, our sample remains consistent with their upper boundary. The observed small planet occurrence likewise drops off at higher metallicities; this is likely due to the higher probability of forming large planets around metal-rich stars, which inhibit the formation of small planets or dynamically eject them \citep[][and references therein]{Zhu2021}.

The contour plot in Fig. \ref{fig:mass-metallicity} highlights a tendency for the more massive planets to be hosted by the more metal-rich stars. A simple linear regression with the \texttt{SciPy} package \citep{Scipy} yields a correlation with high statistical significance, with $p-value = 1.01\times10^{-5}$. However, both the masses and metallicities have significant error bars. We therefore use the \texttt{linmix} package \citep{Kelly2007}\footnote{Ported to Python at \url{https://github.com/jmeyers314/linmix}} to perform a linear regression taking into account the errors, using a Gaussian mixture model and MCMC sampling. The resulting median fit, which is shown in Fig. \ref{fig:mass-metallicity}, is indistinguishable from that derived with \texttt{SciPy}. Taking the median and standard deviation of the MCMC samples as the values and errors of the coefficients, our fit is given by:
\begin{equation}
    \mathrm{Mass = (8.6\pm2.0)\times[Fe/H]+(8.15\pm0.38)}.
\end{equation}
This trend is consistent with our initial hypothesis, that more massive planets would preferentially be found around more metallic stars.

Both the PlanetS catalogue metallicities, and those for the WG3 sample, are taken from the planet discovery and characterization papers, and are thus inhomogeneous. We therefore query the SWEET-Cat catalogue of homogenous stellar parameters \citep{Santos2013, Sousa2021} to obtain uniformly derived metallicities for our sample. Only three stars (TOI-713, TOI-214, and TOI-283) are not found in SWEET-Cat. We also verify the source of the metallicities; 70\% of the WG3 sample and 63\% of the PlanetS sample have metallicities homogenously derived from archival spectra, while the rest are from the literature.
Repeating the linmix modelling with the SWEET-Cat metallicities, our new median fit is given by:
\begin{equation}
    \mathrm{Mass = (7.2\pm1.9)\times[Fe/H]+(7.91\pm0.40)}.
\end{equation}
This is indistinguishable from the previous fit, showing that our population-level results are robust to inhomogeneities in the individual metallicities. Likewise, we test whether the $\mathrm{\approx40\,M_\oplus}$ planet from the PlanetS sample (TOI-849 b, \citealt{Armstrong2020}) is driving the trend, finding that our fits are robust to its removal. 

A tentative three-dimensional correlation between mass, metallicity, and period was previously presented by \cite{Sousa2019, Sousa2021}. As our sample consists only of transiting planets, we are heavily biased towards short periods, so we do not perform an analogous three-dimensional fit. Comparing the multiplicative coefficients on the metallicity, $a \times \mathrm{[Fe/H]}$, our value of $a = 8.6\pm2.0$ is lower than the $a =13.4\pm0.4$ found by \cite{Sousa2021}; however, the overall conclusion that masses increase with higher metallicity remains.

\subsection{Comparing planet masses to typical protoplanetary disk masses}

Planets form in protoplanetary disks, and so their masses - and, for multiplanetary systems, the sum of the planets' masses - cannot exceed that of the disk they formed in. This seemingly obvious statement has in fact been the source of a large quantity of works, as comparing disk masses and planetary masses is a non-trivial endeavour: the exoplanet population is biased in complex ways, and determining disk masses from continuum observations requires several assumptions on factors such as temperature, dust-to-gas-ratio, and dust optical depth \citep[e.g.][]{Manara2023}. Recent studies suggest planets must either form early, in the more massive Class 0 and I disks \citep{Tychoniec2020}, or with formation efficiencies close to 100\% if they form in Class II disks \citep{Mulders2021}.

\begin{figure}[htb!]
    \centering
    \includegraphics[width=\linewidth]{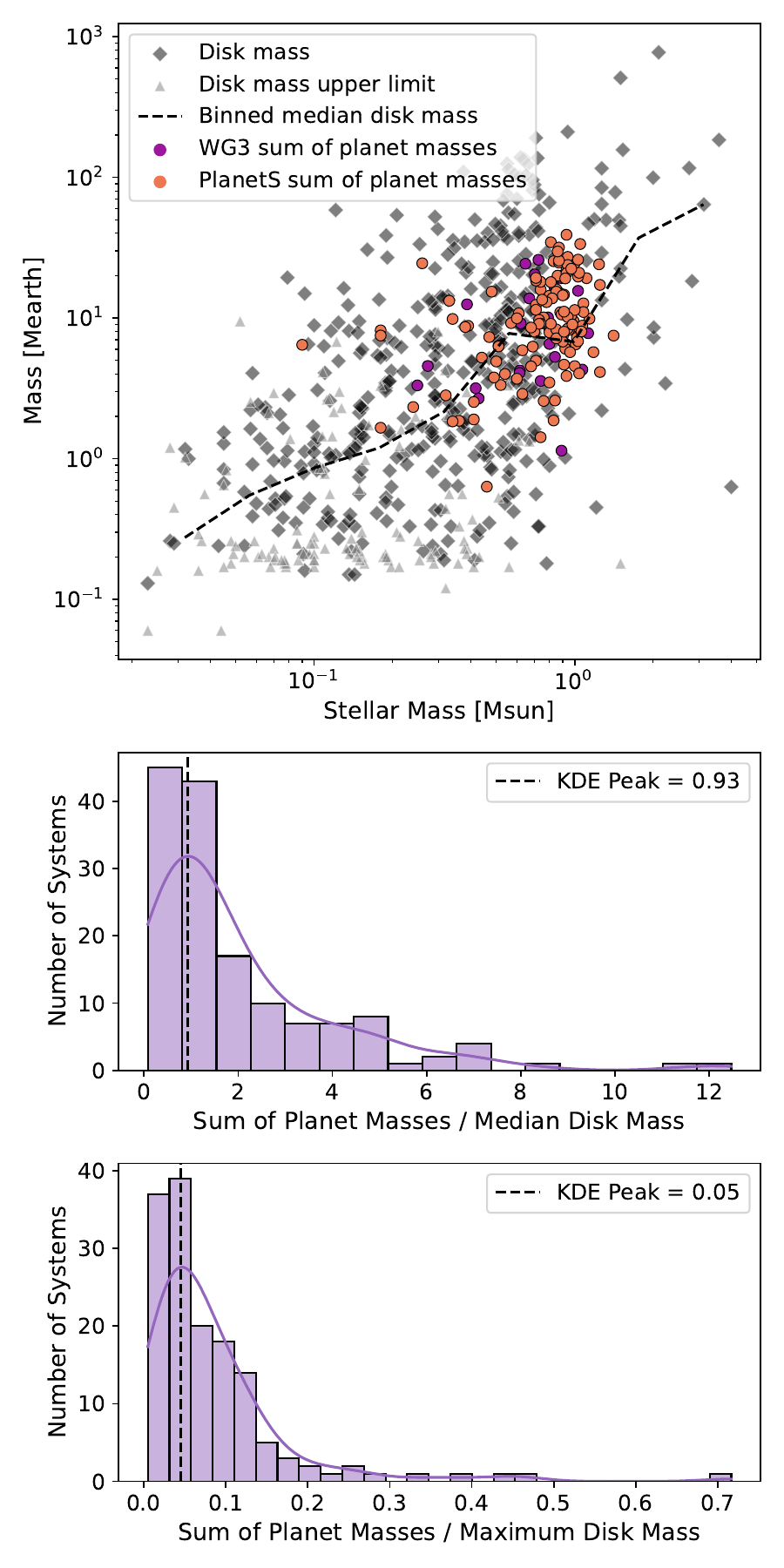}
    \caption{Planet and protoplanetary disk mass comparison. Top: Stellar mass versus sum of planet masses for the WG3 sample (purple points) and small planets from the PlanetS catalogue (orange points), and versus protoplanetary disk masses (grey squares) and disk mass upper limits (light grey triangles). The black dashed line shows the median disk mass over ten stellar mass bins. Middle: sum of planet masses to median disk mass ratio histogram and KDE, where the median disk mass is computed for each planet in a $0.1\,M_\odot$ bin around the mass of the planet's host star, including disk upper mass limits. Bottom: sum of planet masses to maximum disk mass ratio histogram and KDE, using the same binning.}
    \label{fig:disks-fig}
\end{figure}

To compare the masses of the rocky planets that constitute the combined WG3 and PlanetS sample presented in this paper to the likely mass of the disk in which they formed, we obtain a large sample of observationally determined Class II protoplanetary disk masses, together with the masses of their host stars, from \cite{Manara2023}. The top panel of Fig. \ref{fig:disks-fig} shows the sum of planet masses for each system, and the protoplanetary disk masses, as a function of stellar masses. The middle and bottom panels show the ratio between the sum of planet masses and the median and maximum disk masses respectively, with these computed in a $0.1\,M_\odot$ bin around the mass of the planet's host star. We include disk upper mass limits in this computation, as these correspond to disks that are not detected at mm wavelengths but whose existence is inferred from other observations such as IR excess. To exclude them would then bias our median masses towards higher masses. The summed planet masses generally represent only $\sim5\%$ of the maximum disk mass for their stellar type, but are overall very similar to the median disk masses. This suggests they must either form with a very high efficiency, form preferentially in more massive disks, or form earlier in the disk's lifetime. Likewise, the detected planets represent only a lower limit on the total planetary mass in the system; there could be additional planets on wide or misaligned orbits that are not detectable with the current data, increasing the total planet mass and further strengthening our argument.

It is also worth noting that the disks in the sample of \cite{Manara2023} come from low-mass and consequently not particularly irradiated star-forming regions, which are not the norm for Galactic star-formation environments \citep{Lada2003, Fatuzzo2008, Winter2022}. We generally do not know the formation environments of our present-day planet-hosting stars. If they formed in higher-mass regions, which are more densely populated and contain more OB stars, the strong UV radiation environment is expected to have caused strong, rapid photo-evaporation of the protoplanetary disks \citep{Storzer1999, Scally2001, Haworth2018, Haworth2023}, with the effect being particularly pronounced for low-mass disks \citep{Qiao2026}. This further supports early planet formation and/or preferential formation in more massive disks.

\subsection{Peas in a pod?}

Many works in recent years have searched for similarities between planets in multi-planet systems, a pattern that has become known as "peas in a pod" \citep[e.g. ][]{Millholland2017, Hobson2017, Weiss2018, Otegi2022}. The WG3 sample includes 21 multi-planet systems, although for some of these we have only mass upper limits, or only a lower limit of $\mathrm{m \sin{i}}$ for non-transiting planets in the system. Nevertheless, we use these systems, together with a further 114 multi-planet systems from the PlanetS catalogue, to explore potential similarities between their planets. We also note the important caveat that these analyses assume no intermediate planets have been missed.

\begin{figure}[htb!]
    \centering
    \includegraphics[width=\linewidth]{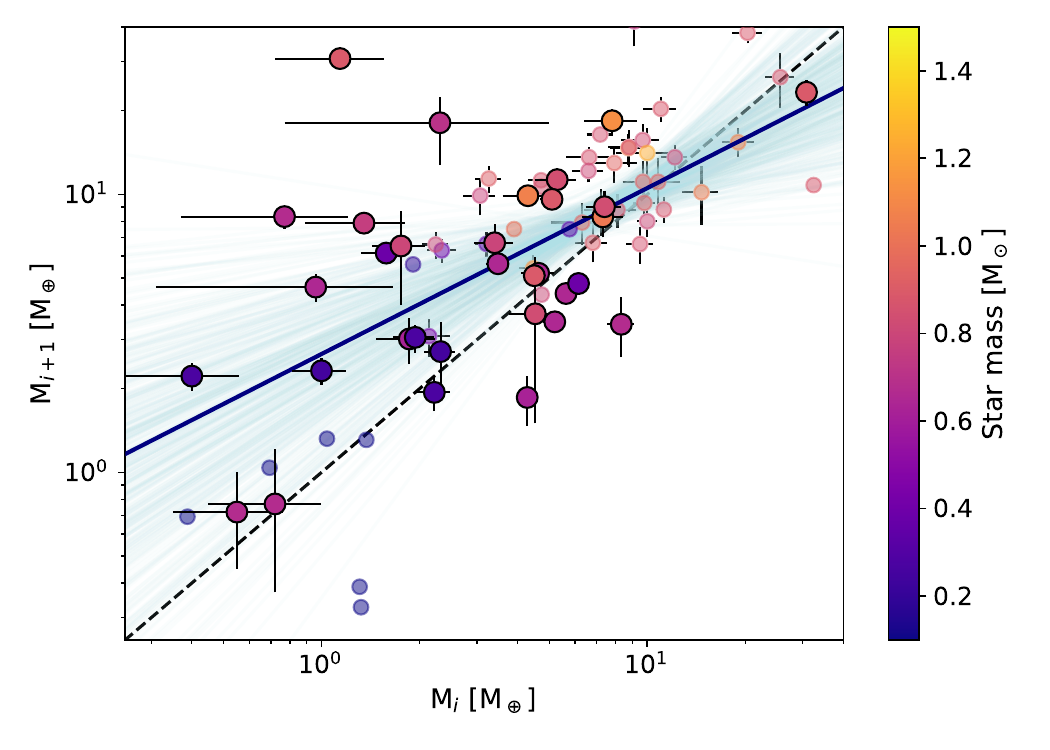}
    \includegraphics[width=\linewidth]{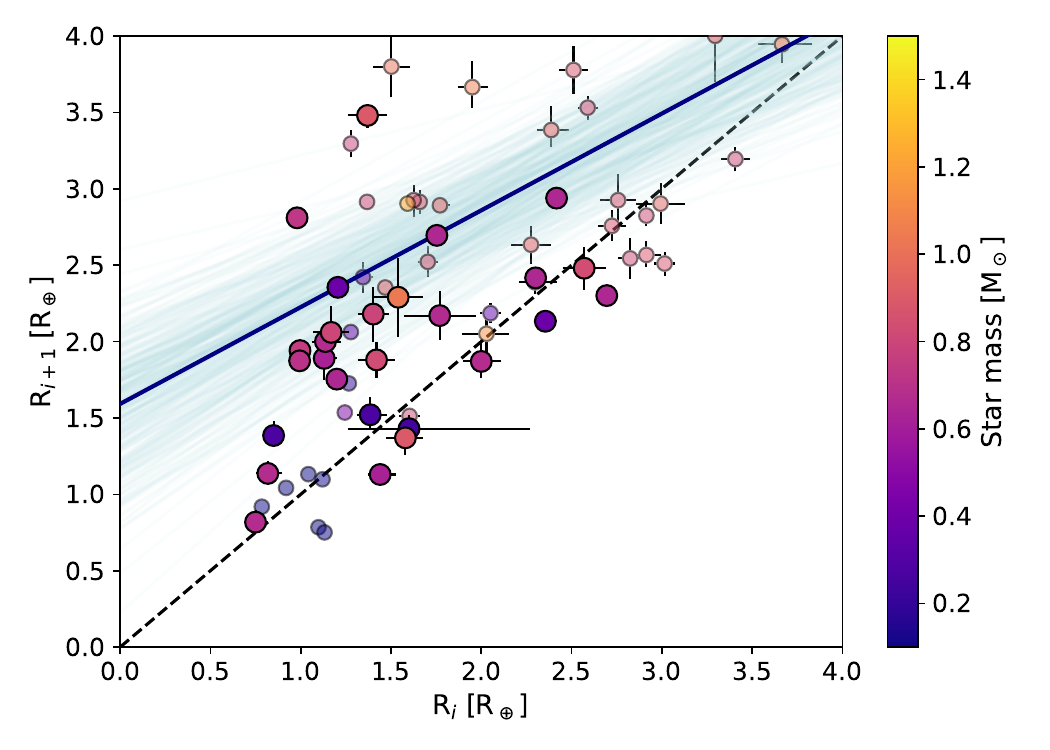}
    \caption{Mass and radius comparison between pairs of consecutive planets. Top [Bottom]: Masses [radii] of pairs of consecutive planets within a multi-planet system, for the WG3 sample (solid circles) and the PlanetS catalogue (semi-transparent circles), coloured by stellar host mass. The dashed grey line shows the 1:1 relationship. The translucent light blue lines show random draws from the \texttt{linmmix} MCMC sampling, and the solid navy line shows the median model.}
    \label{fig:mass-i-i+1}
\end{figure}

Figure \ref{fig:mass-i-i+1} shows masses and radii for pairs of adjacent planets within the same system. We find that both for the WG3 sample and the PlanetS catalogue sample, the outer planets of adjacent pairs tend to be larger and more massive. For the mass, this may be a consequence of the general mass-period-metallicity relationship found by \cite{Sousa2019, Sousa2021}; alternatively, the period-mass part of that correlation may arise from the inclusion of systems with multiple planets. We use the \texttt{linmix} package to perform linear regression taking into account the errors, using a Gaussian mixture model and MCMC sampling. Both masses and radii may grow more similar for larger/more massive inner planets, though the dispersion in the mass fit is large.

\begin{figure*}[htb!]
    \centering
    \includegraphics[width=0.4\textwidth]{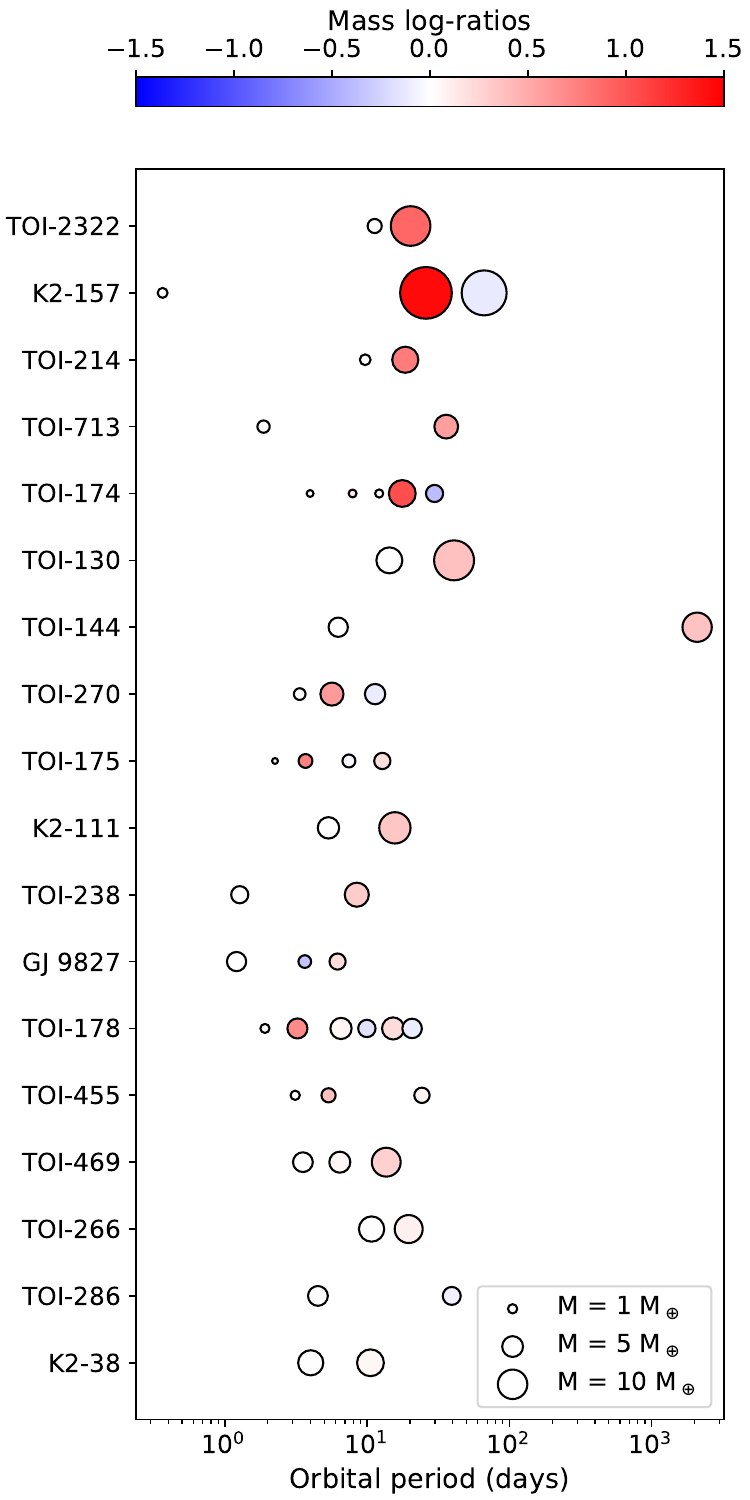}
    \includegraphics[width=0.4\textwidth]{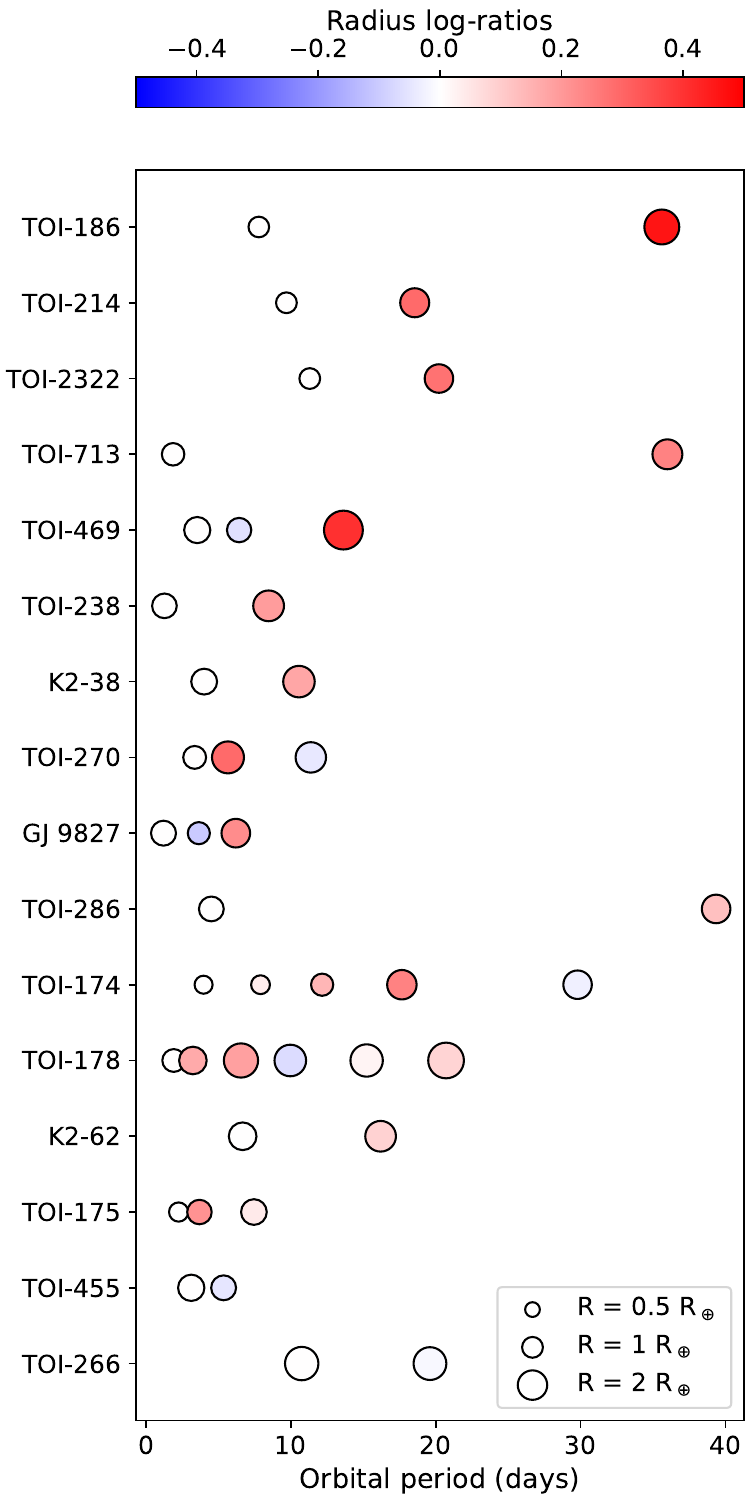}
    \caption{Multiplanetary systems from the WG3 sample. Left (respectively right): planets scaled by mass (radius) and coloured by mass (radius) log-ratio to the adjacent inner planet, ordered by decreasing mass (radius) distance metric as defined in \cite{Otegi2022}. Note that some systems appear in only one plot, due to lack of masses or radii, and that the colour scales differ as the planets are generally more similar in radius than mass.}
    \label{fig:rad-mass-distance}
\end{figure*}

We also find that within each system, planets tend to vary more in mass than in radius, as can be seen in Fig. \ref{fig:rad-mass-distance}, which plots the WG3 systems with planets coloured by mass or radius log-ratio. This is to be expected if the planets have similar densities, as $\mathrm{\rho\sim \frac{M}{R^3}}$, and is further supported by a calculation of the mass and radius distance metrics, $\mathcal{D}_M = \sum_{i=1}^{N_{pl}-1} |\log \frac{M_{i+1}}{M_i}|/N_{pl}-1$ and the analogous $\mathcal{D}_R$, as defined by \cite{Otegi2022}. Of the 13 WG3 systems in which we have both mass and radius for at least two planets and so can compute both metrics, 10 (77\%) have larger $\mathcal{D}_M$ than $\mathcal{D}_R$. This agrees with the results of \cite{Otegi2022} on a larger sample. Likewise, we find a mean $\mathcal{D}_M/\mathcal{D}_R$ ratio of 2.8, in agreement with the observation of \cite{Otegi2022} that the values overall lie closer to the $\mathcal{D}_M = 3 \times \mathcal{D}_R$ line than the $\mathcal{D}_M = \mathcal{D}_R$ line, which they interpret as a possible correlation in density.

\begin{figure}[htb!]
    \centering
    \includegraphics[width=\linewidth]{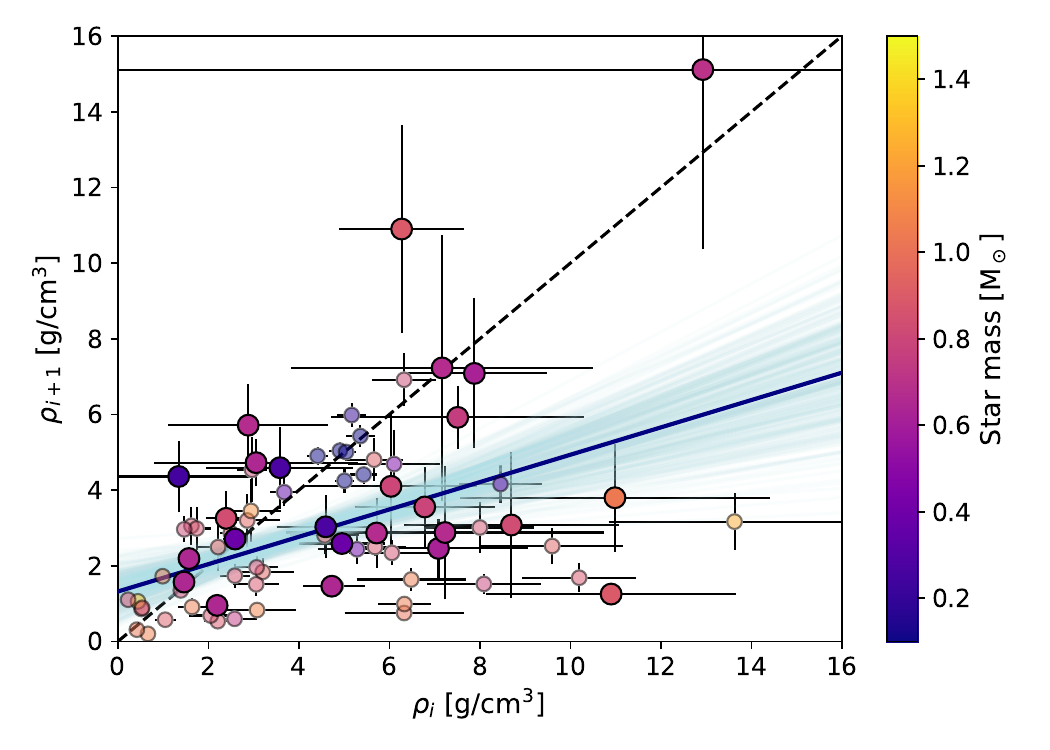}
    \caption{Density of pairs of consecutive planets within a multi-planet system, for the WG3 sample (solid circles) and the PlanetS catalogue (semi-transparent circles), coloured by stellar host mass. The dashed grey line shows the 1:1 relationship. The translucent light blue lines show random draws from the \texttt{linmmix} MCMC sampling, and the solid navy line shows the median model.}
    \label{fig:density-i-i+1}
\end{figure}

Regarding the planetary densities, however, which are plotted in pairs in Fig. \ref{fig:density-i-i+1}, we find for both the WG3 and PlanetS samples that outer planets tend to be less dense than their adjacent inner companions. For the joint sample, we find a significant correlation ($\textrm{p-value} = 8.7\times10^{-5}$). This could reflect a formation in more volatile-rich areas of the protoplanetary disk for the outer planets compared to the inner ones, potentially followed by a disk-driven migration that preserved the planet ordering. Likewise, this could also be an effect of photo-evaporation removing volatiles from the inner planets, particularly as our sample is biased towards short-period planets. 

To conclude, we find that planets orbiting low-mass stars seem to be more similar in mass, radius, and density to their system companions, compared to those orbiting higher-mass stars. This is likely due to the planets around low-mass stars having generally smaller masses and radii, while higher-mass stars host a wider range of planets.

\subsection{Transmission and emission spectroscopy metrics}\label{s:tsm-esm}

\begin{figure}[htb]
    \centering
    \includegraphics[width=\linewidth]{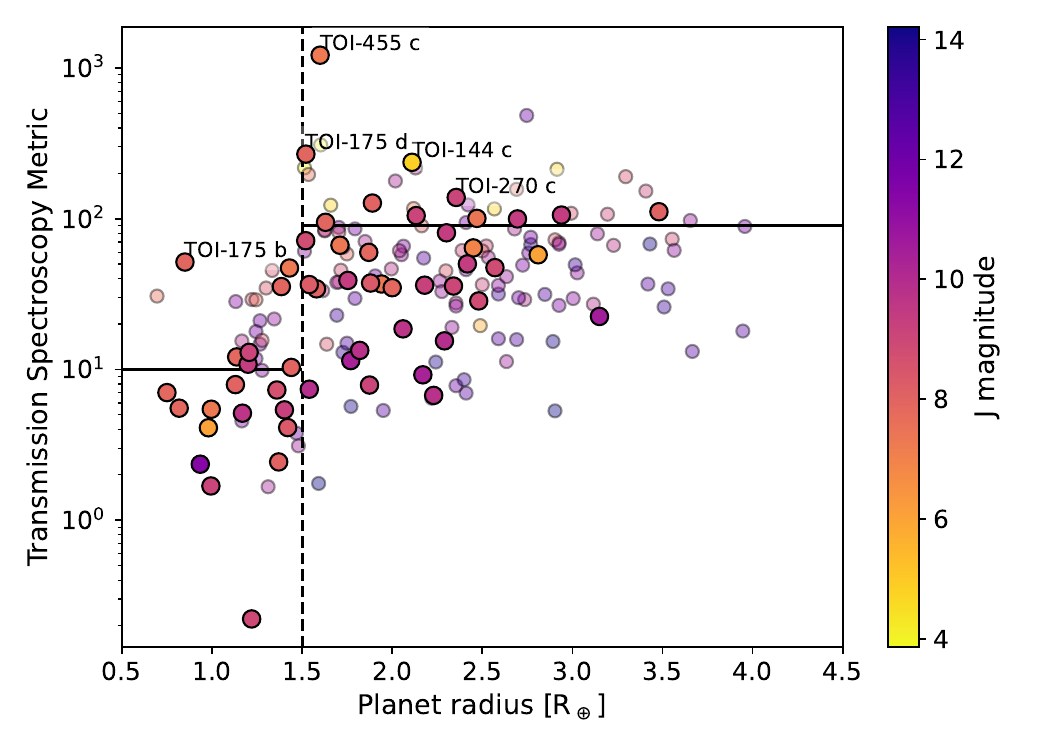}
    \caption{TSM as a function on planet radius for the WG3-confirmed planets (solid circles), with the PlanetS catalogue planets (semi-transparent circles) included for comparison. The points are coloured by stellar J magnitude. The horizontal black lines indicate the two TSM thresholds established by \cite{Kempton2018} for planets with radii below and above $1.5\mathrm{R_\oplus}$ respectively, with the vertical dashed line separating the two groups.}
    \label{fig:TSM_WG3}
\end{figure}

\begin{figure}[htb]
    \centering
    \includegraphics[width=\linewidth]{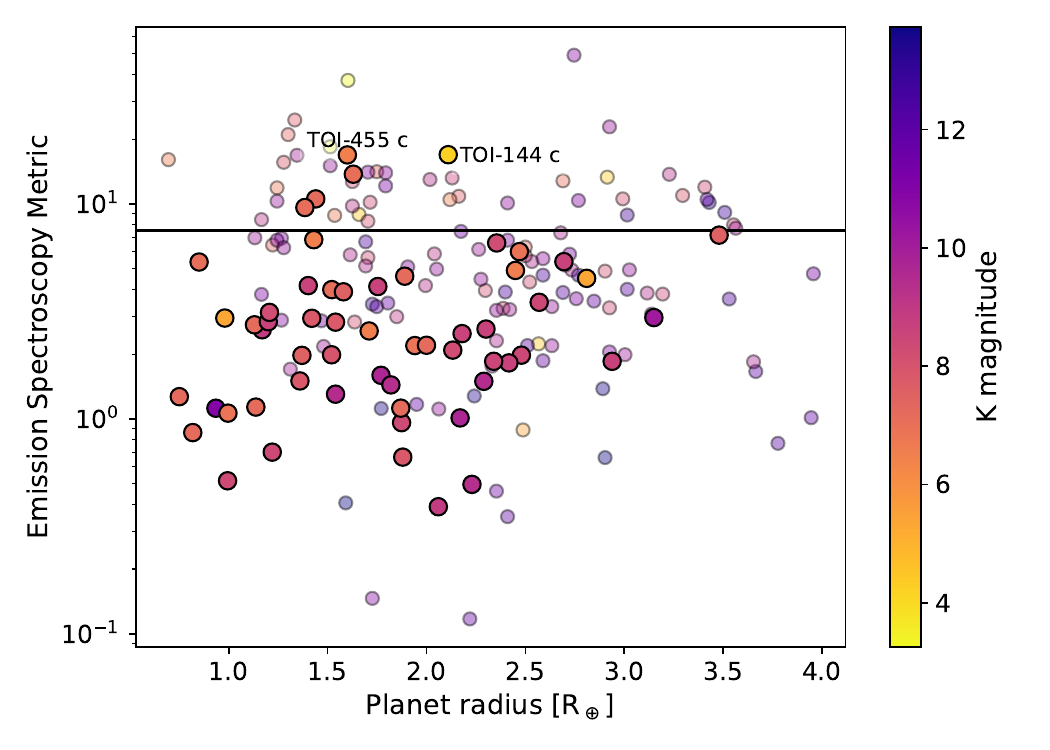}
    \caption{ESM as a function on planet radius for the WG3-confirmed planets (solid circles), with the PlanetS catalogue planets (semi-transparent circles) included for comparison. The points are coloured by stellar K magnitude. The horizontal black line indicates the ESM threshold established by \cite{Kempton2018}.}
    \label{fig:ESM_WG3}
\end{figure}

With the advent of the James Webb Space Telescope \citep[JWST,][]{Gardner2006, Rigby2023} atmospheric studies of super-Earth and sub-Neptune planets have become viable. In this section, we uniformly compute the transmission spectroscopy metric (TSM) and emission spectroscopy metric (ESM) defined by \cite{Kempton2018} for all planets discovered by the WG3, to assess their suitability for atmospheric characterization. Figures \ref{fig:TSM_WG3} and \ref{fig:ESM_WG3} show the values obtained for TSM and ESM respectively for the WG3 planets, together with those from the PlanetS catalogue for comparison. Overall, most small planets with precisely measured masses and radii are not considered suitable for atmospheric characterization by these metrics; however, around 22\% fall above the TSM or ESM thresholds recommended by \cite{Kempton2018}, and around 12\% above both. As reliable masses are crucial to atmospheric modelling, these planets represent valuable targets for atmospheric studies.

Our survey has confirmed some highly promising planets for both transmission and emission spectroscopic characterization, including the best TSM planets in each bin. Tables \ref{tab:TSM_WG3} and \ref{tab:ESM_WG3} list the planets above the thresholds defined by \cite{Kempton2018} for transmission and emission spectroscopy respectively. TOI-455 c \citep[][also known as LTT 1445A c]{Lavie2023} has an extremely high TSM and the second-best ESM, and is the target of an upcoming JWST Director's Discretionary Time (DDT) program \citep{Espinoza2025jwst}, in the framework of the joint JWST/HST Rocky Worlds DDT \footnote{Information available at \url{https://rockyworlds.stsci.edu/index.html}}, for which it was one of the first two targets announced. 
TOI-175 b \citep{Demangeon2021}, meanwhile, has the highest TSM for a planet with $\mathrm{R_p < 1.5\, R_\oplus}$, and has been observed with both HST \citep{Zhou2022} and JWST \citep{Bello2025}, with the latter observations favouring a volcanic atmosphere.

\begin{table}[htb]
\begin{center} 
\caption{TSM values, planet parameters, and J magnitude for WG3 planets above the TSM thresholds.} 
\label{tab:TSM_WG3} 
\centering 
\resizebox{\columnwidth}{!}{
\begin{tabular}{llllll}
\hline \hline
Name & TSM & P [d] & $\mathrm{R_p \, [R_\oplus]}$ & $\mathrm{M_p \, [M_\oplus]}$ & J mag \\ 
\hline
TOI-455 c & 1215.52& 3.123898   & $1.60^{+0.67}_{-0.34}$ & $1.00^{+0.19}_{-0.19}$ & 7.294 \\
TOI-175 d & 268.27 & 7.4507245  & $1.52^{+0.12}_{-0.10}$ & $1.94^{+0.28}_{-0.28}$ & 7.933 \\
TOI-144 c & 236.71 & 6.267852   & $2.11^{+0.05}_{-0.05}$ & $4.30^{+0.70}_{-0.70}$ & 4.87  \\
TOI-270 c & 138.39 & 5.6605731  & $2.36^{+0.06}_{-0.06}$ & $6.15^{+0.37}_{-0.37}$ & 9.099 \\
GJ 9827 d & 127.06 & 6.201812   & $1.89^{+0.16}_{-0.14}$ & $3.02^{+0.58}_{-0.57}$ & 7.984 \\
TOI-469 b & 111.29 & 13.63083   & $3.48^{+0.07}_{-0.08}$ & $9.60^{+0.80}_{-0.80}$ & 8.056 \\
TOI-178 g & 105.76 & 20.71663   & $2.94^{+0.06}_{-0.06}$ & $4.40^{+0.39}_{-0.37}$ & 9.372 \\
TOI-270 d & 104.92 & 11.379573  & $2.13^{+0.06}_{-0.06}$ & $4.78^{+0.43}_{-0.43}$ & 9.099 \\
TOI-178 d & 99.72  & 6.557569   & $2.70^{+0.04}_{-0.05}$ & $5.20^{+0.39}_{-0.43}$ & 9.372 \\
TOI-134 b & 94.67  & 1.40152604 & $1.63^{+0.14}_{-0.14}$ & $4.07^{+0.45}_{-0.45}$ & 7.941 \\
\hline
TOI-175 b & 51.61  & 2.2531136  & $0.85^{+0.06}_{-0.05}$ & $0.40^{+0.16}_{-0.15}$ & 7.933 \\
TOI-455 b & 47.11  & 5.358764   & $1.43^{+0.09}_{-0.09}$ & $2.32^{+0.25}_{-0.25}$ & 7.294 \\
TOI-175 c & 35.48  & 3.6906777  & $1.38^{+0.10}_{-0.08}$ & $2.22^{+0.26}_{-0.25}$ & 7.933 \\
TOI-270 b & 12.99  & 3.3601538  & $1.21^{+0.04}_{-0.04}$ & $1.58^{+0.26}_{-0.26}$ & 9.099 \\
TOI-174 f & 12.12  & 12.1621839 & $1.14^{+0.08}_{-0.08}$ & $0.77^{+0.44}_{-0.40}$ & 7.865 \\
TOI-178 b & 10.84  & 1.9145601  & $1.20^{+0.04}_{-0.04}$ & $0.96^{+0.70}_{-0.65}$ & 9.372 \\
GJ 9827 b & 10.32  & 1.208974   & $1.44^{+0.09}_{-0.07}$ & $4.28^{+0.35}_{-0.33}$ & 7.984 \\
\hline
\end{tabular}
}
\tablefoot{\textit{Top}: Planets with $\mathrm{R_p \geq 1.5\, R_\oplus}$. \textit{Bottom}: Planets with $\mathrm{R_p < 1.5\, R_\oplus}$.}
\end{center}
\end{table}

\begin{table}[htb]
\begin{center} 
\caption{ESM values, planet parameters, and K magnitude for WG3 planets above the ESM threshold.} 
\label{tab:ESM_WG3} 
\centering 
\resizebox{\columnwidth}{!}{
\begin{tabular}{llllll}
\hline \hline
Name & ESM & P [d] & $\mathrm{R_p \, [R_\oplus]}$ & $\mathrm{M_p \, [M_\oplus]}$ & K mag\\
\hline
TOI-144 c & 16.95 & 6.267852   & $2.11^{+0.05}_{-0.05}$ & $4.30^{+0.70}_{-0.70}$ & 4.241 \\
TOI-455 c & 16.91 & 3.123898   & $1.60^{+0.67}_{-0.34}$ & $1.00^{+0.19}_{-0.19}$ & 6.496 \\
TOI-134 b & 13.75 & 1.40152604 & $1.63^{+0.14}_{-0.14}$ & $4.07^{+0.45}_{-0.45}$ & 7.082 \\
GJ 9827 b & 10.56 & 1.208974   & $1.44^{+0.09}_{-0.07}$ & $4.28^{+0.35}_{-0.33}$ & 7.193 \\
TOI-175 c & 9.62  & 3.6906777  & $1.38^{+0.10}_{-0.08}$ & $2.22^{+0.26}_{-0.25}$ & 7.101 \\
\hline
\end{tabular}
}
\end{center}
\end{table}

\subsection{Impact of the observing strategy}

As a follow-up program rather than a blind survey, the WG3 observations were conducted in a way that prioritized planet confirmation. For each star, a handful of RVs were taken, from which emerging signals matching the transiting candidate period were searched for and the activity levels assessed. High-cadence RV campaigns were then performed on the most promising targets. 

\begin{figure}
    \centering
    \includegraphics[width=\linewidth]{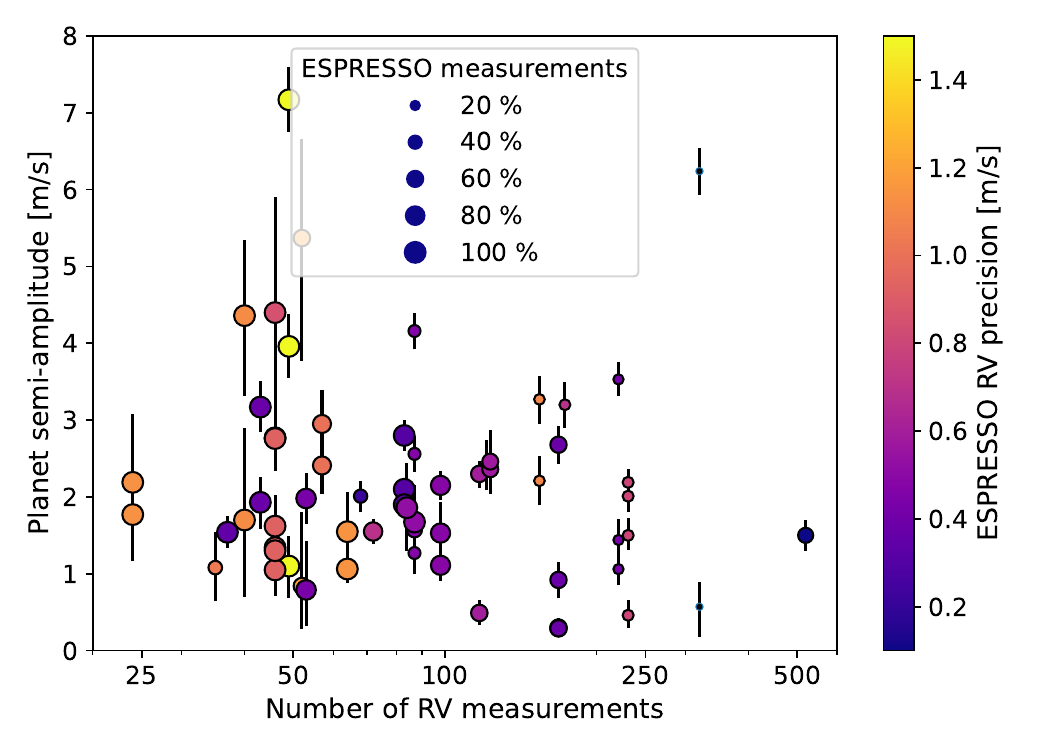}
    \caption{Semi-amplitude of the WG3 planets versus number of RV measurements used to confirm them. The points are coloured by the ESPRESSO RV precision and scaled by the percentage of ESPRESSO measurements.}
    \label{fig:obs-strategy}
\end{figure}

To explore the impact of this observing strategy on the sample of confirmed planets, we plot in Fig. \ref{fig:obs-strategy} the semi-amplitude of the planets versus the number of total RV measurements used to confirm them, showing also the ESPRESSO RV precision and ESPRESSO measurements as a percentage of the total RVs. An important result is that there appears to be a minimum threshold of $\sim75$ measurements required in order to reach semi-amplitudes significantly below the $\mathrm{1\,m \, s^{-1}}$ level. In the context of the upcoming PLATO mission, \citep[][launch expected Q1 2027]{Rauer2025}, which aims specifically to characterize Earth-twins for which the RV semi-amplitudes will range from $\mathrm{\sim1\,m \, s^{-1}}$ for M dwarf hosts to $\mathrm{\sim0.1\,m \, s^{-1}}$ for sun-like stars, this is valuable information for planning follow-up campaigns.

There is also a clear and natural tendency to prioritise targets where better RV precision can be reached, as shown by the colour scale.
As the number of RV measurements increases, the error bars on the semi-amplitudes generally decrease across the left half of the figure, in line with expectations - as these targets have both more measurements and a better measurement precision - up to around 100 measurements. In the right half of the figure, however, we find some systems with hundreds of measurements and large error bars on the semi-amplitudes. This can be explained by the general decrease in the percentage of total RVs constituted by the ESPRESSO measurements. The inclusion of many RVs from less precise instruments, and frequently from multiple instruments with concomitant multiple inter-instrument offsets to be fitted, is then driving up the uncertainties on the semi-amplitudes.

Likewise, we see that the percentage of total RVs constituted by the ESPRESSO measurements tends to decrease with growing number of total RV measurements. Obtaining hundreds of measurements on a single target is a costly investment, that even a GTO will hesitate to make. With the smallest planets requiring large amounts of RVs for confirmation, this highlights the importance of co-operation between different surveys, as has been done, for instance, in the context of TESS follow-up. An important caveat on this is that for the smallest planets, ESPRESSO-like RV precision is necessary, which limits the co-operation possibilities.

\subsection{Stellar activity characterization}

Stellar activity is a well-known confounding factor in exoplanet RV surveys, with active regions on the stellar surface producing spurious RV signals that can mimic or mask true planetary signals \citep[e.g.][and references therein]{Dumusque2014, Dumusque2018}. Transit follow-up surveys are privileged in this regard, as the additional transit information can enable the separation of planetary and activity signals even when the periods are similar. This has been crucial to the analysis of three systems from the WG3 \citep[][; Barros et al. in prep]{Hobson2024, Hobson2025}.

The ESPRESSO pipeline provides numerous activity indicators, as described in \ref{s:ESPRESSO-RVs}. The question then arises of which indicators are the most effective for our sample. To study this, we selected the WG3 targets with at least fifty observations, to allow for a sufficient number of data points for quasi-periodic activity signals to emerge clearly. This sample consists of 18 stars. For these stars, we generated periodograms of all the activity indicators, and inspected the periodograms for significant signals. For each star, we recorded the indicator with the lowest FAP, as a proxy for the most effective indicator. For just over half the sample (10 targets) the FWHM proved to be the most effective indicator, with the remainder being fairly evenly distributed among the other indicators. This is consistent with previous findings on Proxima Centauri \cite{Faria2022}, which was intensively observed in the blind survey GTO sub-programme, and for which the FWHM shows the strongest activity signals.
We also compared the signals seen in the activity indicators to predicted rotation periods using the $\log R'_{\rm hk}-P_{rot}$ relation from \cite{Suarez2015}. We find an overall good agreement between the predicted rotation periods and the strongest signals in the activity indicator periodograms, though for some targets the first harmonic of the predicted rotation period shows more power than the expected rotation period itself. This has previously been observed in both simulations \citep{Boisse2011} and HARPS observational data \citep{Suarez2017}, and is dependent on the configuration of the active regions and the inclination of the stellar axis. 

\begin{figure}
    \centering
    \includegraphics[width=\linewidth]{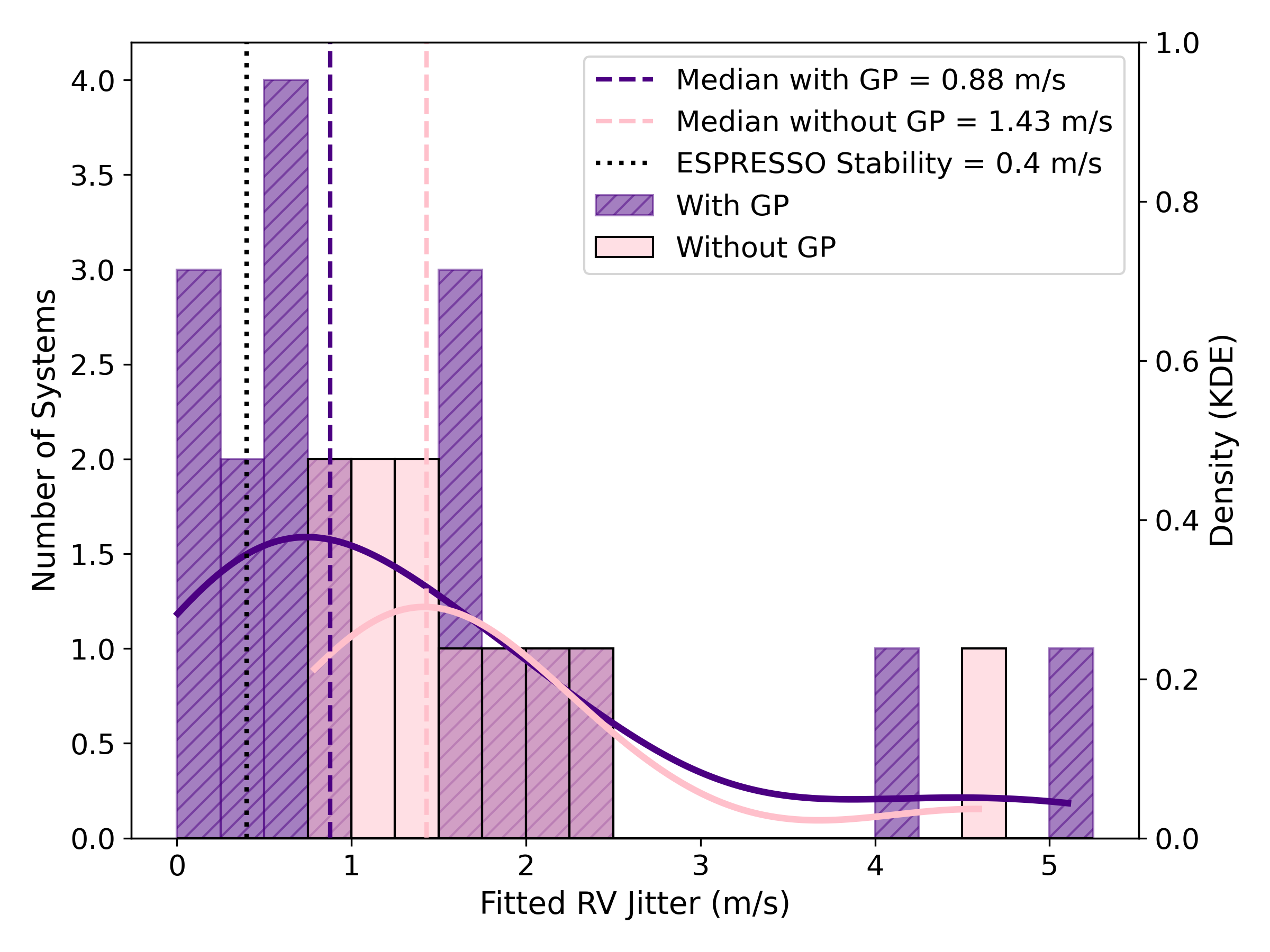}
    \caption{Histograms and KDEs (solid lines) of the fitted RV jitter for the WG3 systems that were modelled without additional GP activity modelling (semi-transparent pink), and with GP modelling (semi-transparent hatched purple). The dashed vertical pink and purple lines show the median RV jitter for systems without and with GP modelling respectively. The dotted black line shows ESPRESSO's long-term RV-precision.}
    \label{fig:jitter}
\end{figure}

Another way to look at the impact of stellar activity on our survey is to examine the fitted RV jitter. This is a white noise term typically included in RV modelling to account for residual stellar effects such as oscillation and granulation noise, instrumental effects, and unmodelled planets. The RV jitter for each of the systems characterised by the WG3 is listed in Table \ref{tab:espresso-summary}. To study this jitter, we divide the systems into two groups: those where a GP was used to model the stellar activity, which should reduce the amount of residual activity-induced RV variation (18 systems); and those without activity modelling (11 systems). Figure \ref{fig:jitter} shows the RV jitter distributions for each group. It is clear that adding a GP model tends to result in a substantial decrease of the fitted RV jitter, as expected. However, given ESPRESSO's excellent long-term RV stability of $\mathrm{40\,cm\, s^{-1}}$ compared to the median with-GP RV jitter of $\mathrm{89\,cm\, s^{-1}}$, it is apparent that there is still unaccounted-for stellar-activity-induced RV variation remaining even after GP modelling. Stellar activity is thus clearly the greatest current obstacle to detecting and characterizing the smallest planets.

\section{Conclusions}\label{s:conclusions}

We have presented an overview of the WG3 sub-program of the ESPRESSO GTO, dedicated to the follow-up, confirmation, and characterization of small transiting planets from K2 and TESS. We have studied the sample of characterized planets, placing them in the context of the overall planet population. Our main results are:
\begin{itemize}
    \item In the $\mathrm{50-200\,S_\oplus}$ insolation range, there is a tentative rocky-to-volatile transition threshold mass at $\mathrm{\sim6\,M_\oplus}$, with planets below this mass being rocky and planets above this mass being volatile-rich.
    \item In the high-insolation regime, we find a population of high-mass rocky planets ($\mathrm{\sim6-10\,M_\oplus}$) that are not observed at lower insolations, suggesting these planets are stripped cores.
    \item Planets below the radius valley have a narrow mass distribution with a sharp cut-off at $\mathrm{\sim10M_\oplus}$, while planets above the valley have a broad distribution spanning the full mass range. The mass distributions intersect at $\mathrm{\sim6\,M_\oplus}$, in line with the proposed rocky-to-volatile threshold.
    \item Planetary mass correlates with metallicity; we find more massive planets around more metal-rich stars.
    \item Planetary masses are comparable to the median protoplanetary disk mass in their stellar mass range, suggesting these rocky planets form either in massive disks, with high formation efficiency, or very early in the disk's lifetime.
    \item In multiplanetary systems, outer planets tend to be more massive, bigger, and less dense than their inner companions.
    \item The ESPRESSO WG3 sub-program has provided many valuable targets for atmospheric characterization.
    \item The results from our observing strategy are discussed, and can serve as a guide for the follow-up of the upcoming PLATO mission. Generally, at least $\approx75$ measurements are required for measuring semi-amplitudes below the $\mathrm{\approx50\,cm \, s^{-1}}$ level.
    \item For ESPRESSO data, the FWHM is generally the most reliable activity indicator. The fitted residual RV jitter suggests that even with activity modelling via GPs, there is significant remaining uncharacterised RV noise induced by stellar activity.
\end{itemize}

\begin{acknowledgements}
We thank the Swiss National Science Foundation (SNSF) and the Geneva University for their continuous support to our planet low-mass companion search programmes. This work has been carried out within the framework of the National Centre of Competence in Research PlanetS supported by the Swiss National Science Foundation under grants 51NF40\_182901 and 51NF40\_205606. The authors acknowledge the financial support of the SNSF. 
This publication makes use of The Data \& Analysis Center for Exoplanets (DACE), which is a facility based at the University of Geneva (CH) dedicated to extrasolar planets data visualisation, exchange and analysis. DACE is a platform of the Swiss National Centre of Competence in Research (NCCR) PlanetS, federating the Swiss expertise in Exoplanet research. The DACE platform is available at \url{https://dace.unige.ch}.
Funding for the TESS mission is provided by NASA's Science Mission Directorate. 
This paper made use of data collected by the TESS mission and are publicly available from the Mikulski Archive for Space Telescopes (MAST) operated by the Space Telescope Science Institute (STScI). 
We acknowledge the use of public TESS data from pipelines at the TESS Science Office and at the TESS Science Processing Operations Center. 
Resources supporting this work were provided by the NASA High-End Computing (HEC) Program through the NASA Advanced Supercomputing (NAS) Division at Ames Research Center for the production of the SPOC data products.
Based on observations collected at the European Southern Observatory under ESO programmes 0102.C-0456, 106.21M2, 108.2254, 110.24CD, 1102.C-0744, 1102.C-0958, 1104.C-0350, and 112.25BG.
This research made use of Lightkurve, a Python package for Kepler and TESS data analysis \citep{Lightkurve}.
JIGH, ASM, CAP and RR acknowledge financial support from the Spanish Ministry of Science, Innovation and Universities (MICIU) project PID2023-149982NB-I00.
Funded/Co-funded by the European Union (ERC, FIERCE, 101052347). Views and opinions expressed are however those of the author(s) only and do not necessarily reflect those of the European Union or the European Research Council. Neither the European Union nor the granting authority can be held responsible for them. This work was supported by FCT - Fundação para a Ciência e a Tecnologia through national funds by grants reference UID/04434/2025.
X. D acknowledges the support from the the Swiss National Science Foundation under the grant SPECTRE (No 200021\_215200).
This work was financed by Portuguese funds through FCT (Funda\c c\~ao para a Ci\^encia e a Tecnologia) in the framework of the project 2022.04048.PTDC (Phi in the Sky, DOI 10.54499/2022.04048.PTDC). CJM also acknowledges FCT and POCH/FSE (EC) support through Investigador FCT Contract 2021.01214.CEECIND/CP1658/CT0001 (DOI 10.54499/2021.01214.CEECIND/CP1658/CT0001).
A.S.M. acknowledges financial support from the Spanish Ministry of Science and Innovation (MICINN) project PID2020-117493GB-I00.
S.G.S. acknowledges support from FCT through FCT contract nr. CEECIND/00826/2018 and POPH/FSE (EC).
We acknowledge financial support from the Agencia Estatal de Investigaci\'on of the Ministerio de Ciencia e Innovaci\'on MCIN/AEI/10.13039/501100011033 and the ERDF “A way of making Europe” through projects PID2021-125627OB-C32 and PID2024-158486OB-C32. 
This work was financed by FCT - Funda\c{c}\~ao para a Ci\^encia e a Tecnologia  under projects UIDB/04434/2020 DOI:10.54499/UIDB/04434/2020, UIDP/04434/2020 DOI: 10.54499/UIDP/04434/2020, PTDC/FIS-AST/4862/2020, UID/04434/2025.
J.L.-B. is funded by the Spanish grants PID2023-150468NB-I00 and CNS2023-144309 from the Ministry of Science, Innovation and Universities (MCIN/AEI/10.13039/501100011033).
Pedro Figueira acknowledges financial support from the Severo Ochoa grant CEX2021-001131-S funded by MCIN/AEI/10.13039/501100011033. Pedro Figueira is also funded by the European Union (ERC, THIRSTEE, 101164189). Views and opinions expressed are however those of the author(s) only and do not necessarily reflect those of the European Union or the European Research Council. Neither the European Union nor the granting authority can be held responsible for them.
O.D.S.D. acknowledges support from e-CHEOPS (PEA No 4000142255).
\end{acknowledgements}

\bibliographystyle{aa}
\bibliography{biblio} 

\begin{thebibliography}{146}
\expandafter\ifx\csname natexlab\endcsname\relax\def\natexlab#1{#1}\fi

\bibitem[{{Akana Murphy} {et~al.}(2023){Akana Murphy}, {Batalha}, {Scarsdale}, {Isaacson}, {Ciardi}, {Gonzales}, {Giacalone}, {Twicken}, {Dattilo}, {Fetherolf}, {Rubenzahl}, {Crossfield}, {Dressing}, {Fulton}, {Howard}, {Huber}, {Kane}, {Petigura}, {Robertson}, {Roy}, {Weiss}, {Beard}, {Chontos}, {Dai}, {Rice}, {Van Zandt}, {Lubin}, {Blunt}, {Polanski}, {Behmard}, {Dalba}, {Hill}, {Rosenthal}, {Brinkman}, {Mayo}, {Turtelboom}, {Angelo}, {Mo{\v{c}}nik}, {MacDougall}, {Pidhorodetska}, {Tyler}, {Kosiarek}, {Holcomb}, {Louden}, {Hirsch}, {Gilbert}, {Anderson}, \& {Valenti}}]{Akana2023}
{Akana Murphy}, J.~M., {Batalha}, N.~M., {Scarsdale}, N., {et~al.} 2023, \aj, 166, 153

\bibitem[{{Ambikasaran} {et~al.}(2015){Ambikasaran}, {Foreman-Mackey}, {Greengard}, {Hogg}, \& {O'Neil}}]{Ambikasaran2015}
{Ambikasaran}, S., {Foreman-Mackey}, D., {Greengard}, L., {Hogg}, D.~W., \& {O'Neil}, M. 2015, IEEE Transactions on Pattern Analysis and Machine Intelligence, 38, 252

\bibitem[{{Antoniadis-Karnavas} {et~al.}(2020){Antoniadis-Karnavas}, {Sousa}, {Delgado-Mena}, {Santos}, {Teixeira}, \& {Neves}}]{Antoniadis2020}
{Antoniadis-Karnavas}, A., {Sousa}, S.~G., {Delgado-Mena}, E., {et~al.} 2020, \aap, 636, A9

\bibitem[{{Armstrong} {et~al.}(2020){Armstrong}, {Lopez}, {Adibekyan}, {Booth}, {Bryant}, {Collins}, {Deleuil}, {Emsenhuber}, {Huang}, {King}, {Lillo-Box}, {Lissauer}, {Matthews}, {Mousis}, {Nielsen}, {Osborn}, {Otegi}, {Santos}, {Sousa}, {Stassun}, {Veras}, {Ziegler}, {Acton}, {Almenara}, {Anderson}, {Barrado}, {Barros}, {Bayliss}, {Belardi}, {Bouchy}, {Brice{\~n}o}, {Brogi}, {Brown}, {Burleigh}, {Casewell}, {Chaushev}, {Ciardi}, {Collins}, {Col{\'o}n}, {Cooke}, {Crossfield}, {D{\'\i}az}, {Delgado Mena}, {Demangeon}, {Dorn}, {Dumusque}, {Eigm{\"u}ller}, {Fausnaugh}, {Figueira}, {Gan}, {Gandhi}, {Gill}, {Gonzales}, {Goad}, {G{\"u}nther}, {Helled}, {Hojjatpanah}, {Howell}, {Jackman}, {Jenkins}, {Jenkins}, {Jensen}, {Kennedy}, {Latham}, {Law}, {Lendl}, {Lozovsky}, {Mann}, {Moyano}, {McCormac}, {Meru}, {Mordasini}, {Osborn}, {Pollacco}, {Queloz}, {Raynard}, {Ricker}, {Rowden}, {Santerne}, {Schlieder}, {Seager}, {Sha}, {Tan}, {Tilbrook}, {Ting}, {Udry}, {Vanderspek}, {Watson}, {West}, {Wilson}, {Winn},
  {Wheatley}, {Villasenor}, {Vines}, \& {Zhan}}]{Armstrong2020}
{Armstrong}, D.~J., {Lopez}, T.~A., {Adibekyan}, V., {et~al.} 2020, \nat, 583, 39

\bibitem[{{Armstrong} {et~al.}(2025){Armstrong}, {Osborn}, {Burn}, {Venturini}, {Adibekyan}, {Bonfanti}, {Burt}, {Collins}, {Delgado Mena}, {Hadjigeorghiou}, {Howell}, {Quinn}, {Sousa}, {Keniger}, {Barrado}, {Barros}, {Bayliss}, {Bouchy}, {Castro-Gonz{\'a}lez}, {Collins}, {Conti}, {Crossfield}, {Diaz}, {Dumusque}, {Feng}, {Lester}, {Lillo-Box}, {Matson}, {Matthews}, {Mordasini}, {Murgas}, {Osborn}, {Palle}, {Santos}, {Schwarz}, {Silva}, {Stassun}, {Str{\o}m}, {Tan}, {Teske}, {Wang}, \& {Wheatley}}]{Armstrong2025}
{Armstrong}, D.~J., {Osborn}, A., {Burn}, R., {et~al.} 2025, \mnras, 537, 3175

\bibitem[{{Astropy Collaboration} {et~al.}(2022){Astropy Collaboration}, {Price-Whelan}, {Lim}, {Earl}, {Starkman}, {Bradley}, {Shupe}, {Patil}, {Corrales}, {Brasseur}, {N{\"o}the}, {Donath}, {Tollerud}, {Morris}, {Ginsburg}, {Vaher}, {Weaver}, {Tocknell}, {Jamieson}, {van Kerkwijk}, {Robitaille}, {Merry}, {Bachetti}, {G{\"u}nther}, {Aldcroft}, {Alvarado-Montes}, {Archibald}, {B{\'o}di}, {Bapat}, {Barentsen}, {Baz{\'a}n}, {Biswas}, {Boquien}, {Burke}, {Cara}, {Cara}, {Conroy}, {Conseil}, {Craig}, {Cross}, {Cruz}, {D'Eugenio}, {Dencheva}, {Devillepoix}, {Dietrich}, {Eigenbrot}, {Erben}, {Ferreira}, {Foreman-Mackey}, {Fox}, {Freij}, {Garg}, {Geda}, {Glattly}, {Gondhalekar}, {Gordon}, {Grant}, {Greenfield}, {Groener}, {Guest}, {Gurovich}, {Handberg}, {Hart}, {Hatfield-Dodds}, {Homeier}, {Hosseinzadeh}, {Jenness}, {Jones}, {Joseph}, {Kalmbach}, {Karamehmetoglu}, {Ka{\l}uszy{\'n}ski}, {Kelley}, {Kern}, {Kerzendorf}, {Koch}, {Kulumani}, {Lee}, {Ly}, {Ma}, {MacBride}, {Maljaars}, {Muna}, {Murphy}, {Norman},
  {O'Steen}, {Oman}, {Pacifici}, {Pascual}, {Pascual-Granado}, {Patil}, {Perren}, {Pickering}, {Rastogi}, {Roulston}, {Ryan}, {Rykoff}, {Sabater}, {Sakurikar}, {Salgado}, {Sanghi}, {Saunders}, {Savchenko}, {Schwardt}, {Seifert-Eckert}, {Shih}, {Jain}, {Shukla}, {Sick}, {Simpson}, {Singanamalla}, {Singer}, {Singhal}, {Sinha}, {Sip{\H{o}}cz}, {Spitler}, {Stansby}, {Streicher}, {{\v{S}}umak}, {Swinbank}, {Taranu}, {Tewary}, {Tremblay}, {Val-Borro}, {Van Kooten}, {Vasovi{\'c}}, {Verma}, {de Miranda Cardoso}, {Williams}, {Wilson}, {Winkel}, {Wood-Vasey}, {Xue}, {Yoachim}, {Zhang}, {Zonca}, \& {Astropy Project Contributors}}]{astropy}
{Astropy Collaboration}, {Price-Whelan}, A.~M., {Lim}, P.~L., {et~al.} 2022, \apj, 935, 167

\bibitem[{{Baranne} {et~al.}(1996){Baranne}, {Queloz}, {Mayor}, {Adrianzyk}, {Knispel}, {Kohler}, {Lacroix}, {Meunier}, {Rimbaud}, \& {Vin}}]{Baranne1996}
{Baranne}, A., {Queloz}, D., {Mayor}, M., {et~al.} 1996, \aaps, 119, 373

\bibitem[{{Barros} {et~al.}(2022){Barros}, {Demangeon}, {Alibert}, {Leleu}, {Adibekyan}, {Lovis}, {Bossini}, {Sousa}, {Hara}, {Bouchy}, {Lavie}, {Rodrigues}, {Gomes da Silva}, {Lillo-Box}, {Pepe}, {Tabernero}, {Zapatero Osorio}, {Sozzetti}, {Su{\'a}rez Mascare{\~n}o}, {Micela}, {Allende Prieto}, {Cristiani}, {Damasso}, {Di Marcantonio}, {Ehrenreich}, {Faria}, {Figueira}, {Gonz{\'a}lez Hern{\'a}ndez}, {Jenkins}, {Lo Curto}, {Martins}, {Micela}, {Nunes}, {Pall{\'e}}, {Santos}, {Rebolo}, {Seager}, {Twicken}, {Udry}, {Vanderspek}, \& {Winn}}]{Barros2022}
{Barros}, S.~C.~C., {Demangeon}, O.~D.~S., {Alibert}, Y., {et~al.} 2022, \aap, 665, A154

\bibitem[{{Baumeister} \& {Tosi}(2023)}]{Baumeister2023}
{Baumeister}, P. \& {Tosi}, N. 2023, \aap, 676, A106

\bibitem[{{Bean} {et~al.}(2021){Bean}, {Raymond}, \& {Owen}}]{Bean2021}
{Bean}, J.~L., {Raymond}, S.~N., \& {Owen}, J.~E. 2021, Journal of Geophysical Research (Planets), 126, e06639

\bibitem[{{Bello-Arufe} {et~al.}(2025){Bello-Arufe}, {Damiano}, {Bennett}, {Hu}, {Welbanks}, {MacDonald}, {Seligman}, {Sing}, {Tokadjian}, {Oza}, \& {Yang}}]{Bello2025}
{Bello-Arufe}, A., {Damiano}, M., {Bennett}, K.~A., {et~al.} 2025, \apjl, 980, L26

\bibitem[{{Boisse} {et~al.}(2011){Boisse}, {Bouchy}, {H{\'e}brard}, {Bonfils}, {Santos}, \& {Vauclair}}]{Boisse2011}
{Boisse}, I., {Bouchy}, F., {H{\'e}brard}, G., {et~al.} 2011, \aap, 528, A4

\bibitem[{{Boley} {et~al.}(2024){Boley}, {Christiansen}, {Zink}, {Hardegree-Ullman}, {Lee}, {Hopkins}, {Wang}, {Fernandes}, {Bergsten}, \& {Bhure}}]{Boley2024}
{Boley}, K.~M., {Christiansen}, J.~L., {Zink}, J., {et~al.} 2024, \aj, 168, 128

\bibitem[{{Bonfils} {et~al.}(2007){Bonfils}, {Mayor}, {Delfosse}, {Forveille}, {Gillon}, {Perrier}, {Udry}, {Bouchy}, {Lovis}, {Pepe}, {Queloz}, {Santos}, \& {Bertaux}}]{Bonfils2007}
{Bonfils}, X., {Mayor}, M., {Delfosse}, X., {et~al.} 2007, \aap, 474, 293

\bibitem[{{Borucki} {et~al.}(2010){Borucki}, {Koch}, {Basri}, {Batalha}, {Brown}, {Caldwell}, {Caldwell}, {Christensen-Dalsgaard}, {Cochran}, {DeVore}, {Dunham}, {Dupree}, {Gautier}, {Geary}, {Gilliland}, {Gould}, {Howell}, {Jenkins}, {Kondo}, {Latham}, {Marcy}, {Meibom}, {Kjeldsen}, {Lissauer}, {Monet}, {Morrison}, {Sasselov}, {Tarter}, {Boss}, {Brownlee}, {Owen}, {Buzasi}, {Charbonneau}, {Doyle}, {Fortney}, {Ford}, {Holman}, {Seager}, {Steffen}, {Welsh}, {Rowe}, {Anderson}, {Buchhave}, {Ciardi}, {Walkowicz}, {Sherry}, {Horch}, {Isaacson}, {Everett}, {Fischer}, {Torres}, {Johnson}, {Endl}, {MacQueen}, {Bryson}, {Dotson}, {Haas}, {Kolodziejczak}, {Van Cleve}, {Chandrasekaran}, {Twicken}, {Quintana}, {Clarke}, {Allen}, {Li}, {Wu}, {Tenenbaum}, {Verner}, {Bruhweiler}, {Barnes}, \& {Prsa}}]{Borucki2010}
{Borucki}, W.~J., {Koch}, D., {Basri}, G., {et~al.} 2010, Science, 327, 977

\bibitem[{{Brahm} {et~al.}(2019){Brahm}, {Espinoza}, {Jord{\'a}n}, {Henning}, {Sarkis}, {Jones}, {D{\'\i}az}, {Jenkins}, {Vanzi}, {Zapata}, {Petrovich}, {Kossakowski}, {Rabus}, {Rojas}, \& {Torres}}]{Brahm2019PARSEC}
{Brahm}, R., {Espinoza}, N., {Jord{\'a}n}, A., {et~al.} 2019, \aj, 158, 45

\bibitem[{{Bressan} {et~al.}(2012){Bressan}, {Marigo}, {Girardi}, {Salasnich}, {Dal Cero}, {Rubele}, \& {Nanni}}]{Bressan2012}
{Bressan}, A., {Marigo}, P., {Girardi}, L., {et~al.} 2012, \mnras, 427, 127

\bibitem[{{Burn} {et~al.}(2024){Burn}, {Mordasini}, {Mishra}, {Haldemann}, {Venturini}, {Emsenhuber}, \& {Henning}}]{Burn2024}
{Burn}, R., {Mordasini}, C., {Mishra}, L., {et~al.} 2024, Nature Astronomy [\eprint[arXiv]{2401.04380}]

\bibitem[{{Castro-Gonz{\'a}lez} {et~al.}(2025){Castro-Gonz{\'a}lez}, {Bouchy}, {Correia}, {Sozzetti}, {Lillo-Box}, {Figueira}, {Lavie}, {Lovis}, {Hobson}, {Sousa}, {Adibekyan}, {Standing}, {Hara}, {Barrado}, {Silva}, {Bourrier}, {Korth}, {Santos}, {Damasso}, {Zapatero Osorio}, {Rodrigues}, {Alibert}, {Barros}, {Cristiani}, {Di Marcantonio}, {Gonz{\'a}lez Hern{\'a}ndez}, {Lo Curto}, {Martins}, {Nunes}, {Palle}, {Pepe}, {Su{\'a}rez Mascare{\~n}o}, \& {Tabernero}}]{Castro2025}
{Castro-Gonz{\'a}lez}, A., {Bouchy}, F., {Correia}, A.~C.~M., {et~al.} 2025, \aap, 699, A344

\bibitem[{{Castro-Gonz{\'a}lez} {et~al.}(2023){Castro-Gonz{\'a}lez}, {Demangeon}, {Lillo-Box}, {Lovis}, {Lavie}, {Adibekyan}, {Acu{\~n}a}, {Deleuil}, {Aguichine}, {Zapatero Osorio}, {Tabernero}, {Davoult}, {Alibert}, {Santos}, {Sousa}, {Antoniadis-Karnavas}, {Borsa}, {Winn}, {Allende Prieto}, {Figueira}, {Jenkins}, {Sozzetti}, {Damasso}, {Silva}, {Astudillo-Defru}, {Barros}, {Bonfils}, {Cristiani}, {Di Marcantonio}, {Gonz{\'a}lez Hern{\'a}ndez}, {Curto}, {Martins}, {Nunes}, {Palle}, {Pepe}, {Seager}, \& {Su{\'a}rez Mascare{\~n}o}}]{Castro2023}
{Castro-Gonz{\'a}lez}, A., {Demangeon}, O.~D.~S., {Lillo-Box}, J., {et~al.} 2023, \aap, 675, A52

\bibitem[{{Castro-Gonz{\'a}lez} {et~al.}(2022){Castro-Gonz{\'a}lez}, {D{\'\i}ez Alonso}, {Men{\'e}ndez Blanco}, {Livingston}, {de Leon}, {Lillo-Box}, {Korth}, {Fern{\'a}ndez Men{\'e}ndez}, {Recio}, {Izquierdo-Ruiz}, {Coya Lozano}, {Garc{\'\i}a de la Cuesta}, {G{\'o}mez Hern{\'a}ndez}, {Vidal Blanco}, {Hevia D{\'\i}az}, {Pardo Silva}, {P{\'e}rez Acevedo}, {Polancos Ruiz}, {Padilla Tijer{\'\i}n}, {V{\'a}zquez Garc{\'\i}a}, {Su{\'a}rez G{\'o}mez}, {Garc{\'\i}a Riesgo}, {Gonz{\'a}lez Guti{\'e}rrez}, {Bonavera}, {Gonz{\'a}lez-Nuevo}, {Rodr{\'\i}guez Pereira}, {S{\'a}nchez Lasheras}, {S{\'a}nchez Rodr{\'\i}guez}, {Mu{\~n}iz}, {Santos Rodr{\'\i}guez}, \& {de Cos Juez}}]{Castro2022}
{Castro-Gonz{\'a}lez}, A., {D{\'\i}ez Alonso}, E., {Men{\'e}ndez Blanco}, J., {et~al.} 2022, \mnras, 509, 1075

\bibitem[{{Chen} {et~al.}(2025){Chen}, {Mordasini}, {Emsenhuber}, {Burn}, {Xie}, \& {Zhou}}]{Chen2025}
{Chen}, D.-C., {Mordasini}, C., {Emsenhuber}, A., {et~al.} 2025, arXiv e-prints, arXiv:2507.09874

\bibitem[{{Cincunegui} {et~al.}(2007){Cincunegui}, {D{\'\i}az}, \& {Mauas}}]{Cincunegui2007}
{Cincunegui}, C., {D{\'\i}az}, R.~F., \& {Mauas}, P.~J.~D. 2007, \aap, 469, 309

\bibitem[{{Courcol} {et~al.}(2016){Courcol}, {Bouchy}, \& {Deleuil}}]{Courcol2016}
{Courcol}, B., {Bouchy}, F., \& {Deleuil}, M. 2016, \mnras, 461, 1841

\bibitem[{{Crane} {et~al.}(2010){Crane}, {Shectman}, {Butler}, {Thompson}, {Birk}, {Jones}, \& {Burley}}]{Crane2010}
{Crane}, J.~D., {Shectman}, S.~A., {Butler}, R.~P., {et~al.} 2010, in Society of Photo-Optical Instrumentation Engineers (SPIE) Conference Series, Vol. 7735, Ground-based and Airborne Instrumentation for Astronomy III, ed. I.~S. {McLean}, S.~K. {Ramsay}, \& H.~{Takami}, 773553

\bibitem[{{Damasso} {et~al.}(2023){Damasso}, {Rodrigues}, {Castro-Gonz{\'a}lez}, {Lavie}, {Davoult}, {Zapatero Osorio}, {Dou}, {Sousa}, {Owen}, {Sossi}, {Adibekyan}, {Osborn}, {Leinhardt}, {Alibert}, {Lovis}, {Delgado Mena}, {Sozzetti}, {Barros}, {Bossini}, {Ziegler}, {Ciardi}, {Matthews}, {Carter}, {Lillo-Box}, {Su{\'a}rez Mascare{\~n}o}, {Cristiani}, {Pepe}, {Rebolo}, {Santos}, {Allende Prieto}, {Benatti}, {Bouchy}, {Brice{\~n}o}, {Di Marcantonio}, {D'Odorico}, {Dumusque}, {Egger}, {Ehrenreich}, {Faria}, {Figueira}, {G{\'e}nova Santos}, {Gonzales}, {Gonz{\'a}lez Hern{\'a}ndez}, {Law}, {Lo Curto}, {Mann}, {Martins}, {Mehner}, {Micela}, {Molaro}, {Nunes}, {Palle}, {Poretti}, {Schlieder}, \& {Udry}}]{Damasso2023}
{Damasso}, M., {Rodrigues}, J., {Castro-Gonz{\'a}lez}, A., {et~al.} 2023, \aap, 679, A33

\bibitem[{{Damasso} {et~al.}(2020){Damasso}, {Sozzetti}, {Lovis}, {Barros}, {Sousa}, {Demangeon}, {Faria}, {Lillo-Box}, {Cristiani}, {Pepe}, {Rebolo}, {Santos}, {Zapatero Osorio}, {Gonz{\'a}lez Hern{\'a}ndez}, {Amate}, {Pasquini}, {Zerbi}, {Adibekyan}, {Abreu}, {Affolter}, {Alibert}, {Aliverti}, {Allart}, {Allende Prieto}, {{\'A}lvarez}, {Alves}, {Avila}, {Baldini}, {Bandy}, {Benz}, {Bianco}, {Borsa}, {Bossini}, {Bourrier}, {Bouchy}, {Broeg}, {Cabral}, {Calderone}, {Cirami}, {Coelho}, {Conconi}, {Coretti}, {Cumani}, {Cupani}, {D'Odorico}, {Deiries}, {Dekker}, {Delabre}, {Di Marcantonio}, {Dumusque}, {Ehrenreich}, {Figueira}, {Fragoso}, {Genolet}, {Genoni}, {G{\'e}nova Santos}, {Hughes}, {Iwert}, {Kerber}, {Knudstrup}, {Landoni}, {Lavie}, {Lizon}, {Lo Curto}, {Maire}, {Martins}, {M{\'e}gevand}, {Mehner}, {Micela}, {Modigliani}, {Molaro}, {Monteiro}, {Monteiro}, {Moschetti}, {Mueller}, {Murphy}, {Nunes}, {Oggioni}, {Oliveira}, {Oshagh}, {Pall{\'e}}, {Pariani}, {Poretti}, {Rasilla}, {Rebord{\~a}o}, {Redaelli},
  {Riva}, {Santana Tschudi}, {Santin}, {Santos}, {S{\'e}gransan}, {Schmidt}, {Segovia}, {Sosnowska}, {Span{\`o}}, {Su{\'a}rez Mascare{\~n}o}, {Tabernero}, {Tenegi}, {Udry}, \& {Zanutta}}]{Damasso2020}
{Damasso}, M., {Sozzetti}, A., {Lovis}, C., {et~al.} 2020, \aap, 642, A31

\bibitem[{{Demangeon} {et~al.}(2021){Demangeon}, {Zapatero Osorio}, {Alibert}, {Barros}, {Adibekyan}, {Tabernero}, {Antoniadis-Karnavas}, {Camacho}, {Su{\'a}rez Mascare{\~n}o}, {Oshagh}, {Micela}, {Sousa}, {Lovis}, {Pepe}, {Rebolo}, {Cristiani}, {Santos}, {Allart}, {Allende Prieto}, {Bossini}, {Bouchy}, {Cabral}, {Damasso}, {Di Marcantonio}, {D'Odorico}, {Ehrenreich}, {Faria}, {Figueira}, {G{\'e}nova Santos}, {Haldemann}, {Hara}, {Gonz{\'a}lez Hern{\'a}ndez}, {Lavie}, {Lillo-Box}, {Lo Curto}, {Martins}, {M{\'e}gevand}, {Mehner}, {Molaro}, {Nunes}, {Pall{\'e}}, {Pasquini}, {Poretti}, {Sozzetti}, \& {Udry}}]{Demangeon2021}
{Demangeon}, O.~D.~S., {Zapatero Osorio}, M.~R., {Alibert}, Y., {et~al.} 2021, \aap, 653, A41

\bibitem[{{D{\'\i}az} {et~al.}(2007){D{\'\i}az}, {Cincunegui}, \& {Mauas}}]{Diaz2007}
{D{\'\i}az}, R.~F., {Cincunegui}, C., \& {Mauas}, P. J.~D. 2007, \mnras, 378, 1007

\bibitem[{{Dragomir} {et~al.}(2019){Dragomir}, {Teske}, {G{\"u}nther}, {S{\'e}gransan}, {Burt}, {Huang}, {Vanderburg}, {Matthews}, {Dumusque}, {Stassun}, {Pepper}, {Ricker}, {Vanderspek}, {Latham}, {Seager}, {Winn}, {Jenkins}, {Beatty}, {Bouchy}, {Brown}, {Butler}, {Ciardi}, {Crane}, {Eastman}, {Fossati}, {Francis}, {Fulton}, {Gaudi}, {Goeke}, {James}, {Klaus}, {Kuhn}, {Lovis}, {Lund}, {McDermott}, {Paegert}, {Pepe}, {Rodriguez}, {Sha}, {Shectman}, {Shporer}, {Siverd}, {Garcia Soto}, {Stevens}, {Twicken}, {Udry}, {Villanueva}, {Wang}, {Wohler}, {Yao}, \& {Zhan}}]{Dragomir2019}
{Dragomir}, D., {Teske}, J., {G{\"u}nther}, M.~N., {et~al.} 2019, \apjl, 875, L7

\bibitem[{{Dumusque}(2018)}]{Dumusque2018}
{Dumusque}, X. 2018, \aap, 620, A47

\bibitem[{{Dumusque} {et~al.}(2014){Dumusque}, {Boisse}, \& {Santos}}]{Dumusque2014}
{Dumusque}, X., {Boisse}, I., \& {Santos}, N.~C. 2014, \apj, 796, 132

\bibitem[{{Eastman} {et~al.}(2013){Eastman}, {Gaudi}, \& {Agol}}]{Eastman2013}
{Eastman}, J., {Gaudi}, B.~S., \& {Agol}, E. 2013, \pasp, 125, 83

\bibitem[{{Espinoza} {et~al.}(2016){Espinoza}, {Brahm}, {Jord{\'a}n}, {Jenkins}, {Rojas}, {Jofr{\'e}}, {M{\"a}dler}, {Rabus}, {Chanam{\'e}}, {Pantoja}, {Soto}, {Morzinski}, {Males}, {Ward-Duong}, \& {Close}}]{Espinoza2016}
{Espinoza}, N., {Brahm}, R., {Jord{\'a}n}, A., {et~al.} 2016, \apj, 830, 43

\bibitem[{{Espinoza} \& {Diamond-Lowe}(2025)}]{Espinoza2025jwst}
{Espinoza}, N. \& {Diamond-Lowe}, H. 2025, {Rocky Worlds DDT: JWST Observations of LTT 1445 A c}, JWST Proposal. Cycle 3, ID. \#9234

\bibitem[{{Espinoza} {et~al.}(2019){Espinoza}, {Kossakowski}, \& {Brahm}}]{Espinoza2019juliet}
{Espinoza}, N., {Kossakowski}, D., \& {Brahm}, R. 2019, \mnras, 490, 2262

\bibitem[{{Faria} {et~al.}(2022){Faria}, {Su{\'a}rez Mascare{\~n}o}, {Figueira}, {Silva}, {Damasso}, {Demangeon}, {Pepe}, {Santos}, {Rebolo}, {Cristiani}, {Adibekyan}, {Alibert}, {Allart}, {Barros}, {Cabral}, {D'Odorico}, {Di Marcantonio}, {Dumusque}, {Ehrenreich}, {Gonz{\'a}lez Hern{\'a}ndez}, {Hara}, {Lillo-Box}, {Lo Curto}, {Lovis}, {Martins}, {M{\'e}gevand}, {Mehner}, {Micela}, {Molaro}, {Nunes}, {Pall{\'e}}, {Poretti}, {Sousa}, {Sozzetti}, {Tabernero}, {Udry}, \& {Zapatero Osorio}}]{Faria2022}
{Faria}, J.~P., {Su{\'a}rez Mascare{\~n}o}, A., {Figueira}, P., {et~al.} 2022, \aap, 658, A115

\bibitem[{{Fatuzzo} \& {Adams}(2008)}]{Fatuzzo2008}
{Fatuzzo}, M. \& {Adams}, F.~C. 2008, \apj, 675, 1361

\bibitem[{{Figueira} {et~al.}(2025){Figueira}, {Faria}, {Silva}, {Castro-Gonz{\'a}lez}, {Gomes da Silva}, {Sousa}, {Bossini}, {Zapatero-Osorio}, {Balsalobre-Ruza}, {Lillo-Box}, {Tabernero}, {Adibekyan}, {Allart}, {Benatti}, {Bouchy}, {Cabral}, {Cristiani}, {Dumusque}, {Gonz{\'a}lez-Hern{\'a}ndez}, {Hara}, {Lo Curto}, {Lovis}, {Mehner}, {Molaro}, {Pepe}, {Santos}, {S{\'e}gransan}, {Sosnowska}, {Rebolo}, {Mascare{\~n}o}, {Sozzetti}, {Udry}, \& {Wehbe}}]{Figueira2025}
{Figueira}, P., {Faria}, J.~P., {Silva}, A.~M., {et~al.} 2025, \aap, 700, A174

\bibitem[{{Foreman-Mackey} {et~al.}(2017){Foreman-Mackey}, {Agol}, {Ambikasaran}, \& {Angus}}]{Foreman-Mackey2017}
{Foreman-Mackey}, D., {Agol}, E., {Ambikasaran}, S., \& {Angus}, R. 2017, \aj, 154, 220

\bibitem[{{Fortney} {et~al.}(2007){Fortney}, {Marley}, \& {Barnes}}]{Fortney2007}
{Fortney}, J.~J., {Marley}, M.~S., \& {Barnes}, J.~W. 2007, \apj, 659, 1661

\bibitem[{{Fressin} {et~al.}(2013){Fressin}, {Torres}, {Charbonneau}, {Bryson}, {Christiansen}, {Dressing}, {Jenkins}, {Walkowicz}, \& {Batalha}}]{Fressin2013}
{Fressin}, F., {Torres}, G., {Charbonneau}, D., {et~al.} 2013, \apj, 766, 81

\bibitem[{{Fulton} {et~al.}(2018){Fulton}, {Petigura}, {Blunt}, \& {Sinukoff}}]{Fulton2018}
{Fulton}, B.~J., {Petigura}, E.~A., {Blunt}, S., \& {Sinukoff}, E. 2018, \pasp, 130, 044504

\bibitem[{{Fulton} {et~al.}(2017){Fulton}, {Petigura}, {Howard}, {Isaacson}, {Marcy}, {Cargile}, {Hebb}, {Weiss}, {Johnson}, {Morton}, {Sinukoff}, {Crossfield}, \& {Hirsch}}]{Fulton2017}
{Fulton}, B.~J., {Petigura}, E.~A., {Howard}, A.~W., {et~al.} 2017, \aj, 154, 109

\bibitem[{{Gaia Collaboration} {et~al.}(2016){Gaia Collaboration}, {Prusti}, {de Bruijne}, {Brown}, {Vallenari}, {Babusiaux}, {Bailer-Jones}, {Bastian}, {Biermann}, {Evans}, {Eyer}, {Jansen}, {Jordi}, {Klioner}, {Lammers}, {Lindegren}, {Luri}, {Mignard}, {Milligan}, {Panem}, {Poinsignon}, {Pourbaix}, {Randich}, {Sarri}, {Sartoretti}, {Siddiqui}, {Soubiran}, {Valette}, {van Leeuwen}, {Walton}, {Aerts}, {Arenou}, {Cropper}, {Drimmel}, {H{\o}g}, {Katz}, {Lattanzi}, {O'Mullane}, {Grebel}, {Holland}, {Huc}, {Passot}, {Bramante}, {Cacciari}, {Casta{\~n}eda}, {Chaoul}, {Cheek}, {De Angeli}, {Fabricius}, {Guerra}, {Hern{\'a}ndez}, {Jean-Antoine-Piccolo}, {Masana}, {Messineo}, {Mowlavi}, {Nienartowicz}, {Ord{\'o}{\~n}ez-Blanco}, {Panuzzo}, {Portell}, {Richards}, {Riello}, {Seabroke}, {Tanga}, {Th{\'e}venin}, {Torra}, {Els}, {Gracia-Abril}, {Comoretto}, {Garcia-Reinaldos}, {Lock}, {Mercier}, {Altmann}, {Andrae}, {Astraatmadja}, {Bellas-Velidis}, {Benson}, {Berthier}, {Blomme}, {Busso}, {Carry}, {Cellino}, {Clementini},
  {Cowell}, {Creevey}, {Cuypers}, {Davidson}, {De Ridder}, {de Torres}, {Delchambre}, {Dell'Oro}, {Ducourant}, {Fr{\'e}mat}, {Garc{\'\i}a-Torres}, {Gosset}, {Halbwachs}, {Hambly}, {Harrison}, {Hauser}, {Hestroffer}, {Hodgkin}, {Huckle}, {Hutton}, {Jasniewicz}, {Jordan}, {Kontizas}, {Korn}, {Lanzafame}, {Manteiga}, {Moitinho}, {Muinonen}, {Osinde}, {Pancino}, {Pauwels}, {Petit}, {Recio-Blanco}, {Robin}, {Sarro}, {Siopis}, {Smith}, {Smith}, {Sozzetti}, {Thuillot}, {van Reeven}, {Viala}, {Abbas}, {Abreu Aramburu}, {Accart}, {Aguado}, {Allan}, {Allasia}, {Altavilla}, {{\'A}lvarez}, {Alves}, {Anderson}, {Andrei}, {Anglada Varela}, {Antiche}, {Antoja}, {Ant{\'o}n}, {Arcay}, {Atzei}, {Ayache}, {Bach}, {Baker}, {Balaguer-N{\'u}{\~n}ez}, {Barache}, {Barata}, {Barbier}, {Barblan}, {Baroni}, {Barrado y Navascu{\'e}s}, {Barros}, {Barstow}, {Becciani}, {Bellazzini}, {Bellei}, {Bello Garc{\'\i}a}, {Belokurov}, {Bendjoya}, {Berihuete}, {Bianchi}, {Bienaym{\'e}}, {Billebaud}, {Blagorodnova}, {Blanco-Cuaresma}, {Boch},
  {Bombrun}, {Borrachero}, {Bouquillon}, {Bourda}, {Bouy}, {Bragaglia}, {Breddels}, {Brouillet}, {Br{\"u}semeister}, {Bucciarelli}, {Budnik}, {Burgess}, {Burgon}, {Burlacu}, {Busonero}, {Buzzi}, {Caffau}, {Cambras}, {Campbell}, {Cancelliere}, {Cantat-Gaudin}, {Carlucci}, {Carrasco}, {Castellani}, {Charlot}, {Charnas}, {Charvet}, {Chassat}, {Chiavassa}, {Clotet}, {Cocozza}, {Collins}, {Collins}, {Costigan}, {Crifo}, {Cross}, {Crosta}, {Crowley}, {Dafonte}, {Damerdji}, {Dapergolas}, {David}, {David}, {De Cat}, {de Felice}, {de Laverny}, {De Luise}, {De March}, {de Martino}, {de Souza}, {Debosscher}, {del Pozo}, {Delbo}, {Delgado}, {Delgado}, {di Marco}, {Di Matteo}, {Diakite}, {Distefano}, {Dolding}, {Dos Anjos}, {Drazinos}, {Dur{\'a}n}, {Dzigan}, {Ecale}, {Edvardsson}, {Enke}, {Erdmann}, {Escolar}, {Espina}, {Evans}, {Eynard Bontemps}, {Fabre}, {Fabrizio}, {Faigler}, {Falc{\~a}o}, {Farr{\`a}s Casas}, {Faye}, {Federici}, {Fedorets}, {Fern{\'a}ndez-Hern{\'a}ndez}, {Fernique}, {Fienga}, {Figueras}, {Filippi},
  {Findeisen}, {Fonti}, {Fouesneau}, {Fraile}, {Fraser}, {Fuchs}, {Furnell}, {Gai}, {Galleti}, {Galluccio}, {Garabato}, {Garc{\'\i}a-Sedano}, {Gar{\'e}}, {Garofalo}, {Garralda}, {Gavras}, {Gerssen}, {Geyer}, {Gilmore}, {Girona}, {Giuffrida}, {Gomes}, {Gonz{\'a}lez-Marcos}, {Gonz{\'a}lez-N{\'u}{\~n}ez}, {Gonz{\'a}lez-Vidal}, {Granvik}, {Guerrier}, {Guillout}, {Guiraud}, {G{\'u}rpide}, {Guti{\'e}rrez-S{\'a}nchez}, {Guy}, {Haigron}, {Hatzidimitriou}, {Haywood}, {Heiter}, {Helmi}, {Hobbs}, {Hofmann}, {Holl}, {Holland }, {Hunt}, {Hypki}, {Icardi}, {Irwin}, {Jevardat de Fombelle}, {Jofr{\'e}}, {Jonker}, {Jorissen}, {Julbe}, {Karampelas}, {Kochoska}, {Kohley}, {Kolenberg}, {Kontizas}, {Koposov}, {Kordopatis}, {Koubsky}, {Kowalczyk}, {Krone-Martins}, {Kudryashova}, {Kull}, {Bachchan}, {Lacoste-Seris}, {Lanza}, {Lavigne}, {Le Poncin-Lafitte}, {Lebreton}, {Lebzelter}, {Leccia}, {Leclerc}, {Lecoeur-Taibi}, {Lemaitre}, {Lenhardt}, {Leroux}, {Liao}, {Licata}, {Lindstr{\o}m}, {Lister}, {Livanou}, {Lobel}, {L{\"o}ffler},
  {L{\'o}pez}, {Lopez-Lozano}, {Lorenz}, {Loureiro}, {MacDonald}, {Magalh{\~a}es Fernandes}, {Managau}, {Mann}, {Mantelet}, {Marchal}, {Marchant}, {Marconi}, {Marie}, {Marinoni}, {Marrese}, {Marschalk{\'o}}, {Marshall}, {Mart{\'\i}n-Fleitas}, {Martino}, {Mary}, {Matijevi{\v{c}}}, {Mazeh}, {McMillan}, {Messina}, {Mestre}, {Michalik}, {Millar}, {Miranda}, {Molina}, {Molinaro}, {Molinaro}, {Moln{\'a}r}, {Moniez}, {Montegriffo}, {Monteiro}, {Mor}, {Mora}, {Morbidelli}, {Morel}, {Morgenthaler}, {Morley}, {Morris}, {Mulone}, {Muraveva}, {Musella}, {Narbonne}, {Nelemans}, {Nicastro}, {Noval}, {Ord{\'e}novic}, {Ordieres-Mer{\'e}}, {Osborne}, {Pagani}, {Pagano}, {Pailler}, {Palacin}, {Palaversa}, {Parsons}, {Paulsen}, {Pecoraro}, {Pedrosa}, {Pentik{\"a}inen}, {Pereira}, {Pichon}, {Piersimoni}, {Pineau}, {Plachy}, {Plum}, {Poujoulet}, {Pr{\v{s}}a}, {Pulone}, {Ragaini}, {Rago}, {Rambaux}, {Ramos-Lerate}, {Ranalli}, {Rauw}, {Read}, {Regibo}, {Renk}, {Reyl{\'e}}, {Ribeiro}, {Rimoldini}, {Ripepi}, {Riva}, {Rixon},
  {Roelens}, {Romero-G{\'o}mez}, {Rowell}, {Royer}, {Rudolph}, {Ruiz-Dern}, {Sadowski}, {Sagrist{\`a} Sell{\'e}s}, {Sahlmann}, {Salgado}, {Salguero}, {Sarasso}, {Savietto}, {Schnorhk}, {Schultheis}, {Sciacca}, {Segol}, {Segovia}, {Segransan}, {Serpell}, {Shih}, {Smareglia}, {Smart}, {Smith}, {Solano}, {Solitro}, {Sordo}, {Soria Nieto}, {Souchay}, {Spagna}, {Spoto}, {Stampa}, {Steele}, {Steidelm{\"u}ller}, {Stephenson}, {Stoev}, {Suess}, {S{\"u}veges}, {Surdej}, {Szabados}, {Szegedi-Elek}, {Tapiador}, {Taris}, {Tauran}, {Taylor}, {Teixeira}, {Terrett}, {Tingley}, {Trager}, {Turon}, {Ulla}, {Utrilla}, {Valentini}, {van Elteren}, {Van Hemelryck}, {van Leeuwen}, {Varadi}, {Vecchiato}, {Veljanoski}, {Via}, {Vicente}, {Vogt}, {Voss}, {Votruba}, {Voutsinas}, {Walmsley}, {Weiler}, {Weingrill}, {Werner}, {Wevers}, {Whitehead}, {Wyrzykowski}, {Yoldas}, {{\v{Z}}erjal}, {Zucker}, {Zurbach}, {Zwitter}, {Alecu}, {Allen}, {Allende Prieto}, {Amorim}, {Anglada-Escud{\'e}}, {Arsenijevic}, {Azaz}, {Balm}, {Beck}, {Bernstein},
  {Bigot}, {Bijaoui}, {Blasco}, {Bonfigli}, {Bono}, {Boudreault}, {Bressan}, {Brown}, {Brunet}, {Bunclark}, {Buonanno}, {Butkevich}, {Carret}, {Carrion}, {Chemin}, {Ch{\'e}reau}, {Corcione}, {Darmigny}, {de Boer}, {de Teodoro}, {de Zeeuw}, {Delle Luche}, {Domingues}, {Dubath}, {Fodor}, {Fr{\'e}zouls}, {Fries}, {Fustes}, {Fyfe}, {Gallardo}, {Gallegos}, {Gardiol}, {Gebran}, {Gomboc}, {G{\'o}mez}, {Grux}, {Gueguen}, {Heyrovsky}, {Hoar}, {Iannicola}, {Isasi Parache}, {Janotto}, {Joliet}, {Jonckheere}, {Keil}, {Kim}, {Klagyivik}, {Klar}, {Knude}, {Kochukhov}, {Kolka}, {Kos}, {Kutka}, {Lainey}, {LeBouquin}, {Liu}, {Loreggia}, {Makarov}, {Marseille}, {Martayan}, {Martinez-Rubi}, {Massart}, {Meynadier}, {Mignot}, {Munari}, {Nguyen}, {Nordlander}, {Ocvirk}, {O'Flaherty}, {Olias Sanz}, {Ortiz}, {Osorio}, {Oszkiewicz}, {Ouzounis}, {Palmer}, {Park}, {Pasquato}, {Peltzer}, {Peralta}, {P{\'e}turaud}, {Pieniluoma}, {Pigozzi}, {Poels}, {Prat}, {Prod'homme}, {Raison}, {Rebordao}, {Risquez}, {Rocca-Volmerange}, {Rosen},
  {Ruiz-Fuertes}, {Russo}, {Sembay}, {Serraller Vizcaino}, {Short}, {Siebert}, {Silva}, {Sinachopoulos}, {Slezak}, {Soffel}, {Sosnowska}, {Strai{\v{z}}ys}, {ter Linden}, {Terrell}, {Theil}, {Tiede}, {Troisi}, {Tsalmantza}, {Tur}, {Vaccari}, {Vachier}, {Valles}, {Van Hamme}, {Veltz}, {Virtanen}, {Wallut}, {Wichmann}, {Wilkinson}, {Ziaeepour}, \& {Zschocke}}]{GAIA2016}
{Gaia Collaboration}, {Prusti}, T., {de Bruijne}, J.~H.~J., {et~al.} 2016, \aap, 595, A1

\bibitem[{{Gaia Collaboration} {et~al.}(2022){Gaia Collaboration}, {Vallenari}, {Brown}, {Prusti}, {de Bruijne}, {Arenou}, {Babusiaux}, {Biermann}, {Creevey}, {Ducourant}, {Evans}, {Eyer}, {Guerra}, {Hutton}, {Jordi}, {Klioner}, {Lammers}, {Lindegren}, {Luri}, {Mignard}, {Panem}, {Pourbaix}, {Randich}, {Sartoretti}, {Soubiran}, {Tanga}, {Walton}, {Bailer-Jones}, {Bastian}, {Drimmel}, {Jansen}, {Katz}, {Lattanzi}, {van Leeuwen}, {Bakker}, {Cacciari}, {Casta{\~n}eda}, {De Angeli}, {Fabricius}, {Fouesneau}, {Fr{\'e}mat}, {Galluccio}, {Guerrier}, {Heiter}, {Masana}, {Messineo}, {Mowlavi}, {Nicolas}, {Nienartowicz}, {Pailler}, {Panuzzo}, {Riclet}, {Roux}, {Seabroke}, {Sordo{\o}rcit}, {Th{\'e}venin}, {Gracia-Abril}, {Portell}, {Teyssier}, {Altmann}, {Andrae}, {Audard}, {Bellas-Velidis}, {Benson}, {Berthier}, {Blomme}, {Burgess}, {Busonero}, {Busso}, {C{\'a}novas}, {Carry}, {Cellino}, {Cheek}, {Clementini}, {Damerdji}, {Davidson}, {de Teodoro}, {Nu{\~n}ez Campos}, {Delchambre}, {Dell'Oro}, {Esquej},
  {Fern{\'a}ndez-Hern{\'a}ndez}, {Fraile}, {Garabato}, {Garc{\'\i}a-Lario}, {Gosset}, {Haigron}, {Halbwachs}, {Hambly}, {Harrison}, {Hern{\'a}ndez}, {Hestroffer}, {Hodgkin}, {Holl}, {Jan{\ss}en}, {Jevardat de Fombelle}, {Jordan}, {Krone-Martins}, {Lanzafame}, {L{\"o}ffler}, {Marchal}, {Marrese}, {Moitinho}, {Muinonen}, {Osborne}, {Pancino}, {Pauwels}, {Recio-Blanco}, {Reyl{\'e}}, {Riello}, {Rimoldini}, {Roegiers}, {Rybizki}, {Sarro}, {Siopis}, {Smith}, {Sozzetti}, {Utrilla}, {van Leeuwen}, {Abbas}, {{\'A}brah{\'a}m}, {Abreu Aramburu}, {Aerts}, {Aguado}, {Ajaj}, {Aldea-Montero}, {Altavilla}, {{\'A}lvarez}, {Alves}, {Anders}, {Anderson}, {Anglada Varela}, {Antoja}, {Baines}, {Baker}, {Balaguer-N{\'u}{\~n}ez}, {Balbinot}, {Balog}, {Barache}, {Barbato}, {Barros}, {Barstow}, {Bartolom{\'e}}, {Bassilana}, {Bauchet}, {Becciani}, {Bellazzini}, {Berihuete}, {Bernet}, {Bertone}, {Bianchi}, {Binnenfeld}, {Blanco-Cuaresma}, {Blazere}, {Boch}, {Bombrun}, {Bossini}, {Bouquillon}, {Bragaglia}, {Bramante}, {Breedt},
  {Bressan}, {Brouillet}, {Brugaletta}, {Bucciarelli}, {Burlacu}, {Butkevich}, {Buzzi}, {Caffau}, {Cancelliere}, {Cantat-Gaudin}, {Carballo}, {Carlucci}, {Carnerero}, {Carrasco}, {Casamiquela}, {Castellani}, {Castro-Ginard}, {Chaoul}, {Charlot}, {Chemin}, {Chiaramida}, {Chiavassa}, {Chornay}, {Comoretto}, {Contursi}, {Cooper}, {Cornez}, {Cowell}, {Crifo}, {Cropper}, {Crosta}, {Crowley}, {Dafonte}, {Dapergolas}, {David}, {David}, {de Laverny}, {De Luise}, {De March}, {De Ridder}, {de Souza}, {de Torres}, {del Peloso}, {del Pozo}, {Delbo}, {Delgado}, {Delisle}, {Demouchy}, {Dharmawardena}, {Di Matteo}, {Diakite}, {Diener}, {Distefano}, {Dolding}, {Edvardsson}, {Enke}, {Fabre}, {Fabrizio}, {Faigler}, {Fedorets}, {Fernique}, {Fienga}, {Figueras}, {Fournier}, {Fouron}, {Fragkoudi}, {Gai}, {Garcia-Gutierrez}, {Garcia-Reinaldos}, {Garc{\'\i}a-Torres}, {Garofalo}, {Gavel}, {Gavras}, {Gerlach}, {Geyer}, {Giacobbe}, {Gilmore}, {Girona}, {Giuffrida}, {Gomel}, {Gomez}, {Gonz{\'a}lez-N{\'u}{\~n}ez},
  {Gonz{\'a}lez-Santamar{\'\i}a}, {Gonz{\'a}lez-Vidal}, {Granvik}, {Guillout}, {Guiraud}, {Guti{\'e}rrez-S{\'a}nchez}, {Guy}, {Hatzidimitriou}, {Hauser}, {Haywood}, {Helmer}, {Helmi}, {Sarmiento}, {Hidalgo}, {Hilger}, {H{\l}adczuk}, {Hobbs}, {Holland}, {Huckle}, {Jardine}, {Jasniewicz}, {Jean-Antoine Piccolo}, {Jim{\'e}nez-Arranz}, {Jorissen}, {Juaristi Campillo}, {Julbe}, {Karbevska}, {Kervella}, {Khanna}, {Kontizas}, {Kordopatis}, {Korn}, {K{\'o}sp{\'a}l}, {Kostrzewa-Rutkowska}, {Kruszy{\'n}ska}, {Kun}, {Laizeau}, {Lambert}, {Lanza}, {Lasne}, {Le Campion}, {Lebreton}, {Lebzelter}, {Leccia}, {Leclerc}, {Lecoeur-Taibi}, {Liao}, {Licata}, {Lindstr{\o}m}, {Lister}, {Livanou}, {Lobel}, {Lorca}, {Loup}, {Madrero Pardo}, {Magdaleno Romeo}, {Managau}, {Mann}, {Manteiga}, {Marchant}, {Marconi}, {Marcos}, {Marcos Santos}, {Mar{\'\i}n Pina}, {Marinoni}, {Marocco}, {Marshall}, {Polo}, {Mart{\'\i}n-Fleitas}, {Marton}, {Mary}, {Masip}, {Massari}, {Mastrobuono-Battisti}, {Mazeh}, {McMillan}, {Messina}, {Michalik},
  {Millar}, {Mints}, {Molina}, {Molinaro}, {Moln{\'a}r}, {Monari}, {Mongui{\'o}}, {Montegriffo}, {Montero}, {Mor}, {Mora}, {Morbidelli}, {Morel}, {Morris}, {Muraveva}, {Murphy}, {Musella}, {Nagy}, {Noval}, {Oca{\~n}a}, {Ogden}, {Ordenovic}, {Osinde}, {Pagani}, {Pagano}, {Palaversa}, {Palicio}, {Pallas-Quintela}, {Panahi}, {Payne-Wardenaar}, {Pe{\~n}alosa Esteller}, {Penttil{\"a}}, {Pichon}, {Piersimoni}, {Pineau}, {Plachy}, {Plum}, {Poggio}, {Pr{\v{s}}a}, {Pulone}, {Racero}, {Ragaini}, {Rainer}, {Raiteri}, {Rambaux}, {Ramos}, {Ramos-Lerate}, {Re Fiorentin}, {Regibo}, {Richards}, {Rios Diaz}, {Ripepi}, {Riva}, {Rix}, {Rixon}, {Robichon}, {Robin}, {Robin}, {Roelens}, {Rogues}, {Rohrbasser}, {Romero-G{\'o}mez}, {Rowell}, {Royer}, {Ruz Mieres}, {Rybicki}, {Sadowski}, {S{\'a}ez N{\'u}{\~n}ez}, {Sagrist{\`a} Sell{\'e}s}, {Sahlmann}, {Salguero}, {Samaras}, {Sanchez Gimenez}, {Sanna}, {Santove{\~n}a}, {Sarasso}, {Schultheis}, {Sciacca}, {Segol}, {Segovia}, {S{\'e}gransan}, {Semeux}, {Shahaf}, {Siddiqui}, {Siebert},
  {Siltala}, {Silvelo}, {Slezak}, {Slezak}, {Smart}, {Snaith}, {Solano}, {Solitro}, {Souami}, {Souchay}, {Spagna}, {Spina}, {Spoto}, {Steele}, {Steidelm{\"u}ller}, {Stephenson}, {S{\"u}veges}, {Surdej}, {Szabados}, {Szegedi-Elek}, {Taris}, {Taylo}, {Teixeira}, {Tolomei}, {Tonello}, {Torra}, {Torra}, {Torralba Elipe}, {Trabucchi}, {Tsounis}, {Turon}, {Ulla}, {Unger}, {Vaillant}, {van Dillen}, {van Reeven}, {Vanel}, {Vecchiato}, {Viala}, {Vicente}, {Voutsinas}, {Weiler}, {Wevers}, {Wyrzykowski}, {Yoldas}, {Yvard}, {Zhao}, {Zorec}, {Zucker}, \& {Zwitter}}]{GaiaDR3}
{Gaia Collaboration}, {Vallenari}, A., {Brown}, A.~G.~A., {et~al.} 2022, arXiv e-prints, arXiv:2208.00211

\bibitem[{{Gardner} {et~al.}(2006){Gardner}, {Mather}, {Clampin}, {Doyon}, {Greenhouse}, {Hammel}, {Hutchings}, {Jakobsen}, {Lilly}, {Long}, {Lunine}, {McCaughrean}, {Mountain}, {Nella}, {Rieke}, {Rieke}, {Rix}, {Smith}, {Sonneborn}, {Stiavelli}, {Stockman}, {Windhorst}, \& {Wright}}]{Gardner2006}
{Gardner}, J.~P., {Mather}, J.~C., {Clampin}, M., {et~al.} 2006, \ssr, 123, 485

\bibitem[{{Ginsburg} {et~al.}(2019){Ginsburg}, {Sip{\H o}cz}, {Brasseur}, {Cowperthwaite}, {Craig}, {Deil}, {Guillochon}, {Guzman}, {Liedtke}, {Lian Lim}, {Lockhart}, {Mommert}, {Morris}, {Norman}, {Parikh}, {Persson}, {Robitaille}, {Segovia}, {Singer}, {Tollerud}, {de Val-Borro}, {Valtchanov}, {Woillez}, {The Astroquery collaboration}, \& {a subset of the astropy collaboration}}]{astroquery}
{Ginsburg}, A., {Sip{\H o}cz}, B.~M., {Brasseur}, C.~E., {et~al.} 2019, \aj, 157, 98

\bibitem[{{Gonzalez}(1997)}]{Gonzalez1997}
{Gonzalez}, G. 1997, \mnras, 285, 403

\bibitem[{{Guerrero} {et~al.}(2021){Guerrero}, {Seager}, {Huang}, {Vanderburg}, {Garcia Soto}, {Mireles}, {Hesse}, {Fong}, {Glidden}, {Shporer}, {Latham}, {Collins}, {Quinn}, {Burt}, {Dragomir}, {Crossfield}, {Vanderspek}, {Fausnaugh}, {Burke}, {Ricker}, {Daylan}, {Essack}, {G{\"u}nther}, {Osborn}, {Pepper}, {Rowden}, {Sha}, {Villanueva}, {Yahalomi}, {Yu}, {Ballard}, {Batalha}, {Berardo}, {Chontos}, {Dittmann}, {Esquerdo}, {Mikal-Evans}, {Jayaraman}, {Krishnamurthy}, {Louie}, {Mehrle}, {Niraula}, {Rackham}, {Rodriguez}, {Rowden}, {Sousa-Silva}, {Watanabe}, {Wong}, {Zhan}, {Zivanovic}, {Christiansen}, {Ciardi}, {Swain}, {Lund}, {Mullally}, {Fleming}, {Rodriguez}, {Boyd}, {Quintana}, {Barclay}, {Col{\'o}n}, {Rinehart}, {Schlieder}, {Clampin}, {Jenkins}, {Twicken}, {Caldwell}, {Coughlin}, {Henze}, {Lissauer}, {Morris}, {Rose}, {Smith}, {Tenenbaum}, {Ting}, {Wohler}, {Bakos}, {Bean}, {Berta-Thompson}, {Bieryla}, {Bouma}, {Buchhave}, {Butler}, {Charbonneau}, {Doty}, {Ge}, {Holman}, {Howard}, {Kaltenegger}, {Kane},
  {Kjeldsen}, {Kreidberg}, {Lin}, {Minsky}, {Narita}, {Paegert}, {P{\'a}l}, {Palle}, {Sasselov}, {Spencer}, {Sozzetti}, {Stassun}, {Torres}, {Udry}, \& {Winn}}]{Guerrero2021}
{Guerrero}, N.~M., {Seager}, S., {Huang}, C.~X., {et~al.} 2021, \apjs, 254, 39

\bibitem[{{Haworth} {et~al.}(2018){Haworth}, {Clarke}, {Rahman}, {Winter}, \& {Facchini}}]{Haworth2018}
{Haworth}, T.~J., {Clarke}, C.~J., {Rahman}, W., {Winter}, A.~J., \& {Facchini}, S. 2018, \mnras, 481, 452

\bibitem[{{Haworth} {et~al.}(2023){Haworth}, {Coleman}, {Qiao}, {Sellek}, \& {Askari}}]{Haworth2023}
{Haworth}, T.~J., {Coleman}, G. A.~L., {Qiao}, L., {Sellek}, A.~D., \& {Askari}, K. 2023, \mnras, 526, 4315

\bibitem[{{Haywood} {et~al.}(2014){Haywood}, {Collier Cameron}, {Queloz}, {Barros}, {Deleuil}, {Fares}, {Gillon}, {Lanza}, {Lovis}, {Moutou}, {Pepe}, {Pollacco}, {Santerne}, {S{\'e}gransan}, \& {Unruh}}]{Haywood2014}
{Haywood}, R.~D., {Collier Cameron}, A., {Queloz}, D., {et~al.} 2014, \mnras, 443, 2517

\bibitem[{{Heller} \& {Armstrong}(2014)}]{Heller2014}
{Heller}, R. \& {Armstrong}, J. 2014, Astrobiology, 14, 50

\bibitem[{{Ho} \& {Van Eylen}(2023)}]{Ho2023}
{Ho}, C. S.~K. \& {Van Eylen}, V. 2023, \mnras, 519, 4056

\bibitem[{{Hobson} {et~al.}(2024){Hobson}, {Bouchy}, {Lavie}, {Lovis}, {Adibekyan}, {Allende Prieto}, {Alibert}, {Barros}, {Castro-Gonz{\'a}lez}, {Cristiani}, {D'Odorico}, {Damasso}, {Di Marcantonio}, {Dumusque}, {Ehrenreich}, {Figueira}, {G{\'e}nova Santos}, {Gilbert}, {Gonz{\'a}lez Hern{\'a}ndez}, {Lillo-Box}, {Lo Curto}, {Martins}, {Mehner}, {Micela}, {Molaro}, {Nunes}, {Palle}, {Pepe}, {Rebolo}, {Rodrigues}, {Santos}, {Sousa}, {Sozzetti}, {Su{\'a}rez Mascare{\~n}o}, {Tabernero}, {Udry}, {Zapatero Osorio}, {Armstrong}, {Ciardi}, {Collins}, {Collins}, {Everett}, {Gandolfi}, {Howell}, {Jenkins}, {Kielkopf}, {Livingston}, {Lund}, {Mireles}, {Ricker}, {Schwarz}, {Seager}, {Tan}, {Ting}, \& {Winn}}]{Hobson2024}
{Hobson}, M.~J., {Bouchy}, F., {Lavie}, B., {et~al.} 2024, \aap, 688, A216

\bibitem[{{Hobson} \& {Gomez}(2017)}]{Hobson2017}
{Hobson}, M.~J. \& {Gomez}, M. 2017, \na, 55, 1

\bibitem[{{Hobson} {et~al.}(2025){Hobson}, {Su{\'a}rez Mascare{\~n}o}, {Lovis}, {Bouchy}, {Lavie}, {Cretignier}, {Silva}, {Sousa}, {Tabernero}, {Adibekyan}, {Allende Prieto}, {Alibert}, {Barros}, {Castro-Gonz{\'a}lez}, {Collins}, {Cristiani}, {D'Odorico}, {Damasso}, {Dragomir}, {Dumusque}, {Ehrenreich}, {Figueira}, {G{\'e}nova Santos}, {Goeke}, {Gonz{\'a}lez Hern{\'a}ndez}, {Hesse}, {Lillo-Box}, {Lo Curto}, {Martins}, {Mehner}, {Micela}, {Molaro}, {Nunes}, {Palle}, {Passegger}, {Pepe}, {Rebolo}, {Rodrigues}, {Santos}, {Sozzetti}, {Tofflemire}, {Udry}, {Watkins}, {Zapatero Osorio}, \& {Ziegler}}]{Hobson2025}
{Hobson}, M.~J., {Su{\'a}rez Mascare{\~n}o}, A., {Lovis}, C., {et~al.} 2025, \aap, 702, A32

\bibitem[{{H{\o}g} {et~al.}(2000){H{\o}g}, {Fabricius}, {Makarov}, {Urban}, {Corbin}, {Wycoff}, {Bastian}, {Schwekendiek}, \& {Wicenec}}]{Tycho-2}
{H{\o}g}, E., {Fabricius}, C., {Makarov}, V.~V., {et~al.} 2000, \aap, 355, L27

\bibitem[{{Howell} {et~al.}(2014){Howell}, {Sobeck}, {Haas}, {Still}, {Barclay}, {Mullally}, {Troeltzsch}, {Aigrain}, {Bryson}, {Caldwell}, {Chaplin}, {Cochran}, {Huber}, {Marcy}, {Miglio}, {Najita}, {Smith}, {Twicken}, \& {Fortney}}]{Howell2014}
{Howell}, S.~B., {Sobeck}, C., {Haas}, M., {et~al.} 2014, \pasp, 126, 398

\bibitem[{{Ida} \& {Lin}(2004)}]{Ida2004}
{Ida}, S. \& {Lin}, D.~N.~C. 2004, \apj, 616, 567

\bibitem[{{Jao} {et~al.}(2011){Jao}, {Henry}, {Subasavage}, {Winters}, {Riedel}, \& {Ianna}}]{Jao2011}
{Jao}, W.-C., {Henry}, T.~J., {Subasavage}, J.~P., {et~al.} 2011, \aj, 141, 117

\bibitem[{{Jenkins}(2002)}]{Jenkins2002}
{Jenkins}, J.~M. 2002, \apj, 575, 493

\bibitem[{{Jenkins} {et~al.}(2010){Jenkins}, {Chandrasekaran}, {McCauliff}, {Caldwell}, {Tenenbaum}, {Li}, {Klaus}, {Cote}, \& {Middour}}]{Jenkins2010}
{Jenkins}, J.~M., {Chandrasekaran}, H., {McCauliff}, S.~D., {et~al.} 2010, in Society of Photo-Optical Instrumentation Engineers (SPIE) Conference Series, Vol. 7740, Software and Cyberinfrastructure for Astronomy, ed. N.~M. {Radziwill} \& A.~{Bridger}, 77400D

\bibitem[{{Jenkins} {et~al.}(2020){Jenkins}, {Tenenbaum}, {Seader}, {Burke}, {McCauliff}, {Smith}, {Twicken}, \& {Chandrasekaran}}]{Jenkins2020}
{Jenkins}, J.~M., {Tenenbaum}, P., {Seader}, S., {et~al.} 2020, {Kepler Data Processing Handbook: Transiting Planet Search}, Kepler Science Document KSCI-19081-003

\bibitem[{{Jenkins} {et~al.}(2016){Jenkins}, {Twicken}, {McCauliff}, {Campbell}, {Sanderfer}, {Lung}, {Mansouri-Samani}, {Girouard}, {Tenenbaum}, {Klaus}, {Smith}, {Caldwell}, {Chacon}, {Henze}, {Heiges}, {Latham}, {Morgan}, {Swade}, {Rinehart}, \& {Vanderspek}}]{Jenkins2016}
{Jenkins}, J.~M., {Twicken}, J.~D., {McCauliff}, S., {et~al.} 2016, in Society of Photo-Optical Instrumentation Engineers (SPIE) Conference Series, Vol. 9913, Software and Cyberinfrastructure for Astronomy IV, ed. G.~{Chiozzi} \& J.~C. {Guzman}, 99133E

\bibitem[{{Kar} {et~al.}(2024){Kar}, {Henry}, {Couperus}, {Vrijmoet}, \& {Jao}}]{Kar2024}
{Kar}, A., {Henry}, T.~J., {Couperus}, A.~A., {Vrijmoet}, E.~H., \& {Jao}, W.-C. 2024, \aj, 167, 196

\bibitem[{{Kelly}(2007)}]{Kelly2007}
{Kelly}, B.~C. 2007, \apj, 665, 1489

\bibitem[{{Kempton} {et~al.}(2018){Kempton}, {Bean}, {Louie}, {Deming}, {Koll}, {Mansfield}, {Christiansen}, {L{\'o}pez-Morales}, {Swain}, {Zellem}, {Ballard}, {Barclay}, {Barstow}, {Batalha}, {Beatty}, {Berta-Thompson}, {Birkby}, {Buchhave}, {Charbonneau}, {Cowan}, {Crossfield}, {de Val-Borro}, {Doyon}, {Dragomir}, {Gaidos}, {Heng}, {Hu}, {Kane}, {Kreidberg}, {Mallonn}, {Morley}, {Narita}, {Nascimbeni}, {Pall{\'e}}, {Quintana}, {Rauscher}, {Seager}, {Shkolnik}, {Sing}, {Sozzetti}, {Stassun}, {Valenti}, \& {von Essen}}]{Kempton2018}
{Kempton}, E. M.~R., {Bean}, J.~L., {Louie}, D.~R., {et~al.} 2018, \pasp, 130, 114401

\bibitem[{{Kipping}(2013)}]{Kipping2013}
{Kipping}, D.~M. 2013, \mnras, 435, 2152

\bibitem[{{Koen} {et~al.}(2010){Koen}, {Kilkenny}, {van Wyk}, \& {Marang}}]{Koen2010}
{Koen}, C., {Kilkenny}, D., {van Wyk}, F., \& {Marang}, F. 2010, \mnras, 403, 1949

\bibitem[{{Kreidberg}(2015)}]{Kreidberg2015}
{Kreidberg}, L. 2015, \pasp, 127, 1161

\bibitem[{{Lada} \& {Lada}(2003)}]{Lada2003}
{Lada}, C.~J. \& {Lada}, E.~A. 2003, \araa, 41, 57

\bibitem[{{Lavie} {et~al.}(2023){Lavie}, {Bouchy}, {Lovis}, {Zapatero Osorio}, {Deline}, {Barros}, {Figueira}, {Sozzetti}, {Gonz{\'a}lez Hern{\'a}ndez}, {Lillo-Box}, {Rodrigues}, {Mehner}, {Damasso}, {Adibekyan}, {Alibert}, {Allende Prieto}, {Cristiani}, {D'Odorico}, {Di Marcantonio}, {Ehrenreich}, {G{\'e}nova Santos}, {Lo Curto}, {Martins}, {Micela}, {Molaro}, {Nunes}, {Palle}, {Pepe}, {Poretti}, {Rebolo}, {Santos}, {Sousa}, {Su{\'a}rez Mascare{\~n}o}, {Tabrenero}, \& {Udry}}]{Lavie2023}
{Lavie}, B., {Bouchy}, F., {Lovis}, C., {et~al.} 2023, \aap, 673, A69

\bibitem[{{Leleu} {et~al.}(2021){Leleu}, {Alibert}, {Hara}, {Hooton}, {Wilson}, {Robutel}, {Delisle}, {Laskar}, {Hoyer}, {Lovis}, {Bryant}, {Ducrot}, {Cabrera}, {Delrez}, {Acton}, {Adibekyan}, {Allart}, {Allende Prieto}, {Alonso}, {Alves}, {Anderson}, {Angerhausen}, {Anglada Escud{\'e}}, {Asquier}, {Barrado}, {Barros}, {Baumjohann}, {Bayliss}, {Beck}, {Beck}, {Bekkelien}, {Benz}, {Billot}, {Bonfanti}, {Bonfils}, {Bouchy}, {Bourrier}, {Bou{\'e}}, {Brandeker}, {Broeg}, {Buder}, {Burdanov}, {Burleigh}, {B{\'a}rczy}, {Cameron}, {Chamberlain}, {Charnoz}, {Cooke}, {Corral Van Damme}, {Correia}, {Cristiani}, {Damasso}, {Davies}, {Deleuil}, {Demangeon}, {Demory}, {Di Marcantonio}, {Di Persio}, {Dumusque}, {Ehrenreich}, {Erikson}, {Figueira}, {Fortier}, {Fossati}, {Fridlund}, {Futyan}, {Gandolfi}, {Garc{\'\i}a Mu{\~n}oz}, {Garcia}, {Gill}, {Gillen}, {Gillon}, {Goad}, {Gonz{\'a}lez Hern{\'a}ndez}, {Guedel}, {G{\"u}nther}, {Haldemann}, {Henderson}, {Heng}, {Hogan}, {Isaak}, {Jehin}, {Jenkins}, {Jord{\'a}n}, {Kiss},
  {Kristiansen}, {Lam}, {Lavie}, {Lecavelier des Etangs}, {Lendl}, {Lillo-Box}, {Lo Curto}, {Magrin}, {Martins}, {Maxted}, {McCormac}, {Mehner}, {Micela}, {Molaro}, {Moyano}, {Murray}, {Nascimbeni}, {Nunes}, {Olofsson}, {Osborn}, {Oshagh}, {Ottensamer}, {Pagano}, {Pall{\'e}}, {Pedersen}, {Pepe}, {Persson}, {Peter}, {Piotto}, {Polenta}, {Pollacco}, {Poretti}, {Pozuelos}, {Queloz}, {Ragazzoni}, {Rando}, {Ratti}, {Rauer}, {Raynard}, {Rebolo}, {Reimers}, {Ribas}, {Santos}, {Scandariato}, {Schneider}, {Sebastian}, {Sestovic}, {Simon}, {Smith}, {Sousa}, {Sozzetti}, {Steller}, {Su{\'a}rez Mascare{\~n}o}, {Szab{\'o}}, {S{\'e}gransan}, {Thomas}, {Thompson}, {Tilbrook}, {Triaud}, {Turner}, {Udry}, {Van Grootel}, {Venus}, {Verrecchia}, {Vines}, {Walton}, {West}, {Wheatley}, {Wolter}, \& {Zapatero Osorio}}]{Leleu2021}
{Leleu}, A., {Alibert}, Y., {Hara}, N.~C., {et~al.} 2021, \aap, 649, A26

\bibitem[{{Leleu} {et~al.}(2024){Leleu}, {Delisle}, {Delrez}, {Bryant}, {Brandeker}, {Osborn}, {Hara}, {Wilson}, {Billot}, {Lendl}, {Ehrenreich}, {Chakraborty}, {G{\"u}nther}, {Hooton}, {Alibert}, {Alonso}, {Alves}, {Anderson}, {Apergis}, {Armstrong}, {B{\'a}rczy}, {Barrado Navascues}, {Barros}, {Battley}, {Baumjohann}, {Bayliss}, {Beck}, {Benz}, {Borsato}, {Broeg}, {Burleigh}, {Casewell}, {Collier Cameron}, {Correia}, {Csizmadia}, {Cubillos}, {Davies}, {Deleuil}, {Deline}, {Demangeon}, {Demory}, {Derekas}, {Edwards}, {Erikson}, {Fortier}, {Fossati}, {Fridlund}, {Gandolfi}, {Gazeas}, {Gillen}, {Gillon}, {Goad}, {G{\"u}del}, {Hawthorn}, {Heitzmann}, {Helling}, {Isaak}, {Jenkins}, {Jenkins}, {Kendall}, {Kiss}, {Korth}, {Lam}, {Laskar}, {Latham}, {Lecavelier des Etangs}, {Magrin}, {Maxted}, {McCormac}, {Mordasini}, {Moyano}, {Nascimbeni}, {Olofsson}, {Osborn}, {Ottensamer}, {Pagano}, {Pall{\'e}}, {Peter}, {Piotto}, {Pollacco}, {Queloz}, {Ragazzoni}, {Rando}, {Rauer}, {Ribas}, {Ricker}, {Saha}, {Santos},
  {Scandariato}, {Seager}, {S{\'e}gransan}, {Simon}, {Smith}, {Sousa}, {Stalport}, {Sulis}, {Szab{\'o}}, {Udry}, {Ulmer-Moll}, {Van Grootel}, {Vanderspek}, {Venturini}, {Villaver}, {Vin{\'e}s}, {Walton}, {West}, {Wheatley}, {Winn}, \& {Zivave}}]{Leleu2024}
{Leleu}, A., {Delisle}, J.-B., {Delrez}, L., {et~al.} 2024, \aap, 688, A211

\bibitem[{{Lightkurve Collaboration} {et~al.}(2018){Lightkurve Collaboration}, {Cardoso}, {Hedges}, {Gully-Santiago}, {Saunders}, {Cody}, {Barclay}, {Hall}, {Sagear}, {Turtelboom}, {Zhang}, {Tzanidakis}, {Mighell}, {Coughlin}, {Bell}, {Berta-Thompson}, {Williams}, {Dotson}, \& {Barentsen}}]{Lightkurve}
{Lightkurve Collaboration}, {Cardoso}, J.~V.~d.~M., {Hedges}, C., {et~al.} 2018, {Lightkurve: Kepler and TESS time series analysis in Python}, Astrophysics Source Code Library

\bibitem[{{Livingston} {et~al.}(2018){Livingston}, {Crossfield}, {Petigura}, {Gonzales}, {Ciardi}, {Beichman}, {Christiansen}, {Dressing}, {Henning}, {Howard}, {Isaacson}, {Fulton}, {Kosiarek}, {Schlieder}, {Sinukoff}, \& {Tamura}}]{Livingston2018}
{Livingston}, J.~H., {Crossfield}, I. J.~M., {Petigura}, E.~A., {et~al.} 2018, \aj, 156, 277

\bibitem[{{Luo} {et~al.}(2024){Luo}, {Dorn}, \& {Deng}}]{Luo2024}
{Luo}, H., {Dorn}, C., \& {Deng}, J. 2024, Nature Astronomy, 8, 1399

\bibitem[{{Luque} \& {Pall{\'e}}(2022)}]{Luque2022}
{Luque}, R. \& {Pall{\'e}}, E. 2022, Science, 377, 1211

\bibitem[{{Manara} {et~al.}(2023){Manara}, {Ansdell}, {Rosotti}, {Hughes}, {Armitage}, {Lodato}, \& {Williams}}]{Manara2023}
{Manara}, C.~F., {Ansdell}, M., {Rosotti}, G.~P., {et~al.} 2023, in Astronomical Society of the Pacific Conference Series, Vol. 534, Protostars and Planets VII, ed. S.~{Inutsuka}, Y.~{Aikawa}, T.~{Muto}, K.~{Tomida}, \& M.~{Tamura}, 539

\bibitem[{{Mayo} {et~al.}(2018){Mayo}, {Vanderburg}, {Latham}, {Bieryla}, {Morton}, {Buchhave}, {Dressing}, {Beichman}, {Berlind}, {Calkins}, {Ciardi}, {Crossfield}, {Esquerdo}, {Everett}, {Gonzales}, {Hirsch}, {Horch}, {Howard}, {Howell}, {Livingston}, {Patel}, {Petigura}, {Schlieder}, {Scott}, {Schumer}, {Sinukoff}, {Teske}, \& {Winters}}]{Mayo2018}
{Mayo}, A.~W., {Vanderburg}, A., {Latham}, D.~W., {et~al.} 2018, \aj, 155, 136

\bibitem[{{Mayor} {et~al.}(2011){Mayor}, {Marmier}, {Lovis}, {Udry}, {S{\'e}gransan}, {Pepe}, {Benz}, {Bertaux}, {Bouchy}, {Dumusque}, {Lo Curto}, {Mordasini}, {Queloz}, \& {Santos}}]{Mayor2011}
{Mayor}, M., {Marmier}, M., {Lovis}, C., {et~al.} 2011, arXiv e-prints, arXiv:1109.2497

\bibitem[{{Mayor} {et~al.}(2003){Mayor}, {Pepe}, {Queloz}, {Bouchy}, {Rupprecht}, {Lo Curto}, {Avila}, {Benz}, {Bertaux}, {Bonfils}, {Dall}, {Dekker}, {Delabre}, {Eckert}, {Fleury}, {Gilliotte}, {Gojak}, {Guzman}, {Kohler}, {Lizon}, {Longinotti}, {Lovis}, {Megevand}, {Pasquini}, {Reyes}, {Sivan}, {Sosnowska}, {Soto}, {Udry}, {van Kesteren}, {Weber}, \& {Weilenmann}}]{Mayor2003}
{Mayor}, M., {Pepe}, F., {Queloz}, D., {et~al.} 2003, The Messenger, 114, 20

\bibitem[{{Millholland} {et~al.}(2017){Millholland}, {Wang}, \& {Laughlin}}]{Millholland2017}
{Millholland}, S., {Wang}, S., \& {Laughlin}, G. 2017, \apjl, 849, L33

\bibitem[{{Mortier} {et~al.}(2012){Mortier}, {Santos}, {Sozzetti}, {Mayor}, {Latham}, {Bonfils}, \& {Udry}}]{Mortier2012}
{Mortier}, A., {Santos}, N.~C., {Sozzetti}, A., {et~al.} 2012, \aap, 543, A45

\bibitem[{{Mortier} {et~al.}(2020){Mortier}, {Zapatero Osorio}, {Malavolta}, {Alibert}, {Rice}, {Lillo-Box}, {Vanderburg}, {Oshagh}, {Buchhave}, {Adibekyan}, {Delgado Mena}, {Lopez-Morales}, {Charbonneau}, {Sousa}, {Lovis}, {Affer}, {Allende Prieto}, {Barros}, {Benatti}, {Bonomo}, {Boschin}, {Bouchy}, {Cabral}, {Collier Cameron}, {Cosentino}, {Cristiani}, {Demangeon}, {Di Marcantonio}, {D'Odorico}, {Dumusque}, {Ehrenreich}, {Figueira}, {Fiorenzano}, {Ghedina}, {Gonz{\'a}lez Hern{\'a}ndez}, {Haldemann}, {Harutyunyan}, {Haywood}, {Latham}, {Lavie}, {Lo Curto}, {Maldonado}, {Manescau}, {Martins}, {Mayor}, {M{\'e}gevand}, {Mehner}, {Micela}, {Molaro}, {Molinari}, {Nunes}, {Pepe}, {Palle}, {Phillips}, {Piotto}, {Pinamonti}, {Poretti}, {Riva}, {Rebolo}, {Santos}, {Sasselov}, {Sozzetti}, {Su{\'a}rez Mascare{\~n}o}, {Udry}, {West}, {Watson}, \& {Wilson}}]{Mortier2020}
{Mortier}, A., {Zapatero Osorio}, M.~R., {Malavolta}, L., {et~al.} 2020, \mnras, 499, 5004

\bibitem[{{Mulders} {et~al.}(2021){Mulders}, {Pascucci}, {Ciesla}, \& {Fernandes}}]{Mulders2021}
{Mulders}, G.~D., {Pascucci}, I., {Ciesla}, F.~J., \& {Fernandes}, R.~B. 2021, \apj, 920, 66

\bibitem[{{Murgas} {et~al.}(2025){Murgas}, {Pall{\'e}}, {Su{\'a}rez Mascare{\~n}o}, {Korth}, {Pozuelos}, {Hobson}, {Lavie}, {Lovis}, {Sousa}, {Bossini}, {Parviainen}, {Castro-Gonz{\'a}lez}, {Adibekyan}, {Allende Prieto}, {Alibert}, {Bouchy}, {Brice{\~n}o}, {Caldwell}, {Ciardi}, {Clark}, {Collins}, {Collins}, {Cristiani}, {Dumusque}, {Ehrenreich}, {Figueira}, {Furlan}, {G{\'e}nova Santos}, {Gnilka}, {Gonz{\'a}lez Hern{\'a}ndez}, {Hartman}, {Howell}, {Jenkins}, {Law}, {Littlefield}, {Lo Curto}, {Mann}, {Martins}, {Mehner}, {Micela}, {Molaro}, {Nunes}, {Pepe}, {Rebolo}, {Relles}, {Santos}, {Scott}, {Seager}, {Sozzetti}, {Udry}, {Watkins}, {Winn}, {Zapatero Osorio}, \& {Ziegler}}]{Murgas2025}
{Murgas}, F., {Pall{\'e}}, E., {Su{\'a}rez Mascare{\~n}o}, A., {et~al.} 2025, \aap, 703, A201

\bibitem[{{Narita} {et~al.}(2017){Narita}, {Hirano}, {Fukui}, {Hori}, {Dai}, {Yu}, {Livingston}, {Ryu}, {Nowak}, {Kuzuhara}, {Sato}, {Takeda}, {Albrecht}, {Kudo}, {Kusakabe}, {Palle}, {Ribas}, {Tamura}, {Van Eylen}, \& {Winn}}]{Narita2017}
{Narita}, N., {Hirano}, T., {Fukui}, A., {et~al.} 2017, \pasj, 69, 29

\bibitem[{{Noyes} {et~al.}(1984){Noyes}, {Hartmann}, {Baliunas}, {Duncan}, \& {Vaughan}}]{Noyes1984}
{Noyes}, R.~W., {Hartmann}, L.~W., {Baliunas}, S.~L., {Duncan}, D.~K., \& {Vaughan}, A.~H. 1984, \apj, 279, 763

\bibitem[{{Otegi} {et~al.}(2020){Otegi}, {Bouchy}, \& {Helled}}]{Otegi2020}
{Otegi}, J.~F., {Bouchy}, F., \& {Helled}, R. 2020, \aap, 634, A43

\bibitem[{{Otegi} {et~al.}(2022){Otegi}, {Helled}, \& {Bouchy}}]{Otegi2022}
{Otegi}, J.~F., {Helled}, R., \& {Bouchy}, F. 2022, \aap, 658, A107

\bibitem[{{Parc} {et~al.}(2024){Parc}, {Bouchy}, {Venturini}, {Dorn}, \& {Helled}}]{Parc2024}
{Parc}, L., {Bouchy}, F., {Venturini}, J., {Dorn}, C., \& {Helled}, R. 2024, \aap, 688, A59

\bibitem[{{Passegger} {et~al.}(2024){Passegger}, {Su{\'a}rez Mascare{\~n}o}, {Allart}, {Gonz{\'a}lez Hern{\'a}ndez}, {Lovis}, {Lavie}, {Silva}, {M{\"u}ller}, {Tabernero}, {Cristiani}, {Pepe}, {Rebolo}, {Santos}, {Adibekyan}, {Alibert}, {Allende Prieto}, {Barros}, {Bouchy}, {Castro-Gonz{\'a}lez}, {D'Odorico}, {Dumusque}, {Di Marcantonio}, {Ehrenreich}, {Figueira}, {G{\'e}nova Santos}, {Lo Curto}, {Martins}, {Mehner}, {Micela}, {Molaro}, {Nari}, {Nunes}, {Pall{\'e}}, {Poretti}, {Rodrigues}, {Sousa}, {Sozzetti}, {Udry}, \& {Zapatero Osorio}}]{Passegger2024}
{Passegger}, V.~M., {Su{\'a}rez Mascare{\~n}o}, A., {Allart}, R., {et~al.} 2024, \aap, 684, A22

\bibitem[{{Patel} \& {Espinoza}(2022)}]{Patel2022}
{Patel}, J.~A. \& {Espinoza}, N. 2022, \aj, 163, 228

\bibitem[{{Pecaut} \& {Mamajek}(2013)}]{Pecaut2013}
{Pecaut}, M.~J. \& {Mamajek}, E.~E. 2013, \apjs, 208, 9

\bibitem[{{Pepe} {et~al.}(2021){Pepe}, {Cristiani}, {Rebolo}, {Santos}, {Dekker}, {Cabral}, {Di Marcantonio}, {Figueira}, {Lo Curto}, {Lovis}, {Mayor}, {M{\'e}gevand}, {Molaro}, {Riva}, {Zapatero Osorio}, {Amate}, {Manescau}, {Pasquini}, {Zerbi}, {Adibekyan}, {Abreu}, {Affolter}, {Alibert}, {Aliverti}, {Allart}, {Allende Prieto}, {{\'A}lvarez}, {Alves}, {Avila}, {Baldini}, {Bandy}, {Barros}, {Benz}, {Bianco}, {Borsa}, {Bourrier}, {Bouchy}, {Broeg}, {Calderone}, {Cirami}, {Coelho}, {Conconi}, {Coretti}, {Cumani}, {Cupani}, {D'Odorico}, {Damasso}, {Deiries}, {Delabre}, {Demangeon}, {Dumusque}, {Ehrenreich}, {Faria}, {Fragoso}, {Genolet}, {Genoni}, {G{\'e}nova Santos}, {Gonz{\'a}lez Hern{\'a}ndez}, {Hughes}, {Iwert}, {Kerber}, {Knudstrup}, {Landoni}, {Lavie}, {Lillo-Box}, {Lizon}, {Maire}, {Martins}, {Mehner}, {Micela}, {Modigliani}, {Monteiro}, {Monteiro}, {Moschetti}, {Murphy}, {Nunes}, {Oggioni}, {Oliveira}, {Oshagh}, {Pall{\'e}}, {Pariani}, {Poretti}, {Rasilla}, {Rebord{\~a}o}, {Redaelli}, {Santana Tschudi},
  {Santin}, {Santos}, {S{\'e}gransan}, {Schmidt}, {Segovia}, {Sosnowska}, {Sozzetti}, {Sousa}, {Span{\`o}}, {Su{\'a}rez Mascare{\~n}o}, {Tabernero}, {Tenegi}, {Udry}, \& {Zanutta}}]{Pepe2021}
{Pepe}, F., {Cristiani}, S., {Rebolo}, R., {et~al.} 2021, \aap, 645, A96

\bibitem[{{Pepe} {et~al.}(2013){Pepe}, {Cristiani}, {Rebolo}, {Santos}, {Dekker}, {M{\'e}gevand}, {Zerbi}, {Cabral}, {Molaro}, {Di Marcantonio}, {Abreu}, {Affolter}, {Aliverti}, {Allende Prieto}, {Amate}, {Avila}, {Baldini}, {Bristow}, {Broeg}, {Cirami}, {Coelho}, {Conconi}, {Coretti}, {Cupani}, {D'Odorico}, {De Caprio}, {Delabre}, {Dorn}, {Figueira}, {Fragoso}, {Galeotta}, {Genolet}, {Gomes}, {Gonz{\'a}lez Hern{\'a}ndez}, {Hughes}, {Iwert}, {Kerber}, {Landoni}, {Lizon}, {Lovis}, {Maire}, {Mannetta}, {Martins}, {Monteiro}, {Oliveira}, {Poretti}, {Rasilla}, {Riva}, {Santana Tschudi}, {Santos}, {Sosnowska}, {Sousa}, {Span{\`o}}, {Tenegi}, {Toso}, {Vanzella}, {Viel}, \& {Zapatero Osorio}}]{Pepe2013}
{Pepe}, F., {Cristiani}, S., {Rebolo}, R., {et~al.} 2013, The Messenger, 153, 6

\bibitem[{{Petigura} {et~al.}(2022){Petigura}, {Rogers}, {Isaacson}, {Owen}, {Kraus}, {Winn}, {MacDougall}, {Howard}, {Fulton}, {Kosiarek}, {Weiss}, {Behmard}, \& {Blunt}}]{Petigura2022}
{Petigura}, E.~A., {Rogers}, J.~G., {Isaacson}, H., {et~al.} 2022, \aj, 163, 179

\bibitem[{{Qiao} {et~al.}(2026){Qiao}, {Coleman}, \& {Haworth}}]{Qiao2026}
{Qiao}, L., {Coleman}, G. A.~L., \& {Haworth}, T.~J. 2026, \mnras, 546, stag034

\bibitem[{{Queloz} {et~al.}(2001){Queloz}, {Henry}, {Sivan}, {Baliunas}, {Beuzit}, {Donahue}, {Mayor}, {Naef}, {Perrier}, \& {Udry}}]{Queloz2001}
{Queloz}, D., {Henry}, G.~W., {Sivan}, J.~P., {et~al.} 2001, \aap, 379, 279

\bibitem[{{Rauer} {et~al.}(2025){Rauer}, {Aerts}, {Cabrera}, {Deleuil}, {Erikson}, {Gizon}, {Goupil}, {Heras}, {Walloschek}, {Lorenzo-Alvarez}, {Marliani}, {Martin-Garcia}, {Mas-Hesse}, {O'Rourke}, {Osborn}, {Pagano}, {Piotto}, {Pollacco}, {Ragazzoni}, {Ramsay}, {Udry}, {Appourchaux}, {Benz}, {Brandeker}, {G{\"u}del}, {Janot-Pacheco}, {Kabath}, {Kjeldsen}, {Min}, {Santos}, {Smith}, {Suarez}, {Werner}, {Aboudan}, {Abreu}, {Acu{\~n}a}, {Adams}, {Adibekyan}, {Affer}, {Agneray}, {Agnor}, {Aguirre B{\o}rsen-Koch}, {Ahmed}, {Aigrain}, {Al-Bahlawan}, {Alcacera Gil}, {Alei}, {Alencar}, {Alexander}, {Alfonso-Garz{\'o}n}, {Alibert}, {Allende Prieto}, {Almeida}, {Alonso Sobrino}, {Altavilla}, {Althaus}, {Alvarez Trujillo}, {Amarsi}, {Ammler-von Eiff}, {Am{\^o}res}, {Andrade}, {Antoniadis-Karnavas}, {Ant{\'o}nio}, {Aparicio del Moral}, {Appolloni}, {Arena}, {Armstrong}, {Aroca Aliaga}, {Asplund}, {Audenaert}, {Auricchio}, {Avelino}, {Baeke}, {Bailli{\'e}}, {Balado}, {Ballber Balaguer{\'o}}, {Balestra}, {Ball}, {Ballans},
  {Ballot}, {Barban}, {Barbary}, {Barbieri}, {Barcel{\'o} Forteza}, {Barker}, {Barklem}, {Barnes}, {Barrado Navascues}, {Barragan}, {Baruteau}, {Basu}, {Baudin}, {Baumeister}, {Bayliss}, {Bazot}, {Beck}, {Belkacem}, {Bellinger}, {Benatti}, {Benomar}, {B{\'e}rard}, {Bergemann}, {Bergomi}, {Bernardo}, {Biazzo}, {Bignamini}, {Bigot}, {Billot}, {Binet}, {Biondi}, {Biondi}, {Birch}, {Bitsch}, {Bluhm Ceballos}, {B{\'o}di}, {Bogn{\'a}r}, {Boisse}, {Bolmont}, {Bonanno}, {Bonavita}, {Bonfanti}, {Bonfils}, {Bonito}, {Bonomo}, {B{\"o}rner}, {Boro Saikia}, {Borreguero Mart{\'\i}n}, {Borsa}, {Borsato}, {Bossini}, {Bouchy}, {Bou{\'e}}, {Boufleur}, {Boumier}, {Bourrier}, {Bowman}, {Bozzo}, {Bradley}, {Bray}, {Bressan}, {Breton}, {Brienza}, {Brito}, {Brogi}, {Brown}, {Brown}, {Brun}, {Bruno}, {Bruns}, {Buchhave}, {Bugnet}, {Buldgen}, {Burgess}, {Busatta}, {Busso}, {Buzasi}, {Caballero}, {Cabral}, {Cabrero Gomez}, {Calderone}, {Cameron}, {Cameron}, {Campante}, {Campos Gestal}, {Canto Martins}, {Cara}, {Carone}, {Carrasco},
  {Casagrande}, {Casewell}, {Cassisi}, {Castellani}, {Castro}, {Catala}, {Catal{\'a}n Fern{\'a}ndez}, {Catelan}, {Cegla}, {Cerruti}, {Cessa}, {Chadid}, {Chaplin}, {Charpinet}, {Chiappini}, {Chiarucci}, {Chiavassa}, {Chinellato}, {Chirulli}, {Christensen-Dalsgaard}, {Church}, {Claret}, {Clarke}, {Claudi}, {Clermont}, {Coelho}, {Coelho}, {Cogato}, {Colom{\'e}}, {Condamin}, {Conde Garc{\'\i}a}, \& {Conseil}}]{Rauer2025}
{Rauer}, H., {Aerts}, C., {Cabrera}, J., {et~al.} 2025, Experimental Astronomy, 59, 26

\bibitem[{{Ricker} {et~al.}(2015){Ricker}, {Winn}, {Vanderspek}, {Latham}, {Bakos}, {Bean}, {Berta-Thompson}, {Brown}, {Buchhave}, {Butler}, {Butler}, {Chaplin}, {Charbonneau}, {Christensen-Dalsgaard}, {Clampin}, {Deming}, {Doty}, {De Lee}, {Dressing}, {Dunham}, {Endl}, {Fressin}, {Ge}, {Henning}, {Holman}, {Howard}, {Ida}, {Jenkins}, {Jernigan}, {Johnson}, {Kaltenegger}, {Kawai}, {Kjeldsen}, {Laughlin}, {Levine}, {Lin}, {Lissauer}, {MacQueen}, {Marcy}, {McCullough}, {Morton}, {Narita}, {Paegert}, {Palle}, {Pepe}, {Pepper}, {Quirrenbach}, {Rinehart}, {Sasselov}, {Sato}, {Seager}, {Sozzetti}, {Stassun}, {Sullivan}, {Szentgyorgyi}, {Torres}, {Udry}, \& {Villasenor}}]{Ricker2015}
{Ricker}, G.~R., {Winn}, J.~N., {Vanderspek}, R., {et~al.} 2015, Journal of Astronomical Telescopes, Instruments, and Systems, 1, 014003

\bibitem[{{Rigby} {et~al.}(2023){Rigby}, {Perrin}, {McElwain}, {Kimble}, {Friedman}, {Lallo}, {Doyon}, {Feinberg}, {Ferruit}, {Glasse}, {Rieke}, {Rieke}, {Wright}, {Willott}, {Colon}, {Milam}, {Neff}, {Stark}, {Valenti}, {Abell}, {Abney}, {Abul-Huda}, {Acton}, {Adams}, {Adler}, {Aguilar}, {Ahmed}, {Albert}, {Alberts}, {Aldridge}, {Allen}, {Altenburg}, {{\'A}lvarez-M{\'a}rquez}, {Alves de Oliveira}, {Andersen}, {Anderson}, {Anderson}, {Argyriou}, {Armstrong}, {Arribas}, {Artigau}, {Arvai}, {Atkinson}, {Bacon}, {Bair}, {Banks}, {Barrientes}, {Barringer}, {Bartosik}, {Bast}, {Baudoz}, {Beatty}, {Bechtold}, {Beck}, {Bergeron}, {Bergkoetter}, {Bhatawdekar}, {Birkmann}, {Blazek}, {Blome}, {Boccaletti}, {B{\"o}ker}, {Boia}, {Bonaventura}, {Bond}, {Bosley}, {Boucarut}, {Bourque}, {Bouwman}, {Bower}, {Bowers}, {Boyer}, {Bradley}, {Brady}, {Braun}, {Breda}, {Bresnahan}, {Bright}, {Britt}, {Bromenschenkel}, {Brooks}, {Brooks}, {Brown}, {Brown}, {Brown}, {Bunker}, {Burger}, {Bushouse}, {Cale}, {Cameron}, {Cameron},
  {Canipe}, {Caplinger}, {Caputo}, {Cara}, {Carey}, {Carniani}, {Carrasquilla}, {Carruthers}, {Case}, {Catherine}, {Chance}, {Chapman}, {Charlot}, {Charlow}, {Chayer}, {Chen}, {Cherinka}, {Chichester}, {Chilton}, {Chonis}, {Clampin}, {Clark}, {Clark}, {Coe}, {Coleman}, {Comber}, {Comeau}, {Connolly}, {Cooper}, {Cooper}, {Coppock}, {Correnti}, {Cossou}, {Coulais}, {Coyle}, {Cracraft}, {Curti}, {Cuturic}, {Davis}, {Davis}, {Dean}, {DeLisa}, {deMeester}, {Dencheva}, {Dencheva}, {DePasquale}, {Deschenes}, {Hunor Detre}, {Diaz}, {Dicken}, {DiFelice}, {Dillman}, {Dixon}, {Doggett}, {Donaldson}, {Douglas}, {DuPrie}, {Dupuis}, {Durning}, {Easmin}, {Eck}, {Edeani}, {Egami}, {Ehrenwinkler}, {Eisenhamer}, {Eisenhower}, {Elie}, {Elliott}, {Elliott}, {Ellis}, {Engesser}, {Espinoza}, {Etienne}, {Etxaluze}, {Falini}, {Feeney}, {Ferry}, {Filippazzo}, {Fincham}, {Fix}, {Flagey}, {Florian}, {Flynn}, {Fontanella}, {Ford}, {Forshay}, {Fox}, {Franz}, {Fu}, {Fullerton}, {Galkin}, {Galyer}, {Garc{\'\i}a Mar{\'\i}n}, {Gardner},
  {Gardner}, {Garland}, {Garrett}, {Gasman}, {Gaspar}, {Gaudreau}, {Gauthier}, {Geers}, {Geithner}, {Gennaro}, {Giardino}, {Girard}, {Giuliano}, {Glassmire}, \& {Glauser}}]{Rigby2023}
{Rigby}, J., {Perrin}, M., {McElwain}, M., {et~al.} 2023, \pasp, 135, 048001

\bibitem[{{Rodrigues} {et~al.}(2025){Rodrigues}, {Barros}, {Santos}, {Davoult}, {Attia}, {Castro-Gonz{\'a}lez}, {Sousa}, {Demangeon}, {Hobson}, {Bossini}, {Ziegler}, {Faria}, {Adibekyan}, {Lovis}, {Lavie}, {Damasso}, {Silva}, {Su{\'a}rez Mascare{\~n}o}, {Pepe}, {Bouchy}, {Alibert}, {Gonz{\'a}lez Hern{\'a}ndez}, {Sozzetti}, {Allende Prieto}, {Cristiani}, {Palle}, {D'Odorico}, {Ehrenreich}, {Figueira}, {Stassun}, {G{\'e}nova Santos}, {Lo Curto}, {Martins}, {Mehner}, {Micela}, {Molaro}, {Nunes}, {Poretti}, {Rebolo}, {Udry}, \& {Zapatero Osorio}}]{Rodrigues2025}
{Rodrigues}, J., {Barros}, S.~C.~C., {Santos}, N.~C., {et~al.} 2025, \aap, 695, A237

\bibitem[{{Rogers}(2015)}]{Rogers2015}
{Rogers}, L.~A. 2015, \apj, 801, 41

\bibitem[{{Santos} {et~al.}(2001){Santos}, {Israelian}, \& {Mayor}}]{Santos2001}
{Santos}, N.~C., {Israelian}, G., \& {Mayor}, M. 2001, \aap, 373, 1019

\bibitem[{{Santos} {et~al.}(2013){Santos}, {Sousa}, {Mortier}, {Neves}, {Adibekyan}, {Tsantaki}, {Delgado Mena}, {Bonfils}, {Israelian}, {Mayor}, \& {Udry}}]{Santos2013}
{Santos}, N.~C., {Sousa}, S.~G., {Mortier}, A., {et~al.} 2013, \aap, 556, A150

\bibitem[{{Scally} \& {Clarke}(2001)}]{Scally2001}
{Scally}, A. \& {Clarke}, C. 2001, \mnras, 325, 449

\bibitem[{{Skrutskie} {et~al.}(2006){Skrutskie}, {Cutri}, {Stiening}, {Weinberg}, {Schneider}, {Carpenter}, {Beichman}, {Capps}, {Chester}, {Elias}, {Huchra}, {Liebert}, {Lonsdale}, {Monet}, {Price}, {Seitzer}, {Jarrett}, {Kirkpatrick}, {Gizis}, {Howard}, {Evans}, {Fowler}, {Fullmer}, {Hurt}, {Light}, {Kopan}, {Marsh}, {McCallon}, {Tam}, {Van Dyk}, \& {Wheelock}}]{2MASS}
{Skrutskie}, M.~F., {Cutri}, R.~M., {Stiening}, R., {et~al.} 2006, \aj, 131, 1163

\bibitem[{{Smith} {et~al.}(2012){Smith}, {Stumpe}, {Van Cleve}, {Jenkins}, {Barclay}, {Fanelli}, {Girouard}, {Kolodziejczak}, {McCauliff}, {Morris}, \& {Twicken}}]{Smith2012}
{Smith}, J.~C., {Stumpe}, M.~C., {Van Cleve}, J.~E., {et~al.} 2012, \pasp, 124, 1000

\bibitem[{{Sousa}(2014)}]{Sousa2014}
{Sousa}, S.~G. 2014, in Determination of Atmospheric Parameters of B, 297--310

\bibitem[{{Sousa} {et~al.}(2021){Sousa}, {Adibekyan}, {Delgado-Mena}, {Santos}, {Rojas-Ayala}, {Soares}, {Legoinha}, {Ulmer-Moll}, {Camacho}, {Barros}, {Demangeon}, {Hoyer}, {Israelian}, {Mortier}, {Tsantaki}, \& {Monteiro}}]{Sousa2021}
{Sousa}, S.~G., {Adibekyan}, V., {Delgado-Mena}, E., {et~al.} 2021, \aap, 656, A53

\bibitem[{{Sousa} {et~al.}(2019){Sousa}, {Adibekyan}, {Santos}, {Mortier}, {Barros}, {Delgado-Mena}, {Demangeon}, {Israelian}, {Faria}, {Figueira}, {Rojas-Ayala}, {Tsantaki}, {Andreasen}, {Brand{\~a}o}, {Ferreira}, {Montalto}, \& {Santerne}}]{Sousa2019}
{Sousa}, S.~G., {Adibekyan}, V., {Santos}, N.~C., {et~al.} 2019, \mnras, 485, 3981

\bibitem[{{Sozzetti} {et~al.}(2021){Sozzetti}, {Damasso}, {Bonomo}, {Alibert}, {Sousa}, {Adibekyan}, {Zapatero Osorio}, {Gonz{\'a}lez Hern{\'a}ndez}, {Barros}, {Lillo-Box}, {Stassun}, {Winn}, {Cristiani}, {Pepe}, {Rebolo}, {Santos}, {Allart}, {Barclay}, {Bouchy}, {Cabral}, {Ciardi}, {Di Marcantonio}, {D'Odorico}, {Ehrenreich}, {Fasnaugh}, {Figueira}, {Haldemann}, {Jenkins}, {Latham}, {Lavie}, {Lo Curto}, {Lovis}, {Martins}, {M{\'e}gevand}, {Mehner}, {Micela}, {Molaro}, {Nunes}, {Oshagh}, {Otegi}, {Pall{\'e}}, {Poretti}, {Ricker}, {Rodriguez}, {Seager}, {Su{\'a}rez Mascare{\~n}o}, {Twicken}, \& {Udry}}]{Sozzetti2021}
{Sozzetti}, A., {Damasso}, M., {Bonomo}, A.~S., {et~al.} 2021, \aap, 648, A75

\bibitem[{{Speagle}(2020)}]{Speagle2020}
{Speagle}, J.~S. 2020, \mnras, 493, 3132

\bibitem[{{Stassun}(2019)}]{Stassun2019}
{Stassun}, K.~G. 2019, VizieR Online Data Catalog, IV/38

\bibitem[{{St{\"o}rzer} \& {Hollenbach}(1999)}]{Storzer1999}
{St{\"o}rzer}, H. \& {Hollenbach}, D. 1999, \apj, 515, 669

\bibitem[{{Stumpe} {et~al.}(2014){Stumpe}, {Smith}, {Catanzarite}, {Van Cleve}, {Jenkins}, {Twicken}, \& {Girouard}}]{Stumpe2014}
{Stumpe}, M.~C., {Smith}, J.~C., {Catanzarite}, J.~H., {et~al.} 2014, \pasp, 126, 100

\bibitem[{{Stumpe} {et~al.}(2012){Stumpe}, {Smith}, {Van Cleve}, {Twicken}, {Barclay}, {Fanelli}, {Girouard}, {Jenkins}, {Kolodziejczak}, {McCauliff}, \& {Morris}}]{Stumpe2012}
{Stumpe}, M.~C., {Smith}, J.~C., {Van Cleve}, J.~E., {et~al.} 2012, \pasp, 124, 985

\bibitem[{{Su{\'a}rez Mascare{\~n}o} {et~al.}(2024){Su{\'a}rez Mascare{\~n}o}, {Passegger}, {Gonz{\'a}lez Hern{\'a}ndez}, {Armstrong}, {Nielsen}, {Lovis}, {Lavie}, {Sousa}, {Silva}, {Allart}, {Rebolo}, {Pepe}, {Santos}, {Cristiani}, {Sozzetti}, {Zapatero Osorio}, {Tabernero}, {Dumusque}, {Udry}, {Adibekyan}, {Allende Prieto}, {Alibert}, {Barros}, {Bouchy}, {Castro-Gonz{\'a}lez}, {Collins}, {Damasso}, {D'Odorico}, {Demangeon}, {Di Marcantonio}, {Ehrenreich}, {Hadjigeorghiou}, {Hara}, {Hawthorn}, {Jenkins}, {Lillo-Box}, {Lo Curto}, {Martins}, {Mehner}, {Micela}, {Molaro}, {Nunes}, {Nari}, {Osborn}, {Pall{\'e}}, {Ricker}, {Rodrigues}, {Rowden}, {Seager}, {Stefanov}, {Str{\o}m}, {Villase{\~n}or}, {Watkins}, {Winn}, {Wohler}, \& {Zambelli}}]{Suarez2024}
{Su{\'a}rez Mascare{\~n}o}, A., {Passegger}, V.~M., {Gonz{\'a}lez Hern{\'a}ndez}, J.~I., {et~al.} 2024, arXiv e-prints, arXiv:2402.04113

\bibitem[{{Su{\'a}rez Mascare{\~n}o} {et~al.}(2015){Su{\'a}rez Mascare{\~n}o}, {Rebolo}, {Gonz{\'a}lez Hern{\'a}ndez}, \& {Esposito}}]{Suarez2015}
{Su{\'a}rez Mascare{\~n}o}, A., {Rebolo}, R., {Gonz{\'a}lez Hern{\'a}ndez}, J.~I., \& {Esposito}, M. 2015, \mnras, 452, 2745

\bibitem[{{Su{\'a}rez Mascare{\~n}o} {et~al.}(2017){Su{\'a}rez Mascare{\~n}o}, {Rebolo}, {Gonz{\'a}lez Hern{\'a}ndez}, \& {Esposito}}]{Suarez2017}
{Su{\'a}rez Mascare{\~n}o}, A., {Rebolo}, R., {Gonz{\'a}lez Hern{\'a}ndez}, J.~I., \& {Esposito}, M. 2017, \mnras, 468, 4772

\bibitem[{{Tabernero} {et~al.}(2022){Tabernero}, {Marfil}, {Montes}, \& {Gonz{\'a}lez Hern{\'a}ndez}}]{Tabernero2022}
{Tabernero}, H.~M., {Marfil}, E., {Montes}, D., \& {Gonz{\'a}lez Hern{\'a}ndez}, J.~I. 2022, \aap, 657, A66

\bibitem[{{Tayar} {et~al.}(2022){Tayar}, {Claytor}, {Huber}, \& {van Saders}}]{Tayar2022}
{Tayar}, J., {Claytor}, Z.~R., {Huber}, D., \& {van Saders}, J. 2022, \apj, 927, 31

\bibitem[{{Toledo-Padr{\'o}n} {et~al.}(2020){Toledo-Padr{\'o}n}, {Lovis}, {Su{\'a}rez Mascare{\~n}o}, {Barros}, {Gonz{\'a}lez Hern{\'a}ndez}, {Sozzetti}, {Bouchy}, {Zapatero Osorio}, {Rebolo}, {Cristiani}, {Pepe}, {Santos}, {Sousa}, {Tabernero}, {Lillo-Box}, {Bossini}, {Adibekyan}, {Allart}, {Damasso}, {D'Odorico}, {Figueira}, {Lavie}, {Lo Curto}, {Mehner}, {Micela}, {Modigliani}, {Nunes}, {Pall{\'e}}, {Abreu}, {Affolter}, {Alibert}, {Aliverti}, {Allende Prieto}, {Alves}, {Amate}, {Avila}, {Baldini}, {Bandy}, {Benatti}, {Benz}, {Bianco}, {Broeg}, {Cabral}, {Calderone}, {Cirami}, {Coelho}, {Conconi}, {Coretti}, {Cumani}, {Cupani}, {Deiries}, {Dekker}, {Delabre}, {Demangeon}, {Di Marcantonio}, {Ehrenreich}, {Fragoso}, {Genolet}, {Genoni}, {G{\'e}nova Santos}, {Hughes}, {Iwert}, {Knudstrup}, {Landoni}, {Lizon}, {Maire}, {Manescau}, {Martins}, {M{\'e}gevand}, {Molaro}, {Monteiro}, {Monteiro}, {Moschetti}, {Mueller}, {Oggioni}, {Oliveira}, {Oshagh}, {Pariani}, {Pasquini}, {Poretti}, {Rasilla}, {Redaelli}, {Riva},
  {Santana Tschudi}, {Santin}, {Santos}, {Segovia}, {Sosnowska}, {Span{\`o}}, {Tenegi}, {Udry}, {Zanutta}, \& {Zerbi}}]{Toledo2020}
{Toledo-Padr{\'o}n}, B., {Lovis}, C., {Su{\'a}rez Mascare{\~n}o}, A., {et~al.} 2020, \aap, 641, A92

\bibitem[{{Trifonov} {et~al.}(2019){Trifonov}, {Rybizki}, \& {K{\"u}rster}}]{Trifonov2019}
{Trifonov}, T., {Rybizki}, J., \& {K{\"u}rster}, M. 2019, \aap, 622, L7

\bibitem[{{Tychoniec} {et~al.}(2020){Tychoniec}, {Manara}, {Rosotti}, {van Dishoeck}, {Cridland}, {Hsieh}, {Murillo}, {Segura-Cox}, {van Terwisga}, \& {Tobin}}]{Tychoniec2020}
{Tychoniec}, {\L}., {Manara}, C.~F., {Rosotti}, G.~P., {et~al.} 2020, \aap, 640, A19

\bibitem[{{Udry} \& {Santos}(2007)}]{Udry2007}
{Udry}, S. \& {Santos}, N.~C. 2007, \araa, 45, 397

\bibitem[{{Valencia} {et~al.}(2007){Valencia}, {Sasselov}, \& {O'Connell}}]{Valencia2007}
{Valencia}, D., {Sasselov}, D.~D., \& {O'Connell}, R.~J. 2007, \apj, 656, 545

\bibitem[{{Van Eylen} {et~al.}(2018){Van Eylen}, {Agentoft}, {Lundkvist}, {Kjeldsen}, {Owen}, {Fulton}, {Petigura}, \& {Snellen}}]{VanEylen2018}
{Van Eylen}, V., {Agentoft}, C., {Lundkvist}, M.~S., {et~al.} 2018, \mnras, 479, 4786

\bibitem[{{Van Eylen} {et~al.}(2021){Van Eylen}, {Astudillo-Defru}, {Bonfils}, {Livingston}, {Hirano}, {Luque}, {Lam}, {Justesen}, {Winn}, {Gandolfi}, {Nowak}, {Palle}, {Albrecht}, {Dai}, {Campos Estrada}, {Owen}, {Foreman-Mackey}, {Fridlund}, {Korth}, {Mathur}, {Forveille}, {Mikal-Evans}, {Osborne}, {Ho}, {Almenara}, {Artigau}, {Barrag{\'a}n}, {Barros}, {Bouchy}, {Cabrera}, {Caldwell}, {Charbonneau}, {Chaturvedi}, {Cochran}, {Csizmadia}, {Damasso}, {Delfosse}, {De Medeiros}, {D{\'\i}az}, {Doyon}, {Esposito}, {F{\H{u}}r{\'e}sz}, {Figueira}, {Georgieva}, {Goffo}, {Grziwa}, {Guenther}, {Hatzes}, {Jenkins}, {Kabath}, {Knudstrup}, {Latham}, {Lavie}, {Lovis}, {Mennickent}, {Mullally}, {Murgas}, {Narita}, {Pepe}, {Persson}, {Redfield}, {Ricker}, {Santos}, {Seager}, {Serrano}, {Smith}, {Su{\'a}rez Mascare{\~n}o}, {Subjak}, {Twicken}, {Udry}, {Vanderspek}, \& {Zapatero Osorio}}]{VanEylen2021}
{Van Eylen}, V., {Astudillo-Defru}, N., {Bonfils}, X., {et~al.} 2021, \mnras, 507, 2154

\bibitem[{{Vaughan} {et~al.}(1978){Vaughan}, {Preston}, \& {Wilson}}]{Vaughan1978}
{Vaughan}, A.~H., {Preston}, G.~W., \& {Wilson}, O.~C. 1978, \pasp, 90, 267

\bibitem[{{Venturini} {et~al.}(2020){Venturini}, {Guilera}, {Haldemann}, {Ronco}, \& {Mordasini}}]{Venturini2020}
{Venturini}, J., {Guilera}, O.~M., {Haldemann}, J., {Ronco}, M.~P., \& {Mordasini}, C. 2020, \aap, 643, L1

\bibitem[{{Venturini} {et~al.}(2024){Venturini}, {Ronco}, {Guilera}, {Haldemann}, {Mordasini}, \& {Miller Bertolami}}]{Venturini2024}
{Venturini}, J., {Ronco}, M.~P., {Guilera}, O.~M., {et~al.} 2024, arXiv e-prints, arXiv:2404.01967

\bibitem[{Virtanen {et~al.}(2020)Virtanen, Gommers, Oliphant, Haberland, Reddy, Cournapeau, Burovski, Peterson, Weckesser, Bright, {van der Walt}, Brett, Wilson, Millman, Mayorov, Nelson, Jones, Kern, Larson, Carey, Polat, Feng, Moore, {VanderPlas}, Laxalde, Perktold, Cimrman, Henriksen, Quintero, Harris, Archibald, Ribeiro, Pedregosa, {van Mulbregt}, \& {SciPy 1.0 Contributors}}]{Scipy}
Virtanen, P., Gommers, R., Oliphant, T.~E., {et~al.} 2020, Nature Methods, 17, 261

\bibitem[{{Vogt} {et~al.}(1994){Vogt}, {Allen}, {Bigelow}, {Bresee}, {Brown}, {Cantrall}, {Conrad}, {Couture}, {Delaney}, {Epps}, {Hilyard}, {Hilyard}, {Horn}, {Jern}, {Kanto}, {Keane}, {Kibrick}, {Lewis}, {Osborne}, {Pardeilhan}, {Pfister}, {Ricketts}, {Robinson}, {Stover}, {Tucker}, {Ward}, \& {Wei}}]{Vogt1994}
{Vogt}, S.~S., {Allen}, S.~L., {Bigelow}, B.~C., {et~al.} 1994, in Society of Photo-Optical Instrumentation Engineers (SPIE) Conference Series, Vol. 2198, Instrumentation in Astronomy VIII, ed. D.~L. {Crawford} \& E.~R. {Craine}, 362

\bibitem[{{Weiss} {et~al.}(2018){Weiss}, {Marcy}, {Petigura}, {Fulton}, {Howard}, {Winn}, {Isaacson}, {Morton}, {Hirsch}, {Sinukoff}, {Cumming}, {Hebb}, \& {Cargile}}]{Weiss2018}
{Weiss}, L.~M., {Marcy}, G.~W., {Petigura}, E.~A., {et~al.} 2018, \aj, 155, 48

\bibitem[{{Wenger} {et~al.}(2000){Wenger}, {Ochsenbein}, {Egret}, {Dubois}, {Bonnarel}, {Borde}, {Genova}, {Jasniewicz}, {Lalo{\"e}}, {Lesteven}, \& {Monier}}]{Wenger2000}
{Wenger}, M., {Ochsenbein}, F., {Egret}, D., {et~al.} 2000, \aaps, 143, 9

\bibitem[{{Winter} \& {Haworth}(2022)}]{Winter2022}
{Winter}, A.~J. \& {Haworth}, T.~J. 2022, European Physical Journal Plus, 137, 1132

\bibitem[{{Zacharias} {et~al.}(2013){Zacharias}, {Finch}, {Girard}, {Henden}, {Bartlett}, {Monet}, \& {Zacharias}}]{Zacharias2013}
{Zacharias}, N., {Finch}, C.~T., {Girard}, T.~M., {et~al.} 2013, \aj, 145, 44

\bibitem[{{Zapatero Osorio} {et~al.}(2026){Zapatero Osorio}, {Tabernero}, {Su{\'a}rez Mascare{\~n}o}, {Sousa}, {Pall{\'e}}, {Lovis}, {Lavie}, {Nari}, {Sozzetti}, {Pepe}, {Rebolo}, {Santos}, {Cristiani}, {Alibert}, {Astudillo-Defru}, {Bouchy}, {Castro-Gonz{\'a}lez}, {Damasso}, {Di Marcantonio}, {Gonz{\'a}lez Hern{\'a}ndez}, {Lillo-Box}, {Martins}, {Rodrigues}, \& {Hobson}}]{Zapatero2026}
{Zapatero Osorio}, M.~R., {Tabernero}, H., {Su{\'a}rez Mascare{\~n}o}, A., {et~al.} 2026, \aap, 706, A166

\bibitem[{{Zeng} {et~al.}(2019){Zeng}, {Jacobsen}, {Sasselov}, {Petaev}, {Vanderburg}, {Lopez-Morales}, {Perez-Mercader}, {Mattsson}, {Li}, {Heising}, {Bonomo}, {Damasso}, {Berger}, {Cao}, {Levi}, \& {Wordsworth}}]{Zeng2019}
{Zeng}, L., {Jacobsen}, S.~B., {Sasselov}, D.~D., {et~al.} 2019, Proceedings of the National Academy of Science, 116, 9723

\bibitem[{{Zhou} {et~al.}(2022){Zhou}, {Ma}, {Wang}, \& {Zhu}}]{Zhou2022}
{Zhou}, L., {Ma}, B., {Wang}, Y., \& {Zhu}, Y. 2022, \aj, 164, 203

\bibitem[{{Zhu} \& {Dong}(2021)}]{Zhu2021}
{Zhu}, W. \& {Dong}, S. 2021, \araa, 59, 291

\end{thebibliography}

\appendix

\section{Stellar parameters for stars with few spectra}

In this appendix, we present the atmospheric stellar parameters for WG3 targets with less than ten ESPRESSO spectra.

\begin{table}[htb]
\caption{Stellar parameters for WG3 targets with less than 10 spectra.}
\label{tab:starparams_fewmeas}
\resizebox{\columnwidth}{!}{  
\begin{tabular}{llllll}
\hline \hline
Name & V [mag]$^a$ & $\mathrm{T_{eff}}$ [K] & Fe/H [dex] & $\log{g}$ [dex] & V$_{\rm broad}$  [km s$^{-1}$]  \\
\hline
TOI-1203 & 8.59$^\dagger$ & \TeffTOIonetwozerothree & \loggTOIonetwozerothree & \FeHTOIonetwozerothree & \VbroadTOIonetwozerothree \\ 
TOI-1233 & 9.24$^\dagger$ & \TeffTOIonetwothreethree & \loggTOIonetwothreethree & \FeHTOIonetwothreethree & \VbroadTOIonetwothreethree \\ 
TOI-696 & 12.482$^\star$ & \TeffTOIsixninesix & \loggTOIsixninesix & \FeHTOIsixninesix & \VbroadTOIsixninesix \\ 
K2-155 & 12.806$^\star$ & \TeffKtwoonefivefive & \loggKtwoonefivefive & \FeHKtwoonefivefive & \VbroadKtwoonefivefive \\ 
TOI-438 & 10.24$^\dagger$ & \TeffTOIfourthreeeight & \loggTOIfourthreeeight & \FeHTOIfourthreeeight & \VbroadTOIfourthreeeight \\ 
TOI-1130 & 11.59$^\dagger$ & \TeffTOIoneonethreezero & \loggTOIoneonethreezero & \FeHTOIoneonethreezero & \VbroadTOIoneonethreezero \\ 
TOI-871 & 10.57$^\dagger$ & \TeffTOIeightsevenone & \loggTOIeightsevenone & \FeHTOIeightsevenone & \VbroadTOIeightsevenone \\ 
TOI-1063 & 9.80$^\dagger$ & \TeffTOIonezerosixthree & \loggTOIonezerosixthree & \FeHTOIonezerosixthree & \VbroadTOIonezerosixthree \\ 
TOI-1078 & 12.20$^\sharp$ & \TeffTOIonezeroseveneight & \loggTOIonezeroseveneight & \FeHTOIonezeroseveneight & \VbroadTOIonezeroseveneight \\ 
TOI-704 & 12.00$^\dagger$ & \TeffTOIsevenzerofour & \loggTOIsevenzerofour & \FeHTOIsevenzerofour & \VbroadTOIsevenzerofour \\ 
TOI-1097 & 9.31$^\dagger$ & \TeffTOIonezeronineseven & \loggTOIonezeronineseven & \FeHTOIonezeronineseven & \VbroadTOIonezeronineseven \\ 
TOI-1230  & 10.25$^\dagger$ & \TeffTOIonetwothreezero & \loggTOIonetwothreezero & \FeHTOIonetwothreezero & \VbroadTOIonetwothreezero \\ 
TOI-1468 & 13.124$^\star$ & \TeffTOIonefoursixeight & \loggTOIonefoursixeight & \FeHTOIonefoursixeight & \VbroadTOIonefoursixeight \\ 
TOI-1058 & 10.81$^\dagger$  & \TeffTOIonezerofiveeight & \loggTOIonezerofiveeight & \FeHTOIonezerofiveeight & \VbroadTOIonezerofiveeight \\ 
TOI-910 & 12.69$^\lozenge$ & \TeffTOInineonezero & \loggTOInineonezero & \FeHTOInineonezero & \VbroadTOInineonezero \\ 
\hline
\end{tabular}
}
\tablefoot{$^a$: V magnitude sources: $\dagger$: the Tycho-2 Catalogue \citep{Tycho-2}, $\star$: The Fourth US Naval Observatory CCD Astrograph Catalog \citep[UCAC4,][]{Zacharias2013}, $\sharp$: \cite{Kar2024}, $\lozenge$: \cite{Jao2011}.}
\end{table}

\section{Juliet fit parameters and plots}

In this appendix, we present the fitted and derived parameters for the \texttt{juliet} fits described in Section \ref{s:analysis}. We also present the RV and photometric plots.

\begin{table}[htp] 
\begin{center} 
\caption{Prior and posterior planetary parameter distributions obtained with \texttt{juliet} for K2-184, for the $1e\_gp$ model.} 
\label{tab:K2-184} 
\centering 
\resizebox{\columnwidth}{!}{
\begin{tabular}{lll} 
\hline  \hline 
Parameter & Prior & Posterior \\ 
\hline 
$\mathrm{\mu_{ESPRESSO18}}$ \dotfill [$\mathrm{m \, s^{-1}}$] & $\mathcal{U}(-33353.8,-33335.1)$ & \ktwooneeightfourmuESPRESSOeighteen \\
$\mathrm{\sigma_{w,ESPRESSO18}}$ \dotfill [$\mathrm{m \, s^{-1}}$] & $\mathcal{J}(0.001,10)$ & \ktwooneeightfoursigmawESPRESSOeighteen \\
$\mathrm{\mu_{ESPRESSO19}}$ \dotfill [$\mathrm{m \, s^{-1}}$] & $\mathcal{U}(-33347.8,-33333.0)$ & \ktwooneeightfourmuESPRESSOnineteen \\
$\mathrm{\sigma_{w,ESPRESSO19}}$ \dotfill [$\mathrm{m \, s^{-1}}$] & $\mathcal{J}(0.001,10)$ & \ktwooneeightfoursigmawESPRESSOnineteen \\
$\mathrm{P_{1}}$ \dotfill [d] & $\mathcal{N}(16.978,0.001)$ & \ktwooneeightfourPpone \\
$\mathrm{t_{0,1}}$ \dotfill [BJD] & $\mathcal{N}(2457148.53,0.01)$ & \ktwooneeightfourtzeropone \\
$\mathrm{\sqrt{e}\sin\omega_{1}}$ \dotfill & $\mathcal{U}(-1,1)$ & \ktwooneeightfoursesinomegapone \\
$\mathrm{\sqrt{e}\cos\omega_{1}}$ \dotfill & $\mathcal{U}(-1,1)$ & \ktwooneeightfoursecosomegapone \\
$\mathrm{K_{1}}$ \dotfill & $\mathcal{U}(0,10)$ & \ktwooneeightfourKpone \\
$\mathrm{\rho_{}}$ \dotfill  [$\mathrm{kg \, m^{-3}}$] & $\mathcal{J}(100,10000)$ & \ktwooneeightfourrho \\
$\mathrm{p_{1}}$ \dotfill & $\mathcal{N}(0.02,0.1)$ & \ktwooneeightfourppone \\
$\mathrm{b_{1}}$ \dotfill & $\mathcal{N}(0.42,0.1)$ & \ktwooneeightfourbpone \\
$\mathrm{q1_{K2-5,K2-18}}$ \dotfill & $\mathcal{U}(0,1)$ & \ktwooneeightfourqoneKtwofiveKtwoeighteen \\
$\mathrm{q2_{K2-5,K2-18}}$ \dotfill & $\mathcal{U}(0,1)$ & \ktwooneeightfourqtwoKtwofiveKtwoeighteen \\
$\mathrm{\sigma_{GP,rv}}$ \dotfill & $\mathcal{TN}(11.1,7.8,0,1000)$ & \ktwooneeightfourGPsigmarv \\
$\mathrm{\alpha_{GP,rv}}$ \dotfill & $\mathcal{TN}(0.008,0.029,0,1000)$ & \ktwooneeightfourGPalpharv \\
$\mathrm{\Gamma_{GP,rv}}$ \dotfill & $\mathcal{TN}(2.1,4.0,0,1000)$ & \ktwooneeightfourGPGammarv \\
$\mathrm{Prot_{GP,rv}}$ \dotfill & $\mathcal{N}(53.1,22.1)$ & \ktwooneeightfourGPProtrv \\
$\mathrm{m_{flux,K2-5}}$ \dotfill & $\mathcal{N}(0,0.1)$ & \ktwooneeightfourmfluxKtwofive \\
$\mathrm{\sigma_{w,K2-5}}$ \dotfill & $\mathcal{J}(0.1,1000)$ & \ktwooneeightfoursigmawKtwofive \\
$\mathrm{\sigma_{GP,K2-5}}$ \dotfill & $\mathcal{N}(3.55\times10^{-5},0.18\times10^{-5})$ & \ktwooneeightfourGPsigmaKtwofive \\
$\mathrm{\rho_{GP,K2-5}}$ \dotfill & $\mathcal{N}(0.142,0.019)$ & \ktwooneeightfourGPrhoKtwofive \\
$\mathrm{m_{flux,K2-18}}$ \dotfill & $\mathcal{N}(0,0.1)$ & \ktwooneeightfourmfluxKtwoeighteen \\
$\mathrm{\sigma_{w,K2-18}}$ \dotfill & $\mathcal{J}(0.1,1000)$ & \ktwooneeightfoursigmawKtwoeighteen \\
$\mathrm{\sigma_{GP,K2-18}}$ \dotfill & $\mathcal{N}(4.67\times10^{-5},0.28\times10^{-5})$ & \ktwooneeightfourGPsigmaKtwoeighteen \\
$\mathrm{\rho_{GP,K2-18}}$ \dotfill & $\mathcal{N}(0.094,0.014)$ & \ktwooneeightfourGPrhoKtwoeighteen \\
$\mathrm{m_{dilution,K2-5}}$ \dotfill & $\mathrm{fixed}$ & \ktwooneeightfourmdilutionKtwofive \\
$\mathrm{m_{dilution,K2-18}}$ \dotfill & $\mathrm{fixed}$ & \ktwooneeightfourmdilutionKtwoeighteen \\
\hline 
$\mathrm{a_{1}}$ \dotfill [au] & $-$ & \ktwooneeightyfourab \\ 
$\mathrm{e_{1}}$ \dotfill & $-$ & \ktwooneeightfoureccpone \\
$\mathrm{\omega_{1}}$ \dotfill [$\degr$]& $-$ & \ktwooneeightfouromegapone \\
$\mathrm{i_{1}}$ \dotfill [$\degr$] & $-$ & \ktwooneeightyfourincb \\ 
$\mathrm{T_{14,1}}$ \dotfill [h] & $-$ & \ktwooneeightyfourTdurb \\ 
$\mathrm{M_{1}}$ \dotfill [$\mathrm{M_\oplus}$] & $-$ & \ktwooneeightyfourmassb \\ 
$\mathrm{R_{1}}$ \dotfill [$\mathrm{R_\oplus}$] & $-$ & \ktwooneeightyfourradb \\ 
$\mathrm{\rho_{1}}$ \dotfill [$\mathrm{g \, cm^{-3}}$] & $-$ & \ktwooneeightyfourrhoplb \\ 
$\mathrm{T_{eq,1}}$ \dotfill [K] & $-$ & \ktwooneeightyfourteqb \\ 
$\mathrm{S_{1}}$ \dotfill [$\mathrm{S_\oplus}$] & $-$ & \ktwooneeightyfourSb \\ 
$\mathrm{\log Z}$ \dotfill  & $-$ & 47499.2 $\pm$ 0.5 \\
\hline 
\end{tabular} 
} 
\tablefoot{\textit{Top}: Fitted parameters. \textit{Bottom}: derived orbital parameters and physical parameters.}
\end{center} 
\end{table} 

\begin{figure}[hbt]
    \centering
    \includegraphics[width=.5\textwidth]{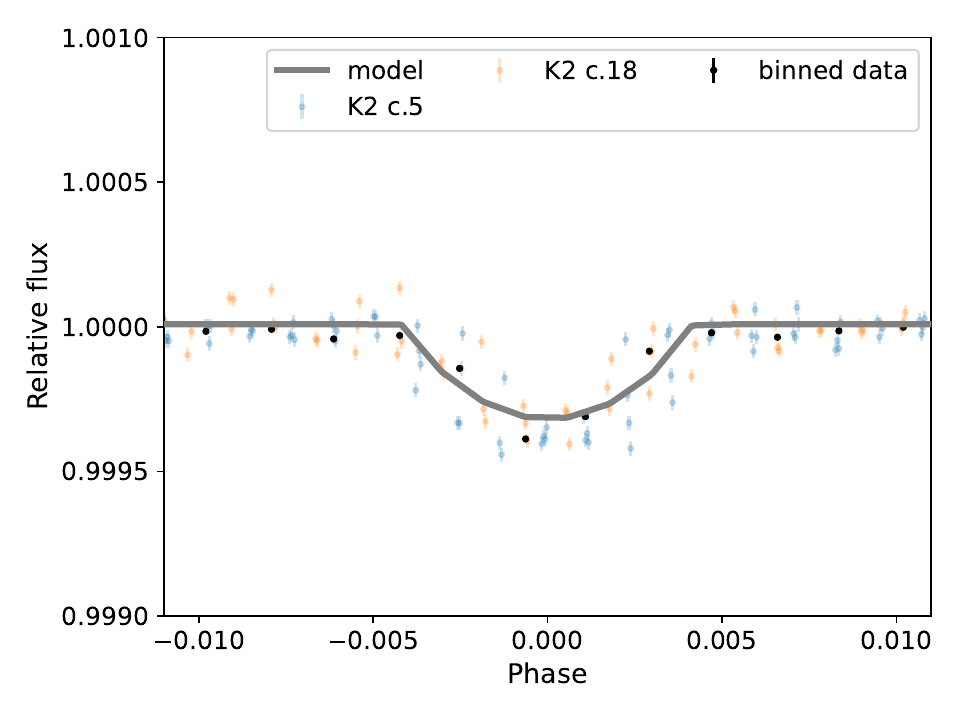}
    \caption{Phase-folded K2 data with the median $1e\_gp$ model (grey line) for K2-184 b. The points show the data from K2 campaign 5 (blue), K2 campaign 18 (orange), and binned data (black).}
    \label{fig:K2-184_K2}
\end{figure}

\begin{figure}[htb]
    \centering
    \includegraphics[width=.45\textwidth]{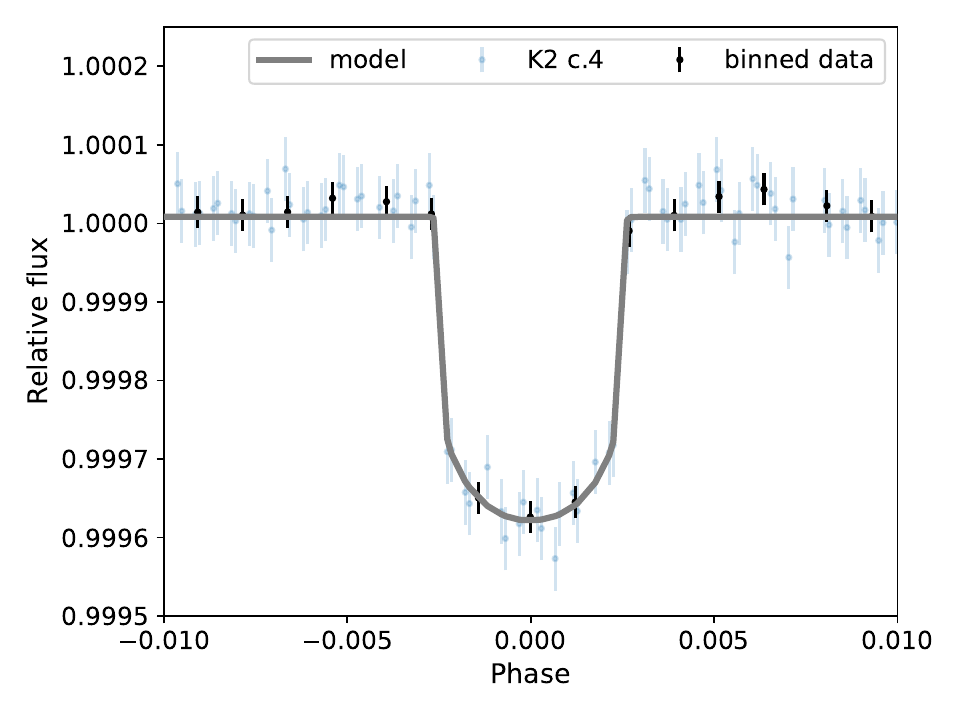}
    \caption{Phase-folded K2 data with the median $1c$ model (grey line) for BD+20594 b. The points show the data from K2 campaign 4 (blue) and binned data (black).}
    \label{fig:BD+20594_K2_stacked}
\end{figure}

\begin{figure*}
    \centering
    \includegraphics[width=.9\textwidth]{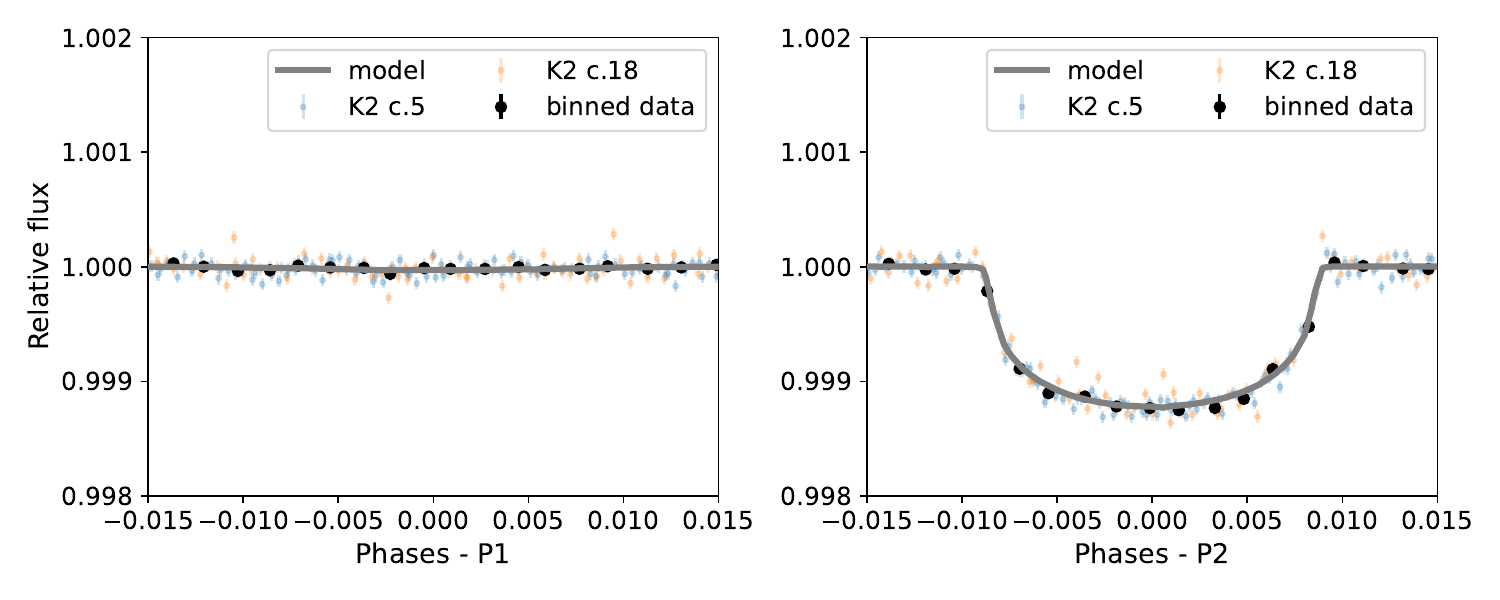}
    \caption{Stacked phase-folded K2 data with the median $2c$ model (grey line), for K2-105 b (left) and K2-105 c (right). The points show the data from K2 campaign 5 (blue) and campaign 18 (orange), and binned data (black).}
    \label{fig:K2-105_TESS_stacked}
\end{figure*}

\begin{figure*}
    \centering
    \includegraphics[width=.9\textwidth]{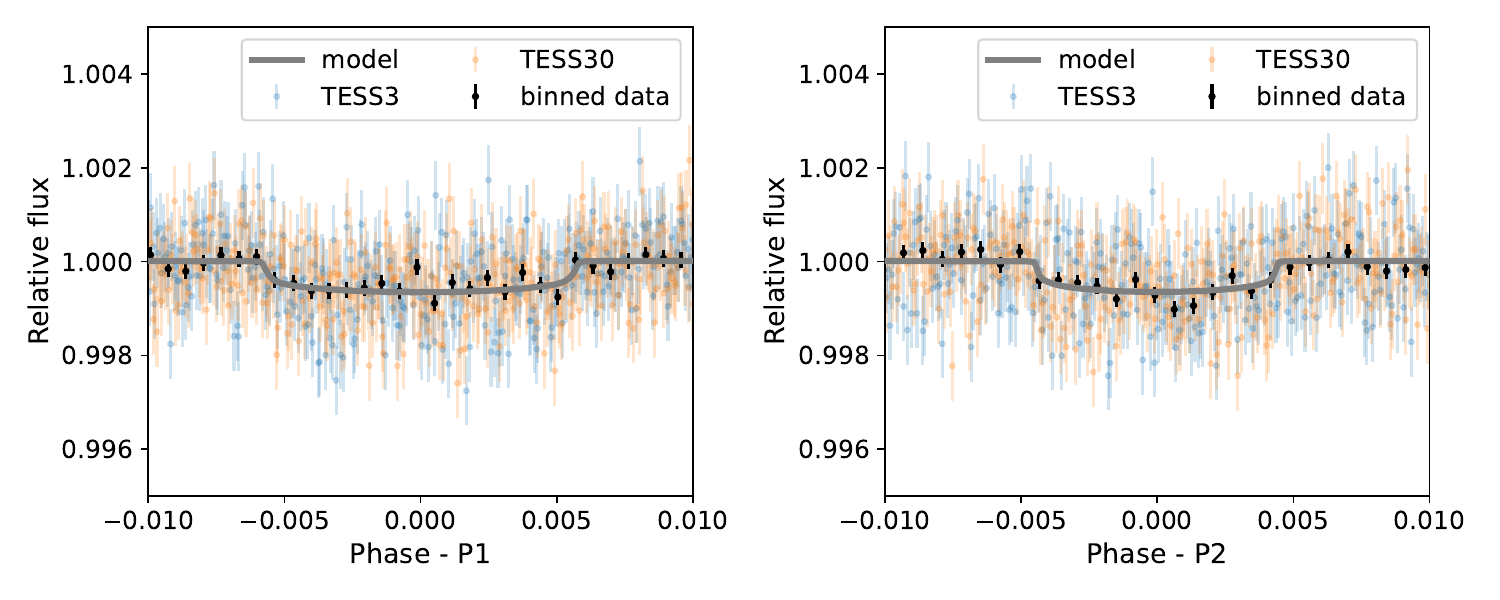}
    \caption{Stacked phase-folded PDCSAP TESS data with the median $2c$ model (grey line), for TOI-266 b (left) and TOI-266 c (right). The points show the data from TESS sector 3 (blue) and sector 30 (orange), and binned data (black).}
    \label{fig:TOI-266_TESS_stacked}
\end{figure*}

\begin{figure*}
    \centering
    \includegraphics[width=.9\textwidth]{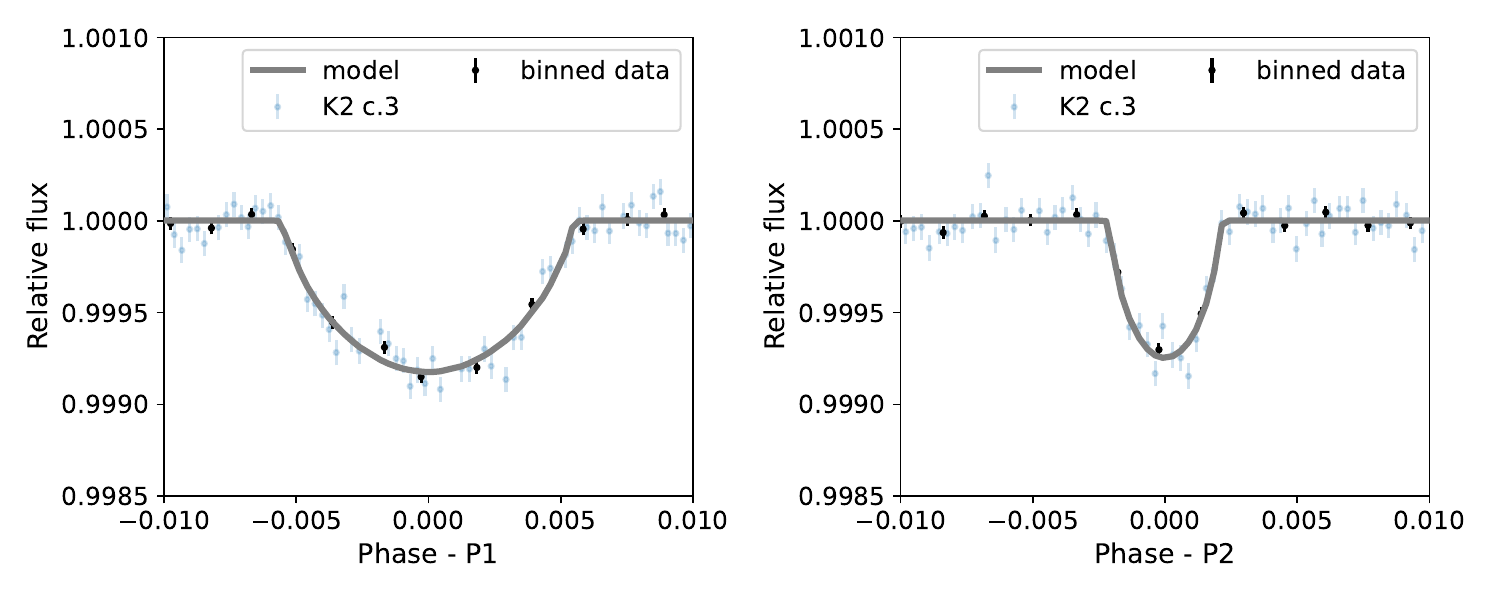}
    \caption{Phase-folded K2 data with the median $2c$ model (grey line), for K2-62 b (left) and K2-62 c (right). The points show the data from K2 campaign 3 (blue) and binned data (black).}
    \label{fig:K2-62_K2_stacked}
\end{figure*}

\begin{figure*}
    \centering
    \includegraphics[width=.9\textwidth]{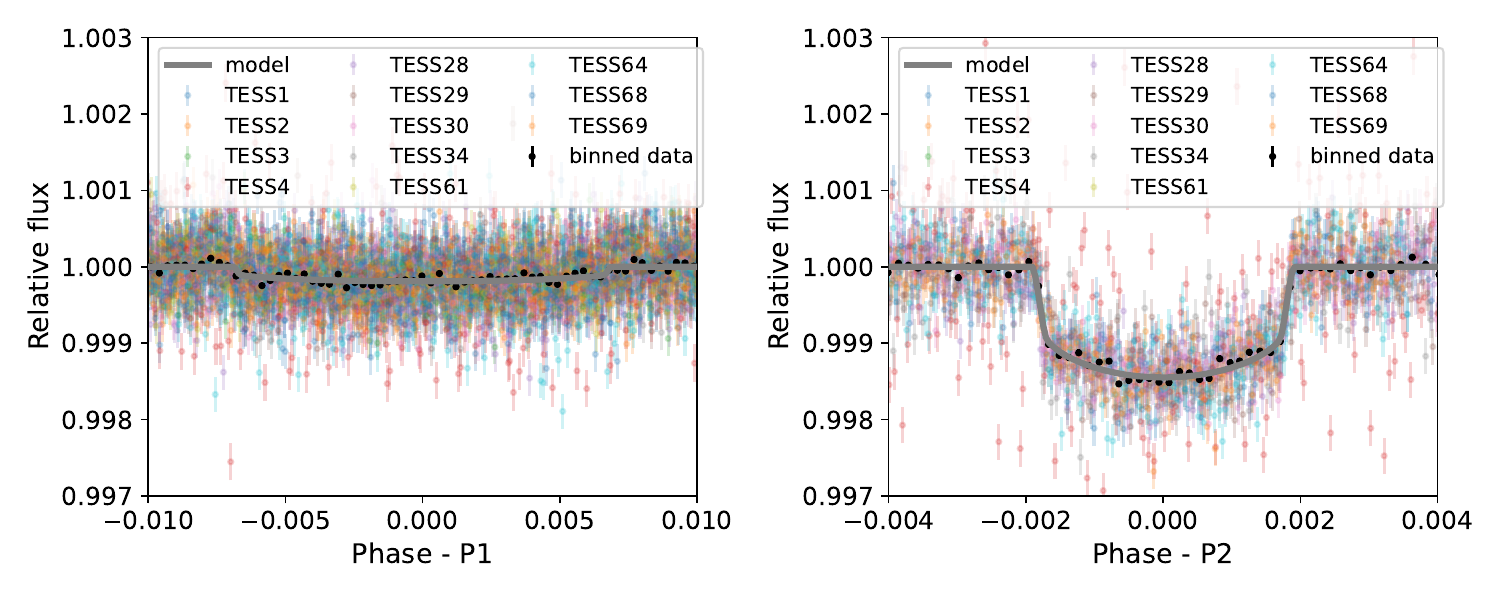}
    \caption{Phase-folded TESS data with the median $2e$ model (grey line), for TOI-186 b (left) and TOI-186 c (right). The points show the data from the different TESS sectors (colours) and binned data (black).}
    \label{fig:TOI-186_TESS_stacked}
\end{figure*}

\begin{figure*}[bht]
    \centering
    \includegraphics[width=.9\textwidth]{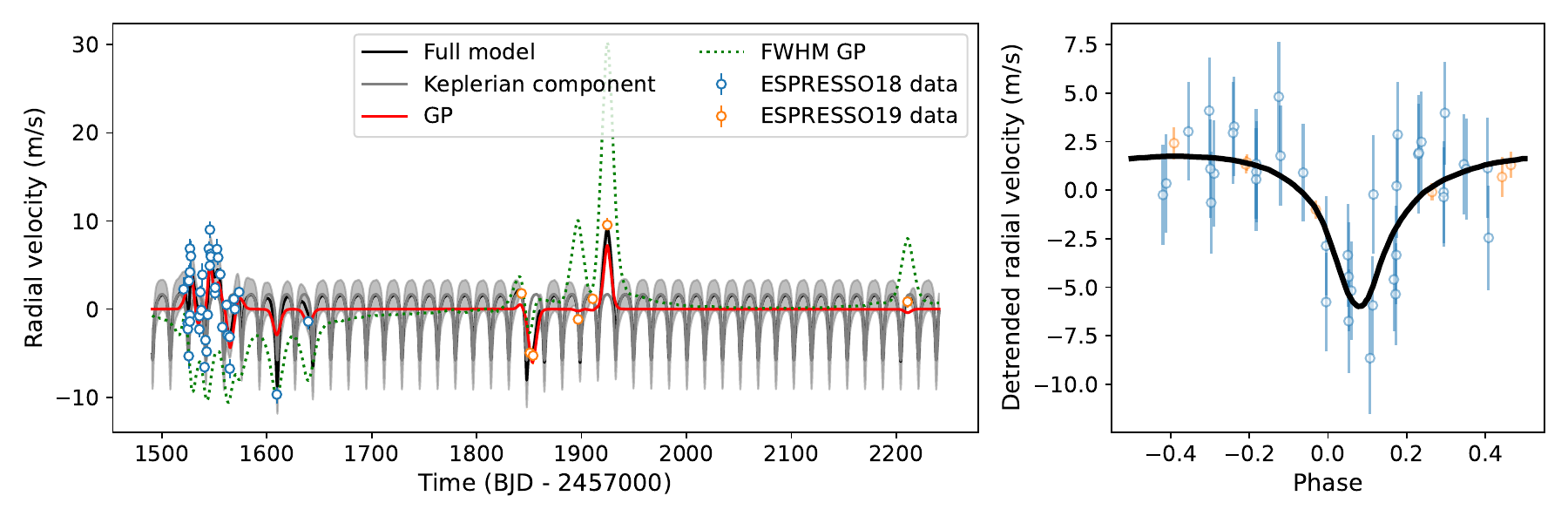}
    \caption{Left: ESPRESSO RVs before (blue dots) and after (orange dots) the fibre link replacement and median model (black) for the $1e\_gp$ model for K2-184 b. The error bars show the RV errors and jitter added in quadrature. The instrumental systemic velocities have been subtracted. Right: phase-folded RVs and median Keplerian model.}
    \label{fig:K2-184_RV}
\end{figure*}

\begin{figure*}
    \centering
    \includegraphics[width=.9\textwidth]{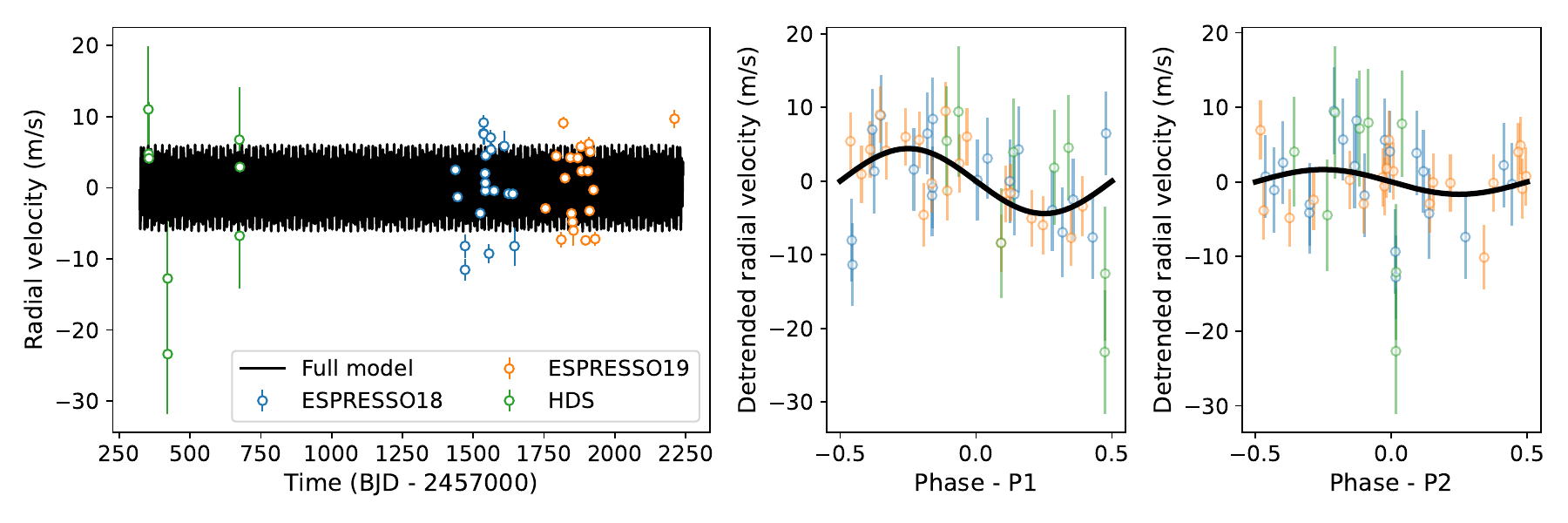}
    \caption{Left: ESPRESSO RVs before (blue dots) and after (orange dots) the fibre link replacement, HDS RVs (green dots), and median model (black) for the $2c$ model for K2-105. The error bars show the RV errors and jitter added in quadrature. The instrumental systemic velocities have been subtracted. Centre and right: phase-folded RVs and median Keplerian model for K2-105 b (centre) and K2-105 c (right).}
    \label{fig:K2-105_RVs}
\end{figure*}

\begin{figure*}
    \centering
    \includegraphics[width=.9\textwidth]{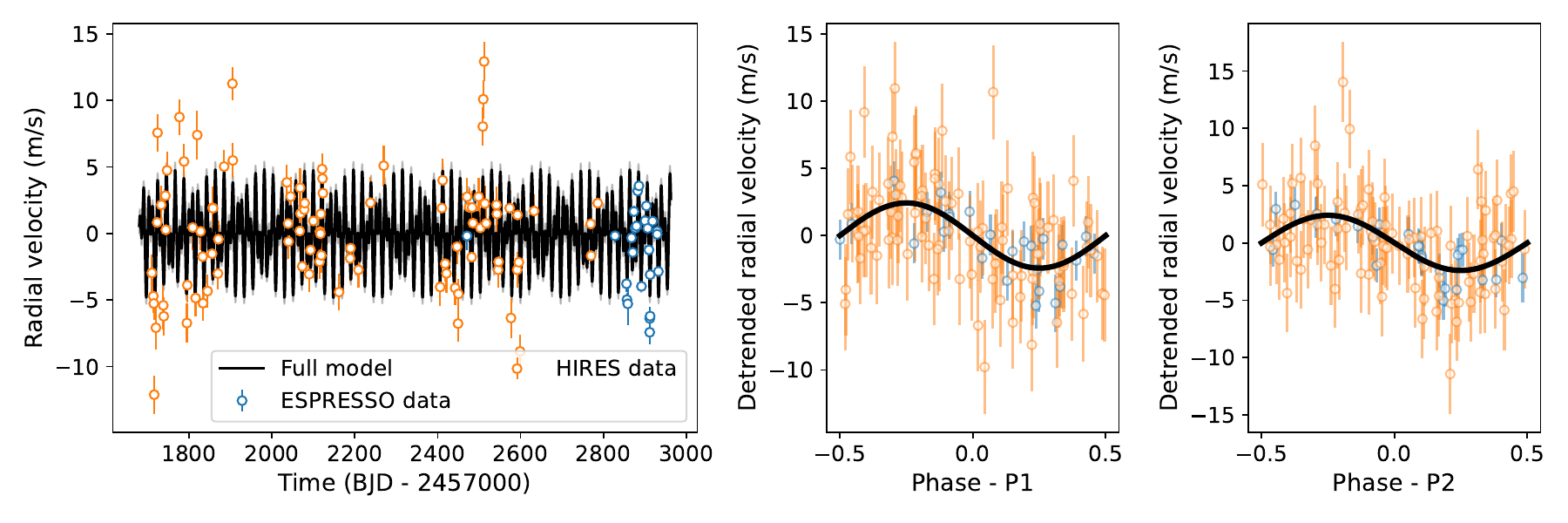}
    \caption{Left: ESPRESSO RVs (blue dots), HIRES RVs (orange dots), and median model (black) for the $2c$ model for TOI-266. The error bars show the RV errors and jitter added in quadrature. The instrumental systemic velocities have been subtracted. Centre and right: phase-folded RVs and median Keplerian model for for TOI-266 b (centre) and TOI-266 c (right).}
    \label{fig:TOI-266_RVs}
\end{figure*}

\begin{figure*}
    \centering
    \includegraphics[width=.9\textwidth]{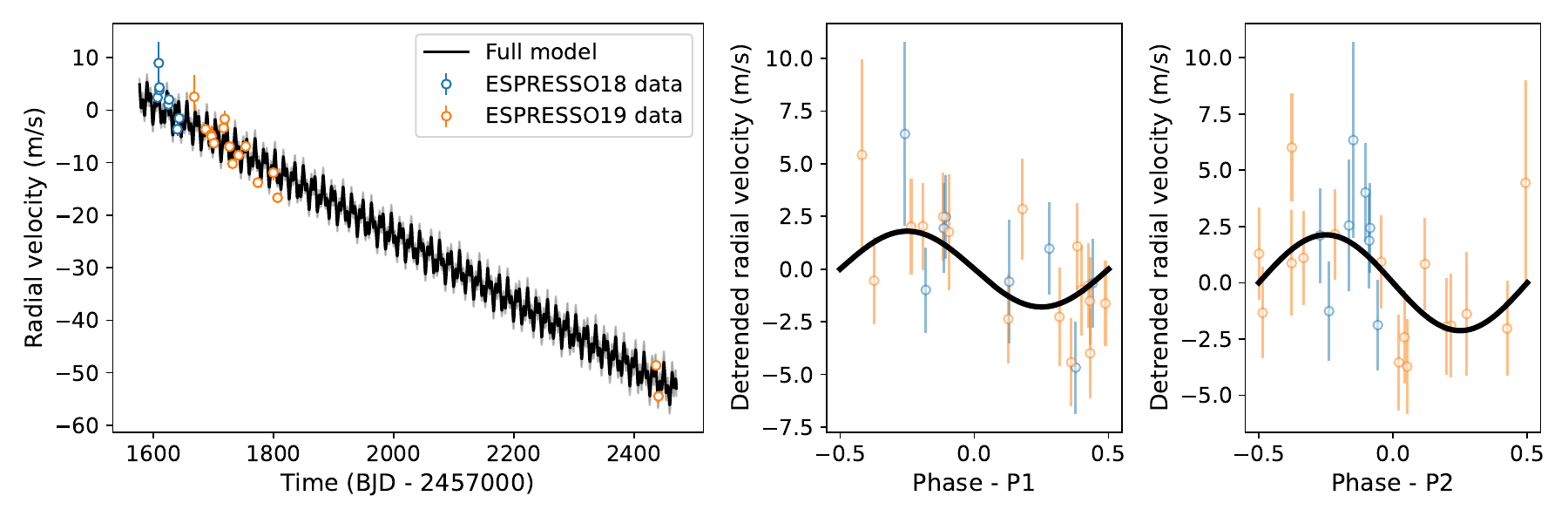}
    \caption{Left: ESPRESSO RVs before (blue dots) and after (orange dots) the fibre link replacement, and median model (black) for the $2c$ model for K2-62. The error bars show the RV errors and jitter added in quadrature. The instrumental systemic velocities have been subtracted. Centre and right: phase-folded RVs and median Keplerian model for K2-62 b (centre) and K2-62 c (right).}
    \label{fig:K2-62_RVs}
\end{figure*}

\begin{figure*}
    \centering
    \includegraphics[width=.9\textwidth]{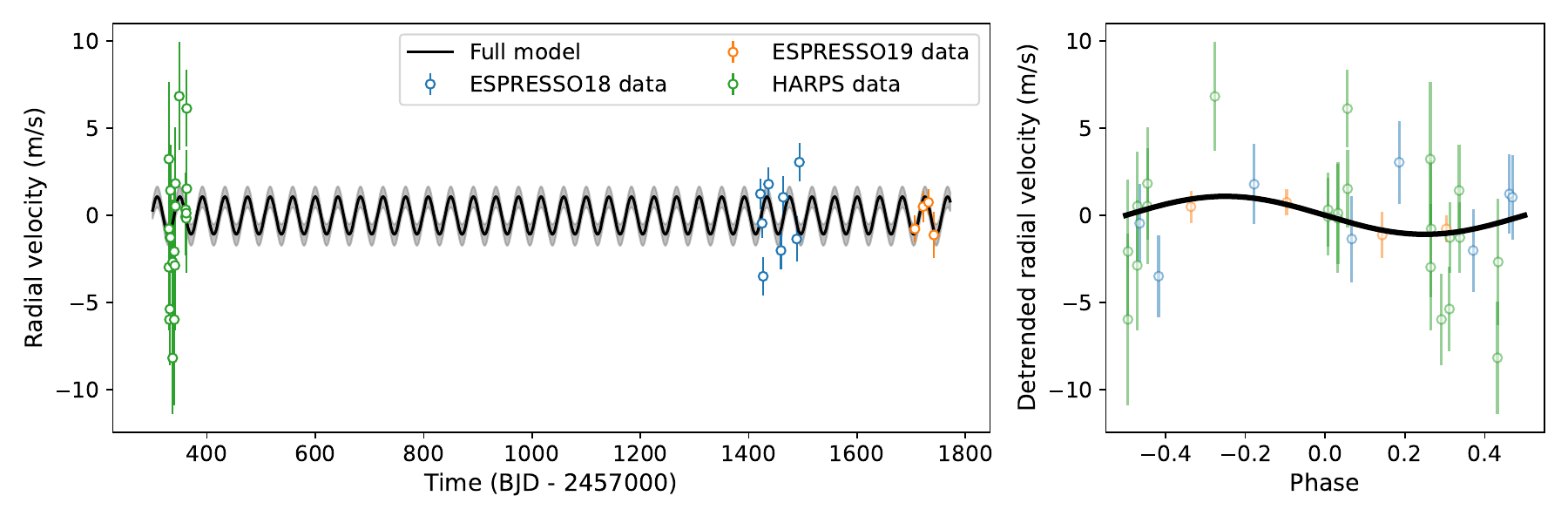}
    \caption{Left: ESPRESSO RVs before (blue dots) and after (orange dots) the fibre link replacement, HARPS RVs (green dots), and median model (black) for the $1c$ model for BD+20594 b. The error bars show the RV errors and jitter added in quadrature. The instrumental systemic velocities have been subtracted. Right: phase-folded RVs and median Keplerian model.}
    \label{fig:BD+20594_RV}
\end{figure*}

\clearpage

\begin{figure*}[hbt!]
    \centering
    \includegraphics[width=.9\textwidth]{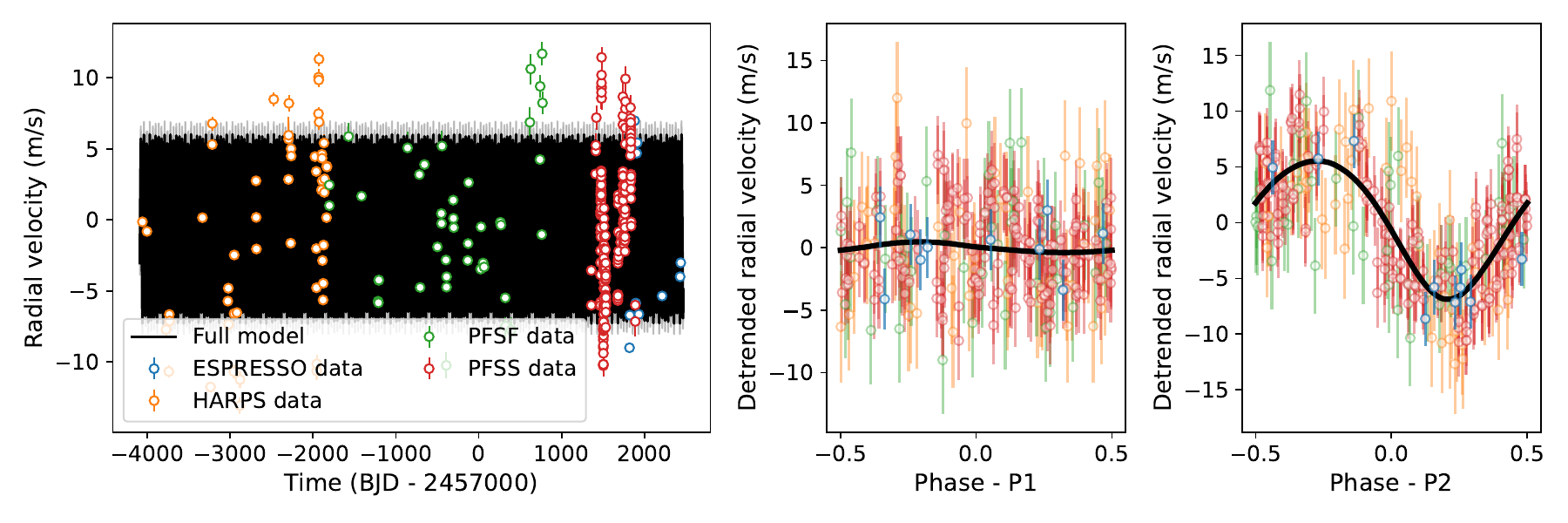}
    \caption{Left: ESPRESSO RVs (blue dots), HARPS RVs (orange dots), PFS RVs before (green dots) and after (red dots) the 2018 detector and slit change, and median model (black) for the $2e$ model for TOI-186. The error bars show the RV errors and jitter added in quadrature. The instrumental systemic velocities have been subtracted. Centre and right: phase-folded RVs and median Keplerian model for TOI-186 b (centre) and TOI-186 c (right).}
    \label{fig:TOI-186_RVs}
\end{figure*}

\begin{table}[bht] 
\begin{center} 
\caption{Prior and posterior planetary parameter distributions obtained with \texttt{juliet} for K2-105, for the $2c$ model on the full dataset.} 
\label{tab:K2-105} 
\centering 
\resizebox{\columnwidth}{!}{
\begin{tabular}{lll} 
\hline  \hline 
Parameter & Prior & Posterior \\ 
\hline 
$\mathrm{\mu_{ESPRESSO18}}$ \dotfill [$\mathrm{m \, s^{-1}}$] & $\mathcal{U}(-32546.2,-32525.5)$ & \ktwooneofivemuESPRESSOeighteen \\
$\mathrm{\sigma_{w,ESPRESSO18}}$ \dotfill [$\mathrm{m \, s^{-1}}$] & $\mathcal{J}(0.001,10)$ & \ktwooneofivesigmawESPRESSOeighteen \\
$\mathrm{\mu_{ESPRESSO19}}$ \dotfill [$\mathrm{m \, s^{-1}}$] & $\mathcal{U}(-32540.1,-32523.0)$ & \ktwooneofivemuESPRESSOnineteen \\
$\mathrm{\sigma_{w,ESPRESSO19}}$ \dotfill [$\mathrm{m \, s^{-1}}$] & $\mathcal{J}(0.001,10)$ & \ktwooneofivesigmawESPRESSOnineteen \\
$\mathrm{\mu_{Subaru}}$ \dotfill [$\mathrm{m \, s^{-1}}$] & $\mathcal{U}(-23.06,11.33)$ & \ktwooneofivemuSubaru \\
$\mathrm{\sigma_{w,Subaru}}$ \dotfill [$\mathrm{m \, s^{-1}}$] & $\mathcal{J}(0.001,10)$ & \ktwooneofivesigmawSubaru \\
$\mathrm{P_{1}}$ \dotfill [d] & $\mathcal{N}(5.022,0.001)$ & \ktwooneofivePpone \\
$\mathrm{t_{0,1}}$ \dotfill [BJD] & $\mathcal{N}(2458437.97,0.01)$ & \ktwooneofivetzeropone \\
$\mathrm{P_{2}}$ \dotfill [d] & $\mathcal{N}(8.267,0.001)$ & \ktwooneofivePptwo \\
$\mathrm{t_{0,2}}$ \dotfill [BJD] & $\mathcal{N}(2457147.99,0.01)$ & \ktwooneofivetzeroptwo \\
$\mathrm{K_{1}}$ \dotfill [$\mathrm{m \, s^{-1}}$] & $\mathcal{U}(0,10)$ & \ktwooneofiveKpone \\
$\mathrm{K_{2}}$ \dotfill [$\mathrm{m \, s^{-1}}$] & $\mathcal{U}(0,10)$ & \ktwooneofiveKptwo \\
$\mathrm{p_{1}}$ \dotfill & $\mathcal{U}(0,0.05)$ & \ktwooneofiveppone \\
$\mathrm{b_{1}}$ \dotfill & $\mathcal{U}(0,2)$ & \ktwooneofivebpone \\
$\mathrm{\rho_{}}$ \dotfill [$\mathrm{kg \, m^{-3}}$] & $\mathcal{J}(100,1000)$ & \ktwooneofiverho \\
$\mathrm{p_{2}}$ \dotfill & $\mathcal{N}(0.0339,0.1)$ & \ktwooneofivepptwo \\
$\mathrm{b_{2}}$ \dotfill & $\mathcal{N}(0.31,0.1)$ & \ktwooneofivebptwo \\
$\mathrm{q1_{K2-5,K2-18}}$ \dotfill & $\mathcal{U}(0,1)$ & \ktwooneofiveqoneKtwofiveKtwoeighteen \\
$\mathrm{q2_{K2-5,K2-18}}$ \dotfill & $\mathcal{U}(0,1)$ & \ktwooneofiveqtwoKtwofiveKtwoeighteen \\
$\mathrm{m_{flux,K2-5}}$ \dotfill & $\mathcal{N}(0,0.1)$ & \ktwooneofivemfluxKtwofive \\
$\mathrm{\sigma_{w,K2-5}}$ \dotfill & $\mathcal{J}(0.1,1000)$ & \ktwooneofivesigmawKtwofive \\
$\mathrm{\sigma_{GP,K2-5}}$ \dotfill & $\mathcal{N}(21.67\times10^{-5},0.73\times10^{-5})$ & \ktwooneofiveGPsigmaKtwofive \\
$\mathrm{\rho_{GP,K2-5}}$ \dotfill & $\mathcal{N}(0.0439,0.0041)$ & \ktwooneofiveGPrhoKtwofive \\
$\mathrm{m_{flux,K2-18}}$ \dotfill & $\mathcal{N}(0,0.1)$ & \ktwooneofivemfluxKtwoeighteen \\
$\mathrm{\sigma_{w,K2-18}}$ \dotfill & $\mathcal{J}(0.1,1000)$ & \ktwooneofivesigmawKtwoeighteen \\
$\mathrm{\sigma_{GP,K2-18}}$ \dotfill & $\mathcal{N}(23.5\times10^{-5},1.0\times10^{-5})$ & \ktwooneofiveGPsigmaKtwoeighteen \\
$\mathrm{\rho_{GP,K2-18}}$ \dotfill & $\mathcal{N}(0.0928,0.0059)$ & \ktwooneofiveGPrhoKtwoeighteen \\
$\mathrm{e_{1}}$ \dotfill & $\mathrm{fixed}$ & \ktwooneofiveeccpone \\
$\mathrm{\omega_{1}}$ \dotfill [$\degr$] & $\mathrm{fixed}$ & \ktwooneofiveomegapone \\
$\mathrm{e_{2}}$ \dotfill & $\mathrm{fixed}$ & \ktwooneofiveeccptwo \\
$\mathrm{\omega_{2}}$ \dotfill [$\degr$] & $\mathrm{fixed}$ & \ktwooneofiveomegaptwo \\
$\mathrm{m_{dilution,K2-5}}$ \dotfill & $\mathrm{fixed}$ & \ktwooneofivemdilutionKtwofive \\
$\mathrm{m_{dilution,K2-18}}$ \dotfill & $\mathrm{fixed}$ & \ktwooneofivemdilutionKtwoeighteen \\
\hline 
$\mathrm{a_{1}}$ \dotfill [au] & $-$ & \ktwooneofiveab \\ 
$\mathrm{i_{1}}$ \dotfill [$\degr$] & $-$ & $-$ \\ 
$\mathrm{T_{14,1}}$ \dotfill [h] & $-$ & $-$ \\ 
$\mathrm{M_{1}}$ \dotfill [$\mathrm{M_\oplus}$] & $-$ & \ktwooneofivemassb \\ 
$\mathrm{R_{1}}$ \dotfill [$\mathrm{R_\oplus}$] & $-$ & $-$ \\  
$\mathrm{\rho_{1}}$ \dotfill [$\mathrm{g \, cm^{-3}}$] & $-$ & $-$ \\ 
$\mathrm{T_{eq,1}}$ \dotfill [K] & $-$ & \ktwooneofiveteqb \\ 
$\mathrm{S_{1}}$ \dotfill [$\mathrm{S_\oplus}$] & $-$ & \ktwooneofiveSb \\ 
$\mathrm{a_{2}}$ \dotfill [au] & $-$ & \ktwooneofiveac \\ 
$\mathrm{i_{2}}$ \dotfill [$\degr$] & $-$ & \ktwooneofiveincc \\ 
$\mathrm{T_{14,2}}$ \dotfill [h] & $-$ & \ktwooneofiveTdurc \\ 
$\mathrm{M_{2}}$ \dotfill [$\mathrm{M_\oplus}$] & $-$ & \ktwooneofivemassc \\ 
$\mathrm{R_{2}}$ \dotfill [$\mathrm{R_\oplus}$] & $-$ & \ktwooneofiveradc \\ 
$\mathrm{\rho_{2}}$ \dotfill [$\mathrm{g \, cm^{-3}}$] & $-$ & \ktwooneofiverhoplc \\ 
$\mathrm{T_{eq,2}}$ \dotfill [K] & $-$ & \ktwooneofiveteqc \\ 
$\mathrm{S_{2}}$ \dotfill [$\mathrm{S_\oplus}$] & $-$ & \ktwooneofiveSc \\ 
$\mathrm{\log Z}$ \dotfill  & $-$ & 41266.4 $\pm$ 0.5 \\
\hline 
\end{tabular} 
} 
\tablefoot{\textit{Top}: Fitted parameters. \textit{Bottom}: derived orbital parameters and physical parameters.}
\end{center} 
\end{table} 

\begin{table}[bhtp] 
\begin{center} 
\caption{Prior and posterior planetary parameter distributions obtained with \texttt{juliet} for TOI-266, for the $2c$ model with all data.} 
\label{tab:TOI-266} 
\centering 
\resizebox{\columnwidth}{!}{
\begin{tabular}{lll} 
\hline  \hline 
Parameter & Prior & Posterior \\ 
\hline 
$\mathrm{\mu_{ESPRESSO}}$ \dotfill [$\mathrm{m \, s^{-1}}$] & $\mathcal{U}(41398.7,41409.7)$ & \toitwosixsixmuESPRESSO \\
$\mathrm{\sigma_{w,ESPRESSO}}$  \dotfill [$\mathrm{m \, s^{-1}}$] & $\mathcal{J}(0.001,10)$ & \toitwosixsixsigmawESPRESSO \\
$\mathrm{\mu_{HIRES}}$  \dotfill [$\mathrm{m \, s^{-1}}$] & $\mathcal{U}(-13.53,11.50)$ & \toitwosixsixmuHIRES \\
$\mathrm{\sigma_{w,HIRES}}$  \dotfill [$\mathrm{m \, s^{-1}}$] & $\mathcal{J}(0.001,10)$ & \toitwosixsixsigmawHIRES \\
$\mathrm{P_{1}}$ \dotfill [d] & $\mathcal{N}(10.751,0.001)$ & \toitwosixsixPpone \\
$\mathrm{t_{0,1}}$ \dotfill [BJD] & $\mathcal{N}(2458393.09,0.01)$ & \toitwosixsixtzeropone \\
$\mathrm{K_{1}}$ \dotfill [$\mathrm{m \, s^{-1}}$] & $\mathcal{U}(0.0,10)$ & \toitwosixsixKpone \\
$\mathrm{\rho_{}}$ \dotfill [$\mathrm{kg \, m^{-3}}$] & $\mathcal{J}(100,10000)$ & \toitwosixsixrho \\
$\mathrm{p_{1}}$ \dotfill & $\mathcal{N}(0.0242,0.1)$ & \toitwosixsixppone \\
$\mathrm{b_{1}}$ \dotfill & $\mathcal{N}(0.52,0.1)$ & \toitwosixsixbpone \\
$\mathrm{P_{2}}$ \dotfill [d] & $\mathcal{N}(19.606,0.001)$ & \toitwosixsixPptwo \\
$\mathrm{t_{0,2}}$ \dotfill [BJD] & $\mathcal{N}(2458398.29,0.01)$ & \toitwosixsixtzeroptwo \\
$\mathrm{K_{2}}$ \dotfill [$\mathrm{m \, s^{-1}}$] & $\mathcal{U}(0,10)$ & \toitwosixsixKptwo \\
$\mathrm{p_{2}}$ \dotfill & $\mathcal{N}(0.0241,0.1)$ & \toitwosixsixpptwo \\
$\mathrm{b_{2}}$ \dotfill & $\mathcal{N}(0.4,0.1)$ & \toitwosixsixbptwo \\
$\mathrm{q_{1,TESS}}$ \dotfill & $\mathcal{U}(0.0,1.0)$ & \toitwosixsixqoneTESS \\
$\mathrm{q_{2,TESS}}$ \dotfill & $\mathcal{U}(0.0,1.0)$ & \toitwosixsixqtwoTESS \\
$\mathrm{m_{flux,TESS3}}$ \dotfill & $\mathcal{N}(0.0,0.1)$ & \toitwosixsixmfluxTESSthree \\
$\mathrm{\sigma_{w,TESS3}}$ \dotfill & $\mathcal{J}(0.1,1000.0)$ & \toitwosixsixsigmawTESSthree \\
$\mathrm{\sigma_{GP,TESS3}}$ \dotfill & $\mathcal{N}(1.6\times10^{-5},4.1\times10^{-5})$ & \toitwosixsixGPsigmaTESSthree \\
$\mathrm{\rho_{GP,TESS3}}$ \dotfill & $\mathcal{N}(6.1,287.5)$ & \toitwosixsixGPrhoTESSthree \\
$\mathrm{m_{flux,TESS30}}$ \dotfill & $\mathcal{N}(0,0.1)$ & \toitwosixsixmfluxTESSthirty \\
$\mathrm{\sigma_{w,TESS30}}$ \dotfill & $\mathcal{J}(0.1,1000)$ & \toitwosixsixsigmawTESSthirty \\
$\mathrm{\sigma_{GP,TESS30}}$ \dotfill & $\mathcal{N}(10.4\times10^{-5},1.1\times10^{-5})$ & \toitwosixsixGPsigmaTESSthirty \\
$\mathrm{\rho_{GP,TESS30}}$ \dotfill & $\mathcal{N}(0.112,0.036)$ & \toitwosixsixGPrhoTESSthirty \\
$\mathrm{e_{1}}$ \dotfill & $\mathrm{fixed}$ & \toitwosixsixeccpone \\
$\mathrm{\omega_{1}}$ \dotfill [$\degr$] & $\mathrm{fixed}$ & \toitwosixsixomegapone \\
$\mathrm{e_{2}}$ \dotfill & $\mathrm{fixed}$ & \toitwosixsixeccptwo \\
$\mathrm{\omega_{2}}$ \dotfill [$\degr$] & $\mathrm{fixed}$ & \toitwosixsixomegaptwo \\
\hline 
$\mathrm{a_{1}}$ \dotfill [au] & $-$ & \toitwosixtysixab \\ 
$\mathrm{i_{1}}$ \dotfill [$\degr$] & $-$ & \toitwosixtysixincb \\ 
$\mathrm{T_{14,1}}$ \dotfill [h] & $-$ & \toitwosixtysixTdurb \\ 
$\mathrm{M_{1}}$ \dotfill [$\mathrm{M_\oplus}$] & $-$ & \toitwosixtysixmassb \\ 
$\mathrm{R_{1}}$ \dotfill [$\mathrm{R_\oplus}$] & $-$ & \toitwosixtysixradb \\ 
$\mathrm{\rho_{1}}$ \dotfill [$\mathrm{g \, cm^{-3}}$] & $-$ & \toitwosixtysixrhoplb \\ 
$\mathrm{T_{eq,1}}$ \dotfill [K] & $-$ & \toitwosixtysixteqb \\ 
$\mathrm{S_{1}}$ \dotfill [$\mathrm{S_\oplus}$] & $-$ & \toitwosixtysixSb \\ 
$\mathrm{a_{2}}$ \dotfill [au] & $-$ & \toitwosixtysixac \\ 
$\mathrm{i_{2}}$ \dotfill [$\degr$] & $-$ & \toitwosixtysixincc \\ 
$\mathrm{T_{14,2}}$ \dotfill [h] & $-$ & \toitwosixtysixTdurc \\ 
$\mathrm{M_{2}}$ \dotfill [$\mathrm{M_\oplus}$] & $-$ & \toitwosixtysixmassc \\ 
$\mathrm{R_{2}}$ \dotfill [$\mathrm{R_\oplus}$] & $-$ & \toitwosixtysixradc \\ 
$\mathrm{\rho_{2}}$ \dotfill [$\mathrm{g \, cm^{-3}}$] & $-$ & \toitwosixtysixrhoplc \\ 
$\mathrm{T_{eq,2}}$ \dotfill [K] & $-$ & \toitwosixtysixteqc \\ 
$\mathrm{S_{2}}$ \dotfill [$\mathrm{S_\oplus}$] & $-$ & \toitwosixtysixSc \\ 
$\mathrm{\log Z}$ \dotfill  & $-$ & 165320.3 $\pm$ 0.5 \\
\hline 
\end{tabular} 
} 
\tablefoot{\textit{Top}: Fitted parameters. \textit{Bottom}: derived orbital parameters and physical parameters.}
\end{center} 
\end{table}

\begin{table}[bht] 
\begin{center} 
\caption{Prior and posterior planetary parameter distributions obtained with \texttt{juliet} for K2-62, for the $2c$ model.} 
\label{tab:K2-62} 
\centering 
\resizebox{\columnwidth}{!}{ 
\begin{tabular}{lll} 
\hline  \hline 
Parameter & Prior & Posterior \\ 
\hline 
$\mathrm{\mu_{ESPRESSO18}}$ \dotfill [$\mathrm{m \, s^{-1}}$] & $\mathcal{U}(4360.0,4372.6)$ & \ktwosixtytwomuESPRESSOeighteen \\
$\mathrm{\sigma_{w,ESPRESSO18}}$ \dotfill [$\mathrm{m \, s^{-1}}$] & $\mathcal{J}(0.001,10)$ & \ktwosixtytwosigmawESPRESSOeighteen \\
$\mathrm{\mu_{ESPRESSO19}}$ \dotfill [$\mathrm{m \, s^{-1}}$] & $\mathcal{U}(4305.4,4362.4)$ & \ktwosixtytwomuESPRESSOnineteen \\
$\mathrm{\sigma_{w,ESPRESSO19}}$ \dotfill [$\mathrm{m \, s^{-1}}$] & $\mathcal{J}(0.001,10)$ & \ktwosixtytwosigmawESPRESSOnineteen \\
$\mathrm{P_{1}}$ \dotfill [d] & $\mathcal{N}(6.672,0.001)$ & \ktwosixtytwoPpone \\
$\mathrm{t_{0,1}}$ \dotfill [BJD] & $\mathcal{N}(2456982.69,0.01)$ & \ktwosixtytwotzeropone \\
$\mathrm{K_{1}}$ \dotfill [$\mathrm{m \, s^{-1}}$] & $\mathcal{U}(0,10)$ & \ktwosixtytwoKpone \\
$\mathrm{\rho_{}}$ \dotfill [$\mathrm{kg \, m^{-3}}$] & $\mathcal{J}(100,10000)$ & \ktwosixtytworho \\
$\mathrm{p_{1}}$ \dotfill & $\mathcal{N}(0.027,0.1)$ & \ktwosixtytwoppone \\
$\mathrm{b_{1}}$ \dotfill & $\mathcal{U}(0,1)$ & \ktwosixtytwobpone \\
$\mathrm{P_{2}}$ \dotfill [d] & $\mathcal{N}(16.197,0.001)$ & \ktwosixtytwoPptwo \\
$\mathrm{t_{0,2}}$ \dotfill [BJD] & $\mathcal{N}(2456991.54,0.01)$ & \ktwosixtytwotzeroptwo \\
$\mathrm{K_{2}}$ \dotfill [$\mathrm{m \, s^{-1}}$] & $\mathcal{U}(0,10)$ & \ktwosixtytwoKptwo \\
$\mathrm{p_{2}}$ \dotfill & $\mathcal{N}(0.027,0.1)$ & \ktwosixtytwopptwo \\
$\mathrm{b_{2}}$ \dotfill & $\mathcal{U}(0,1)$ & \ktwosixtytwobptwo \\
$\mathrm{q_{1,K2}}$ \dotfill & $\mathcal{U}(0,1)$ & \ktwosixtytwoqoneKtwothree \\
$\mathrm{q_{2,K2}}$ \dotfill & $\mathcal{U}(0,1)$ & \ktwosixtytwoqtwoKtwothree \\
$\mathrm{rv_{intercept}}$ \dotfill [$\mathrm{m \, s^{-1}}$] & $\mathcal{U}(-100,100)$ & \ktwosixtytworvintercept \\
$\mathrm{rv_{slope}}$ \dotfill [$\mathrm{m \, s^{-1}}$] & $\mathcal{U}(-100,100)$ & \ktwosixtytworvslope \\
$\mathrm{m_{flux,K2-3}}$ \dotfill & $\mathcal{N}(0,0.1)$ & \ktwosixtytwomfluxKtwothree \\
$\mathrm{\sigma_{w,K2-3}}$ \dotfill & $\mathcal{J}(0.1,1000)$ & \ktwosixtytwosigmawKtwothree \\
$\mathrm{\sigma_{GP,K2-3}}$ \dotfill & $\mathcal{N}(5.6\times10^{-5},0.3\times10^{-5})$ & \ktwosixtytwoGPsigmaKtwothree \\
$\mathrm{\rho_{GP,K2-3}}$ \dotfill & $\mathcal{N}(0.161,0.023)$ & \ktwosixtytwoGPrhoKtwothree \\
$\mathrm{e_{1}}$ \dotfill & $\mathrm{fixed}$ & \ktwosixtytwoeccpone \\
$\mathrm{\omega_{1}}$ \dotfill [$\degr$] & $\mathrm{fixed}$ & \ktwosixtytwoomegapone \\
$\mathrm{e_{2}}$ \dotfill & $\mathrm{fixed}$ & \ktwosixtytwoeccptwo \\
$\mathrm{\omega_{2}}$ \dotfill  [$\degr$] & $\mathrm{fixed}$ & \ktwosixtytwoomegaptwo \\
$\mathrm{m_{dilution,K2-3}}$ \dotfill & $\mathrm{fixed}$ & \ktwosixtytwomdilutionKtwothree \\
\hline 
$\mathrm{a_{1}}$ \dotfill [au] & $-$ & \ktwosixtytwoab \\ 
$\mathrm{i_{1}}$ \dotfill [$\degr$] & $-$ & \ktwosixtytwoincb \\ 
$\mathrm{T_{14,1}}$ \dotfill [h] & $-$ & \ktwosixtytwoTdurb \\ 
$\mathrm{M_{1}}$ \dotfill [$\mathrm{M_\oplus}$] & $-$ & \ktwosixtytwomassb \\ 
$\mathrm{R_{1}}$ \dotfill [$\mathrm{R_\oplus}$] & $-$ & \ktwosixtytworadb \\ 
$\mathrm{\rho_{1}}$ \dotfill [$\mathrm{g \, cm^{-3}}$] & $-$ & \ktwosixtytworhoplb \\ 
$\mathrm{T_{eq,1}}$ \dotfill [K] & $-$ & \ktwosixtytwoteqb \\ 
$\mathrm{S_{1}}$ \dotfill [$\mathrm{S_\oplus}$] & $-$ & \ktwosixtytwoSb \\ 
$\mathrm{a_{2}}$ \dotfill [au] & $-$ & \ktwosixtytwoac \\ 
$\mathrm{i_{2}}$ \dotfill [$\degr$] & $-$ & \ktwosixtytwoincc \\ 
$\mathrm{T_{14,2}}$ \dotfill [h] & $-$ & \ktwosixtytwoTdurc \\ 
$\mathrm{M_{2}}$ \dotfill [$\mathrm{M_\oplus}$] & $-$ & \ktwosixtytwomassc \\ 
$\mathrm{R_{2}}$ \dotfill [$\mathrm{R_\oplus}$] & $-$ & \ktwosixtytworadc \\ 
$\mathrm{\rho_{2}}$ \dotfill [$\mathrm{g \, cm^{-3}}$] & $-$ & \ktwosixtytworhoplc \\ 
$\mathrm{T_{eq,2}}$ \dotfill [K] & $-$ & \ktwosixtytwoteqc \\ 
$\mathrm{S_{2}}$ \dotfill [$\mathrm{S_\oplus}$] & $-$ & \ktwosixtytwoSc \\ 
$\mathrm{\log Z}$ \dotfill  & $-$ & 24602.0 $\pm$ 0.6 \\
\hline 
\end{tabular} 
} 
\tablefoot{\textit{Top}: Fitted parameters. \textit{Bottom}: derived orbital parameters and physical parameters.}
\end{center} 
\end{table} 

\begin{table}[bht] 
\begin{center} 
\caption{Prior and posterior planetary parameter distributions obtained with \texttt{juliet} for BD+20954, for the $1c$ model on the joint dataset.} 
\label{tab:BD+20954} 
\centering 
\resizebox{\columnwidth}{!}{
\begin{tabular}{lll} 
\hline  \hline 
Parameter & Prior & Posterior \\ 
\hline 
$\mathrm{\mu_{ESPRESSO18}}$ \dotfill [$\mathrm{m \, s^{-1}}$] & $\mathcal{U}(-20494.4,-20489.6)$ & \bdmuESPRESSOeighteen \\
$\mathrm{\sigma_{w,ESPRESSO18}}$ \dotfill [$\mathrm{m \, s^{-1}}$] & $\mathcal{J}(0.001,10)$ & \bdsigmawESPRESSOeighteen \\
$\mathrm{\mu_{ESPRESSO19}}$ \dotfill [$\mathrm{m \, s^{-1}}$] & $\mathcal{U}(-20496.0,-20494.1)$ & \bdmuESPRESSOnineteen \\
$\mathrm{\sigma_{w,ESPRESSO19}}$ \dotfill [$\mathrm{m \, s^{-1}}$] & $\mathcal{J}(0.001,10)$ & \bdsigmawESPRESSOnineteen \\
$\mathrm{\mu_{HARPS}}$ \dotfill [$\mathrm{m \, s^{-1}}$] & $\mathcal{U}(-20345.3,-20330.3)$ & \bdmuHARPS \\
$\mathrm{\sigma_{w,HARPS}}$ \dotfill [$\mathrm{m \, s^{-1}}$] & $\mathcal{J}(0.001,10)$ & \bdsigmawHARPS \\
$\mathrm{P_{1}}$ \dotfill [d] & $\mathcal{N}(41.686,0.001)$ & \bdPpone \\
$\mathrm{t_{0,1}}$ \dotfill [BJD] & $\mathcal{N}(2457151.90,0.01)$ & \bdtzeropone \\
$\mathrm{K_{1}}$ \dotfill [$\mathrm{m \, s^{-1}}$] & $\mathcal{U}(0,10)$ & \bdKpone \\
$\mathrm{\rho_{}}$ \dotfill [$\mathrm{kg \, m^{-3}}$] & $\mathcal{J}(100,10000)$ & \bdrho \\
$\mathrm{p_{1}}$ \dotfill & $\mathcal{U}(0.005,0.05)$ & \bdppone \\
$\mathrm{b_{1}}$ \dotfill & $\mathcal{U}(0,1)$ & \bdbpone \\
$\mathrm{q_{1,K2}}$ \dotfill & $\mathcal{U}(0,1)$ & \bdqoneKtwofour \\
$\mathrm{q_{2,K2}}$ \dotfill & $\mathcal{U}(0,1)$ & \bdqtwoKtwofour \\
$\mathrm{m_{flux,K2-4}}$ \dotfill & $\mathcal{N}(0,0.1)$ & \bdmfluxKtwofour \\
$\mathrm{\sigma_{w,K2-4}}$ \dotfill & $\mathcal{J}(0.1,1000)$ & \bdsigmawKtwofour \\
$\mathrm{\sigma_{GP,K2-4}}$ \dotfill & $\mathcal{N}(7.31\times10^{-5},0.14\times10^{-5})$ & \bdGPsigmaKtwofour \\
$\mathrm{\rho_{GP,K2-4}}$ \dotfill & $\mathcal{N}(0.01561,0.00068)$ & \bdGPrhoKtwofour \\
$\mathrm{e_{1}}$ \dotfill & $\mathrm{fixed}$ & \bdeccpone \\
$\mathrm{\omega_{1}}$ \dotfill [$\degr$] & $\mathrm{fixed}$ & \bdomegapone \\
$\mathrm{m_{dilution,K2-4}}$ \dotfill & $\mathrm{fixed}$ & \bdmdilutionKtwofour \\
\hline 
$\mathrm{a_{1}}$ \dotfill [au] & $-$ & \bdab \\ 
$\mathrm{i_{1}}$ \dotfill [$\degr$] & $-$ & \bdincb \\ 
$\mathrm{T_{14,1}}$ \dotfill [h] & $-$ & \bdTdurb \\ 
$\mathrm{M_{1}}$ \dotfill [$\mathrm{M_\oplus}$] & $-$ & \bdmassb \\ 
$\mathrm{R_{1}}$ \dotfill [$\mathrm{R_\oplus}$] & $-$ & \bdradb \\ 
$\mathrm{\rho_{1}}$ \dotfill [$\mathrm{g \, cm^{-3}}$] & $-$ & \bdrhoplb \\ 
$\mathrm{T_{eq,1}}$ \dotfill [K] & $-$ & \bdteqb \\ 
$\mathrm{S_{1}}$ \dotfill [$\mathrm{S_\oplus}$] & $-$ & \bdSb \\ 
$\mathrm{\log Z}$ \dotfill  & $-$ & 25990.0 $\pm$ 0.4 \\
\hline 
\end{tabular} 
} 
\tablefoot{\textit{Top}: Fitted parameters. \textit{Bottom}: derived orbital parameters and physical parameters.}
\end{center} 
\end{table}

\begin{table}[bhtp] 
\begin{center} 
\caption{Prior and posterior planetary parameter distributions obtained with \texttt{juliet} for TOI-186, for the $2e$ model on the joint dataset.} 
\label{tab:TOI-186} 
\centering 
\resizebox{\columnwidth}{!}{ 
\begin{tabular}{lll} 
\hline  \hline 
Parameter & Prior & Posterior \\ 
\hline 
$\mathrm{\mu_{ESPRESSO}}$ \dotfill [$\mathrm{m \, s^{-1}}$] & $\mathcal{U}(59595.4,59611.4)$ & \toioneeightysixmuESPRESSO \\
$\mathrm{\sigma_{w,ESPRESSO}}$ \dotfill [$\mathrm{m \, s^{-1}}$] & $\mathcal{J}(0.001,10)$ & \toioneeightysixsigmawESPRESSO \\
$\mathrm{\mu_{HARPS}}$ \dotfill [$\mathrm{m \, s^{-1}}$] & $\mathcal{U}(-13.2,11.2)$ & \toioneeightysixmuHARPS \\
$\mathrm{\sigma_{w,HARPS}}$ \dotfill [$\mathrm{m \, s^{-1}}$] & $\mathcal{J}(0.001,10)$ & \toioneeightysixsigmawHARPS \\
$\mathrm{\mu_{PFSF}}$ \dotfill [$\mathrm{m \, s^{-1}}$] & $\mathcal{U}(-10.1,11.8)$ & \toioneeightysixmuPFSF \\
$\mathrm{\sigma_{w,PFSF}}$ \dotfill [$\mathrm{m \, s^{-1}}$] & $\mathcal{J}(0.001,10)$ & \toioneeightysixsigmawPFSF \\
$\mathrm{\mu_{PFSS}}$ \dotfill [$\mathrm{m \, s^{-1}}$] & $\mathcal{U}(-9.7,12.0)$ & \toioneeightysixmuPFSS \\
$\mathrm{\sigma_{w,PFSS}}$ \dotfill [$\mathrm{m \, s^{-1}}$] & $\mathcal{J}(0.001,10)$ & \toioneeightysixsigmawPFSS \\
$\mathrm{P_{1}}$ \dotfill [d] & $\mathcal{N}(7.790,0.001)$ & \toioneeightysixPpone \\
$\mathrm{t_{0,1}}$ \dotfill [BJD] & $\mathcal{N}(2458371.23,0.01)$ & \toioneeightysixtzeropone \\
$\mathrm{\sqrt{e}\sin\omega_{1}}$ \dotfill & $\mathcal{U}(-1,1)$ & \toioneeightysixsesinomegapone \\
$\mathrm{\sqrt{e}\cos\omega_{1}}$ \dotfill & $\mathcal{U}(-1,1)$ & \toioneeightysixsecosomegapone \\
$\mathrm{e_{1}}$ \dotfill & $-$ & \toioneeightysixeccpone \\
$\mathrm{\omega_{1}}$ \dotfill [$\degr$] & $-$ & \toioneeightysixomegapone \\
$\mathrm{K_{1}}$ \dotfill [$\mathrm{m \, s^{-1}}$] & $\mathcal{U}(0,10)$ & \toioneeightysixKpone \\
$\mathrm{\rho_{}}$ \dotfill [$\mathrm{kg \, m^{-3}}$] & $\mathcal{J}(100,10000)$ & \toioneeightysixrho \\
$\mathrm{p_{1}}$ \dotfill & $\mathcal{N}(0.012,0.1)$ & \toioneeightysixppone \\
$\mathrm{b_{1}}$ \dotfill & $\mathcal{N}(0.42,0.1)$ & \toioneeightysixbpone \\
$\mathrm{P_{2}}$ \dotfill [d] & $\mathcal{N}(35.613,0.001)$ & \toioneeightysixPptwo \\
$\mathrm{t_{0,2}}$ \dotfill [BJD] & $\mathcal{N}(2458385.93,0.01)$ & \toioneeightysixtzeroptwo \\
$\mathrm{\sqrt{e}\sin\omega_{2}}$ \dotfill & $\mathcal{U}(-1,1)$ & \toioneeightysixsesinomegaptwo \\
$\mathrm{\sqrt{e}\cos\omega_{2}}$ \dotfill & $\mathcal{U}(-1,1)$ & \toioneeightysixsecosomegaptwo \\
$\mathrm{e_{2}}$ \dotfill & $-$ & \toioneeightysixeccptwo \\
$\mathrm{\omega_{2}}$ \dotfill [$\degr$] & $-$ & \toioneeightysixomegaptwo \\
$\mathrm{K_{2}}$ \dotfill [$\mathrm{m \, s^{-1}}$] & $\mathcal{U}(0,10)$ & \toioneeightysixKptwo \\
$\mathrm{p_{2}}$ \dotfill & $\mathcal{N}(0.024,0.1)$ & \toioneeightysixpptwo \\
$\mathrm{b_{2}}$ \dotfill & $\mathcal{N}(0.59,0.1)$ & \toioneeightysixbptwo \\
$\mathrm{q_{1,TESS}}$ \dotfill & $\mathcal{U}(0,1)$ & \toioneeightysixqoneTESS \\
$\mathrm{q_{2,TESS}}$ \dotfill & $\mathcal{U}(0,1)$ & \toioneeightysixqtwoTESS \\
\hline 
$\mathrm{a_{1}}$ \dotfill [au] & $-$ & \toioneeightysixab \\ 
$\mathrm{i_{1}}$ \dotfill [$\degr$] & $-$ & \toioneeightysixincb \\ 
$\mathrm{T_{14,1}}$ \dotfill [h] & $-$ & \toioneeightysixTdurb \\ 
$\mathrm{M_{1}}$ \dotfill [$\mathrm{M_\oplus}$] & $-$ & \toioneeightysixmassb \\ 
$\mathrm{R_{1}}$ \dotfill [$\mathrm{R_\oplus}$] & $-$ & \toioneeightysixradb \\ 
$\mathrm{\rho_{1}}$ \dotfill [$\mathrm{g \, cm^{-3}}$] & $-$ & \toioneeightysixrhoplb \\ 
$\mathrm{T_{eq,1}}$ \dotfill [K] & $-$ & \toioneeightysixteqb \\ 
$\mathrm{S_{1}}$ \dotfill [$\mathrm{S_\oplus}$] & $-$ & \toioneeightysixSb \\ 
$\mathrm{a_{2}}$ \dotfill [au] & $-$ & \toioneeightysixac \\ 
$\mathrm{i_{2}}$ \dotfill [$\degr$] & $-$ & \toioneeightysixincc \\ 
$\mathrm{T_{14,2}}$ \dotfill [h] & $-$ & \toioneeightysixTdurc \\ 
$\mathrm{M_{2}}$ \dotfill [$\mathrm{M_\oplus}$] & $-$ & \toioneeightysixmassc \\ 
$\mathrm{R_{2}}$ \dotfill [$\mathrm{R_\oplus}$] & $-$ & \toioneeightysixradc \\ 
$\mathrm{\rho_{2}}$ \dotfill [$\mathrm{g \, cm^{-3}}$] & $-$ & \toioneeightysixrhoplc \\ 
$\mathrm{T_{eq,2}}$ \dotfill [K] & $-$ & \toioneeightysixteqc \\ 
$\mathrm{S_{2}}$ \dotfill [$\mathrm{S_\oplus}$] & $-$ & \toioneeightysixSc \\ 
$\mathrm{\log Z}$ \dotfill  & $-$ & 61324.3 $\pm$ 0.6 \\
\hline
\end{tabular}  
}
\tablefoot{\textit{Top}: Fitted parameters. \textit{Bottom}: derived orbital parameters and physical parameters.}
\end{center} 
\end{table}

\clearpage
\onecolumn

\section{ExoMDN model output plots}

In this section, we present the results of the ExoMDN modelling described in section \ref{s:composition}.

\begin{figure*}[htb]
    \centering
    \includegraphics[width=.7\textwidth]{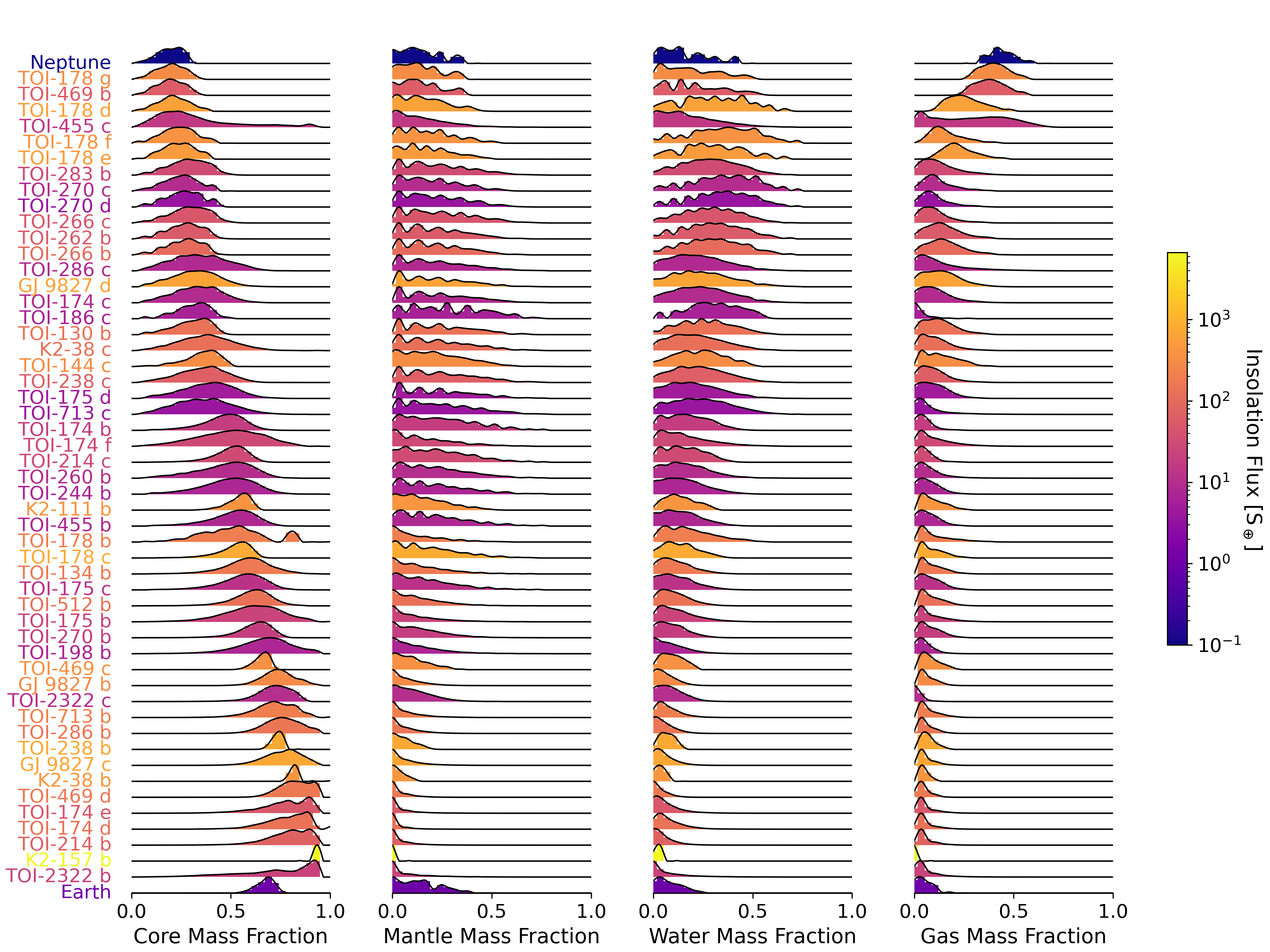}
    \caption{Normalized histograms of the core, mantle, water and gas mass fraction posteriors obtained with ExoMDN for the WG3 sample. The planets are ordered by core mass fraction and labelled. We include the Earth and Neptune at the bottom and top respectively as comparison points.}
    \label{fig:ExoMDN_mass}
\end{figure*}

\begin{figure*}[hbt]
    \centering
    \includegraphics[width=.7\textwidth]{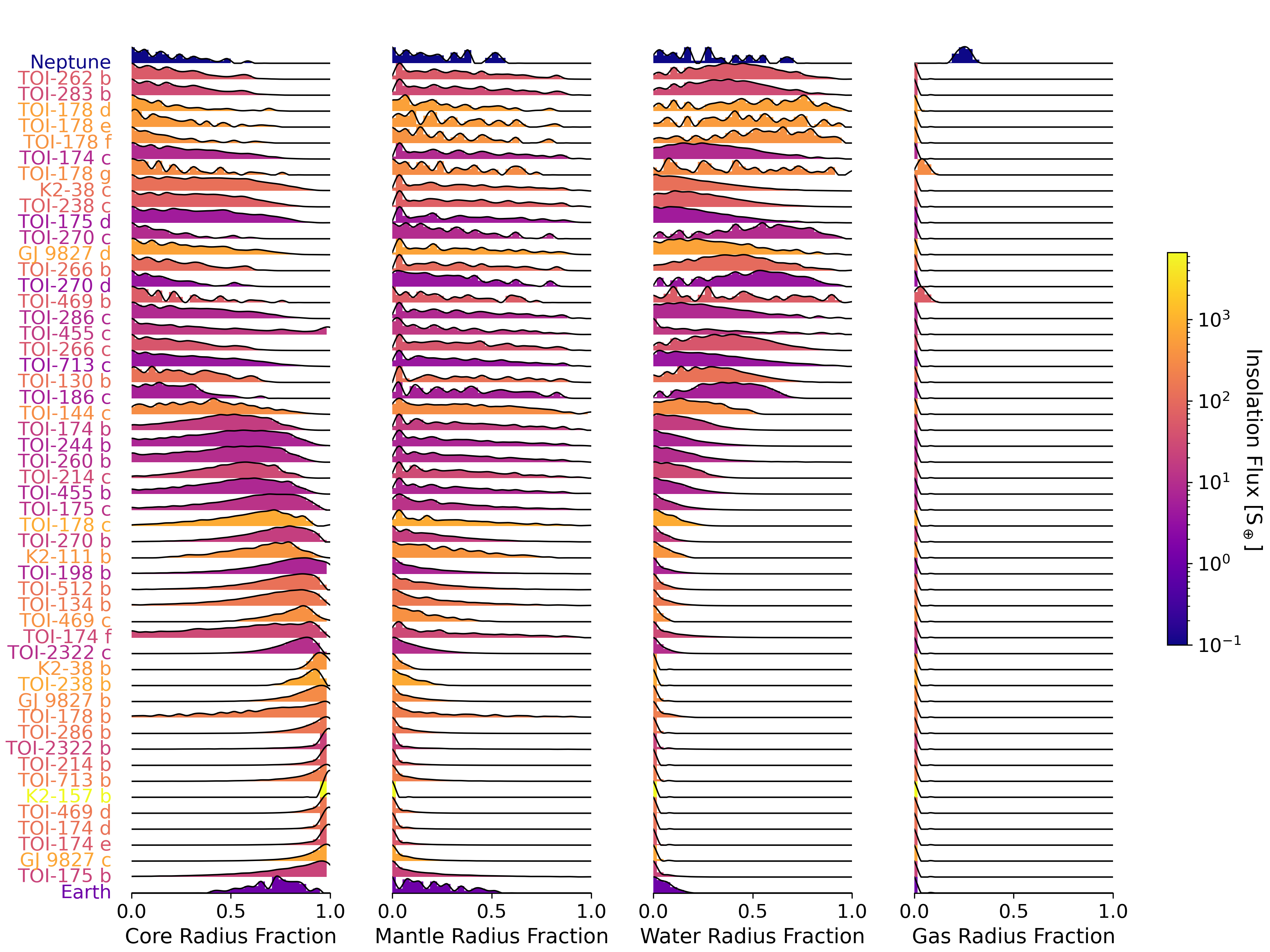}
    \caption{Normalized histograms of the core, mantle, water and gas radius fraction posteriors obtained with ExoMDN for the WG3 sample. The planets are ordered by core radius fraction. We include the Earth and Neptune at the bottom and top respectively as comparison points.}
    \label{fig:ExoMDN_radius}
\end{figure*}

\end{document}